\documentclass[journal]{IEEEtran}
\onecolumn
\usepackage{amsmath,amstext,amsfonts,amssymb,amsthm} 
\usepackage{graphicx,color} 
\usepackage[latin1]{inputenc}   
\usepackage[T1]{fontenc} 
\usepackage[english]{babel} 
\usepackage{hyperref}
\usepackage{bm}

\def \A {\mathbf{A}}
\def \Acal{\mathcal{A}}
\def \a {\mathbf{a}}

\def \b {\mathbf{b}}

\def \Ecal {\mathcal{E}}
\def \e {\mathbf{e}}

\def \f {\mathbf{f}}
\def \G {\mathbf{G}}

\def \i {\mathbf{i}}

\def \Ocal{\mathcal{O}}

\def \Q {\mathbf{Q}}

\def \R {\mathbf{R}}

\def \Scal {\mathcal{S}}
\def \T {\mathbf{T}}
\def \t {\mathbf{t}}

\def \det {\mathrm{det} \ }
\def \Tr {\mathrm{Tr}\,}
\def \tr {\mathrm{tr}\,}

\def \dist {\mathrm{dist}}

\renewcommand{\Im}{\mathrm{Im}}

\newcommand{\supp}{\mathrm{supp}}

\newcommand{\Var}{\mathrm{Var}}

\def\tr{\mathrm{Tr}}
\def\supp{\mathrm{Supp}}
\def\var{\mathbf{Var}}

\def\ex{\mathbb{E}}
\def\im{\mathrm{Im}}

\newcommand{\bs}{\boldsymbol}
\newtheorem{corollary}{Corollary}
\numberwithin{corollary}{section}
\newtheorem{theorem}{Theorem}
\numberwithin{theorem}{section}

\numberwithin{definition}{section}
\newtheorem{lemma}{Lemma}
\numberwithin{lemma}{section}
\newtheorem{remark}{Remark}
\numberwithin{remark}{section}

\numberwithin{claim}{section}
\newtheorem{proposition}{Proposition}
\numberwithin{proposition}{section}
\newtheorem{assumption}{Assumption}
\numberwithin{assumption}{section}

\numberwithin{conjecture}{section}

\numberwithin{property}{section}
\numberwithin{equation}{section}

\begin{document}

\title{On the largest singular values of certain large random matrices with application to the estimation of the minimal dimension of the state-space representations of high-dimensional time series.}

\author
{
 D. Tieplova, 
  P. Loubaton, \IEEEmembership{Fellow, IEEE},
  \thanks
  {
    Daria Tieplova was with the Department of Theoretical Physics, Institute for Low Temperature Physics and Engineering, Kharkiv, Ukraine and with Laboratoire d'Informatique Gaspard Monge, UMR 8049, Universit\'e Paris-Est Marne la Vall\'ee, 5 Bd. Descartes, Cit\'e Descartes, 77454 Marne la Vall\'ee Cedex 2, France. She is now with the Department of Statistics and Actuarial Science,  Run Run Shaw Building, The University of Hong Kong, Pokfulam Road, Hong Kong,  daria.tieplova@u-pem.fr
  }
  \thanks
  {
  Philippe Loubaton is with Laboratoire d'Informatique Gaspard Monge, UMR 8049,  Universit\'e Paris-Est Marne la Vall\'ee, 5 Bd. Descartes, Cit\'e Descartes, 77454 Marne la Vall\'ee Cedex 2 , France, philippe.loubaton@u-pem.fr
  }
   \thanks
  {
    This work was partially supported by  Labex B\'ezout, funded by ANR, project ANR-10-LABX-58, and by ANR project HIDITSA, ANR-17-CE40-0003 
  }
}  

\maketitle
\date{}

\begin{abstract}
This paper is devoted to the estimation of the minimal dimension P of the state-space realizations of a high-dimensional time series y, defined as a noisy version (the noise is white and Gaussian) of a useful signal with low rank rational spectral density, in the high-dimensional asymptotic regime where the number of available samples N and the dimension of the time series M converge towards infinity at the same rate. In the classical low-dimensional regime, P is estimated as the number of significant singular values of the  empirical autocovariance matrix between the past and the future of y, or as the number of significant  estimated canonical correlation coefficients between the past and the future of y. Generalizing large random matrix methods developed in the past to analyze classical spiked models, the behaviour of the above singular values and canonical correlation coefficients is studied in the high-dimensional regime. It is proved that they are smaller than certain thresholds depending on the statistics of the noise, except a finite number of outliers that are due to the useful signal. The number of singular values of the sample autocovariance matrix above the threshold is evaluated, is shown to be almost independent from P in general, and cannot therefore be used to estimate P accurately. In contrast, the number s of canonical correlation coefficients larger than the corresponding threshold is shown to be less than or equal to P, and explicit conditions under which it is equal to P are provided. Under the corresponding assumptions, s is thus a consistent estimate of P in the high-dimensional regime. The core of the paper is the development of the necessary large random matrix tools. 
\end{abstract}

    \section{Introduction}
\subsection{The addressed problem and the results.}
\label{subsec:addressed-problem}

Due to the spectacular development of data acquisition devices and sensor networks, it becomes very common to be faced with
high-dimensional time series in various fields such as digital communications, environmental sensing, electroencephalography, analysis of financial datas, industrial monitoring, .... In this context, it is not always possible to collect a large enough number of observations to perform statistical inference because the durations of the signals are limited and/or because their statistics are not time-invariant over large enough temporal windows. As a result, fundamental inference schemes do not behave as in the classical low-dimensional regimes. This stimulated 
considerably in the ten past years the development of new statistical approaches aiming at mitigating
the above mentioned difficulties. \\

In particular, a number of works proposed to use large random matrix theory in the context of high-dimensional statistical signal processing, traditionally modelled by a double asymptotic regime in which the dimension $M$ of the time series and the sample size $N$ both converge towards $+\infty$. These contributions addressed, among others, detection or estimation problems in the context of the so-called narrow band array processing model, also known in the statistical community as the linear factor model. The  $M$--dimensional observation $(y_n)_{n=1, \ldots, N}$ is a noisy version of a useful signal $(u_n)_{n \in \mathbb{Z}}$ that can be written as 
$u_n = H s_n$  where $(s_n)_{n \in \mathbb{Z}}$ is a $K$-dimensional non observable 
signal and $H$ is a $M \times K$ unknown (or partially unknown) deterministic matrix. In this context, some relevant informations have to be inferred on the useful signal $(u_n)_{n \in \mathbb{Z}}$ from the available 
samples $y_1, \ldots, y_N$, e.g. estimation of $K$, of the column space of $H$, the non zero eigenvalues 
and associated eigenvectors of the covariance matrix $R_u = \mathbb{E}(u_n u_n^{*})$ when $(s_n)_{n \in \mathbb{Z}}$is assumed to be a stationary sequence,....The $M \times N$ observed matrix 
$Y$ collecting the $N$ observations appears as the sum of a full rank random matrix due to the additive 
noise with the $M \times N$ rank $K$ matrix $U = H S$ where $U$ and $S$ 
are defined in the same way than $Y$. In this context, a number of detection and estimation schemes are based on functionals of the empirical "spatial" covariance matrix 
$\hat{R}_y = \frac{Y Y^{*}}{N}$. In the traditional low dimensional regime where $M$ is fixed while $N \rightarrow +\infty$,  $\hat{R}_y$ behaves as the true covariance matrix $R_y = \mathbb{E}(y_n y_n^{*})$ 
of the observation, and this allows to evaluate quite easily the behaviour of the various algorithms. The main 
difficulty of the high-dimensional regime follows from the well known observation that, when $M$ and $N$ converge towards $+\infty$ at the same rate, 
then $\hat{R}_y$ is not a good estimator of $R_y$ in the sense that the spectral norm $\| \hat{R}_y - R_y \|$ of the estimation error does not converge towards $0$. However, when the rank $K$ does not scale with 
$M$ and $N$, an assumption which in practice means that $\frac{K}{M}$ is small enough, large random matrix theory results related to the so-called spiked models, characterizing, among others, the eigenvalue distribution and the $K$ 
largest eigenvalues and related eigenvectors of $\hat{R}_y$ (see  e.g. \cite{baik-benarous-peche-spike}, \cite{baik-silverstein-spike}, \cite{benaych-nadakuditi-ann-math}, \cite{benaych-rao-2}, \cite{paul-2007}), allow to 
evaluate the behaviour of the relevant functionals of $\hat{R}_y$, to analyze the performance of the traditional schemes, and, sometimes, to propose improved algorithms (see e.g. \cite{bianchi-debbah-maida-najim-ieeeit-2011}, \cite{couillet-pascal-silverstein-jmva-2015}, \cite{couillet-jmva-2015} 
\cite{loubaton-2}, \cite{krichtman-nadler-2009}, \cite{nadakuditi-edelman-2008}, 
\cite{nadakuditi-silverstein-2010}, \cite{hachem-1}, \cite{vallet-1}, \cite{hachem-2}). In particular, provided the $K$ non zero eigenvalues of $R_u$ are larger than a certain threshold depending on the noise statistics, then, under certain extra assumptions, $K$ can 
be estimated consistently as the number of "significant" eigenvalues of $\hat{R}_y$. \\

In this work, we consider the more general context where the useful signal $(u_n)_{n \in \mathbb{Z}}$ coincides with the output of a $K$ inputs / $M$ outputs filter, $K < M$, with unknown causal and causally invertible rational transfer function $H(z)$ driven by a $K$ dimensional non observable sequence $(i_n)_{n \in \mathbb{Z}}$ verifying 
$\mathbb{E}(i_{n+k} i_n^{*}) = I_K \delta_k$, which, necessarily, coincides with a normalized version of the innovation sequence of $u$ defined as the prediction error 
$u_n - u_n/\mathrm{sp}(u_{n-k}, k \geq 1)$. Normalized version of the innovation  means that for each $n$, the components of $i_n$ represent an orthonormal basis of the $K$--dimensional space generated by the components of $u_n - u_n/\mathrm{sp}(u_{n-k}, k \geq 1)$. We remark that, for each frequency $f \in [-1/2, 1/2]$, the spectral density of $u$ is a rank $K < M$ matrix, except if $e^{2 i \pi f}$ is a zero of  $H(z)$.  In the following, we denote by $P$ the Mac-Millan degree of $H(z)$, i.e. the minimal dimension of the state-space representations of $H(z) = D + C(zI -A)^{-1}B$ where $A$ is a $P \times P$ matrix with spectral radius $\rho(A)<1$  and where $C,B,D$ are $M \times P, P \times K, M \times M$ matrices respectively. It is well known (see e.g. 
\cite{kailath-1980}, \cite{vanoverschee-demoor}, \cite{lindquist-picci-book-2015}, Appendix A) that the minimality of the state-space representation of $H(z)$ is equivalent to $(C,A)$ observable and $(A,B)$ commandable. We recall 
that $(C,A)$ observable means that for each $L \geq P$, the $ML \times P$ observability matrix $\mathcal{O}^{(L)}$ of $(C,A)$ defined by 
\begin{equation}
\label{eq:def-OL}
\mathcal{O}^{(L)} = \left( \begin{array}{c} C \\ C A \\ \vdots \\  C A^{L-1} \end{array} \right)
\end{equation}
is a rank $P$ matrix, while, similarly, $(A,B)$ is commandable if the $P \times ML$ commandability matrix 
$\mathcal{C}^{(L)}$ of $(A,B)$ defined by
\begin{equation}
\label{eq:def-CL}
\mathcal{C}^{(L)} = \left(  A^{L-1} B,  A^{L-2} B, \ldots,  B \right).
\end{equation}
is rank $P$ as well. Then, $u_n$ can be represented in the state-space form as 
\begin{equation}
\label{eq:state-space}
x_{n+1}  =  A x_n + B i_n, \;  u_n  = C x_n + D i_n,
\end{equation}
where the $P$-dimensional Markovian sequence $(x_n)_{n \in \mathbb{Z}}$ is called the state-space sequence associated to (\ref{eq:state-space}). Moreover, we assume that the observed $M$--dimensional multivariate time series $(y_n)_{n \in \mathbb{Z}}$ is given by
\begin{equation}
\label{eq:dynamic-factor-model}
y_n = u_n  + v_n,
\end{equation}
where $(v_n)_{n \in \mathbb{Z}}$ is a complex Gaussian "noise" term such that $\mathbb{E}(v_{n+k} v_n^{*}) = R \delta_k$ for some 
unknown positive definite matrix $R$. $(v_n)_{n \in \mathbb{Z}}$ is of course independent from the useful signal $u$.\\

The estimation of the (minimal) dimension $P$ of the state-space 
representation (\ref{eq:state-space}) from $N$ avalaible samples $y_1, \ldots, y_N$ is an important 
problem of multivariate time series analysis in that estimating $P$ first allows to address the estimation of matrices $C$ and $A$, as well as of matrices $B, D$ and $R$, at least if the three later 
matrices are identifiable. In the standard asymptotic regime where $N \rightarrow +\infty$ while 
$M$ remains fixed, standard estimation procedures are based on the following well known ingredients (the reader is referred e.g. to \cite{vanoverschee-demoor} and  \cite{lindquist-picci-book-2015} and the references therein). First, as 
$(v_n)_{n \in \mathbb{Z}}$ is an uncorrelated sequence, the autocovariance sequence $(R_{y,k})_{k \in \mathbb{Z}}$ defined by $R_{y,k} = \mathbb{E}(y_{n+k}y_n^{*})$ verifies $R_{y,k} = R_{u,k} = \mathbb{E}(u_{n+k}u_n^{*})$ for each $k \neq 0$. Next, the autocovariance sequence $R_u$ of $u$ can be represented as
\begin{equation}
\label{eq:expre-autocovariance}
R_{u,k}  = C A^{k-1} G
\end{equation}
for each $k \geq 1$, where matrix $G$ coincides with $G=\mathbb{E}(x_{n+1}u_n^{*})$, which is also equal to 
$G= \mathbb{E}(x_{n+1}y_n^{*})$ because signals $u$ and $v$ are uncorrelated. Moreover, the pair $(A,G)$ is commandable, and every triple $(A',C',G')$ of $P \times P, M \times P, P \times M$ matrices for which (\ref{eq:expre-autocovariance}) holds can be obtained from $(A,C,G)$ by a similarity transform. If we define the autocovariance matrix $R^{L}_{f|p,u}$
between the past and the future of $u$ as
\begin{equation}
\label{eq:def-autocovariance}
R^{L}_{f|p,u}  = \mathbb{E} \left[ \left( \begin{array}{c} u_{n+L} \\  u_{n+L+1} \\ \vdots \\ u_{n+2L-1} \end{array} \right) \left(
u_n^{*}, u_{n+1}^{*}, \ldots, u_{n+L-1}^{*} \right) \right]
\end{equation}
then, it holds that
\begin{equation}
\label{eq:factorisation-hankel}
R^{(L)}_{f|p,u} = \mathcal{O}^{(L)} \, \mathcal{C}^{(L)},
\end{equation}
where $\mathcal{O}^{(L)}$ is the observability matrix of the pair $(C,A)$  and $\mathcal{C}^{(L)}$ represents the commandability matrix of $(A,G)$. 
For each $L \geq P$, matrices $\mathcal{O}^{(L)}$ and $ \mathcal{C}^{(L)}$ are full rank, so that the rank of $R^{(L)}_{f|p,u}$ remains equal to $P$, and each minimal rank factorization of $R^{(L)}_{f|p,u}$
can be written as (\ref{eq:factorisation-hankel}) for some particular triple $(A,C,G)$.
As matrix $R^{L}_{f|p,y}$ defined in the same way than $R^{(L)}_{f|p,u}$ coincides with $R^{(L)}_{f|p,u}$, 
we deduce from the above properties that $P$ coincides with the rank of $R^{L}_{f|p,y}$ for each integer $L \geq P$. 
Moreover, a particular pair $(C,A)$ can be identified from any minimal rank factorisation of  $R^{L}_{f|p,y}$. 

In order to estimate $P$ from the available samples $y_1, \ldots, y_N$, a standard approach is to estimate  $P$ as the number of
"significant" singular values of the empirical estimate $\hat{R}^{L}_{f|p,y}$ of the true matrix $R^{L}_{f|p,y} = R^{L}_{f|p,u}$ defined by
$$
\hat{R}^{L}_{f|p,y} = \frac{Y_{f,N} Y_{p,N}^{*}}{N},
$$
where matrices $Y_{f,N}$ and $Y_{p,N}$ are defined as
\begin{equation}
\label{eq:def-intro-Yp}
Y_{p,N} =  \left( \begin{array}{ccccc} y_1 & y_2 & \ldots & y_{N-1} & y_N \\
y_2 & y_3 & \ldots & y_N & y_{N+1} \\
\vdots & \vdots & \vdots & \vdots & \vdots \\
\vdots & \vdots & \vdots & \vdots & \vdots \\
y_L & y_{L+1} & \ldots & y_{N+L-2} & y_{N+L-1} \\
\end{array} \right)
\end{equation}
and 
\begin{equation}
\label{eq:def-intro-Yf}
Y_{f,N} =   \left( \begin{array}{ccccc} y_{L+1} & y_{L+2} & \ldots & y_{N-1+L} & y_{N+L} \\
y_{L+2} & y_{L+3} & \ldots & y_{N+L} & y_{N+L+1} \\
\vdots & \vdots & \vdots & \vdots & \vdots \\
\vdots & \vdots & \vdots & \vdots & \vdots \\
y_{2L} & y_{2L+1} & \ldots & y_{N+2L-2} & y_{N+2L-1} \\
\end{array} \right).
\end{equation}
We note that the samples $(y_{N+l})_{l=1, \ldots, 2L-1}$ are supposed to be available 
while we have assumed that only the first $N$ samples are observed. In order to 
simplify the presentation, this end effect is neglected. We also notice that a pair $(C,A)$ 
can also be estimated from the truncated singular value decomposition of $\hat{R}^{L}_{f|p,y}$ (see \cite{vanoverschee-demoor} and \cite{ciuso-picci-2004} for a statistical analysis of the corresponding estimates). This approach provides consistent estimates of $P,C,A$ when $N \rightarrow +\infty$ while $M$, $K, P$ and $L$ are fixed because in this context, $\| \hat{R}^{L}_{f|p,y} - R^{L}_{f|p,y} \| \rightarrow 0$. 
\\


Another way to estimate $P$ is to resort to the canonical analysis of the observation $y$. In particular, $P$ coincides with the number of non zero canonical correlation coefficients between the spaces ${\cal Y}_{p,L}$ and ${\cal Y}_{f,L}$ generated respectively by the components of $y_{n+k}, k= 0,  \ldots, L-1$ and $y_{n+k}, k= L, \ldots, 2L-1$ for any $L \geq P$. We recall that these coefficients are defined as the singular values of matrix 
$(R^{L}_{y})^{-1/2} R^{L}_{f|p,y}  (R^{L}_{y})^{-1/2}$ where $R^{L}_{y}$ represents the covariance matrix 
of the $ML$--dimensional vector $(y_n^{T}, \ldots, y_{n+L-1}^{T})^{T}$.  
In order to estimate $P$ from the $N$ avalaible observations $y_1, \ldots, y_N$, a standard solution is to estimate the canonical correlation coefficients between ${\cal Y}_{p,L}$ and ${\cal Y}_{f,L}$ by the
canonical correlation coefficients between the row spaces of matrices $Y_{p,N}$ and $Y_{f,N}$ defined by (\ref{eq:def-intro-Yp}) and (\ref{eq:def-intro-Yf})
respectively, and to estimate $P$ as the number of significant coefficients, i.e. as the number of significant
singular values of matrix $(\hat{R}^{L}_{f,y})^{-1/2} \hat{R}^{L}_{f|p,y} (\hat{R}^{L}_{p,y})^{-1/2}$, or equivalently as the number of significant eigenvalues of $(\hat{R}^{L}_{f,y})^{-1/2} \hat{R}^{L}_{f|p,y} (\hat{R}^{L}_{p,y})^{-1} \hat{R}_{f|p,y}^{L*} (\hat{R}^{L}_{f,y})^{-1/2} $. Here, matrices $\hat{R}^{L}_{f,y}$ and  $\hat{R}^{L}_{p,y}$
are defined by  $\hat{R}^{L}_{f,y} = \frac{Y_{f,N}Y_{f,N}^{*}}{N}$ and  $\hat{R}^{L}_{p,y} = \frac{Y_{p,N}Y_{p,N}^{*}}{N}$ respectively. In the standard low-dimensional regime $N \rightarrow +\infty$ and $M,K,P,L$ are fixed, it holds that $\| \hat{R}^{L}_{i,y} - R^{L}_{y} \| \rightarrow 0$ for $i=p,f$ as well as 
$\| \hat{R}^{L}_{f|p,y} - R^{L}_{f|p,y} \| \rightarrow 0$. This immediately leads to the conclusion that this approach provides consistent estimates of $P$. We again refer to  \cite{vanoverschee-demoor} and  \cite{lindquist-picci-book-2015} and the references therein. \\

If $M$ is large and that the sample size $N$ cannot be arbitrarily larger than $M$, the ratio $ML/N$ may not be small enough to make reliable the above statistical analysis, in the sense that it cannot be expected that $ \hat{R}^{L}_{f|p,y}$ and $\hat{R}^{L}_{i,y}$, $i=p,f$ are close enough in the spectral norm sense from the true matrices $R^{L}_{f|p,y}$ and $ R^{L}_{y}$ respectively. It is thus relevant to study the behaviour of the above estimators of $P$ in asymptotic regimes where $M$ and $N$ both converge towards $+\infty$ in such a way that $c_N = \frac{ML}{N}$ converges towards a non zero constant $c_*$. In this context,  matrix $\hat{R}^{L}_{f|p,y}$ is no
longer a consistent estimate of the true matrix $R^{(L)}_{f|p,y}$ in the spectral norm sense. Therefore, the singular values of
$\hat{R}^{(L)}_{f|p,y}$ have no reasons to behave as those of $R^{(L)}_{f|p,y}$, and the same conclusion holds for matrices $(\hat{R}^{L}_{f,y})^{-1/2} \hat{R}^{L}_{f|p,y} (\hat{R}^{L}_{p,y})^{-1/2}$ and 
$(R^{L}_{y})^{-1/2} R^{L}_{f|p,y}  (R^{L}_{y})^{-1/2}$. Thus, it appears of fundamental interest to evaluate the behaviour of the singular values of $\hat{R}^{L}_{f|p,y}$ and $(\hat{R}^{L}_{f,y})^{-1/2} \hat{R}^{L}_{f|p,y} (\hat{R}^{L}_{p,y})^{-1/2}$, and to study whether the largest singular values still allow to estimate $P$ consistently, at least if the power of the useful signal $u$ and the non zero singular values of $R^{L}_{f|p,u}$ or the non zero canonical correlation coefficients between the spaces $\mathcal{U}_{p,L}$ and $\mathcal{U}_{f,L}$ are large enough. \\

In this paper, we address these problems when the integers $K$ and $P$ do not scale with $M$ and $N$, 
and thus remain fixed integers. This in practice means that the following results are likely to 
be useful when the rank $K$ of the spectral density of $u$ is much smaller than $M$, and when 
$P$ is small enough compared to $M$ and $N$. As $P$ is supposed to be a fixed integer, the integer 
$L \geq P$ will also be assumed to remain fixed when $M$ and $N$ converge towards $+\infty$. As explained below, the assumption $K,P,L$ remain fixed implies that the matrices $\hat{R}^{L}_{f|p,y}$ 
and $(\hat{R}^{L}_{f,y})^{-1/2} \hat{R}^{L}_{f|p,y} (\hat{R}^{L}_{p,y})^{-1/2}$ are low rank perturbations 
of the random matrices  $\hat{R}^{L}_{f|p,v}$ and $(\hat{R}^{L}_{f,v})^{-1/2} \hat{R}^{L}_{f|p,v} (\hat{R}^{L}_{p,v})^{-1/2}$ built from the noise samples $v_1, \ldots, v_N$ instead of $y_1, \ldots, y_N$. It is thus in principle 
possible to use the perturbation techniques developed in \cite{benaych-nadakuditi-ann-math}, \cite{benaych-rao-2}, \cite{paul-2007}. However, the random matrix models that come into play in this paper are considerably more 
complicated than in \cite{benaych-nadakuditi-ann-math}, \cite{benaych-rao-2}, \cite{paul-2007}. Thus, the following 
results cannot be considered as direct consequences of \cite{benaych-nadakuditi-ann-math}, \cite{benaych-rao-2}, \cite{paul-2007}. \\

We first evaluate in Section \ref{sec:autocov_signal} the behaviour of the largest singular values of $\hat{R}^{L}_{f|p,y}$, or equivalently of the largest eigenvalues of $\hat{R}^{L}_{f|p,y}\hat{R}^{L*}_{f|p,y}$ and take benefit of the results in \cite{loubaton-tieplova-2020} in which the asymptotic behaviour of the eigenvalues of $\hat{R}^{L}_{f|p,v}\hat{R}^{L*}_{f|p,v}$ is characterized. Introducing some extra assumptions, we deduce from \cite{loubaton-tieplova-2020} that for each $\epsilon > 0$, almost surely, for $N$ large enough, all the eigenvalues $\hat{R}^{L}_{f|p,v}\hat{R}^{L*}_{f|p,v}$ are less than $x_{+,*} + \epsilon$ for a certain $x_{+,*} > 0$.  Using the perturbation techniques developed in 
\cite{benaych-rao-2} and \cite{paul-2007}, we obtain that the number of eigenvalues of $\hat{R}^{L}_{f|p,y}\hat{R}^{L*}_{f|p,y}$ that may escape from the interval $[0, x_{+,*}]$ is between $0$ and $2r$ where $r$ represents the rank of the covariance matrix $R^{(L)}_u$ of the vector $(u_n^{T}, \ldots, u_{n+L-1}^{T})^{T}$. When $P=1$ and $R = \sigma^{2} I$ for some $\sigma^{2}$, for any $r \geq 1$, we indicate how to produce simple examples such that $2r -1$ eigenvalues of $\hat{R}^{L}_{f|p,y}$ escape from  $[0, x_{+,*}]$. This behaviour leads to the conclusion that $P$ cannot be estimated consistently as the number of eigenvalues that are larger than $x_{+,*}$ even if the useful $u$ is powerful enough and the non zero singular values of $R^{(L)}_{f|p,u}$ are large enough. While it would be possible to address the case $c_* \geq 1$, we will assume that $c_* < 1$ to simplify the exposition. Therefore,  $c_N$ verifies $c_N < 1$ for each $N$ large enough . \\

Always under the assumption $c_* < 1$, using the same approach, we then study in Section \ref {sec:cor_coeficients} the largest eigenvalues of $(\hat{R}^{L}_{f,y})^{-1/2} \hat{R}^{L}_{f|p,y} (\hat{R}^{L}_{p,y})^{-1} (\hat{R}^{L}_{f|p,y})^{*} (\hat{R}^{L}_{f,y})^{-1/2}$, which also coincide 
with those of matrix $\Pi_{p,y} \Pi_{f,y}$ where $\Pi_{p,y}$ and 
$\Pi_{f,y}$ represent the orthogonal projection matrices on the spaces generated by the rows of $Y_p$ 
and $Y_f$ respectively. We first study the eigenvalue distribution of $\Pi_{p,v} \Pi_{f,v}$, and 
establish that it converges towards the free multiplicative convolution product of $c_* \delta_1 + (1-c_*) \delta_0$ with itself. Notice that  $c_N \delta_1 + (1-c_N) \delta_0$ coincides with the eigenvalue distribution of matrices $\Pi_{p,v}$ and $\Pi_{f,v}$. We also establish that almost surely, for $N$ large enough, all the eigenvalues of $\Pi_{p,v} \Pi_{f,v}$ lie in a neighbourhood of the support $[0, 4c_*(1-c_*)] \cup \{ 1 \} \mathbf{1}_{c_* > 1/2}$ of its limit distribution. Using the above mentioned perturbation techniques, we establish that if $s$ represents the number of eigenvalues of $\Pi_{p,y} \Pi_{f,y}$ that escape from $[0, 4c_*(1-c_*)] \cup  \{ 1 \}  \mathbf{1}_{c_* > 1/2}$, then, $s \leq P$, and eventually provide the explicit conditions under which $s=P$. These conditions hold if $c_* < 1/2$ and if the power of $u$ and the non zero canonical correlation coefficients between the spaces $\mathcal{U}_{p,L}$ and $\mathcal{U}_{f,L}$ are large enough. These results allow to conclude that, under certain reasonable well defined assumptions, it is possible to estimate $P$ consistently using the largest singular values of $(\hat{R}^{L}_{f,y})^{-1/2} \hat{R}^{L}_{f|p,y} (\hat{R}^{L}_{p,y})^{-1/2}$, 
but that the use of the largest singular values of $ \hat{R}^{L}_{f|p,y} $ appears unreliable. \\

It has hard to explain intuitively why the use of the normalized matrix $(\hat{R}^{L}_{f,y})^{-1/2} \hat{R}^{L}_{f|p,y} (\hat{R}^{L}_{p,y})^{-1/2}$ allows to estimate $P$ consistently under certain assumptions, while this is not the case for matrix $\hat{R}^{L}_{f|p,y}$. We however mention that matrix $(\hat{R}^{L}_{f,v})^{-1/2} \hat{R}^{L}_{f|p,v} (\hat{R}^{L}_{p,v})^{-1/2}$ defined by replacing $y$ by $v$ does not 
depend on the covariance matrix $R$ of the random vectors $(v_n)_{n \in \mathbb{Z}}$, while it is of course not the case 
of matrix $\hat{R}^{L}_{f|p,v}$. This invariance property appears of course attractive, and plays an important role in the following. We also mention that matrix $(\hat{R}^{L}_{f,y})^{-1/2} \hat{R}^{L}_{f|p,y} (\hat{R}^{L}_{p,y})^{-1/2}$ is  connected with the canonical analysis of the time series $y$. Generally speaking, this analysis has well established merits. In particular, it leads to the concept of stochastically balanced state-space realizations which are known to be useful to derive model reduction algorithms (\cite{vanoverschee-demoor} and  \cite{lindquist-picci-book-2015} and the references therein). \\ 

We believe that the main findings of this paper are of potential interest for statistical signal processing and time series analysis researchers. However, the large random matrix models that come into play in this paper are rather 
complicated, and were almost not considered in previous works. Therefore, new random matrix tools have to be developed and a number of technical intermediate results have to be established. In order to improve the readability of this paper, we postpone the most technical steps in the Appendix, and sometimes provide sketches of proof rather than detailed arguments.

\subsection{On the literature.}
We first mention that the problems considered in this paper have connections with the 
"Generalized Dynamic Factor Models" introduced in the econometrics field, see e.g. \cite{forni-hallin-lippi-reichlin-2000}, \cite{forni-hallin-lippi-reichlin-2004}, \cite{deistler-anderson-filler-chen-ejc-2010}. In these works, the observation 
is still given by $y_n = u_n + v_n$ where $u_n =  [H(z)]i_n$ and $v_n$ are called the common component and the idiosyncratic component respectively. $H(z)$ is not assumed in \cite{forni-hallin-lippi-reichlin-2000} and \cite{forni-hallin-lippi-reichlin-2004} to be rational, 
while $v$ is not necessarily an uncorrelated time series. These papers still address estimation 
problems in the asymptotic regime where $M$ and $N$ converge towards $+\infty$, but \cite{forni-hallin-lippi-reichlin-2000}, \cite{forni-hallin-lippi-reichlin-2004}, \cite{deistler-anderson-filler-chen-ejc-2010} assume that the eigenvalues 
of the spectral density matrix of $v$ remain bounded when $M$ and $N$ increase, while the 
$K$ non zero eigenvalues of the spectral density of $u$ converge towards $+\infty$. In this context, 
it appears possible to estimate consistently from the available samples a number of parameters attached to the useful signal $u$. In particular, if $H(z)$ is rational, the estimation of $P$ does not pose any problem (see  \cite{deistler-anderson-filler-chen-ejc-2010} devoted to the case $H(z)$ rational). In contrast, the technical assumptions formulated in the present paper imply that the eigenvalues of 
the spectral densities of $u$ and $v$ are of the same order of magnitude when $M$ and $N$ increase. We refer the reader to \cite{rosuel-et-al-ieeesp-21} for a discussion on the practical relevance of the context of the present paper. Therefore, the solutions developed in  \cite{forni-hallin-lippi-reichlin-2000}, \cite{forni-hallin-lippi-reichlin-2004}, \cite{deistler-anderson-filler-chen-ejc-2010} cannot be used to design consistent 
estimates of $P$ under our assumptions. \\

We next review the existing works that are more directly related to the present paper. The behaviour of the eigenvalues of matrix $R^{L}_{f|p,v}R^{L*}_{f|p,v}$ was studied in 
\cite{loubaton-tieplova-2020}, and we refer to this paper for the various references that 
addressed similar problems when $L=1$, in the non Gaussian case, or when the time series 
$(v_n)_{n \in \mathbb{Z}}$ is possibly correlated in the time domain. 
Apart \cite{li-wang-yao-ann-stat-2016}, we are not aware of any previous work addressing the behaviour of the largest singular values of matrices depending on estimated autocovariance matrices of $y$ at non zero lags in the presence of a low rank useful signal $u$. \cite{li-wang-yao-ann-stat-2016} assumes that 
$v$ is possibly non Gaussian with covariance matrix $R = \sigma^{2} I$, and that the useful signal 
$u$ is given by $u_n = H s_n$ where $H$ is a $M \times K$ matrix verifying $H^{*} H = I_K$ and where the components $(s_{k,n})_{n \in \mathbb{Z}}$ of $(s_n)_{n \in \mathbb{Z}}$ are independent times series. Using the above mentioned perturbation analysis, \cite{li-wang-yao-ann-stat-2016} studies the eigenvalues of $\hat{R}^{1}_{f|p,y} \hat{R}^{1*}_{f|p,y}$ that escape from the interval $[0,x_{+,*}]$ introduced above. 
We notice that if $L=1$, matrix  $\hat{R}^{1}_{f|p,y}$ coincides with the standard estimate of the autocovariance matrix of $y$ at lag $1$. \\

We finally mention that a number of previous works addressed the behaviour of the canonical correlation coefficients between the row spaces of two large random matrices.  However, the underlying random matrix models are simpler than in the present paper. More specifically, the structured random matrices $Y_{p,L}$ and $Y_{f,L}$ as well as $V_{p,L}$ and $V_{f,L}$  are replaced by mutually independent matrices $X_1$ and $X_2$ with i.i.d. elements, a property that is not verified by $Y_{p,L}$, $Y_{f,L}$,  $V_{p,L}$ and $V_{f,L}$. \cite{wachter-1980} addressed the case of $M \times N$ mutually independent complex Gaussian  matrices $X_1$ and $X_2$ with i.i.d. entries, and derived the corresponding limit distribution of the squared canonical correlation coefficients. This is equivalent to evaluating the limit eigenvalue distribution of $\Pi_1 \Pi_2$ where $\Pi_1$ and $\Pi_2$ represent the orthogonal matrices on the row spaces of $X_1$ and $X_2$. We note that the result of \cite{wachter-1980} appears as a trivial consequence of basic free probability theory results (see e.g. \cite{voiculescu-al-1992} \cite{hiai-petz-book}, \cite{mingo-speicher-book}, as well as \cite{tulino-verdu-book} for a more engineering oriented presentation) because under the above hypotheses, $\Pi_1$ and $\Pi_2$ are almost surely asymptotically free.  More recently, \cite{yang-pan-2012} extended this result to the case where $X_1$ and $X_2$ are independent matrices with non Gaussian i.i.d. entries. We also note that \cite{yang-pan-2015} took benefit of this result to propose independence tests between 2 sets of i.i.d. high-dimensional samples. We mention that \cite{bao-hu-pan-zhou-2019} extended the result of \cite{wachter-1980} to the case where $X_1$ and $X_2$ have Gaussian i.i.d. entries, but this time $\mathbb{E} \{\frac{X_1 X_2^{*}}{N}\}$ is a non zero low rank matrix. We finally notice that in \cite{tieplova-loubaton-pastur-2020}, we established the convergence of the eigenvalue distribution of $\Pi_{p,v} \Pi_{f,v}$ by establishing the almost sure freeness of $\Pi_{p,v}$ and  $\Pi_{f,v}$. We however mention that in order to study the largest eigenvalues of $\Pi_{p,y} \Pi_{f,y}$ using perturbation techniques, it is also necessary to evaluate the asymptotic behaviour of the resolvent of $\Pi_{p,v} \Pi_{f,v}$, a more difficult issue that is solved in the present paper. 

\subsection{Assumptions, notations and basic tools.}
\label{ch:notations}
We now introduce the main assumptions, notations and fundamental tools that will be used throughout this paper. 

\textbf{Assumptions}

\begin{itemize}
	\item  We assume that $L$ is a fixed parameter verifying $L \geq P$, and that $M$ and $N$ 
	converge towards $+\infty$ in such a way that 
	\begin{equation}
	\label{eq:asymptotic-regime}
	c_N = \frac{ML}{N} \rightarrow c_*, \, 0 < c_* < 1
	\end{equation}
	This regime will be referred to as $N \rightarrow +\infty$ in the following. In the regime 
	(\ref{eq:asymptotic-regime}), $M$ should be interpreted as an integer $M=M(N)$ depending on $N$. 
	The various matrices we have introduced above thus depend on $N$ and will be 
	denoted $R_N, Y_{f,N},Y_{p,N}, \ldots$. In order to simplify the notations, the dependency 
	w.r.t. $N$ will sometimes be omitted. We notice that the results of Section 
	\ref{sec:autocov_signal} devoted to the study of the largest eigenvalues of
	$\hat{R}^{L}_{f|p,y} \hat{R}^{L*}_{f|p,y}$ could be generalized to the case $c_* \geq  1$, but we prefer to assume $c_* < 1$ in order to 
	simplify the presentation of the corresponding results. It therefore holds that $c_N < 1$ for each $N$ large enough. 
	
	\item The sequence of covariance matrices $(R_N)_{N \geq 1}$ of $M$--dimensional vectors $(v_n)_{n=1, \ldots, N}$ 
	is supposed to verify
	\begin{equation}
	\label{eq:hypothesis-R-bis}
	a \, I \leq R_N \leq b \, I
	\end{equation}
	for each $N$, where $a>0$ and $b>0$ are 2  constants. $\lambda_{1,N} \geq \lambda_{2,N} \geq \ldots \geq \lambda_{M,N}$ represent the eigenvalues of $R_N$ arranged in the decreasing order and $f_{1,N}, \ldots, f_{M,N}$ denote the corresponding eigenvectors. Hypothesis 
	(\ref{eq:hypothesis-R-bis}) is obviously equivalent to $\lambda_{M,N} \geq a$ and $\lambda_{1,N} \leq b$ for 
	each $N$.  
\end{itemize}

\textbf{Notations}

\begin{itemize}
	\item For each $1\le i\le 2L$ and $1\le m\le M$, $\mathbf{f}_i^m$ represents the vector of the canonical basis of $\mathbb{C}^{2ML}$ with 1 at the index $m+(i-1)M$ and zeros elsewhere. In order to simplify the notations, we mention that if 
	$i \leq L$, vector $\mathbf{f}_i^m$ may also represent, depending on the context, the vector of the canonical basis of $\mathbb{C}^{ML}$ with 1 at the index $m+(i-1)M$ and zeros elsewhere. Vector $\mathbf{e}_j$ with $1\le j\le N$ represents the $j$~--th vector of the canonical basis of $\mathbb{C}^N$. 
	
	 \item For each integer $l \geq 1$, we define the $l \times l$ "shift" matrix  $J_l$  as
	 \begin{equation}
	 \label{eq:def-J}
	 (J_l)_{ij} = \delta_{j-(i+1)}.
	 \end{equation}
	 
	 \item $\mathbb{R}^+$ and $\mathbb{R}^{-}$ represent respectively the set of all non-negative numbers and non-positive numbers, and we
	 denote $\mathbb{R}^* \equiv \mathbb{R}\setminus \{0\}$, $\mathbb{R}^{+*}\equiv \mathbb{R}^+\setminus \{0\}$ and $ \mathbb{R}^{-*}\equiv \mathbb{R}^-\setminus \{0\}$. We also define $\mathbb{C}^+ = \{z \in \mathbb{C} : \im(z) > 0\}$. We finally denote by $\rho(z)$ the distance from $z \in \mathbb{C}$ to $\mathbb{R}^{+}$, i.e. 
\begin{equation}
\label{eq:def-rho}
\rho(z) = \mathrm{dist}(z,\mathbb{R}^{+})
\end{equation}
	 
	   \item By a nice constant, we mean a positive deterministic constant which does not depend on the dimensions $M$ and $N$ nor of the complex variable $z$ that appears in the various Stieltjes transforms introduced in this paper. In the following, $\kappa$ will represent a generic nice constant whose value may
	   change from one line to the other. A nice polynomial $P(z)$ is a polynomial whose degree and coefficients
	   are nice constants.
	   
	    \item If $(\alpha_N)_{N \geq 1}$ is a sequence of positive real numbers and if $\Omega$ is a domain of $\mathbb{C}^{+}$, we will say that
	    a sequence of functions $(f_N(z))_{N \geq 1}$ verifies $f_N(z)=\mathcal{O}_z(\alpha_N)$ for $z \in \Omega$ if there exists two nice polynomials $P_1$ and $P_2$ such that $|f_N(z)| \le \alpha_N P_1(|z|)P_2(\frac{1}{|\im z|})$ for each $z \in \Omega$. If $\Omega = \mathbb{C}^{+}$, we will just write
	    $f_N(z)=\mathcal{O}_z(\alpha_N)$ without mentioning the domain. We notice that 
	    if $P_1$, $P_2$ and $Q_1$, $Q_2$ are nice polynomials, then $P_1(|z|)P_2(\frac{1}{|\im z|})+Q_1(|z|)Q_2(\frac{1}{|\im z|})\le (P_1+Q_1)(|z|)(P_2+Q_2)(\frac{1}{|\im z|})$, from which we conclude that if the sequences $(f_{1,N})_{N \geq 1}$ and $(f_{2,N})_{N \geq 1}$ are $\mathcal{O}_z(\alpha_N)$ on $\Omega$, then it also holds $f_{1,N}(z)+f_{2,N}(z)=\mathcal{O}_z(\alpha_N)$ on $\Omega$.

	     \item For any matrix  $A$, $\| A \|$ and $\| A \|_{F}$ represent its spectral norm and Frobenius 
	     norm respectively. The transpose, conjugate, and conjugate
	     transpose of $A$ are respectively denoted by $A^T$ ,$\bar{A}$ and $A^*$. If $A$ is a square matrix, $\mathrm{Im}(A)$ is the Hermitian matrix defined by
	     $\mathrm{Im}(A) = \frac{A - A^{*}}{2i}$. If $A$ and $B$ are Hermitian matrices, $A \ge B$ stands for $A-B$ non-negative definite. 
	     
	     \item $\mathcal{C}_{c}^{\infty}(\mathbb{R},\mathbb{R})$ represents the set of all $\mathcal{C}^{\infty}$ real-valued compactly supported functions defined on $\mathbb{R}$.
	     
	     \item If $\xi$ is a random variable, we denote by $\xi^{\circ}$ the zero mean random variable
	     defined by 
	     \begin{equation}
	     \label{eq:def-chirond}
	     \xi^{\circ} = \xi - \mathbb{E} \xi.	     
	     \end{equation}
\end{itemize}

\textbf{Fundamentals tools}\\

If $n$ is a positive integer, then a $n \times n$ matrix-valued positive measure $\omega$ is a $\sigma$--additive function from the Borel sets of $\mathbb{R}$
onto the set of all positive $n \times n$ matrices (see e.g. \cite{rozanov}, Chapter 1 for more details). If $\omega$ is a  $n \times n$ matrix-valued positive finite \footnote{finite means that $\Tr(\omega(\mathbb{R})) < +\infty$} measure, the Stieltjes transform $S_{\omega}$ of $\omega$ is the function defined for each $z \in \mathbb{C} \setminus \mathrm{Supp}(\omega)$ by 
\begin{equation}
\label{eq:def-stieltjes-matrix}
S_{\omega}(z) = \int \frac{d \omega(\lambda)}{\lambda - z}
\end{equation}
In the following, if $B$ is a Borel set of $\mathbb{R}$, we denote by $\mathcal{S}_n(B)$ the set of all Stieltjes transforms of $n \times n$ matrix-valued positive finite measures carried by $B$. $\mathcal{S}_1(B)$ is denoted  $\mathcal{S}(B)$. We just mention the following useful properties of the elements of $\mathcal{S}_n(\mathbb{R})$ and $\mathcal{S}_n(\mathbb{R}^{+})$: 
if $S \in \mathcal{S}_n(\mathbb{R})$ and if $\omega$ represents its associated 
 $n \times n$ matrix-valued positive finite measure , then, $S$ is analytic on $\mathbb{C} \setminus \mathbb{R}$ and verifies 
\begin{equation}
\label{eq:properties-calS(R)}
\| S(z) \| \leq \frac{\| \omega(\mathbb{R}) \|}{\Im z}, \; \Im S(z) \geq  0
\end{equation}
if $z \in \mathbb{C}^{+}$. Moreover, $\omega(\mathbb{R}) = \lim_{y \rightarrow +\infty} -iy S(iy)$. When the positive matrix $ \omega(\mathbb{R})$ is positive definite, 
$\Im S(z) > 0$ on $\mathbb{C}^{+}$. If $S \in  \mathcal{S}_n(\mathbb{R}^{+})$, then $S$ is analytic 
on $\mathbb{C} - \mathbb{R}^{+}$ and also satisfies 
\begin{equation}
\label{eq:ImzS}
\Im z S(z) \geq 0, \, z \in \mathbb{C}^{+}, \, \| S(z) \| \leq \frac{\| \omega(\mathbb{R}_{+}) \|}{\rho(z)}, 
z \in \mathbb{C} - \mathbb{R}^{+}
\end{equation}
When  $ \omega(\mathbb{R}_{+}) > 0$, we also have $\Im z S(z) > 0$ on $\mathbb{C}^{+}$. 
We refer the reader to Proposition 4.1 in \cite{loubaton-tieplova-2020}
for other useful properties, and for a converse of (\ref{eq:properties-calS(R)}, \ref{eq:ImzS}). We finally 
mention the following immediate properties: 
\begin{equation}
\label{eq:property-zs(z2)}
S \in \mathcal{S}_n(\mathbb{R}^{+}) \implies {\bf S} \in  \mathcal{S}_n(\mathbb{R})
\end{equation}
where ${\bf S}(z)$ is defined for $z \in \mathbb{C}^{+}$ by ${\bf S}(z) = z S(z^{2})$. Moreover, if $\omega$ and $\bs{\omega}$ are the positive matrix-valued measures associated to $S$ and ${\bf S}$, the following equality holds:
\begin{equation}
    \label{eq:mass-zs(z2)}
\omega(\mathbb{R}^{+}) = \bs{\omega}(\mathbb{R})   
\end{equation}
If $A$ is a $n \times n$  matrix, the resolvent of $A$ is defined as the matrix-valued function $Q_{A}$ defined on $\mathbb{C} - \{ \lambda_1(A), \ldots, \lambda_n(A) \}$ by 
\begin{equation}
\label{eq:def-resolvente-A}
Q_{A}(z) = \left( A - z I \right)^{-1}
\end{equation}
If $A$ is Hermitian, it is clear that $Q_A$ coincides with the Stieltjes transform of the $n \times n$ positive matrix-valued measure $\omega_{A}$ given by 
$$
\omega_A = \sum_{k=1}^{n} \delta_{\lambda_k(A)} f_k(A) f_k(A)^{*}
$$
where $(f_k(A))_{k=1, \ldots, n}$ represent the eigenvectors of $A$. We notice that $\omega_{A}(\mathbb{R}) = I$, so that (\ref{eq:properties-calS(R)}) leads to 
\begin{equation}
\label{eq:upper-bound-norm-QA}
\| Q_A(z) \| \leq \frac{1}{\Im z}
\end{equation}
on $\mathbb{C}^{+}$ and 
\begin{equation}
\label{eq:upper-bound-norm-QA-Apositive}
\| Q_A(z) \| \leq \frac{1}{\rho(z)}
\end{equation}
on $\mathbb{C} - \mathbb{R}^{+}$ if $A \geq 0$. We also mention that $Q_A$ satisfies the "resolvent identity"
\begin{equation}
\label{eq:resolvent-identity-QA}
I + zQ_A(z) = Q_A(z) \, A = A \, Q_A(z)
\end{equation}
for each $z$. If $\nu_{A} = \frac{1}{n} \sum_{k=1}^{n} \delta_{\lambda_k(A)}$ represents the empirical 
eigenvalue distribution of $A$, 
$\frac{1}{n} \Tr Q_A(z)$ is the Stieltjes transform of $\nu_{A}$. \\

We recall that if $(A_N)_{N \geq 1}$ is a sequence of $N \times N$ Hermitian (possibly random) matrices, 
a convenient way to study the behaviour of the sequence of probability measures $(\nu_{A_N})_{N \geq 1}$ when $N \rightarrow +\infty$
is to study the asymptotic behaviour of the corresponding Stieltjes transforms 
$S_{\nu_{A_N}}(z) = \frac{1}{N} \Tr(Q_{A_N}(z))$ because the weak convergence of sequence
$(\nu_{A_N})_{N \geq 1}$ towards a probability measure $\nu_*$ is equivalent to the convergence of 
$S_{\nu_{A_N}}(z)$ towards the Stieltjes transform of $\nu_*$ for each $z \in \mathbb{C}^{+}$. This explains
why Stieltjes transforms and resolvents play an important role in large random matrix theory. We refer 
the reader to e.g. \cite{bai-silverstein-book}, \cite{pastur-shcherbina-book}, \cite{bai-yao-book}. See also \cite{couillet-debbah-book} and \cite{tulino-verdu-book} for more engineering oriented books. \\

We also recall  Montel's theorem (see e.g. \cite{conway-book}), also called the Normal Family Theorem, which is frequently used in the large random matrix literature. If $(s_N(z))_{N \geq 1}$ is a sequence 
of functions that are holomorphic on a domain $\Omega$, and such that, for each compact set $\mathcal{K} \subset \Omega$, $\sup_{N \geq 1} \sup_{z \in \mathcal{K}} |s_N(z)| < +\infty$, then it is possible to extract from 
$(s_N(z))_{N \geq 1}$ a subsequence converging uniformly on each compact subset of $\Omega$ 
towards a function $s_*(z)$ holomorphic on $\Omega$. Note in particular that if for each $N \geq 1$, $s_N$ is the Stieltjes transform of a probability measure, then $(s_N(z))_{N \geq 1}$ verifies the above 
assumptions for $\Omega = \mathbb{C}^{+}$ because $|s_N(z)| \leq \frac{1}{\Im z}$ on $\mathbb{C}^{+}$ 
for each $N \geq 1$. \\

In this paper, we will consider frequently $2n \times 2n$  matrices $\A$ given by 
$$
\A = \left( \begin{array}{cc} 0 & B \\ C & 0 \end{array} \right)
$$
where $B$ and $C$ are $n \times n$ matrices. The resolvent $\Q_{\A}$ of $\A$ is given by
\begin{equation}
\label{eq:expre-Qlinearization}
\Q_{\A}(z) = \left( \begin{array}{cc} z Q_{BC}(z^{2}) & Q_{BC}(z^{2}) \, B \\ C \,  Q_{BC}(z^{2}) & z Q_{CB}(z^{2}) \end{array} \right)
\end{equation}
If the eigenvalues of $BC$ are real and positive, the eigenvalues of 
$\A$ are the $\pm \left(\sqrt{\lambda_k(BC)}\right)_{k=1, \ldots, n}$. \\

We finally recall the two Gaussian tools that will be used in the sequel in order to 
evaluate the asymptotic behaviour of certain resolvents. 
\begin{proposition}
	\label{prop:integration-by-parts}
	\textbf{(Integration by parts formula.)}
	Let $\xi=[\xi_1,\ldots,\xi_K]^T$ be a complex Gaussian random vector such that $\ex\{\xi\}=0$, $\ex\{\xi\xi^T\}=0$ and $\ex\{\xi\xi^*\}=\Omega$. If $\Gamma:(\xi)\mapsto\Gamma(\xi,\bar{\xi})$ is a $\mathcal{C}^1$ complex function polynomially bounded together with its derivatives, then 
	\begin{align}\label{integr}
	\ex\{\xi_i\Gamma(\xi)\}=\sum_{k=1}^{K}\Omega_{ik}\ex\left\{\dfrac{\partial\Gamma(\xi)}{\partial\bar{\xi}_k}\right\}.
	\end{align}
\end{proposition}

\begin{proposition}
	\label{prop:poincare}
	\textbf{(Poincar\'e-Nash inequality.)}
	Let $\xi=[\xi_1,\ldots,\xi_K]^T$ be a complex Gaussian random vector such that $\ex\{\xi\}=0$, $\ex\{\xi\xi^T\}=0$ and $\ex\{\xi\xi^*\}=\Omega$. If $\Gamma:(\xi)\mapsto\Gamma(\xi,\bar{\xi})$ is a $\mathcal{C}^1$ complex function polynomially bounded together with its derivatives, then, noting $\nabla_{\xi}\Gamma=[\frac{\partial\Gamma}{\partial\xi_1},\ldots,\frac{\partial\Gamma}{\partial\xi_K}]^T$ and $\nabla_{\bar{\xi}}\Gamma=[\frac{\partial\Gamma}{\partial\bar{\xi}_1},\ldots,\frac{\partial\Gamma}{\partial\bar{\xi}_K}]^T$
	\begin{align}\label{p-n}
	\var\{\Gamma(\xi)\}\le\ex\left\{\nabla_{\xi}\Gamma(\xi)^T\Omega\overline{\nabla_{\xi}\Gamma(\xi)}\right\}+\ex\left\{\nabla_{\bar{\xi}}\Gamma(\xi)^*\Omega\nabla_{\bar{\xi}}\Gamma(\xi)\right\}.
	\end{align}
\end{proposition}

The combination of these two tools was first proposed in \cite{pastur-2005}, see also \cite{pastur-shcherbina-book}
for an exhaustive reference. We also mention \cite{hachem-khorunzhy-loubaton-najim-pastur-2008} in which 
Propositions \ref{prop:integration-by-parts} and \ref{prop:poincare} are used in order to study the capacity 
of large MIMO channels.

 \section{The largest singular values of the empirical autocovariance matrix.}
\label{sec:autocov_signal}
\subsection{Review of the zero signal case.} 
In this paragraph, we briefly present the useful results from \cite{loubaton-tieplova-2020} concerning the study of the singular values of matrix $\hat{R}^{L}_{f|p,v}$,
or equivalently of the eigenvalues of  $\hat{R}^{L}_{f|p,v}  (\hat{R}^{L}_{f|p,v})^{*}$. All along
Section \ref{sec:autocov_signal}, we will denote by $W_{p,N}$ and $W_{f,N}$ the $ML \times N$ normalized matrices defined by 
\begin{equation}
\label{eq:def-Wpf-intro}
W_{p,N} = \frac{1}{\sqrt{N}}V_{p,N}, \, W_{f,N} = \frac{1}{\sqrt{N}}V_{f,N}
\end{equation}
and by $W_N$ the $2ML \times N$ matrix given by 
\begin{equation}
\label{eq:def-W-intro}
W_N = \left( \begin{array}{c} W_{p,N} \\ W_{f,N} \end{array} \right)
\end{equation}
We first mention that matrices $W_{i,N}$, $i=p,f$ verify the following property. 
\begin{equation}
  \label{eq:Wi-bounded}
  \mbox{There exists a nice constant $\kappa$ such that, almost surely, for each $N$ large enough}, \; \| W_{i,N} \| < \kappa
  \end{equation}
If $R_N$ was equal to $I_M$, this property would be an immediate consequence of Theorem 1.1 in \cite{L:15}. In the general
case, it is an immediate consequence of Eq. (3.1) in \cite{loubaton-tieplova-2020} and of (\ref{eq:hypothesis-R-bis}). \\

In order to study the asymptotic behaviour of the eigenvalues of $W_{f,N} W_{p,N}^{*}W_{p,N} W_{f,N}^{*}$, \cite{loubaton-tieplova-2020} studied the behaviour of the resolvent, denoted $Q_{N,W}(z)$, of the $ML \times ML$ matrix $W_{f,N} W_{p,N}^{*}W_{p,N} W_{f,N}^{*}$, i.e. 
\begin{equation}
\label{eq:def-QNW}
Q_{N,W}(z) = \left( W_{f,N} W_{p,N}^{*}W_{p,N} W_{f,N}^{*} - z I \right)^{-1}
\end{equation}
The entries of $Q_{N,W}$ are easily seen to concentrate almost surely around their mathematical expectations. Therefore, it is sufficient to study the behaviour of $\mathbb{E}(Q_{N,W}(z))$ using Propositions \ref{prop:integration-by-parts} and \ref{prop:poincare}. As the entries of $W_{f,N}W_{p,N}^{*}W_{p,N}W_{f,N}^{*}$ are bi-quadratic functions of the entries of $W_N$, the Gaussian calculations that allow to evaluate 
$\mathbb{E}(Q_{N,W}(z))$ are very complicated. Therefore,  \cite{loubaton-tieplova-2020} used the well-known linearization trick that consists in studying the resolvent ${\bf Q}_{N,W}(z)$ of the $2ML \times 2ML$ hermitized version 
$$
\begin{pmatrix}
0 & W_{f,N} W_{p,N}^*\\
W_{p,N} W_{f,N}^*&0
\end{pmatrix}
$$
Formula (\ref{eq:expre-Qlinearization}) allows eventually to deduce $\mathbb{E}(Q_{N,W}(z))$ from the first diagonal block of $\Q_{N,W}(z)$. This linearization trick will also be used extensively in the present paper.
In the following, every $2ML \times 2ML$ matrix $\G$ such as $\Q_N(z)$ will be written 
\begin{align*}
\mathbf{G}=\begin{pmatrix}
\mathbf{G_{pp}}&\mathbf{G_{pf}}\\
\mathbf{G_{fp}}&\mathbf{G_{ff}}
\end{pmatrix},
\end{align*}  
where the 4 matrices $(\mathbf{G}_{i,j})_{i,j \in {p,f}}$ are $ML \times ML$. Sometimes, the blocks will be denoted $\mathbf{G}(pp)$, $\mathbf{G}(pf)$, ....\\

In order to introduce the main results of \cite{loubaton-tieplova-2020}, we recall Proposition 6.1 in \cite{loubaton-tieplova-2020}: for each $z \in \mathbb{C}^{+}$, the equation 
\begin{equation}
\label{eq:t-autocovariance}
t_N(z)=\dfrac{1}{M}\tr R_N\left(-zI_M-\dfrac{zc_Nt_N(z)}{1- zc_N^2t_N^2(z)}R_N\right)^{-1} 
\end{equation}
has a unique solution for which $t_N(z)$ and $z t_N(z)$ belongs to $\mathbb{C}^{+}$. Moreovoer, $t_N$ 
is the Stieltjes transform of a positive measure $\mu_N$ carried by $\mathbb{R}^{+}$, and the 
$M \times M$ matrix-valued function $T_N(z)$ defined by 
\begin{equation}
\label{eq:def-T-autocovariance}
T_N(z) = - \left(zI_M+ \dfrac{zc_Nt_N(z)}{1- zc_N^2t_N^2(z)}R_N\right)^{-1},
\end{equation}
belongs to $\mathcal{S}_M(\mathbb{R}^{+})$. Its associated positive matrix-valued measure, denoted 
$\nu_N^{T}$, verifies $\nu_N^{T}(\mathbb{R}^{+}) = I$. We also define $\t_N(z)$ and $\T_N(z)$ by 
\begin{equation}
\label{eq:def-bold-t-autocovariance}
\t_N(z) = z t_N(z^{2})
\end{equation}
and 
\begin{equation}
\label{eq:def-bold-T-autocovariance}
\T_N(z) = z T_N(z^{2}) = \left(-zI_M-\dfrac{c_N \t_N(z)}{1-c_N^2 \t_N^2(z)}R_N \right)^{-1}
\end{equation}
which, by (\ref{eq:property-zs(z2)}), belong to $\mathcal{S}(\mathbb{R})$ and $\mathcal{S}_M(\mathbb{R})$ 
respectively. Moreover, the positive matrix-valued measure $\bs{\nu}_N^{\T}$ associated to $\T_N$ verifies $\bs{\nu}_N^{\T}(\mathbb{R}) = \nu_N^{T}(\mathbb{R}^{+}) = I$. Then, the following Proposition can be deduced from the results of \cite{loubaton-tieplova-2020}. 
\begin{proposition}
\label{prop:convergences-stieltjes-transforms-autocovariance}

We consider sequences  of deterministic $ML \times ML$ and $2ML \times 2ML$ matrices 
$(A_N)_{N \geq 1}$ and $(\A_N)_{N \geq 1}$ verifying $\sup_{N} \|A_N\| < +\infty$ and 
$\sup_{N} \|\A_N\| < +\infty$. Then, we have 
\begin{equation}
\label{eq:Q-T-trace}
\frac{1}{ML} \mathrm{Tr} \left( (Q_N(z) - I_L \otimes T_N(z)) A_N \right) \rightarrow 0
\end{equation}
and 
\begin{equation}
\label{eq:boldQ-boldT-trace}
\frac{1}{2ML} \mathrm{Tr} \left( (\Q_N(z) - I_{2L} \otimes \T_N(z)) \A_N \right) \rightarrow 0
\end{equation}
where the convergence holds almost surely and uniformly on the compact subsets of $\mathbb{C} \setminus \mathbb{R}^{+}$ 
and of  $\mathbb{C} \setminus \mathbb{R}$ respectively. Moreover, if $(a_N)_{N \geq 1}$, 
and $(b_N)_{N \geq 1}$ (resp.  $(\a_N)_{N \geq 1}$,  $(\b_N)_{N \geq 1}$) represent sequences of $ML$-dimensional 
(resp. $2ML$-dimensional) deterministic vectors verifying $\sup_{N} \|a_N\| < +\infty$ and $\sup_{N} \|b_N\| < +\infty$
(resp.  $\sup_{N} \|\a_N\| < +\infty$ and $\sup_{N} \|\b_N\| < +\infty$), we also have 
\begin{equation}
\label{eq:Q-T-quadraticform}
a_N^{*} \left( Q_N(z) -  I_M \otimes T_N(z) \right) b_N \rightarrow 0
\end{equation}
and 
\begin{equation}
\label{eq:boldQ-boldT-quadraticform}
\a_N^{*} \left( \Q_N(z) -  I_{2L} \otimes \T_N(z) \right) \b_N \rightarrow 0
\end{equation}
almost surely and uniformly on the compact subsets of $\mathbb{C} \setminus \mathbb{R}^{+}$ and 
$\mathbb{C} \setminus \mathbb{R}$ respectively. 
\end{proposition}
We denote in the following $\hat{\nu}_N$ the empirical eigenvalue distribution of matrix $W_{f,N} W_{p,N}^{*}W_{p,N} W_{f,N}^{*}$. The use of (\ref{eq:Q-T-trace}) for $A_N = I$ leads to the conclusion that if $\nu_N$ represents the probability 
measure defined by 
\begin{equation}
\label{eq:def-nuN}
\nu_N  = \frac{1}{M} \Tr \nu_N^{T}
\end{equation}
then $\hat{\nu}_N - \nu_N \rightarrow 0$ weakly almost surely. Therefore, the empirical eigenvalue distribution $\hat{\nu}_N$ of $W_{f,N} W_{p,N}^{*}W_{p,N} W_{f,N}^{*}$ has a deterministic behaviour when $N \rightarrow +\infty$, and 
measure $\nu_N$ will be referred to as the deterministic equivalent of $\hat{\nu}_N$ in the following. 
\cite{loubaton-tieplova-2020} also characterized the support of $\nu_N$, or equivalently the support of 
$\mu_N$ because Assumption (\ref{eq:hypothesis-R-bis}) implies that $\nu_N$ and $\mu_N$ are absolutely continuous one with respect to each other. For this, the behaviour of 
$t_N(z)$ when $z$ converges towards the real axis is studied in \cite{loubaton-tieplova-2020}. It is shown that for each $x > 0$, the limit of $t_N(z)$ when $z \in \mathbb{C}^{+}$ converges towards $x$ exists and is finite. This limit is still denoted $t_N(x)$ in the following. This property implies that $\mu_N$ and $\nu_N$ are absolutely continuous w.r.t. the Lebesgue measure (see e.g. Theorem 2.1 in \cite{silverstein-choi-1995}). Moreover, it is shown that the corresponding densities converge towards $+\infty$ when $x \rightarrow 0, x > 0$. In order to analyse the common support $\mathcal{S}_N$ of $\mu_N$ and $\nu_N$, the function $w_N(z)$ defined by 
\begin{equation}
\label{eq:def-wN}
w_N(z) = z c_N t_N(z) - \frac{1}{c_N t_N(z)} 
\end{equation}
is introduced. For each $z \in \mathbb{C} - \mathbb{R}^{+}$, $w_N(z)$ 
is solution of the equation $\phi_N(w_N(z)) = z$ where $\phi_N(w)$ is the function defined by 
\begin{equation}
\label{eq:def-phiN}
\phi_N(w) = c_N w^{2} \, \frac{1}{M} \mathrm{Tr} R_N \left( R_N - w I \right)^{-1} \, \left( c_N \, \frac{1}{M} \mathrm{Tr} R_N \left( R_N - w I \right)^{-1} - 1 \right).
\end{equation}
To understand the equation $\phi_N(w_N(z)) = z$, we remark that $T_N(z)$ can be written in terms of $w_N(z)$ as 
\begin{equation}
\label{eq:expre-TN-w}
T_N(z) = \frac{w_N(z)}{z} \, \left( R_N - w_N(z) I \right)^{-1}
\end{equation}
so that $t_N(z) = \frac{1}{M} \Tr R_N T_N(z)$ is equal to 
\begin{equation}
\label{eq:expre-tN-w}
t_N(z) = \frac{w_N(z)}{z} \, \frac{1}{M} \Tr\left(R_N (R_N - w_N(z) I)^{-1} \right)
\end{equation}
Plugging (\ref{eq:expre-tN-w}) into (\ref{eq:def-wN}) leads to $\phi_N(w_N(z)) = z$. 
Moreover, if we define by $w_N(x)$ for $x > 0$ the limit of $w_N(z)$ when $z \rightarrow x, z \in \mathbb{C}^{+}$, the equality $\phi_N(w_N(z)) = z$ is also valid on $\mathbb{R}^{+}$. It is proved that $x \in \mathcal{S}_N^{\circ}$ if and only $\Im(w_N(x)) > 0$ ($\mathcal{S}_N^{\circ}$ represents the interior of $\mathcal{S}_N$) and that $x \in \left(\mathcal{S}_N^{\circ}\right)^{c}$ if and only if $w_N(x)$ is real. Moreover, $w_N'(x) > 0$ for each $x \in \left(\mathcal{S}_N \right)^{c}$. Finally, if $x \in \left(\mathcal{S}_N \right)^{c}$, it holds that 
\begin{equation}
\label{eq:properties-phi-outside-support}
\phi_N(w_N(x))  = x, \; \phi^{'}(w_N(x)) >  0, \; w_N(x) \, \frac{1}{M} \mathrm{Tr} R_N \left( R_N - w_N(x) I \right)^{-1} < 0.
\end{equation}
This property allows to prove that the support $\mathcal{S}_N$ of $\mu_N$ contains $0$, and coincides with the union of intervals whose end points, apart $0$, are the extrema of $\phi_N$ whose arguments verify $\frac{1}{M} \mathrm{Tr} R_N \left( R_N - w I \right)^{-1} < 0$, see Corollary 7.2 in \cite{loubaton-tieplova-2020}. If we denote by $x_{+,N}$ the largest element of $\mathcal{S}_N$, then, $x_{+,N} = \phi_N(w_{+,N})$ where $w_{+,N}> \lambda_{1,N} = \lambda_1(R_N)$ is the largest solution of $\phi_N'(w) = 0$. It is established that $\sup_{N \geq 1} x_{+,N} < +\infty$ and $\sup_{N \geq 1} w_{+,N} < +\infty$. A sufficient condition on the eigenvalues of $R_N$ ensuring that the support of $\mu_N$ is reduced to the single interval $[0,x_{+,N}]$ is formulated (see Lemma 7.7 in  \cite{loubaton-tieplova-2020}). Using the 
Haagerup-Thorbjornsen approach (\cite{HT:05}), it is finally proved that if $d > 0$ verifies $[d, +\infty) \cap \cup_{N \geq N_0} \mathcal{S}_N = \emptyset$  for some integer $N_0$,  almost surely, for $N$ large enough, all the eigenvalues of $W_{f,N}W_{p,N}^{*}W_{p,N}W_{f,N}^{*}$ are smaller than $d$. When $\mathcal{S}_N$ is reduced to $[0,x_{+,N}]$, this property implies that for each $\epsilon > 0$, for each $N$ large enough, then all the eigenvalues of  $W_{f,N}W_{p,N}^{*}W_{p,N}W_{f,N}^{*}$ are smaller than $\sup_{N \geq N_0} x_{+,N} + \epsilon$ where $N_0$ is a large enough integer.

\subsection{Signal model and first assumptions}\label{sec:signal_model}
Now we pass to the case when signal $(u_n)_{n \in \mathbb{Z}}$ is present, and evaluate its influence on the eigenvalues of matrix $\frac{Y_f Y_p^{*}}{N} \left( \frac{Y_f Y_p^{*}}{N} \right)^{*}$. For this, we use a classical approach based on the observation that matrix 
$\frac{Y_f Y_p^{*}}{N}$ is a finite rank perturbation of matrix $\frac{V_f  V_p^{*}}{N}$
due to the noise $(v_n)_{n \in \mathbb{Z}}$. It will be assumed that for each $N$ large enough, the
support $\mathcal{S}_N$ of measure $\mu_N$ associated to $t_N(z)$ is reduced to the single interval
$\mathcal{S}_N = [0, x_{N,+}]$, see Assumption \ref{as:support-mu} below. \\

We recall that the useful signal 
$(u_n)_{n \in \mathbb{Z}}$ is generated by the minimal state-space representation (\ref{eq:state-space}). 
As $M$ is supposed to increase towards $+\infty$, it is first necessary to precise how the parameters 
of (\ref{eq:state-space}) depend on $M$. We formulate the following assumptions: 
\begin{assumption}
	\label{as:state-space-equation}
	\begin{itemize}
		\item   $(i_n)_{n \in \mathbb{Z}}$ is a $K$--dimensional white noise sequence such that 
		$\mathbb{E}(i_n i_n^{*}) = I_K$, and which is independent of $M$ and $N$
		\item The dimension $P$ of the state-space does not scale with $M$ and $N$ and matrices $A$ and $B$
		are independent of $M$ and $N$. 
		\item Matrices $C = C_N$ and $D = D_N$ depend of $M$ and thus on $N$, and are supposed to verify 
		\begin{equation}
		\label{eq:hypotheses-C-D}
		\sup_{N} \|C_N \| < + \infty, \, \sup_{N} \|D_N \| < + \infty
		\end{equation}
	\end{itemize}
\end{assumption}
We recall that $L \geq P$. As a consequence of Assumption \ref{as:state-space-equation}, the $P$--dimensional Markovian signal $(x_n)_{n \in \mathbb{Z}}$ is independent of $M$ and $N$. We define matrix $\mathcal{H}_N$ as the $ML \times KL$ block-Toeplitz matrix defined by 
\begin{equation}
\label{eq:def-H}
\mathcal{H}_N = \left( \begin{array}{ccccc} D_N & 0 & \ldots & \ldots & 0 \\
C_N B & D_N & 0 & \ddots & 0 \\
\vdots & C_N B & \ddots & \ddots & \vdots \\
C_N A^{L-3} B & \ddots & \ddots & \ddots & \vdots \\
C_N A^{L-2} B & C_N A^{L-3} B & \ddots & C_N B & D_N \end{array} \right)
\end{equation}
Then, it is easy to check that the $ML$--dimensional vector $u_n^{L} = (u_n^{T}, \ldots, u_{n+L-1}^{T})^{T}$ 
can be written as
\begin{equation}
\label{eq:expre-uL}
u_n^{L} = \left( \mathcal{O}_N, \mathcal{H}_N \right) \left( \begin{array}{c} x_n \\  i_n^{L}  \end{array} \right)
\end{equation}
where  $i_n^{L}$ is defined as $u_n^{L}$ and where we recall that the observability matrix 
$\mathcal{O}_N$ is defined by (\ref{eq:def-OL}). We formulate the following assumption:
\begin{assumption}
	\label{as:rank-O-H}
	The rank $r \leq P + KL$ of matrix $(\mathcal{O}_N, \mathcal{H}_N)$ remains constant for $N$ large enough.
\end{assumption}
In the following, we denote by  $U_{f,N}$ and $U_{p,N}$ the $ML \times N$ matrices defined as the 
analogues of $Y_{f,N}$ and $Y_{p,N}$
obtained by replacing the $M$--dimensional vectors $(y_n)_{n=1, \ldots, N+2L-1}$ by the $M$--dimensional vectors 
$(u_n)_{n=1, \ldots, N+2L-1}$. We also denote by $R_{u,N}^{L} = \mathbb{E}(u_n^{L} u_n^{L*})$ the covariance 
matrix of $u_{n}^{L}$, and recall that $\mathbb{E}(u_{n+L}^{L} u_n^{L*})$ coincides with $ R_{f|p,N}^{L} = \mathbb{E}(y_{n+L}^{L} y_n^{L*})$. We also recall that $\mathrm{Rank}(R_{f|p,N}^{L}) = P$ for each $L \geq P$ and claim that Assumption \ref{as:rank-O-H} implies that for $N$ large enough, $\mathrm{Rank}(R_{u,N}^{L}) = r$ for each $L \geq P$ . This is because $R_{u,N}^{L}$ is given by 
\begin{equation}
\label{eq:expre-RuNL-OcalH}
R_{u,N}^{L} = \left( \mathcal{O}_N, \mathcal{H}_N \right)  \left( \begin{array}{cc} R_x & 0 \\
0 & I_{KL} \end{array} \right)  \left( \mathcal{O}_N, \mathcal{H}_N \right)^{*}
\end{equation}
where $R_x = \mathbb{E}(x_n x_n^{*})$ coincides with 
$$
R_x = \sum_{k=0}^{\infty} A^{k} B B^{*} A^{*k}
$$
$R_x$ is positive definite because the minimality of the state-space representation (\ref{eq:state-space}) of $u$ 
implies that the pair $(A,B)$ is commandable. Therefore, Assumption \ref{as:rank-O-H} implies 
that $\mathrm{Rank}(R_{u,n}^{L}) = r$  for each $N$ large enough. In the following, we denote by 
\begin{equation}
\label{eq:svd-Ru}
R_{u,N}^{L} = \Theta_N \Delta_N^{2} \Theta_N^{*}
\end{equation}
the eigenvalue / eigenvector decomposition of $R_{u,N}^{L}$ where $\Delta_N^{2} = \mathrm{Diag}( \delta_{1,N}^{2}, \ldots, \delta_{r,N}^{2})$ where $(\delta_{k,N}^{2})_{k=1, \ldots, r}$ are the eigenvalues of $R_{u,N}^{L}$ arranged in the decreasing order and where $\Theta_N$ is the $ML \times r$ orthogonal matrix corresponding to the 
eigenvectors.  \\

We now take benefit of Assumptions \ref{as:state-space-equation} and \ref{as:rank-O-H} to evaluate the properties of matrices $\frac{U_{i,N} U_{i,N}^{*}}{N}$ for $i=p,f$ and $\frac{U_{f,N} U_{p,N}^{*}}{N}$. 
\begin{proposition}
\label{prop:properties-RUhat-Rfphat}
The following convergence result hold: 
\begin{align}
\label{eq:convergence-Ru-UiUi*}
& \left\| \frac{U_{i,N} U_{i,N}^{*}}{N} - R_{u,N}^{L} \right\|  \rightarrow  0  \\
\label{eq:convergence-Rfp-UfUp*}
& \left\| \frac{U_{f,N} U_{p,N}^{*}}{N} - R_{f|p,N}^{L} \right\| \rightarrow 0
\end{align}
for $i=p,f$.
\end{proposition}
{\bf Proof.} 
In the following, we denote by $X_{1,N}$ and $X_{L+1,N}$ 
the $P \times N$ matrices defined by 
\begin{equation}
\label{eq:def-X}
X_{1,N} = (x_1, x_2, \ldots, x_N), \; X_{L+1,N} = (x_{L+1}, x_{L+2}, \ldots, x_{N+L})
\end{equation}
and by $I_{f,N}$ and $I_{p,N}$ the $KL \times N$ matrices defined as the analogues of $Y_{f,N}$ and $Y_{p,N}$
obtained by replacing the $M$--dimensional vectors $(y_n)_{n=1, \ldots, N+2L-1}$ by the $K$--dimensional vectors 
$(i_n)_{n=1, \ldots, N+2L-1}$. It is easy to check that 
\begin{equation}
\label{eq:expre-Up-Uf}
U_{p,N} = \mathcal{O}_N \, X_1 + \mathcal{H}_N \, I_{p,N}, \; U_{f,N} = \mathcal{O}_N \, X_{L+1,N} + \mathcal{H}_N \, I_{f,N}
\end{equation}
As $P, K, L$ remain fixed, matrix 
$$
\frac{1}{N} \left( \begin{array}{c} X_{1,N} \\ I_{p,N} \end{array} \right) \left( \begin{array}{cc} X_{1,N}^{*} & I_{p,N}^{*} \end{array} \right)
$$
converges almost surely towards the covariance matrix of vector $\left( \begin{array}{c} x_n \\  i_n^{L}
\end{array} \right)$, i.e. matrix 
$$
\left( \begin{array}{cc} R_x & 0 \\
0 & I_{KL} \end{array} \right)
$$
As the rank of this matrix is obviously $P + KL$, the same property holds for $\left( \begin{array}{c} X_{1,N} \\ I_{p,N} \end{array} \right)$ for $N$ large enough. Moreover, (\ref{eq:hypotheses-C-D}) implies that 
\begin{equation}
\label{eq:bounded}
\sup_{N} \left\| \left(\mathcal{O}_N, \mathcal{H}_N \right) \right\| < +\infty
\end{equation} 
Using the equation
$$
\frac{U_{p,N}U_{p,N}^{*}}{N} =  \left(\mathcal{O}_N, \mathcal{H}_N \right) \, \frac{1}{N} \left( \begin{array}{c} X_{1,N} \\ I_{p,N} \end{array} \right) \left( \begin{array}{cc} X_{1,N}^{*} & I_{p,N}^{*} \end{array} \right) \, \left(\mathcal{O}_N, \mathcal{H}_N \right)^{*}, 
$$
 (\ref{eq:expre-RuNL-OcalH}) and (\ref{eq:bounded}) imply that
\begin{equation}
\label{eq:convergence-Ru-UpUp*}
\| R_{u,N}^{L} - \frac{U_{p,N} U_{p,N}^{*}}{N} \| \rightarrow 0
\end{equation}
It holds similarly that 
\begin{equation}
\label{eq:convergence-Ru-UfUf*}
\| R_{u,N}^{L} - \frac{U_{f,N} U_{f,N}^{*}}{N} \| \rightarrow 0
\end{equation}
Moreover, the column space of matrices $U_{p,N}$ and $U_{f,N}$ both coincide with the 
$r$--dimensional column space of $(\mathcal{O}_N, \mathcal{H}_N)$ for $N$ large enough. Therefore, 
$\mathrm{Rank}\left( \frac{U_{i,N} U_{i,N}^{*}}{N} \right) = r$ almost surely for $N$ large enough. 
We also remark that
$$
\frac{1}{N} \left( \begin{array}{c} X_{L+1,N} \\ I_{f,N} \end{array} \right) \left(  X_{1,N}^{*}  I_{p,N}^{*}  \right) \rightarrow \mathbb{E} \left[ \left( \begin{array}{c} x_{n+L} \\  i_{n+L}^{L} \end{array} \right)  \left( x_n^{*},   i_n^{L*}  \right) \right] = \left[ \begin{array}{c} \mathbb{E}\left( x_{n+L} (x_n^{*},  i_n^{L*}) \right) \\ 0
\end{array} \right]
$$
Therefore, using (\ref{eq:bounded}), we obtain that 
\begin{equation}
\label{eq:argument-convergence-Rfp-hatRfp}
\left\| \frac{1}{N} \left( \mathcal{O}_N, \mathcal{H}_N \right) \left( \begin{array}{c} X_{L+1,N} \\ I_{f,N} \end{array} \right) \left(  X_{1,N}^{*}  I_{p,N}^{*}  \right)
\left( \begin{array}{c} \mathcal{O}_N^{*} \\ \mathcal{H}_N^{*} \end{array} \right) - 
\left( \mathcal{O}_N, \mathcal{H}_N \right)  \left( \begin{array}{c}\mathbb{E}\left( x_{n+L} \, u_n^{L*} \right) \\ 0 \end{array} \right)  \right\| \rightarrow 0
\end{equation}
because (\ref{eq:expre-uL}) holds.  It is easily seen
that matrix $\mathbb{E}\left( x_{n+L} u_n^{L*} \right)$ coincides with $\mathcal{C}_N = (A^{L-1} G, \ldots, G)$
(we recall that $G = \mathbb{E}(x_{n+1}u_n^*)$, see Paragraph \ref{subsec:addressed-problem}).
Moroever, as $R_{f|p,N}^{L} = \mathbb{E}(u_{n+L}^{L} u_n^{L*})$ is equal to $\mathcal{O}_N \mathcal{C}_N$, 
we obtain that 
$$
\left( \mathcal{O}_N, \mathcal{H}_N \right)  \left( \begin{array}{c}\mathbb{E}\left( x_{n+L} \, u_n^{L*} \right) \\ 0 \end{array} \right) = \mathcal{O}_N \mathcal{C}_N = R_{f|p,N}^{L} 
$$
Therefore, (\ref{eq:argument-convergence-Rfp-hatRfp}) implies that (\ref{eq:convergence-Rfp-UfUp*}) holds.  $\blacksquare$ \\

We introduce the singular value decompositions
of matrices $\frac{U_{p,N}}{\sqrt{N}}$ and $\frac{U_{f,N}}{\sqrt{N}}$:
\begin{equation}
\label{eq:svd-Up-Uf}
\frac{U_{p,N}}{\sqrt{N}} = \Theta_{p,N} \, \Delta_{p,N} \tilde{\Theta}_{p,N}^{*}, \; \frac{U_{f,N}}{\sqrt{N}} = 
\Theta_{f,N} \, \Delta_{f,N} \tilde{\Theta}_{f,N}^{*}
\end{equation}
where $\Theta_{i,N}, \Delta_{i,N}, \tilde{\Theta}_{i,N}$ are $ML \times r$, $r \times r$, $N \times r$ matrices that of course depend on $N$ for $i=p,f$. We deduce from (\ref{eq:convergence-Rfp-UfUp*}) that $\mathrm{Rank}\left( \frac{U_{f,N} U_{p,N}^{*}}{N} \right) = P$ for each $N$ large enough. As $\frac{U_{f,N} U_{p,N}^{*}}{N}$ coincides with $\Theta_{f,N} \Delta_{f,N} \tilde{\Theta}_{f,N}^{*} \tilde{\Theta}_{p,N} \Delta_{p,N} \Theta_{p,N}^{*}$, we obtain that $\mathrm{Rank} \left(\Delta_{f,N} \tilde{\Theta}_{f,N}^{*} \tilde{\Theta}_{p,N} \Delta_{p,N} \right) = P$,  
$\mathrm{Rank} \left(\Delta_{N} \tilde{\Theta}_{f,N}^{*} \tilde{\Theta}_{p,N} \Delta_{N} \right) = P$  and 
that $\mathrm{Rank} \left( \tilde{\Theta}_{f,N}^{*} \tilde{\Theta}_{p,N} \right) = P$ for each $N$ large enough. 
As in the previous works devoted to the study of conventional spiked models (see e.g. \cite{benaych-rao-2}, \cite{chapon-couillet-hachem-mestre}), it is necessary to introduce assumptions concerning the existence of limits of certain terms depending on the statistics of the useful signal $u$. In particular, 
we will need the following assumption. 
	\begin{assumption}
		\label{as:signal}
		$r \times r$ matrices $\Delta_N$ and $\Theta_N^{*} R_{f|p,N}^{L} \Theta_N$ converge towards
		matrices $\Delta_*$ and $\Gamma_*$ respectively. It is moreover assumed that $\Delta_* > 0$.  
	\end{assumption}
We notice that $\mathrm{Rank}(\Gamma_*) = P$. As seen below, the proofs of the main results of this paper appear simpler 
when we assume the following condition
		\begin{equation}
    \label{eq:entries-Delta*-different}
  \delta_{1,*} > \ldots > \delta_{r,*}  
\end{equation}
where $(\delta_{k,*})_{k=1, \ldots, r}$ represent the diagonal entries of $\Delta_*$. 
Therefore, in the following, we will assume that condition (\ref{eq:entries-Delta*-different}) holds, and discuss 
briefly in Sections  \ref{subsec:entries-Delta*} and \ref{subsec:without-condition-section3} below how the results can be extended to the case where some of the diagonal entries of $\Delta_* > 0$ coincide. In order to explain why condition (\ref{eq:entries-Delta*-different}) allows to simplify the following arguments, we establish the following result. 
\begin{proposition}
  \label{prop:convergence-Deltai-Delta-Thetai-Theta}
 For $i=p,f$, matrices $(\Delta_{i,N})_{N \geq 1}$ verify
 \begin{equation}
     \label{eq:convergence-Deltai-Delta}
  \| \Delta_{i,N} - \Delta_N \| \rightarrow 0 \; a.s.  
 \end{equation}
 Moreover, if condition (\ref{eq:entries-Delta*-different}) holds, and if the $r$ left singular vectors  
 $(\theta_{i,N,k})_{k=1, \ldots, r}$ of $\frac{U_{i,N}}{\sqrt{N}}$ are chosen in such a way that 
 $\theta_{N,k}^{*} \theta_{i,N,k}$ is real and positive, then we have 
  \begin{equation}
     \label{eq:convergence-Thetai-Theta}
  \| \Theta_{i,N} - \Theta_N \| \rightarrow 0 \; a.s.
 \end{equation}
\end{proposition}
{\bf Proof.} (\ref{eq:convergence-Deltai-Delta}) is a consequence of (\ref{eq:convergence-Ru-UiUi*}) and of the Weyl inequalities which imply that 
$| \delta^{2}_{i,N,k} - \delta^{2}_{N,k} | \leq \| \frac{U_{i,N}U_{i,N}^{*}}{N} - R_{u,N}^{L} \|$. We thus notice that (\ref{eq:convergence-Deltai-Delta}) holds even  when  Assumption (\ref{as:signal}) is not verified. In order to verify (\ref{eq:convergence-Thetai-Theta}), we first remark that (\ref{eq:convergence-Ru-UiUi*}), 
Assumption (\ref{as:signal}) and (\ref{eq:convergence-Deltai-Delta}) imply that 
\begin{equation}
    \label{eq:convergence-matrices-with-Delta*}
\|  \Theta_{i,N} \Delta_*^{2}  \Theta_{i,N}^{*} - \Theta_{N} \Delta_*^{2}  \Theta_{N}^{*} \| \rightarrow 0
\end{equation}
when $N \rightarrow +\infty$. Condition (\ref{eq:entries-Delta*-different}) implies that the eigenvalues of matrices $\Theta_{i,N} \Delta_*^{2}  \Theta_{i,N}^{*}$ and $\Theta_{N} \Delta_*^{2}  \Theta_{N}^{*}$ have multiplicity 1. Therefore, standard results of perturbation theory of Hermitian matrices lead to the conclusion that 
$$
\| \theta_{i,N,k} \theta_{i,N,k}^{*} - \theta_{N,k} \theta_{N,k}^{*} \| \rightarrow 0
$$
for $k=1, \ldots, r$. This implies that $\| \theta_{i,N,k} - (\theta_{N,k}^{*} \theta_{i,N,k}) \, \theta_{N,k} \| \rightarrow 0$ as well as $|\theta_{N,k}^{*} \theta_{i,N,k}|^{2} \rightarrow 1$. 
The condition $(\theta_{N,k}^{*} \theta_{i,N,k})$ real positive leads to $(\theta_{N,k}^{*} \theta_{i,N,k}) \rightarrow 1$ and to 
$$
\| \theta_{i,N,k} - \theta_{N,k} \| \rightarrow 0
$$
for each $k=1, \ldots, r$. This completes the proof of (\ref{eq:convergence-Thetai-Theta}). 
$\blacksquare$ \\

Condition (\ref{eq:entries-Delta*-different}) allows to replace matrices $\Delta_{i,N}$ and $\Theta_{i,N}$ for $i=p,f$ 
by matrices $\Delta_N$ and $\Theta_N$ up to error terms that converge towards $0$. In particular, 
$\frac{U_{f,N} U_{p,N}^{*}}{N} = \Theta_{f,N} \Delta_{f,N} \tilde{\Theta}_{f,N}^{*} \tilde{\Theta}_{p,N} \Delta_{p,N} \Theta_{p,N}^{*}$
verifies $\| \frac{U_{f,N} U_{p,N}^{*}}{N} - \Theta_N \Delta_N \tilde{\Theta}_{f,N}^{*} \tilde{\Theta}_{p,N} \Delta_{N} \Theta_{N}^{*}\| \rightarrow 0$. We introduce the rank $P$ matrix $\Gamma_N$ given by 
\begin{equation}
\label{eq:def-Gamma}
\Gamma_N = \Delta_{N} \tilde{\Theta}_{f,N}^{*} \tilde{\Theta}_{p,N} \Delta_{N}
\end{equation}
Then, (\ref{eq:convergence-Rfp-UfUp*}) implies that 
\begin{equation}
\label{eq:expre-Rfp-Gamma}
\| R_{f|p,N}^{L} - \Theta_N \Gamma_N \Theta_N^{*} \| \rightarrow 0
\end{equation}
and that, under condition (\ref{eq:entries-Delta*-different}), 
\begin{equation}
    \label{eq:limit-GammaN-under-condition}
    \lim_{N \rightarrow +\infty} \Gamma_N = \Gamma_*
\end{equation}
We notice that if some of the entries of $\Delta_*$ coincide, then (\ref{eq:limit-GammaN-under-condition}) 
does no longer hold. This point will be explained in Section \ref{subsec:entries-Delta*}. 
If we consider the singular value decomposition
\begin{equation}
\label{eq:svd-Gamma}
\Gamma_N = \Upsilon_N \Xi_N \tilde{\Upsilon}_N^{*}
\end{equation}
of matrix $\Gamma_N$, then, (\ref{eq:expre-Rfp-Gamma}) implies that the $P$ non zero singular values of 
$R_{f|p,N}^{L}$ have the same asymptotic behaviour than the $P$ non zero singular values $(\chi_{k,N})_{k=1, \ldots, P}$
of $\Gamma_N$, and converge towards the singular values of matrix $\Gamma_*$. \\

We finally notice that the canonical correlation coefficients between the row spaces of $U_{p,N}$ and $U_{f,N}$, i.e. the singular values of matrix $\tilde{\Theta}_{f,N}^* \tilde{\Theta}_{p,N}$, and 
the canonical correlation coefficients between the  spaces $\mathcal{U}_{p,L}$ and $\mathcal{U}_{f,L}$ generated by the components 
of $(u_{n+L})_{n=0, \ldots, L-1}$ and $(u_{n+L})_{n=L, \ldots, 2L-1}$, i.e. the singular values of matrix 
$\Delta_N^{-1} \Theta_N^{*} R_{f|p,N}^{L} \Theta_N \Delta_N^{-1}$, 
 have the same asymptotic behaviour. For this, we just use 
 (\ref{eq:expre-Rfp-Gamma}), (\ref{eq:limit-GammaN-under-condition}) as well as the convergence of $\Delta_N$ towards 
 $\Delta_* > 0$, and obtain that 
  \begin{equation}
      \label{eq:convergence-coeff-corr-cano-Up-Uf-theorique}
      \| \Delta_N^{-1} \Theta_N^{*} R_{f|p,N}^{L} \Theta_N \Delta_N^{-1} - \tilde{\Theta}_{f,N}^{*} \tilde{\Theta}_{p,N} \| \rightarrow 0
  \end{equation}
\subsection{General approach}
We first briefly explain the general approach that will 
be used in the following to evaluate the behaviour of the eigenvalues of $\frac{Y_{f,N} Y_{p,N}^{*}}{N}\, 
\frac{Y_{p,N} Y_{f,N}^{*}}{N}$. In order to simplify the notations, we denote by $\Sigma_{i,N}$ and $W_{i,N}$ matrices 
$\Sigma_{i,N} = \frac{Y_{i,N}}{\sqrt{N}}$ and $W_{i,N} = \frac{V_{i,N}}{\sqrt{N}}$ for $i=p,f$.  
It is easy to check that 
\begin{equation}
\label{eq:expre-SigmafSigmap*}
\Sigma_f \Sigma_p^{*} = W_f W_p^{*} + (\Theta_f, W_f \tilde{\Theta}_p \Delta_p) 
\left( \begin{array}{cc} \Delta_f  \tilde{\Theta}_f^{*} \tilde{\Theta}_p \Delta_p & I_r \\
I_r & 0 \end{array} \right) \left( \begin{array}{c} \Theta_p^{*} \\ \Delta_f \tilde{\Theta}_f^{*} W_p^{*} \end{array} \right)
\end{equation}
We denote by $\mathcal{A}$ and $\mathcal{B}$ the matrices defined by 
\begin{equation}
\label{eq:defmathcalA}
\mathcal{A} = \left( \Theta_f, W_f \tilde{\Theta}_p \Delta_p  \right)
\end{equation}
and 
\begin{equation}
\label{eq:defmathcalB}
\mathcal{B} = \left( \Theta_p, W_p \tilde{\Theta}_f \Delta_f  \right) \, \left( \begin{array}{cc} \Delta_p  \tilde{\Theta}_p^{*} \tilde{\Theta}_f \Delta_f & I_r \\
I_r & 0 \end{array} \right)
\end{equation}
Then, an easy calculation leads to 
\begin{equation}
\label{eq:expre-matrice-det-1}
\left( \begin{array}{cc} - z \, I & \Sigma_f \Sigma_p^{*} \\ \Sigma_p \Sigma_f^{*} & -z \, I \end{array} \right) = 
\left( \begin{array}{cc} - z \, I & W_f W_p^{*} \\ W_p W_f^{*} & -z \, I \end{array} \right) \, + 
\, \left( \begin{array}{cc} \mathcal{A} & 0 \\ 0 & \mathcal{B} \end{array} \right) \,  \left( \begin{array}{cc} 0  & I_{2r} \\ I_{2r}  & 0  \end{array} \right) \, 
\left( \begin{array}{cc} \mathcal{A}^* & 0 \\ 0 & \mathcal{B}^* \end{array} \right)
\end{equation}
We recall that ${\bf Q}_W(z)$ represents the resolvent of matrix $\left( \begin{array}{cc} 0 & W_f W_p^{*} \\ W_p W_f^{*} & 0 \end{array} \right)$. Consider a positive real number $y$  such that $y$ is not eigenvalue of $\left( \begin{array}{cc} 0 & W_f W_p^{*} \\ W_p W_f^{*} & 0 \end{array} \right)$ for each $N$ large enough (some conditions on such an eigenvalue will be precised below). 
For $z=y$, the left handside of (\ref{eq:expre-matrice-det-1}) can also be written as
\begin{equation}
\label{eq:expre-matrice-det-2}
\left( \begin{array}{cc} - y \, I & \Sigma_f \Sigma_p^{*} \\ \Sigma_p \Sigma_f^{*} & -y \, I \end{array} \right) = 
\left( \begin{array}{cc} - y \, I & W_f W_p^{*} \\ W_p W_f^{*} & -y \, I \end{array} \right) \, \left( I_{2ML} + 
{\bf Q}_W(y) \, \left( \begin{array}{cc} \mathcal{A} & 0 \\ 0 & \mathcal{B} \end{array} \right) \,  \left( \begin{array}{cc} 0  & I_{2r} \\ I_{2r}  & 0  \end{array} \right) \, 
\left( \begin{array}{cc} \mathcal{A}^* & 0 \\ 0 & \mathcal{B}^* \end{array} \right) \right)
\end{equation}
Therefore, $y$ is eigenvalue of $\left( \begin{array}{cc} 0 & \Sigma_f \Sigma_p^{*} \\ \Sigma_p \Sigma_f^{*} & 0 \end{array} \right)$ if and only the determinant of the second term of the right handside of (\ref{eq:expre-matrice-det-2}) vanishes. Using the identity $\mathrm{det}(I + E F) = \mathrm{det}(I + F E)$, we obtain that $y$ is an eigenvalue of
$\left( \begin{array}{cc} 0 & \Sigma_f \Sigma_p^{*} \\ \Sigma_p \Sigma_f^{*} & 0 \end{array} \right)$ if and only
\begin{equation}
\label{eq:annulation-det-1}
\mathrm{det} \left( I_{4r} +  \left( \begin{array}{cc} \mathcal{A}^* & 0 \\ 0 & \mathcal{B}^* \end{array} \right) 
{\bf Q}_W(y) \left( \begin{array}{cc} \mathcal{A} & 0 \\ 0 & \mathcal{B} \end{array} \right)  \left( \begin{array}{cc} 0  & I_{2r} \\ I_{2r}  & 0  \end{array} \right) \right) = 0
\end{equation}
or equivalently if
\begin{equation}
\label{eq:annulation-det-2}
\mathrm{det} \left( I_{4r} + F_N(y) \right) = 0
\end{equation}
where $F_N(z)$ is the $4r \times 4r$ matrix-valued function given by 
\begin{equation}
\label{eq:def-F}
F_N(z) =  \left( \begin{array}{cc} \mathcal{A}^* {\bf Q}_{W,pf}(z) \mathcal{B} &  \mathcal{A}^* {\bf Q}_{W,pp}(z) \mathcal{A} \\ \mathcal{B}^* {\bf Q}_{W,ff}(z) \mathcal{B} &  \mathcal{B}^* {\bf Q}_{W,fp}(z) \mathcal{A} \end{array} \right)
\end{equation}

We will see that under certain technical assumptions, $F_N(y)$ converges towards a deterministic matrix 
$F_*(y)$ and that the solutions of (\ref{eq:annulation-det-2}) converge towards the solutions of the 
deterministic equation $\mathrm{det} \left( I_{4r} + F_*(y) \right) = 0$, which, fortunately, can be analyzed. 

\subsection{New assumptions and their consequences.}\label{subsec:new-assumptions}
We need to distinguish two kinds of extra-assumptions.
\begin{itemize}
	\item Assumptions on the asymptotic behaviour of the eigenvalue distribution of matrix $R_N$. \\
	\begin{assumption}
		\label{as:convergence-alpha}
		If $\omega_N = \frac{1}{M}\sum_{k=1}^{M} \delta_{\lambda_{k,N}}$ is the eigenvalue distribution of matrix $R_N$, it is assumed that
		\begin{equation}
		\label{eq:limit-lambdas}
		\lim_{N \rightarrow +\infty} \lambda_{1,N} = \lambda_{+,*}  \;  \lim_{N \rightarrow +\infty} \lambda_{M,N} = \lambda_{-,*}
		\end{equation}
		We note that $\lambda_{-,*} \geq a > 0$ and $\lambda_{+,*} \leq b$ where $a$ and $b$ are defined by (\ref{eq:hypothesis-R-bis}). Moreover, sequence $(\omega_N)_{N \geq 1}$ is assumed to converge weakly
		towards a probability measure $\omega_*$, which, necessarily, is carried by $[\lambda_{-,*}, \lambda_{+,*}]$
	\end{assumption}
	\begin{assumption}
		\label{as:support-mu}
		It is assumed that for each $N$ large enough, it exists a nice constant $\kappa > 0$ such that the eigenvalues $(\lambda_{k,N})_{k=1, \ldots, M}$ satisfy 
\begin{equation}
    \label{eq:sufficient-condition-support}
    |\lambda_{k,N} - \lambda_{l,N}| \leq \kappa \left( \frac{|k-l|}{M} \right)^{1/2}
  \end{equation}
 for each pair $(k,l)$, $1 \leq k \leq l \leq M$, so that the support $\mathcal{S}_N$ of $\mu_N$ is equal to $\mathcal{S}_N = [0, x_{+,N}]$  (see Lemma 7.7 in  \cite{loubaton-tieplova-2020}). Moreover, we add the following condition: for
		each $N$ large enough, 
		\begin{equation}
		\label{eq:eloignement-lambda1-lambda-k}
		\lambda_{1,N} - \lambda_{k,N} \leq \kappa \, \frac{k-1}{M}
		\end{equation}
		for some nice constant $\kappa$. 
	\end{assumption}
	\item Assumptions on the asymptotic behaviour of matrices depending both of the useful signal and the noise.
	\begin{assumption}
		\label{as:signal-bruit}
		We recall that $(f_{k,N})_{k=1, \ldots, M}$ represent the eigenvectors of matrix $R_N$. We consider the $M \times M$ matrix-valued function positive measure
		$\omega^{R}_N$ defined by
		$$
		\omega^{R}_N = \sum_{k=1}^{M} \delta_{\lambda_{k,N}} \, f_{k,N} f_{k,N}^{*}
		$$
		and introduce the $r \times r$ positive matrix-valued measure $\gamma_N$ defined by         
		\begin{equation}
		\label{eq:def-gamma_N}
		d \gamma_N(\lambda) = \Theta_N^{*} \left(I_L \otimes d \omega_N^{R}(\lambda) \right) \Theta_N
		\end{equation}
		Then it is assumed that the sequence $(\gamma_N)_{N \geq 1}$ converges weakly towards a certain measure $\gamma_*$.       
	\end{assumption}
 \end{itemize}
It is clear that Assumptions \ref{as:convergence-alpha}, \ref{as:support-mu}, \ref{as:signal-bruit} look rather strong (notice however that the assumptions are satisfied when $R_N = \sigma^{2} I$ for some $\sigma^{2} > 0$). This does not limit the usefulness of the results of Section \ref{sec:autocov_signal} because our goal is to establish that, despite the above strong Assumptions, the number of largest eigenvalues of $\hat{R}_{f|p,y} \hat{R}_{f|p,y}^{*}= \Sigma_{f} \Sigma_p^{*} \Sigma_p \Sigma_f^{*}$ that escape from $[0,x_{+,N}]$ is not at all related to $P$. Therefore, the conclusion of the results of Section \ref{sec:autocov_signal} is that, even if strong Assumptions hold, the largest eigenvalues of $\hat{R}_{f|p,y} \hat{R}_{f|p,y}^{*}$ cannot be used to estimate $P$ consistently. \\
 
 We now state some consequences of Assumption \ref{as:convergence-alpha} and Assumption \ref{as:support-mu}, which, in some sense, show that $x_{+,N}, w_{+,N}$, functions $t_N(z), w_N(z)$ and measure $\mu_N$  have,  when $N \rightarrow +\infty$, limits that satisfy the same properties that their finite $N$ equivalents. We recall that $w_{+,N} > 0$ is defined by $w_{+,N} = w_N(x_{+,N})$ and verifies $x_{+,N}=\phi_N(w_{+,N})$, $\phi_N'(w_{+,N}) = 0$ and $w_{+,N} > \lambda_{1,N}$ (we recall that $w_N$ and $\phi_N$ are defined by (\ref{eq:def-wN}) and (\ref{eq:def-phiN})). We omit the proof of the two following Propositions, and refer the reader to the proofs of Proposition 4.1 and Proposition 4.2 in the Thesis \cite{tieplova-thesis}. 

 \begin{proposition}
	\label{prop:consequences-assumption-alpha}
        Sequences $(w_{+,N})_{N \geq 1}$ and $(x_{+,N})_{N \geq 1}$ converge towards finite limits $w_{+,*}$ and $x_{+,*}$ respectively. Moreover, $w_{+,*}$ verifies $w_{+,*} > \lambda_{+,*}$. If $\phi_{*}(w)$ is the function defined on $\mathbb{C} - [\lambda_{-,*}, \lambda_{+,*}]$ by
 		\begin{equation}
 		\label{eq:def-phi*}
 		\phi_*(w) = (c_* w)^{2} \, \left( \int_{\lambda_{-,*}}^{\lambda_{+,*}} \frac{\lambda \, d \omega_*(\lambda)}{
 			w - \lambda} \right)^{2} + c_* w^{2} \,  \int_{\lambda_{-,*}}^{\lambda_{+,*}} \frac{\lambda \, d \omega_*(\lambda)}{
 			w - \lambda}
		\end{equation}
                then, $\phi_N(w) \rightarrow \phi_*(w)$ uniformly on the compact subsets of  $\mathbb{C} - [\lambda_{-,*}, \lambda_{+,*}]$. Moreover, it holds that 
 		\begin{equation}
		\label{eq:relation-1-w+*-x+*}
 		x_{+,*} = \phi_*(w_{+,*})
 		\end{equation}
The sequence $(\mu_N)_{N \geq 1}$ converges weakly towards a probability measure $\mu_*$. The support
 		$\mathcal{S}_*$ of $\mu_*$ is included into $[0,x_{+,*}]$, and the Stieltjes transform $t_*(z)$
 		of $\mu_*$ verifies the equation
 		\begin{equation}
 		\label{eq:canonique-mu*}
 		t_*(z) = \int_{\lambda_{-,*}}^{\lambda_{+,*}} \, \frac{\lambda}{-z(1 + \frac{c_* t_*(z) \, \lambda}{1 - z (c_* t_*(z))^{2}})} \, d\omega_*(\lambda)
		\end{equation}
 		for each $z \in \mathbb{C} - [0, x_{+,*}]$. Moreover, $t_N(z)$ converges uniformly towards $t_*(z)$
                on the compact subsets of $ \mathbb{C} - [0, x_{+,*}]$. If $w_{*}(z)$ is the function defined on $\mathbb{C} - [0, x_{+,*}]$
                by
 		\begin{equation}
 		\label{eq:def-w*}
 		w_{*}(z) = c_* z t_*(z) - \frac{1}{c_* t_*(z)}
 		\end{equation}
 		then, $w_*$ is holomorphic on $\mathbb{C} - [0, x_{+,*}]$ and $w_N(z)$ converges uniformly towards $w_*(z)$
                on the compact subsets of $\mathbb{C} - [0, x_{+,*}]$. $w_*(z)$ satisfies 
 		\begin{equation}
 		\label{eq:phi*(w*)}
 		\phi_*(w_*(z)) = z 
 		\end{equation}
 		for each $z \in \mathbb{C} - [0, x_{+,*}]$. Finally, 
 		\begin{equation}
 		\label{eq:limit-t*(x)}
		\lim_{x \rightarrow x_{+,*}, x > x_{+,*}} t_*(x) \; \mbox{exists, is finite, is still denoted $t_*(x_{+,*})$, and  Eq. (\ref{eq:canonique-mu*}) holds for $z = x_{+,*}$}
 		\end{equation}
 		Moreover, we have    
 		\begin{equation}
 		\label{eq:relation-2-w+*-x+*}
 		w_{+,*} = w_*(x_{+,*})
 		\end{equation}     
 \end{proposition}

We recall that $\nu_{N}^{T}$ is the $M \times M$ matrix-valued positive measure associated to matrix-valued Stieltjes transform
 $T_N(z)$, and introduce for each $N$ the $r \times r$ matrix-valued measure $\beta_N$
 defined by
 \begin{equation}
 \label{eq:def-beta}
 d \beta_N(\lambda) = \Theta_N^{*} \left(I_L \otimes d \nu_N^{T}(\lambda) \right) \Theta_N
 \end{equation}
We notice that $\nu_N^{T}(\mathbb{R}^{+}) = I$ implies that $\beta_N(\mathbb{R}^{+}) = I$. Using the identity (\ref{eq:expre-TN-w}),
we obtain immediately that the Stieltjes transform $T_{\beta_N}(z)$ of $\beta_N$ is given by 
\begin{equation}
\label{eq:expre-TbetaN}
T_{\beta_N}(z) = \frac{w_N(z)}{z} \int \frac{d \gamma_N(\lambda)}{\lambda - w_N(z)}
\end{equation}
 Then, the following result is a consequence of Assumption \ref{as:signal-bruit}. 
 \begin{proposition}
 	\label{prop-consequences-assumptions-signal-bruit}
 	The sequence of measures $(\beta_N)_{N \geq 1}$ converges weakly towards a measure $\beta_*$ whose support is included into $[0,x_{+,*}]$, and which verifies $\beta_*([0,x_{+,*}]) = \beta_*(\mathbb{R}^{+}) = I$. The Stieltjes transform $T_{\beta_{*}}(z)$ of $\beta_*$ is given by
 	\begin{equation}
 	\label{eq:expre-Tbeta*}
 	T_{\beta_*}(z) = \frac{w_*(z)}{z} \, \int_{\lambda_{-,*}}^{\lambda_{+,*}} \frac{d \gamma_*(\lambda)}{\lambda - w_*(z)}
 	\end{equation}
 	for each $z \in \mathbb{C} - [0, x_{+,*}]$. Moreover, it holds that
 	\begin{equation}
	\label{eq:epxre-Tbeta*x+}
 	T_{\beta_*}(x_{+,*}) = \lim_{x \rightarrow x_{+,*}, x > x_{+,*}} T_{\beta_*}(x) = \lim_{N \rightarrow +\infty} T_{\beta_N}(x_{+,N}) = \frac{w_{+,*}}{x_{+,*}} \, \int_{\lambda_{-,*}}^{\lambda_{+,*}} \frac{d \gamma_*(\lambda)}{\lambda - w_{+,*}} 
 	\end{equation}
 \end{proposition}

We finally conclude this paragraph by the following result. 
\begin{proposition}
	\label{prop:support-eigenvalues-extended-matrices}
	Assume that $y > \sqrt{x_{+,*}}$. Then, for each $N$ large enough, $y$ is not eigenvalue of matrix $\left( \begin{array}{cc} 0 & W_{f,N} W_{p,N}^{*} \\ W_{p,N} W_{f,N}^{*} & 0 \end{array} \right)$, and $y^{2}$ is not eigenvalue of $W_f W_p^{*} W_p W_f^{*}$. 
\end{proposition}
{\bf Proof.} As $y > \sqrt{x_{+,*}}$ and that $\lim_{N \rightarrow +\infty} x_{+,N} = x_{+,*}$, it exists $N_0$ such that $y > \sqrt{x_{+,N}}$ and $y^{2} > x_{+,N}$ for each $N \geq N_0$. Therefore, $y^{2}$ does not belong to
$\cup_{N \geq N_0} \mathcal{S}_N$. Theorem 8.1 in \cite{loubaton-tieplova-2020} thus implies that $y^{2}$ and $y$ cannot be one of the eigenvalues of matrices $ W_{f,N} W_{p,N}^{*}W_{p,N} W_{f,N}^{*}$ and 
$\left( \begin{array}{cc} 0 & W_{f,N} W_{p,N}^{*} \\ W_{p,N} W_{f,N}^{*} & 0 \end{array} \right)$ for $N \geq N_0$ 
respectively.  $\blacksquare$

\subsection{Asymptotic behaviour of the eigenvalues of $\Sigma_f \Sigma_p^{*} \Sigma_p \Sigma_f^{*}$.}\label{sec:asym_beh_eigenval}
In this paragraph, we characterize the possible eigenvalues of  $\Sigma_f \Sigma_p^{*} \Sigma_p \Sigma_f^{*}$ that escape
from the interval $[0, x_{+,*}]$. For this, for each $\delta > 0$ small enough, we study the positive eigenvalues of $\left( \begin{array}{cc} 0 & \Sigma_{f,N} \Sigma_{p,N}^{*} \\ \Sigma_{p,N} \Sigma_{f,N}^{*} & 0 \end{array} \right)$ that are almost surely, for $N$ large enough, strictly greater than $\sqrt{x_{+,*} + \delta}$. We first mention that Theorem 8.1 in \cite{loubaton-tieplova-2020} implies that the resolvent $Q_{W}(z)$ 
of $W_{f,N} W_{p,N}^{*} W_{p,N} W_{f,N}^{*}$ and the resolvent 
$\Q_{W}(z)$ of matrix $\left( \begin{array}{cc} 0 & W_{f,N} W_{p,N}^{*} \\ W_{p,N} W_{f,N}^{*} & 0 \end{array} \right)$
are almost surely, for each $N$ large enough, holomorphic in  $\mathbb{C}- [0, x_{+,N}]$ and in $\mathbb{C}- [-\sqrt{x_{+,N}}, \sqrt{x_{+,N}}]$ respectively. Therefore, almost surely, for each $N$ large enough, function $F_N(z)$ defined by (\ref{eq:def-F}) is holomorphic on $\mathbb{C}- [-\sqrt{x_{+,N}}, \sqrt{x_{+,N}}]$. As $\lim_{N \rightarrow +\infty} x_{+,N} = x_{+,*}$, $F_N(z)$ is also holomorphic
on $\mathbb{C}- [-\sqrt{x_{+,*} + \delta}, \sqrt{x_{+,*}+\delta}]$ for each $\delta >0$ for $N$ large enough. \\

We first establish that the sequence of analytic functions $(F_N(z))_{N \geq 1}$ 
almost surely converges uniformly on each compact subset of $\mathbb{C} - [-\sqrt{x_{+,*}}, \sqrt{x_{+,*}}]$ 
towards a deterministic function $F_*(z)$ which is analtyic in $\mathbb{C} - [-\sqrt{x_{+,*}}, \sqrt{x_{+,*}}]$. 
Adapting the stability results of the zeros of certain analytic functions proved in 
\cite{benaych-nadakuditi-ann-math} and \cite{chapon-couillet-hachem-mestre}, we obtain that for $\delta$ small
enough, the solutions of the equation $\mathrm{det}(I + F_N(y)) = 0$, $y > \sqrt{x_{+,*}+\delta}$,  converge towards the 
solutions of the limit equation $\mathrm{det}(I + F_*(y)) = 0$, $y > \sqrt{x_{+,*}}$. \\

In order to study the asymptotic behaviour of $F_N$, we first consider the asymptotic behaviour of matrix $ \mathcal{A}^* {\bf Q}_{W,N}(pf) \mathcal{B}$, which is given by
$$
\mathcal{A}^* {\bf Q}_{W,N}(pf) \mathcal{B} = \left( \begin{array}{c} \Theta_f^{*} \\ \Delta_p \tilde{\Theta}_p^{*} W_f^{*} \end{array}\right) \, {\bf Q}_{W,N}(pf) \, \left( \Theta_p, W_p \tilde{\Theta}_f \Delta_f  \right) \, \left( \begin{array}{cc} \Delta_p  \tilde{\Theta}_p^{*} \tilde{\Theta}_f \Delta_f & I_r \\
I_r & 0 \end{array} \right)
$$
 In order to  study matrix $\mathcal{A}^* {\bf Q}_{W,N}(pf) \mathcal{B}$ when $N \rightarrow +\infty$, it is 
necessary to evaluate the asymptotic behaviour of sesquilinear forms of matrices ${\bf Q}_{W,N}(pf)$, 
$W_f^{*} {\bf Q}_{W,N}(pf)$, ${\bf Q}_{W,N}(pf) W_p$ and $W_f^{*} {\bf Q}_{W,N}(pf) W_p$. The following result holds.
\begin{lemma}
	\label{le:extra-properties}
	For each $z \in \mathbb{C}- [-\sqrt{x_{+,*}}, \sqrt{x_{+,*}}]$ and for each 
	bounded sequences $(a_N, b_N)_{N \geq 1}$ and $(\tilde{a}_N, \tilde{b}_N)$ of $ML$--dimensional and $N$--dimensional deterministic vectors, it holds that 
 	\begin{itemize}
		\item $a_N^{*} \, {\bf Q}_{W,N}(pf) \, b_N \rightarrow 0$ almost surely
 		\item $\tilde{a}_N^{*} \, W_f^{*} \, {\bf Q}_{W,N}(pf) \, b_N \rightarrow 0$ almost surely
		\item $a_N^*  \, {\bf Q}_{W,N}(pf) \, W_p \, \tilde{b}_N \rightarrow 0$ almost surely
		\item $\tilde{a}_N^{*} \, W_f^{*} \, {\bf Q}_{W,N}(pf) \, W_p \, \tilde{b}_N + 
		\frac{(c_N {\bf t}_N(z))^{2}}{1 - (c_N {\bf t}_N(z))^{2}} \, \tilde{a}_N^{*} 
		\tilde{b}_N \rightarrow 0$ almost surely. 
 	\end{itemize}
 	Moreover, the convergence is uniform over each compact subset of $\mathbb{C}- [-\sqrt{x_{+,*}}, \sqrt{x_{+,*}}]$
	and it holds that, almost surely
	\begin{equation}
 	\label{eq:A*QpfB}
 	\mathcal{A}^{*} \, {\bf Q}_{W,N}(pf) \, \mathcal{B} - \left( \begin{array}{cc} 0 & 0 \\ - \frac{(c_N {\bf t}_N(z))^{2}}{1 - (c_N {\bf t}_N(z))^{2}}
	\Gamma_N^{*} & 0 \end {array} \right) \rightarrow 0
 	\end{equation}
	the convergence being uniform on compact subsets of $ \mathbb{C}- [-\sqrt{x_{+,*}}, \sqrt{x_{+,*}}]$. Finally,
	the above properties hold if $a_N,b_N,\tilde{a}_N, \tilde{b}_N$ are random bounded vectors that are 
	independent from the noise sequence $(v_n)_{n \geq 1}$, i.e. from the entries of matrices $(W_N)_{N \geq 1}$.
 \end{lemma}
 {\bf Sketch of proof}. The proof of this result uses ingredients that are very similar to the calculations of Section 5 and Paragraph 6.2 in \cite{loubaton-tieplova-2020}. We therefore only provide a sketch of proof.  When $z \in \mathbb{C}^{+}$, the first item follows from (\ref{eq:boldQ-boldT-trace}) and from the observation that 
$\left(I_{2L} \otimes T_N(z)\right)(pf) = 0$. The convergence for each $z \in  \mathbb{C}- [-\sqrt{x_{+,*}}, \sqrt{x_{+,*}}]$ follows
from the observation that almost surely, for each $\delta > 0$, functions  $(a_N^{*} \, {\bf Q}_{W,N}(pf) \, b_N)$  are analytic on  $ \mathbb{C}- [-\sqrt{x_{+,*}+\delta}, \sqrt{x_{+,*}+\delta}]$ for $N$ large enough. The use of Montel's theorem allows to prove the almost sure convergence for each $z  \in \mathbb{C}- [-\sqrt{x_{+,*}}, \sqrt{x_{+,*}}]$, as well as the uniformity of the convergence on each compact subset of  $\mathbb{C}- [-\sqrt{x_{+,*}}, \sqrt{x_{+,*}}]$. 
 To establish the second and the third item of Lemma \ref{le:extra-properties} when $z \in \mathbb{C}^{+}$, 
we first show that $\mathbb{E}( W_f^{*} \, {\bf Q}_{W,pf} ) = 0$  and $\mathbb{E}({\bf Q}_{W,pf} \, W_p ) = 0$
using the invariance of the distribution of $(v_n)_{n \in \mathbb{Z}}$ under the transformation 
$v_n \rightarrow e^{i n \theta} v_n$ for each $\theta$, and use the Poincar\'e-Nash inequality. We finally prove the uniform convergence on compact subsets of  $\mathbb{C}- [-\sqrt{x_{+,*}}, \sqrt{x_{+,*}}]$ using Montel's theorem. We note that
the sequences of functions defined in item (ii) and (iii) are almost surely bounded on each compact subsets of
$\mathbb{C}- [-\sqrt{x_{+,*}}, \sqrt{x_{+,*}}]$ because matrices $W_f$ and $W_p$ are almost surely bounded, see (\ref{eq:Wi-bounded}). \\

We denote by $\alpha_N(z)$ and ${\bs \alpha}_N(z)$ the functions defined by 
\begin{equation}
\label{eq:def-alphaN}
\alpha_N(z) = \mathbb{E} \left( \frac{1}{ML} \Tr \left[ (I_L \otimes R_N) Q_{W,N}(z) \right] \right) 
\end{equation}
and 
\begin{equation}
\label{eq:def-alphaNbold}
{\bs \alpha}_N(z) = \mathbb{E} \left( \frac{1}{ML} \Tr \left[ (I_{L} \otimes R_N) \Q_{W,N}(pp)(z) \right] \right) 
\end{equation}
We notice that  ${\bs \alpha}_N(z) = z \alpha_N(z^{2})$. 
The proof of the fourth item  of Lemma \ref{le:extra-properties} needs to use the Gaussian calculations of Section 5 in \cite{loubaton-tieplova-2020} to establish that
$$
\tilde{a}_N^{*} \, W_f^{*} \, {\bf Q}_{W,N}(pf) \, W_p \, \tilde{b}_N + 
\frac{(c_N {\bs \alpha}_N(z))^{2}}{1 - (c_N {\bs \alpha}_N(z))^{2}} \, \tilde{a}_N^{*} 
\tilde{b}_N \rightarrow 0 \; a.s.
$$
for each $z \in \mathbb{C}^{+}$. It is proved in Paragraph 5.2 in \cite{loubaton-tieplova-2020} that $\alpha_N(z) - t_N(z) \rightarrow 0$ for each
$z \in \mathbb{C}^{+}$. As ${\bs \alpha}_N(z) = z \alpha_N(z^{2})$ and ${\bf t}_N(z) = z t_N(z^{2})$, this implies that
${\bs \alpha}_N(z) - {\bf t}_N(z) \rightarrow 0$ if $\mathrm{Arg}(z) \in ]0, \pi/2[$. This convergence domain
can be extended to $\mathbb{C}^{+}$ using classical arguments based Montel's theorem. From this, we deduce immediately
that
$$
\frac{(c_N {\bs \alpha}_N(z))^{2}}{1 - (c_N {\bs \alpha}_N(z))^{2}} - \frac{(c_N {\bf t}_N(z))^{2}}{1 - (c_N {\bf t}_N(z))^{2}}
\rightarrow 0
$$
for each $z \in \mathbb{C}^{+}$, and that, for each $z \in \mathbb{C}^{+}$, 
 \begin{equation}
 \label{eq:convergence-fp}
 \tilde{a}_N^{*} \, W_f^{*} \, {\bf Q}_{W,N}(pf) \, W_p \, \tilde{b}_N + 
 \frac{(c_N {\bf t}_N(z))^{2}}{1 - (c_N {\bf t}_N(z))^{2}} \, \tilde{a}_N^{*} 
 \tilde{b}_N \rightarrow 0, \; a.s.
 \end{equation}
Matrices $W_f$ and $W_p$ are almost surely bounded. Therefore, for each $\delta > 0$, $\tilde{a}_N^{*} \, W_f^{*} \, {\bf Q}_{W,N}(pf) \, W_p \, \tilde{b}_N$ and $\frac{(c_N {\bf t}_N(z))^{2}}{1 - (c_N {\bf t}_N(z))^{2}}$ are analytic on $\mathbb{C} - [-\sqrt{x_{+,*}+\delta}, \sqrt{x_{+,*}+\delta}]$ and bounded on each compact subset of $\mathbb{C} - [-\sqrt{x_{+,*}}, \sqrt{x_{+,*}}]$.
Montel's theorem thus implies that (\ref{eq:convergence-fp}) holds for each $z \in \mathbb{C} - [-\sqrt{x_{+,*}}, \sqrt{x_{+,*}}]$.
Moreover, the convergence is uniform on each compact subset of $\mathbb{C} - [-\sqrt{x_{+,*}}, \sqrt{x_{+,*}}]$. \\

We now assume that $a_N,b_N,\tilde{a}_N, \tilde{b}_N$ are random bounded vectors independent from 
the $(v_n)_{n \geq 1}$, and just verify that $a_N^{*} {\bf Q}_{W,N}(pf) b_N \rightarrow 0$ almost surely
still holds. We denote by $(\Omega_{a,b},\mathbb{P}_{a,b})$ and $(\Omega_{v},\mathbb{P}_{v})$ the probability spaces on which $(a_N,b_N)_{N \geq 1}$ and the random variables $(v_n)_{n \geq 1}$ are defined. We consider the event $A$ on which 
$a_N^{*} {\bf Q}_{W,N}(pf) b_N$ does not converge towards zero, and justify that $\mathbb{P}(A) = 0$ where 
$\mathbb{P} = \mathbb{P}_{a,b} \otimes \mathbb{P}_v$.  For each element $\omega_{a,b} \in \Omega_{a,b}$, 
we denote by $A_{\omega_{a,b}}$ the event 
$$
A_{\omega_{a,b}} = \{ \omega_v \in \Omega_{v}, (\omega_{a,b}, \omega_v) \in A \}
$$
Then, the Fubini theorem leads to 
\begin{equation}
\label{eq:fubini}
\mathbb{P}(A) = \int_{\Omega_{a,b}} \mathbb{P}(A_{\omega_{a,b}})  \mathbb{P}_{a,b}(d\omega_{a,b})
\end{equation}
As the sequence of realizations $(a_N(\omega_{a,b}))_{N \geq 1}$ and $(b_N(\omega_{a,b}))_{N \geq 1}$ are bounded vectors, item (i) implies that $a_N^{*}(\omega_{a,b}) {\bf Q}_{W,N}(pf) b_N(\omega_{a,b}) \rightarrow 0$ almost surely, or equivalently, $\mathbb{P}(A_{\omega_{a,b}}) = 0$. (\ref{eq:fubini}) leads to the conclusion that $\mathbb{P}(A) = 0$ as expected. \\

(\ref{eq:A*QpfB}) is an immediate consequence of the statements of items (i) to (iv) and their generalization to the context of random vectors $(a_N,b_N,\tilde{a}_N, \tilde{b}_N)$ (because the columns of 
$\Theta_{i,N}, \tilde{\Theta}_{i,N}$ are bounded random vectors for $i=p,f$ and the entries of $\Delta_{i,N}$  
are bounded random variables), as well as 
of Condition (\ref{eq:entries-Delta*-different}) which implies that
$r \times r$ diagonal matrices $\Delta_{p,N}$ and $\Delta_{f,N}$ (resp. orthogonal $ML \times r$ matrices
$\Theta_{f,N}$ and $\Theta_{p,N}$) have the same asymptotic behaviour than matrix $\Delta_N$ (resp. matrix $\Theta_N$). $ \blacksquare$ \\

Using the same kind of arguments as in the proof of Lemma \ref{le:extra-properties}, it is possible to establish the following result. 
\begin{proposition}
	\label{prop:converge-FN}
	For each $z \in \mathbb{C} - [-\sqrt{x_{+,*}}, \sqrt{x_{+,*}}]$, it holds that
	\begin{equation}
	\label{eq:A*QppA}
	\mathcal{A}^{*} \, {\bf Q}_{W,N}(pp) \, \mathcal{A} - \left(\begin{array}{cc}  -\Theta_N^{*} \left( z I + \frac{c_N {\bf t}_N(z)}{1 - (c_N {\bf t}_N(z))^{2}} I_L \otimes R_N \right)^{-1} \Theta_N & 0 \\ 0 &  \frac{c_N {\bf t}_N(z)}{1 - (c_N {\bf t}_N(z))^{2}} \Delta_N^{2} \end{array} \right) \rightarrow 0 \; a.s.
	\end{equation}
	
	\begin{equation}
	\label{eq:B*QffB}
	\mathcal{B}^{*} \, {\bf Q}_{W,N}(ff) \, \mathcal{B} -  \left( \begin{array}{cc} \Gamma_N^{*} & I \\ I & 0 \end{array} \right) \, \left( \begin{array}{cc}  -\Theta_N^{*} \left( z I + \frac{c_N {\bf t}_N(z)}{1 - (c_N {\bf t}_N(z))^{2}} I_L \otimes R_N \right)^{-1} \Theta_N & 0 \\ I & 0 \end{array} \right) \, \left( \begin{array}{cc} \Gamma_N & I \\ I & 0 \end{array} \right) \rightarrow 0 \; a.s.
	\end{equation}
	
	\begin{equation}
	\label{eq:B*QffA} 
	\mathcal{B}^{*} \, {\bf Q}_{W,N}(ff) \, \mathcal{A} -  \left( \begin{array}{cc} 0 &  - \frac{(c_N {\bf t}_N(z))^{2}}{1 - (c_N {\bf t}_N(z))^{2}} \Gamma_N \\ 0 & 0
	\end{array} \right) \rightarrow 0 \; a.s. 
	\end{equation}
	The convergence is moreover uniform on each compact subset of $ \mathbb{C} - [-\sqrt{x_{+,*}}, \sqrt{x_{+,*}}]$. 
\end{proposition}
Lemma \ref{le:extra-properties} and Proposition \ref{prop:converge-FN} imply that for each $z \in \mathbb{C} - [-\sqrt{x_{+,*}}, \sqrt{x_{+,*}}]$, almost surely, matrix
$F_N(z)$ has the same asymptotic behaviour than the $4r \times 4r$ matrix $F_{d,N}(z)$ defined by
\begin{equation}
\label{eq:def-Fd}
F_{d,N}(z) = \left( \begin{array}{cc} F_{d,N}^{11}(z) & F_{d,N}^{1,2}(z) \\
F_{d,N}^{2,1}(z) &  F_{d,N}^{2,2}(z) \end{array} \right)
\end{equation}
where the $2r \times 2r$ blocks of $F_{d,N}(z)$ are characterized in Lemma \ref{le:extra-properties} 
and in Proposition \ref{prop:converge-FN}. The assumptions formulated
in Paragraph \ref{subsec:new-assumptions} imply that matrix $F_{d,N}(z)$ converges for each $z \in  \mathbb{C} - [-\sqrt{x_{+,*}}, \sqrt{x_{+,*}}]$
towards a limit $F_*(z)$, the convergence being uniform on each compact subset of $\mathbb{C} - [-\sqrt{x_{+,*}}, \sqrt{x_{+,*}}]$. More
precisely, $t_N(z)$ converges towards $t_{*}(z)$ uniformly on each compact subset of $\mathbb{C} - [0, x_{+,*}]$, which implies that
${\bf t}_N(z) = z t_N(z^{2})$ converges  uniformly on each compact subset of $\mathbb{C} - [-\sqrt{x_{+,*}}, \sqrt{x_{+,*}}]$ towards
${\bf t}_*(z) = z t_*(z^{2})$. We notice that matrix $- \left( z I + \frac{c_N {\bf t}_N(z)}{1 - (c_N {\bf t}_N(z))^{2}} I_L \otimes R_N \right)^{-1}$
coincides with matrix $I_L \otimes {\bf T}_N(z) = I_L \otimes z T_N(z^{2}) = z \int_{0}^{x_{+,N}} \frac{I_L \otimes d \nu_N^{T}(\lambda)}{\lambda - z^{2}}$. We denote by ${\bf T}_{\beta_N}(z)$ the 
function defined by ${\bf T}_{\beta_N}(z) = z T_{\beta_N}(z^{2})$, which can also be written as 
$$
\Theta_N^{*} (I_L \otimes {\bf T}_N(z)) \Theta_N = {\bf T}_{\beta_N}(z)
$$
and which, by (\ref{eq:property-zs(z2)}), coincides with the Stieltjes transform of a positive matrix-valued measure carried by $[-\sqrt{x_{+,N}}, \sqrt{x_{+,N}}]$. 
Proposition \ref{prop-consequences-assumptions-signal-bruit} implies that ${\bf T}_{\beta_N}(z)$ converges uniformly on each compact subset of $\mathbb{C} \setminus [-\sqrt{x_{+,*}}, \sqrt{x_{+,*}}]$ towards the $r \times r$ matrix ${\bf T}_{\beta_*}(z)$ defined by
\begin{equation}
\label{eq:def -Tbeta*}
{\bf T}_{\beta_*}(z) = z T_{\beta_*}(z^{2})
\end{equation}
where we recall that $T_{\beta_*}(z) = \int_{0}^{x_{+,*}} \frac{d \beta_*(\lambda)}{\lambda - z}$ is the Stieltjes transform of the positive matrix-valued measure $\beta_*$. ${\bf T}_{\beta_*}$ is an element of $\mathcal{S}_r(\mathbb{R})$, its associated positive measure, denoted $\bs{\beta}_*$, is carried by $[-\sqrt{x_{+,*}}, \sqrt{x_{+,*}}]$, and verifies $\bs{\beta}_*([-\sqrt{x_{+,*}}, \sqrt{x_{+,*}}]) = \bs{\beta}_*(\mathbb{R}) = I$ because 
$\beta_*([0,x_{+,*}]= \beta_*(\mathbb{R}^{+}) = I$ (see (\ref{eq:property-zs(z2)}) and 
(\ref{eq:mass-zs(z2)})). All this imply that 
$$
F_{d,N}^{(1,1)}(z) = \left( \begin{array}{cc} 0 & 0 \\ - \frac{(c_N {\bf t}_N(z))^{2}}{1 - (c_N {\bf t}_N(z))^{2}} 
\Gamma^{*}_N & 0 \end{array} \right) \rightarrow F_{*}^{1,1}(z) = \left( \begin{array}{cc} 0 & 0 \\ - \frac{(c_* {\bf t}_*(z))^{2}}{1 - (c_* {\bf t}_*(z))^{2}} 
\Gamma^{*}_* & 0 \end{array} \right)
$$
$$
F_{d,N}^{(1,2)}(z) = \left( \begin{array}{cc} {\bf T}_{\beta_N}(z) & 0 \\ 0 &  \frac{c_N {\bf t}_N(z)}{1 - (c_N {\bf t}_N(z))^{2}} \Delta_N^{2} \end{array} \right) \rightarrow F_{*}^{1,2}(z) = 
\left( \begin{array}{cc} {\bf T}_{\beta_*}(z) & 0 \\ 0 &  \frac{c_* {\bf t}_*(z)}{1 - (c_* {\bf t}_*(z))^{2}} \Delta_*^{2} \end{array} \right)
$$
$$
F_{d,N}^{2,1}(z) = \left( \begin{array}{cc} \Gamma_N & I \\ I & 0 \end{array} \right) F_{d,N}^{1,2}(z) 
\left( \begin{array}{cc} \Gamma_N^{*} & I \\ I & 0 \end{array} \right) \rightarrow 
F_{*}^{2,1}(z) =  \left( \begin{array}{cc} \Gamma_* & I \\ I & 0 \end{array} \right) F_{*}^{1,2}(z) 
\left( \begin{array}{cc} \Gamma_*^{*} & I \\ I & 0 \end{array} \right)
$$
$$
F_{d,N}^{2,2}(z)  = \left( \begin{array}{cc} 0 &   - \frac{(c_N {\bf t}_N(z))^{2}}{1 - (c_N {\bf t}_N(z))^{2}} 
\Gamma_N \\ 0 & 0 \end{array} \right) \rightarrow F_{*}^{2,2}(z) = \left( \begin{array}{cc} 0 &   - \frac{(c_* {\bf t}_*(z))^{2}}{1 - (c_* {\bf t}_*(z))^{2}} 
\Gamma_* \\ 0 & 0 \end{array} \right)
$$
where we recall that $\Gamma_*$ is defined by Assumption \ref{as:signal}. 
The previous results show that $(F_N(z))_{N \geq 1}$ converge uniformly towards $F_*(z)$ over each compact subset of
$\mathbb{C} - [-\sqrt{x_{+,*}}, \sqrt{x_{+,*}}]$. It is thus reasonable to expect that for $\delta > 0$ small
enough, the solutions of the equation
$\mathrm{det}(I + F_N(y))= 0$ satisfying $y > \sqrt{x_{+,*}+\delta}$ will converge towards the roots of
$\mathrm{det}(I + F_*(y))= 0$ satisfying $y > \sqrt{x_{+,*}}$. \\

We now study the solutions of $\mathrm{det}(I + F_*(y))= 0$, $y > \sqrt{x_{+,*}}$ . For $y > \sqrt{x_{+,*}}$, we express in a more convenient manner the equation $\mathrm{det}(I + F_*(y)) = 0$. 
This equation holds if and only
\begin{equation}
\label{eq:equation-vp-modified-1}
\mathrm{det}\left( \left( \begin{array}{cc} I & 0 \\ 0 & \Omega_* \end{array} \right) \left( I + F_*(y) \right) 
\left( \begin{array}{cc}  \Omega_*^{*} & 0 \\ 0 & I \end{array} \right) \right) = 0
\end{equation}
where
$$
\Omega_* = \left( \begin{array}{cc} \Gamma_* & I \\ I & 0 \end{array} \right)^{-1} =  \left( \begin{array}{cc} 0 & I \\ I & -\Gamma_* \end{array} \right)
$$
The matrix whose determinant vanishes in (\ref{eq:equation-vp-modified-1}) is equal to 
\begin{equation}
\label{eq:matrix-equation-vp-modified-1}
\left( \begin{array}{cccc} 0 & I & {\bf T}_{\beta_*}(z) & 0 \\
I & - \frac{\Gamma_*^{*}}{1 - (c_* {\bf t}_*(z))^{2}} & 0 & \frac{c_N {\bf t}_*(z)}{1 - (c_* {\bf t}_*(z))^{2}} \Delta_*^{2} \\
{\bf T}_{\beta_*}(z) & 0  & 0 & I \\
0 &  \frac{c_* {\bf t}_*(z)}{1 - (c_* {\bf t}_*(z))^{2}} \Delta_*^{2} & I & - \frac{\Gamma_*}{1 - (c_* {\bf t}_*(z))^{2}} \end{array} \right)
\end{equation}
As the lower diagonal $2r \times 2r$  block of this matrix is invertible, its determinant is $0$ 
if and only the determinant of its Schur complement is $0$. After some calculations, we obtain  
that $\mathrm{det}(I + F_*(y)) = 0$ if and only if $\mathrm{det}(I - K_*(y)) = 0$ where 
$K_*(z)$ is the $2r \times 2r$ matrix-valued function defined for each 
$z \in \mathbb{C} - [-\sqrt{x_{+,*}}, \sqrt{x_{+,*}}]$ by 
\begin{equation}
\label{eq:def-G*}
K_*(z) = \left( \begin{array}{cc} \frac{c_* {\bf t}_*(z)}{1 - (c_* {\bf t}_*(z))^{2}} \Delta_*^{2} 
{\bf T}_{\beta_*}(z) & \frac{\Gamma_*^{*}}{1 - (c_* {\bf t}_*(z))^{2}}  \\
\frac{{\bf T}_{\beta_*}(z) \, \Gamma_* {\bf T}_{\beta_*}(z)}{1 - (c_* {\bf t}_*(z))^{2}} & \frac{c_* {\bf t}_*(z)}{1 - (c_* {\bf t}_*(z))^{2}} 
{\bf T}_{\beta_*}(z) \, \Delta_*^{2} \end{array} \right)
\end{equation}
$K_*(z)$ can be factorized as 
$$
K_*(z) = \left( \begin{array}{cc} I & 0 \\ 0 & {\bf T}_{\beta_*}(z) \end{array} \right) 
\left( \begin{array}{cc} \frac{c_* {\bf t}_*(z)}{1 - (c_* {\bf t}_*(z))^{2}} \Delta_*^{2} &
\frac{\Gamma_*^{*}}{1 - (c_* {\bf t}_*(z))^{2}}  \\  \frac{\Gamma_*}{1 - (c_* {\bf t}_*(z))^{2}} &
\frac{c_* {\bf t}_*(z)}{1 - (c_* {\bf t}_*(z))^{2}} \Delta_*^{2} \end{array} \right) 
\left( \begin{array}{cc} {\bf T}_{\beta_*}(z)  & 0 \\ 0 & I \end{array} \right) 
$$
For each $y > \sqrt{x_{+,*}}$,  ${\bf T}_{\beta_*}(y)$ can be written as 
$${\bf T}_{\beta_*}(y) = \int_{-\sqrt{x_{+,*}}}^{\sqrt{x_{+,*}}} \frac{d \bs{\beta}_*(\lambda)}{\lambda- y}$$
and verifies ${\bf T}_{\beta_*}(y) \leq -\frac{1}{\sqrt{x_{+,*}} + y} \, \bs{\beta}_*([-\sqrt{x_{+,*}}, \sqrt{x_{+,*}}]) =  - \frac{I}{\sqrt{x_{+,*}} + y}$
because we recall that $\bs{\beta}_*([-\sqrt{x_{+,*}}, \sqrt{x_{+,*}}]) = \bs{\beta}_*(\mathbb{R}) = I$. 
Therefore, ${\bf T}_{\beta_*}(y)$ is negative definite, and  thus invertible. 
Hence, $\mathrm{det}(I - K_*(y)) = 0$ if and only 
\begin{equation}
\label{eq:equation-vp-modified-2}
\mathrm{det} \left(   \left( \begin{array}{cc} \frac{c_* {\bf t}_*(y)}{1 - (c_* {\bf t}_*(y))^{2}} \Delta_*^{2} &
\frac{\Gamma_*^{*}}{1 - (c_* {\bf t}_*(y))^{2}}  \\  \frac{\Gamma_*}{1 - (c_* {\bf t}_*(y))^{2}} &
\frac{c_* {\bf t}_*(y)}{1 - (c_* {\bf t}_*(y))^{2}} \Delta_*^{2} \end{array} \right) -  \left( \begin{array}{cc} ({\bf T}_{\beta_*}(y))^{-1} & 0 \\ 0 & ({\bf T}_{\beta_*}(y))^{-1} \end{array} \right) \right) = 0
\end{equation}
In the following, we denote by $H_*(z)$ the $2r \times 2r$ matrix-valued function defined
on $\mathbb{C} - [\sqrt{x_{+,*}}, \sqrt{x_{+,*}}]$ by 
\begin{equation}
\label{eq:def-H_lim}
H_*(z) = \left( \begin{array}{cc}   \frac{c_* {\bf t}_*(z))}{1 - (c_* {\bf t}_*(z))^{2}} \Delta_*^{2} - ({\bf T}_{\beta_*}(z))^{-1} &
\frac{\Gamma_*^{*}}{1 - (c_* {\bf t}_*(z))^{2}}  \\  \frac{\Gamma_*}{1 - (c_* {\bf t}_*(z))^{2}} &
\frac{c_* {\bf t}_*(z)}{1 - (c_* {\bf t}_*(z))^{2}} \Delta_*^{2} -  ({\bf T}_{\beta_*}(z))^{-1}  \end{array} \right)
\end{equation}
$H_*(z)$ is of course holomorphic on  $\mathbb{C} - [-\sqrt{x_{+,*}}, \sqrt{x_{+,*}}]$, and the solutions of
$\mathrm{det}(I + F_*(y)) = 0$, $y > \sqrt{x_{+,*}}$, coincide with the solutions of
\begin{equation}
\label{eq:limit-equation-eigenvalues}
\mathrm{det}\left( H_{*}(y) \right) = 0
\end{equation}
where $y > \sqrt{x_{+,*}}$. In order to characterize the roots of (\ref{eq:limit-equation-eigenvalues}), we first establish the following Proposition.
\begin{proposition}
	\label{prop:properties-H*}
	For each $z \in \mathbb{C}^{+}$, $\mathrm{Im}(H_*(z)) > 0$, and function $y \rightarrow H_{*}(y)$ is increasing
	in the sense of the partial order defined on the set of all Hermitian matrices on the interval $[\sqrt{x_{+,*}}, +\infty[$. 
\end{proposition}
{\bf Proof.} As $\bs{\beta}_*(\mathbb{R}) = I$, $\Im({\bf T}_{\beta_*}(z))$ is positive definite for $z \in \mathbb{C}^{+}$ and $\mathrm{Im}\left( ({\bf T}_{\beta_*}(z))^{-1} \right) < 0$. Therefore, in order to
establish that $\mathrm{Im}(H_{*}(z)) > 0$ on $\mathbb{C}^{+}$, it is sufficient to prove that $\mathrm{Im}(H_{*,1}(z)) > 0$ on $\mathbb{C}^{+}$ where
$H_{*,1}(z)$ is the function defined by
$$
H_{*,1}(z) = \left( \begin{array}{cc}   \frac{c_* {\bf t}_*(z)}{1 - (c_* {\bf t}_*(z))^{2}} \Delta_*^{2}  &
\frac{\Gamma_*^{*}}{1 - (c_* {\bf t}_*(z))^{2}}  \\  \frac{\Gamma_*}{1 - (c_* {\bf t}_*(z))^{2}} &
\frac{c_* {\bf t}_*(z)}{1 - (c_* {\bf t}_*(z))^{2}} \Delta_*^{2}   \end{array} \right)
$$
After some calculations, we obtain that
$$
\mathrm{Im}(H_{*,1}(z)) = \frac{1}{|1 - (c_* {\bf t}_*(z))^{2}|^{2}} \, \left( \begin{array}{cc} \mathrm{Im}(c_* t_*(z)) (1+ |c_* t_*(z)|^{2}) \Delta_*^{2} & \mathrm{Im}\left( (c_* t_*(z))^{2} \right) \Gamma_*^{*} \\
\mathrm{Im}\left( (c_* t_*(z))^{2} \right) \Gamma_* & \mathrm{Im}(c_* t_*(z)) (1+ |c_* t_*(z)|^{2}) \Delta_*^{2} \end{array} \right)
$$
It is clear that $\mathrm{Im}(c_* t_*(z)) (1+ |c_* t_*(z)|^{2}) \Delta_*^{2} > 0$. Therefore, $\mathrm{Im}(H_{*,1}(z)) > 0$ if and only if
$$
\mathrm{Im}(c_* t_*(z)) (1+ |c_* t_*(z)|^{2}) \Delta_*^{2} -  \frac{\left[\mathrm{Im}\left( (c_* t_*(z))^{2} \right)\right]^{2}}{ \mathrm{Im}(c_* t_*(z)) (1+ |c_* t_*(z)|^{2})} \, \Gamma_*^{*} \Delta_*^{-2} \Gamma_* > 0
$$
or equivalently, if and only if
\begin{equation}
\label{eq:ImH*-positive}
I - \frac{\left[\mathrm{Im}\left( (c_* t_*(z))^{2} \right)\right]^{2}}{ \left[\mathrm{Im}(c_* t_*(z)) (1+ |c_* t_*(z)|^{2}) \right]^{2}} \, \Delta_*^{-1} \Gamma_*^{*} \Delta_*^{-2} \Gamma_* \Delta_*^{-1} > 0 
\end{equation}
We first claim that $\Delta_*^{-1} \Gamma_*^{*} \Delta_*^{-2} \Gamma_* \Delta_*^{-1} \leq I$. To verify this, we
notice that for each $N$, matrix $\Delta_N^{-1} \Gamma_N^{*} \Delta_N^{-2} \Gamma_N \Delta_N^{-1}$ coincides with
$\tilde{\Theta}_{f,N}^{*} \tilde{\Theta}_{p,N} \tilde{\Theta}_{p,N}^{*} \tilde{\Theta}_{f,N}$ which is less than $I$. Therefore,
$$
\lim_{N \rightarrow +\infty} \Delta_N^{-1} \Gamma_N^{*} \Delta_N^{-2} \Gamma_N \Delta_N^{-1} = \Delta_*^{-1} \Gamma_*^{*} \Delta_*^{-2} \Gamma_* \Delta_*^{-1} \leq I
$$
$\frac{\left[\mathrm{Im}\left( (c_* t_*(z))^{2} \right)\right]^{2}}{ \left[\mathrm{Im}(c_* t_*(z)) (1+ |c_* t_*(z)|^{2}) \right]^{2}}$ is equal to
$$
\frac{\left[\mathrm{Im}\left( (c_* t_*(z))^{2} \right)\right]^{2}}{ \left[\mathrm{Im}(c_* t_*(z)) (1+ |c_* t_*(z)|^{2}) \right]^{2}} =
\frac{ 4 \left[ \mathrm{Re}(c_* t_*(z))\right]^{2}}{(1+ |c_* t_*(z)|^{2})^{2}}
$$
For $z \in \mathbb{C}^{+}$, $\mathrm{Im}(t_*(z)) > 0$. Therefore, it holds that $ \left( \mathrm{Re}(c_* t_*(z))\right)^{2} < |c_* t_*(z)|^{2}$
and that
$$
\frac{\left[\mathrm{Im}\left( (c_* t_*(z))^{2} \right)\right]^{2}}{ \left[\mathrm{Im}(c_* t_*(z)) (1+ |c_* t_*(z)|^{2}) \right]^{2}} <
\frac{ 4 |c_* t_*(z)|^{2}}{  (1+ |c_* t_*(z)|^{2})^{2} } \leq 1
$$
This establishes  (\ref{eq:ImH*-positive}) and $\mathrm{Im}(H_*(z)) > 0$. $\blacksquare$ \\

We now prove that  $y \rightarrow H_{*}(y)$ is increasing on the interval $[\sqrt{x_{+,*}}, +\infty[$. For this, we use the following  representation
of holomorphic matrix-valued functions whose imaginary part is positive on $\mathbb{C}^{+}$ (see e.g. \cite{gesztesy-tsekanovskii}):
\begin{equation}
\label{eq:herglotz}
H_{*}(z) = A + B z + \int \frac{1 + \lambda z}{\lambda -z} \frac{d \sigma(\lambda)}{1 + \lambda^{2}}
\end{equation}
where $A$ is Hermitian, $B \geq 0$ and $\sigma$ is a positive matrix-valued measure for which
$$
\mathrm{Tr} \left( \frac{d \sigma(\lambda)}{1 + \lambda^{2}} \right) < +\infty
$$
$B = \lim_{y \rightarrow +\infty} \frac{H_{*}(iy)}{iy}$ is easily seen to be equal to
$$
B = \lim_{y \rightarrow +\infty} \left( \begin{array}{cc} -\frac{{\bf T}_{\beta_*}(iy)}{iy} & 0 \\ 0 & -\frac{{\bf T}_{\beta_*}(iy)}{iy} \end{array} \right) = I_{2r}
$$
while for any interval $[y_1, y_2]$, it holds that
$$
\sigma([y_1, y_2]) = \frac{1}{\pi} \lim_{\epsilon \rightarrow 0} \int_{y_1}^{y_2} \mathrm{Im}(H_*(y + i \epsilon)) dy
$$
As $\mathrm{Im}(H_{*}(y)) = 0$ if $|y| > \sqrt{x_{+,*}}$, the support of $\sigma$ is included into $[-\sqrt{x_{+,*}}, \sqrt{x_{+,*}}]$. Therefore,
we get immediately from (\ref{eq:herglotz}) that $y \rightarrow H_{*}(y)$ is strictly increasing on $]\sqrt{x_{+,*}}, +\infty]$, i.e.
$H_{*}(y_2) > H_{*}(y_1)$ if $y_2 > y_1$. We also notice that
the last item of Proposition \ref{prop:consequences-assumption-alpha} as well as Proposition \ref{prop-consequences-assumptions-signal-bruit} imply that $\lim_{y \rightarrow \sqrt{x_{+,*}}} H_*(y) = H_*(\sqrt{x_{+,*}})$
exists and is finite. Moreover, it holds that $H_*(\sqrt{x_{+,*}}) < H_*(y)$ for $y > \sqrt{x_{+,*}}$.
\begin{corollary}
	\label{coro:increasing-eigenvalues}
	The eigenvalues (arranged in the decreasing order) $(\lambda_{k,*}(y))_{k=1, \ldots, 2r}$ of matrix $H_*(y)$ are strictly increasing
	functions of $y$ on $[\sqrt{x_{+,*}}, +\infty[$, i.e., for each $k=1, \ldots, 2r$, it holds that
	\begin{equation}
	\label{eq:increasing-eigenvalues}
	\lambda_{k,*}(y_1) < \lambda_{k,*}(y_2) \; \mbox{if $\sqrt{x_{+,*}} \leq y_1 < y_2$}
	\end{equation}
	Moreover,  the number $s$ of solutions of (\ref{eq:limit-equation-eigenvalues}) (taking into account their
	multiplicities) for which $y > \sqrt{x_{+,*}}$ belongs to
	$\{ 0,1, \ldots, 2r \}$, and coincides with the number of strictly negative eigenvalues of matrix $H_{*}(\sqrt{x_{+,*}})$. 
\end{corollary}
{\bf Proof.} We have shown that if $\sqrt{x_{+,*}} \leq y_1 < y_2$, then $H_{*}(y_1) < H_{*}(y_2)$. The Weyl's
inequalities (see e.g. \cite{horn-johnson}, Paragraph 4.3) thus imply that (\ref{eq:increasing-eigenvalues}) holds. Moreover, as matrix $B$ in (\ref{eq:herglotz}) is equal to $I_{2r}$, it is clear that for each $k=1, \ldots, 2r$,
$\lambda_{k,*}(y)$ converges towards $+\infty$ when $y \rightarrow +\infty$. 
For $k=1, \ldots, 2r$, the equation $\lambda_{k,*}(y) = 0$ has thus 1 solution $y > \sqrt{x_{+,*}}$
if $\lambda_{k,*}(x_{+,*}) < 0$ and no solution if $\lambda_{k,*}(x_{+,*}) \geq 0$. (\ref{eq:limit-equation-eigenvalues})
holds if and only if one of the eigenvalues of $H_{*}(y)$ is equal to $0$. Therefore, if we denote by $\tilde{s}$ the number of
positive eigenvalues of $H_{*}(\sqrt{x_{+,*}})$, for $j=1, \ldots, \tilde{s}$,  it must hold that $\lambda_{j,*}(y) > 0$
for $y > \sqrt{x_{+,*}}$. Moroever, $\lambda_{\tilde{s}+1,*}(\sqrt{x_{+,*}}) < 0$ implies that the equation
$\lambda_{\tilde{s}+1,*}(y) = 0$ has a unique solution $y_{1,*} > \sqrt{x_{+,*}}$. Similarly, the equation
$\lambda_{\tilde{s}+2,*}(y) = 0$ has a unique solution denoted $y_{2,*}$. Moreover, as $\lambda_{\tilde{s}+2,*}(y) \leq \lambda_{\tilde{s}+1,*}(y)$
for each $y$, we deduce that $\lambda_{\tilde{s}+2,*}(y_{1,*}) \leq \lambda_{\tilde{s}+1,*}(y_{1,*}) = 0$. If $\lambda_{\tilde{s}+2,*}(y_{1,*}) < 0$,
$y_{2,*}$ must be strictly greater than $y_{1,*}$. As a root of (\ref{eq:limit-equation-eigenvalues}),
$y_{1,*}$ has thus multiplicity 1. If $\lambda_{\tilde{s}+2,*}(y_{1,*}) = 0$,
the multiplicity of $y_{1,*}$ as a root of  (\ref{eq:limit-equation-eigenvalues}) is at least equal to 2. Iterating the process, we obtain that the number of solutions $s$ (taking into account the multiplicities) of (\ref{eq:limit-equation-eigenvalues}) is equal to $s = 2r - \tilde{s}$. Moreover,
solutions $y_{1,*}, \ldots, y_{s,*}$ satisfy $y_{1,*} \leq y_{2,*} \leq \ldots \leq y_{s,*}$. $\blacksquare$ \\

Corollary \ref{coro:increasing-eigenvalues} implies that Eq. $\mathrm{det}(I + F_*(y)) = 0$ has $s$ ($0 \leq s \leq 2r$) solutions $(y_{k,*})_{k=1, \ldots, s}$ strictly greater than $\sqrt{x_{+,*}}$. We recall that, 
almost surely, the sequence of functions $(F_N(z))_{N \geq 1}$ converges uniformly on each compact subset of 
$\mathbb{C} - [-\sqrt{x_{+,*}}, \sqrt{x_{+,*}}]$ towards $F_*(z)$. We now take benefit of the arguments used in
\cite{benaych-nadakuditi-ann-math}, Lemma 6.1 and in the proof of Theorem 2.1 in \cite{chapon-couillet-hachem-mestre}
to derive the following result. 

\begin{corollary}
	\label{coro:final-result}
	For each $\delta > 0$ small enough, almost surely, for $N$ large enough, Eq. $\mathrm{det}(I + F_N(y))=0$ has $s$ solutions $y_{1,N} \leq  y_{2,N} \ldots 
	\leq y_{s,N}$ such that $y_{k,N} > \sqrt{x_{+,*}+\delta}$, and for each $k=1, \ldots, s$, it holds that $\lim_{N \rightarrow +\infty} y_{k,N} = y_{k,*}$. 
\end{corollary}
    {\bf Proof.} We just provide a sketch of proof because we follow the arguments in \cite{benaych-nadakuditi-ann-math}
    and \cite{chapon-couillet-hachem-mestre}. In order to simplify the exposition, we assume that $y_{1,*} < \ldots < y_{s,*}$, but
    the following arguments can be extended immediately to the case where some $(y_{k,*})_{k=1, \ldots, s}$ coincide. We first justify that almost surely, for $N$ large enough, the solutions of
     $\mathrm{det}(I + F_N(y))=0$, $y >  \sqrt{x_{+,*}+\delta}$ are bounded by a nice constant. To verify this, we 
     remark that (\ref{eq:Wi-bounded}) implies that it exists a nice constant $\kappa$ for which, almost surely, 
     $\| \mathcal{A}_N \| \leq \kappa$ and $\| \mathcal{B}_N \| \leq \kappa$ for each $N$ large enough. We recall that
     $\mathcal{A}_N$ and $\mathcal{B}_N$ are defined by (\ref{eq:defmathcalA}) and (\ref{eq:defmathcalB}). Moreover, for $y  >  \sqrt{x_{+,*}+\delta}$,
     the inequality $\| {\bf Q}_W(y) \| \leq \frac{1}{y -  \sqrt{x_{+,*}+\delta}}$ holds. Therefore, matrix $F_N(y)$ verifies
     $\| F_N(y) \| < \frac{\kappa}{y -  \sqrt{x_{+,*}+\delta}}$ for some nice constant $\kappa$, and all the eigenvalues of $F_N(y)$
     satisfy $|\lambda_{j}(F_N(y))| \leq  \frac{\kappa}{y -  \sqrt{x_{+,*}+\delta}}$, for $j=1, \ldots, 2r$. For $y$ larger than a nice constant $y_{max}$,
     $\mathrm{det}(I_{4r} + F_N(y))$ cannot therefore vanish. This implies that almost surely,
     for $N$ large enough, the solutions of $\mathrm{det}(I + F_N(y))=0$, $y >  \sqrt{x_{+,*}+\delta}$ belong to $(\sqrt{x_{+,*}+\delta}, y_{max})$.
     We choose $\delta$ in such a way that $\sqrt{x_{+,*}+\delta} < y_{1,*}$. We consider any open interval $(a_1, a_2)$ such that $(a_1,a_2) \subset( \sqrt{x_{+,*}+\delta}, \max(y_{max}, y_{s,*})+ \delta)$ and $a_i \neq y_{k,*}$ for $i=1,2$ and $k=1, \ldots, s$. Then, using the arguments in the proof of Theorem 2.1 in \cite{chapon-couillet-hachem-mestre},
       we obtain that the equations $\mathrm{det}(I + F_N(y))=0$ and $\mathrm{det}(I + F_*(y)) = 0$ have the same number of solutions located in
       $(a_1,a_2)$. Choosing $(a_1, a_2) = (\sqrt{x_{+,*}+\delta}, \max(y_{max}, y_{s,*})+ \delta)$ leads to the conclusion that
         $\mathrm{det}(I + F_N(y))=0$ has $s$ solutions $y_{1,N}, \ldots, y_{s,N}$ larger than $\sqrt{x_{+,*}+\delta}$. We fix $k \in \{ 1, 2, \ldots, s \}$
         and establish that $y_{k,N} \rightarrow y_{k,*}$. For this, we choose $\epsilon > 0$ arbitrarily small, and choose $(a_1,a_2) = (y_{k,*} - \epsilon, y_{k,*} + \epsilon)$. Then,  \cite{chapon-couillet-hachem-mestre} implies that almost surely, for $N$ large enough, $\mathrm{det}(I + F_N(y))=0$ has 1 solution $y_{k,N}$ in $(y_{k,*} - \epsilon, y_{k,*} + \epsilon)$, and that $|y_{k,N} - y_{k,*}| < \epsilon$. This is equivalent to
         $\lim_{N \rightarrow +\infty} y_{k,N} = y_{k,*}$ as expected. $\blacksquare$. \\
   
\begin{remark}
\label{re:existence-limits}
We notice that the existence of the limits $\Delta_*, \Gamma_*, t_*, \beta_*$ introduced in the various Assumptions of Section \ref{sec:autocov_signal} allows to establish that $F_N(z)$ converges towards the deterministic and independent of $N$ function $F_*(z)$, and to prove
that the solutions of  $\mathrm{det}(I + F_N(y))=0$ larger than $\sqrt{x_{+,*}+\delta}$ converge towards the corresponding solutions of  $\mathrm{det}(I + F_*(y))=0$. If the above limits are not supposed to exist, we can just establish that $F_N(z)$ has the same asymptotic behaviour that the term $F_{d,N}(z)$ introduced in (\ref{eq:def-Fd}). As $F_{d,N}(z)$ depends on $N$, it is not possible to adapt the arguments in the proof of Theorem 2.1 in \cite{chapon-couillet-hachem-mestre} to establish rigorously that  the solutions of  $\mathrm{det}(I + F_N(y))=0$ larger than $\sqrt{x_{+,N}}$ have the same behaviour than the corresponding solutions of  $\mathrm{det}(I + F_{d,N}(y))=0$. However, the existence of  $\Delta_*, \Gamma_*, t_*, \beta_*$ can be considered as purely technical assumptions that allow to derive well founded mathematical results. In particular, even if the limits are not supposed to exist, in practice, for $N$ large enough, the eigenvalues of $\Sigma_{f,N} \Sigma_{p,N}^{*} \Sigma_{p,N} \Sigma_{f,N}^{*}$ that escape from $[0, x_{+,N}]$ should be close from the solutions of $\mathrm{det}(I + F_{d,N}(y))=0$ larger than $x_{+,N}$ in a number of scenarios. However, the derivation of reasonable alternative conditions under which this behaviour holds seems difficult. 
\end{remark}  

We have thus established the Theorem: 
\begin{theorem}
	\label{theo:convergence-spiked-eigenvalues}
         Almost surely, for each $N$ large enough, the $s$ largest eigenvalues $\hat{\lambda}_{1,N} \geq \ldots \geq \hat{\lambda}_{s,N}$ of matrix $\Sigma_{f,N} \Sigma_{p,N}^{*} \Sigma_{p,N} \Sigma_{f,N}^{*}$ escape from the interval $[0, x_{+,*}]$,
	and converge towards $\rho_{1,*} \geq \ldots \geq \rho_{s,*} > x_{+,*}$ defined by $\rho_{k,*}= y_{s +1 - k,*}^{2}$ 
	for $k=1, \ldots, s$. Moreover, for each $\delta > 0$, the eigenvalues $(\hat{\lambda}_{k,N})_{k \geq s+1}$ belong
        to $[0, x_{+,*}+\delta]$. 
\end{theorem}
$s$ and the limit eigenvalues $(\rho_{k,*})_{k=1, \ldots, s}$ depend on the limit distributions $\omega_*$ and 
$\beta_*$ that are rather immaterial. It is thus more appropriate to evaluate the asymptotic behaviour of
the largest eigenvalues of $\Sigma_{f,N} \Sigma_{p,N}^{*} \Sigma_{p,N} \Sigma_{f,N}^{*}$ by using a finite $N$ 
equivalent of $H_*(z)$. We thus define function $H_N(z)$ by 
\begin{equation}
\label{eq:def-HN}
H_N(z) = \left( \begin{array}{cc}   \frac{c_N {\bf t}_N(z))}{1 - (c_N {\bf t}_N(z))^{2}} \Delta_N^{2} - ({\bf T}_{\beta_N}(z))^{-1} &
\frac{\Gamma_N^{*}}{(1 - (c_N {\bf t}_N(z))^{2}}  \\  \frac{\Gamma_N}{(1 - (c_N {\bf t}_N(z))^{2}} &
\frac{c_N {\bf t}_N(z)}{1 - (c_N {\bf t}_N(z))^{2}} \Delta_N^{2} -  ({\bf T}_{\beta_N}(z))^{-1}  \end{array} \right)
\end{equation}
For each $N$ large enough and for each $\delta > 0$, $H_N(z)$ is holomorphic in $\mathbb{C} - [-\sqrt{x_{+,*} + \delta}, \sqrt{x_{+,*} + \delta}]$ and converges uniformly on
each compact subset of  $\mathbb{C} - [-\sqrt{x_{+,*}}, \sqrt{x_{+,*}}]$ towards function $H_*(z)$. Using
again the approach of \cite{benaych-nadakuditi-ann-math} and  \cite{chapon-couillet-hachem-mestre}, we obtain that, for each  
$N$ large enough, the equation $\mathrm{det}(H_N(y)) = 0$ has $s$ solutions $y_{1,N,*} \leq \ldots \leq y_{s,N,*}$ 
strictly larger than $\sqrt{x_{+,N} + \delta}$ for some $\delta > 0$ small enough, and which satisfy $y_{k,N,*} - y_{k,*} \rightarrow 0$ when  $N \rightarrow +\infty$. Moreover, the convergence of $x_{+,N}$ and
$w_{+,N}$ towards $x_{+,*} = \phi_{*}(w_{+,*})$ and $w_{+,*} = w_*(x_{+,*})$ imply that ${\bf t}_N(x_{+,N})$ converge towards ${\bf t}_*(x_{+,*})$.
Therefore, (\ref{eq:epxre-Tbeta*x+}) leads to the following Corollary. 
\begin{corollary}
	\label{coro:final-result-finite-N}
	$H_N(\sqrt{x_{+,N}})$ converges towards $H_{*}(\sqrt{x_{+,*}})$. Moreover, if $\mathrm{det}(H_{*}(\sqrt{x_{+,*}})) \neq 0$, for $N$ large enough, $s$ also coincides with the number of strictly negative eigenvalues of matrix $H_N(\sqrt{x_{+,N}})$. Finally, if we define $\rho_{k,N}$ by $\rho_{k,N}= y_{s+1-k,N,*}^{2}$ for $k=1, \ldots, s$ , then it holds that $\hat{\lambda}_{k,N} - \rho_{k,N} \rightarrow 0$ almost surely. 
\end{corollary}
{\bf Proof.} It just remains to remark that if $0$ is not eigenvalue of $H_*(\sqrt{x_{+,*}})$, then, for each $N$ large 
enough, $s$ is equal to the number of strictly negative eigenvalues of matrix $H_N(\sqrt{x_{+,N}})$. $\blacksquare$ \\

Matrix $H_N(\sqrt{x_{+,N}})$ can be written in a more explicit way, so that $s$ can be evaluated using the following 
alternative formulation. 
\begin{corollary}
\label{coro:evaluation-s-simple}
Define $G_N$ as the $r \times r$ matrix given by 
\begin{equation}
\label{eq:def-G}
G_N = \frac{c_N w_{+,N}}{\sqrt{x_{+,N}}} \frac{1}{M} \mathrm{Tr}(R_N(w_{+,N} I - R_N)^{-1}) \left[ \left( 
\Theta_N^{*} (I_L \otimes (w_{+,N} I - R_N)^{-1}) \Theta_N \right)^{-1} - \Delta_N^{2} \right] 
\end{equation}
Then, if $\mathrm{det}(H_{*}(\sqrt{x_{+,*}})) \neq 0$, for each $N$ large enough, $s$ coincides with the number of strictly negative eigenvalues of the $2r \times 2r$ matrix 
\begin{equation}
\label{eq:matrix-simple-s}
 \left( \begin{array}{cc} G_N &  \Gamma_N^{*} \\ 
\Gamma_N & G_N \end{array} \right)
\end{equation}
\end{corollary}
{\bf Proof.} Writing ${\bf t}_N(z)$ as ${\bf t}_N(z) = z t_N(z^{2})$, and using the expression (\ref{eq:expre-tN-w})  of
$t_N(z)$ in terms of $w_N(z)$, we obtain after some algebra that matrix $H_N(\sqrt{x_{+,N}})$ is given by
\begin{equation}
\label{eq:expre-H(sqrt(x))}
H_N(\sqrt{x_{+,N}}) = \left(1+c_N \frac{1}{M} \mathrm{Tr}(R_N(w_{+,N} I - R_N)^{-1}) \right) \, \left( \begin{array}{cc} G_N &  \Gamma_N^{*} \\ 
\Gamma_N & G_N \end{array} \right)
\end{equation}
As $w_{+,N} > \lambda_{1,N}$, we have $\frac{1}{M} \mathrm{Tr}(R_N(w_{+,N} I - R_N)^{-1} > 0$ 
and $1+c_N \frac{1}{M} \mathrm{Tr}(R_N(w_{+,N} I - R_N)^{-1}) > 0$. 
$s$ thus coincides with the 
number of strictly negative eigenvalues of (\ref{eq:matrix-simple-s}).  $\blacksquare$ 
\subsection{When Condition  (\ref{eq:entries-Delta*-different}) does not hold.}
\label{subsec:entries-Delta*}
We now consider the case where Condition (\ref{eq:entries-Delta*-different}) does not hold, and briefly indicate how the above results have to be modified. For this, we denote by
$k$ the number of different diagonal entries of $\Delta_*$ and by $m_1, \ldots, m_k$ their multiplicities, which also coincide with the multiplicities of the $k$ different eigenvalues of matrices $\Theta_N \Delta_* \Theta_N^{*}$ and $\Theta_{i,N} \Delta_* \Theta_{i,N}^{*}$ for $i=p,f$. If $(\Theta_N(l))_{l=1, \ldots, k}$ and  $(\Theta_{i,N}(l))_{l=1, \ldots, k}$ represent the $ML \times m_l$ matrices for defined by $\Theta_N = (\Theta_N(1), \ldots, \Theta_N(k))$ and  $\Theta_{i,N} = (\Theta_{i,N}(1), \ldots, \Theta_{i,N}(k))$, then, (\ref{eq:convergence-matrices-with-Delta*}) and standard results of 
perturbation theory imply that 
\begin{equation}
    \label{eq:convergence-eigenspaces}
\| \Theta_{i,N}(l) \Theta_{i,N}(l)^{*} - \Theta_{N}(l) \Theta_{N}(l)^{*} \| \rightarrow 0
\end{equation}
for each $l=1, \ldots, k$. We denote by $X_{i,N}(l)$ the $m_l \times m_l$ random matrix defined by $X_{i,N}(l) = \Theta_{N}(l)^{*}  \Theta_{i,N}(l)$, and deduce from (\ref{eq:convergence-eigenspaces}) that
\begin{equation}
    \label{eq:approximation-eigenspace}
\| \Theta_{i,N}(l) -  \Theta_{N}(l) X_{i,N}(l) \| \rightarrow 0    
\end{equation}
as well as 
\begin{equation}
    \label{eq:convergence-XiN*XiN}
X_{i,N}(l)^{*}  X_{i,N}(l) - I_{m_l} \rightarrow 0, \;  X_{i,N}(l)  X_{i,N}(l)^{*} - I_{m_l} \rightarrow 0 
\end{equation}
Therefore, matrix 
$\Theta_{i,N}$ can be replaced up to error terms by matrix $\Theta_N X_{i,N}$ where 
$X_{i,N}$ represents the $r \times r$ block diagonal matrix with diagonal blocks 
$X_{i,N}(1), \ldots, X_{i,N}(k)$. It is useful to notice that the very definition 
of $X_{i,N}$ implies that the equality 
\begin{equation}
    \label{eq:XtimesDelta*}
    X_{i,N} \Delta_* = \Delta_* X_{i,N}
\end{equation}
holds. Another consequence of (\ref{eq:convergence-XiN*XiN}) is related to the asymptotic behaviour 
of matrix $\Gamma_N$. In particular, (\ref{eq:limit-GammaN-under-condition}) does no longer hold, and we rather have 
\begin{equation}
    \label{eq:behaviour-GammaN-condition-fails}
    \Gamma_N - X_{f,N}^{-1} \Gamma_* X_{p,N}^{-*} \rightarrow 0 \; a.s.
\end{equation}
To justify (\ref{eq:behaviour-GammaN-condition-fails}), we notice that (\ref{eq:convergence-Rfp-UfUp*}) leads to 
$$
\Theta_N^{*} R_{f|p,N}^{L} \Theta_N - \Theta_N^{*} \Theta_{f,N} \Delta_N \tilde{\Theta}_{f,N}^{*}  \tilde{\Theta}_{p,N} \Delta_N \Theta_{p,N}^{*} \Theta_N \rightarrow 0
$$
Therefore, (\ref{eq:approximation-eigenspace}) leads to 
$$
X_{f,N} \Gamma_N X_{p,N}^{*} - \Gamma_* \rightarrow 0
$$
and to (\ref{eq:behaviour-GammaN-condition-fails}). Therefore, $\Gamma_N$ does not converge towards a deterministic matrix, and 
rather behaves as the random matrix $X_{f,N}^{-1} \Gamma_* X_{p,N}^{-*}$. Moreover, 
the reader may check that the convergence results (\ref{eq:A*QppA}) and (\ref{eq:B*QffB}) have to be modified as follows: in (\ref{eq:A*QppA}), matrix 
$$
\Theta_N^{*} \left( z I + \frac{c_N {\bf t}_N(z)}{1 - (c_N {\bf t}_N(z))^{2}} I_L \otimes R_N \right)^{-1} \Theta_N = {\bf T}_{\beta_N}(z)
$$
has to be replaced by $X_{f,N}^{*} {\bf T}_{\beta_N}(z) X_{f,N}$ while in (\ref{eq:B*QffB}), 
$ {\bf T}_{\beta_N}(z)$ has to be exchanged with $X_{p,N}^{*} {\bf T}_{\beta_N}(z) X_{p,N}$. 
Matrix $F_{d,N}$ is thus modified. The modified matrix, still denoted $F_{d,N}(z)$, does no longer converge towards matrix $F_*(z)$ introduced after Proposition \ref{prop:converge-FN}, but appears to have almost surely the same asymptotic behaviour than the random matrix $F_{*,N}(z)$ obtained by replacing $\Gamma_N$ by $X_{f,N}^{-1} \Gamma_* X_{p,N}^{-*}$, and ${\bf T}_{\beta_*}$ 
in the definitions of $F_{*}^{1,2}(z)$ and  $F_{*}^{2,1}(z)$ 
by  $X_{f,N}^{*} {\bf T}_{\beta_*}(z) X_{f,N}$ and $X_{p,N}^{*} {\bf T}_{\beta_*}(z) X_{p,N}$
respectively. However, after some algebra, it is easily seen that $\mathrm{det}(I + F_{*,N}(z))=0$  if and only if
$\mathrm{det}(I - K_{*,N}(z))=0$, where $K_{*,N}$ is defined by
\begin{equation}
\label{eq:def-K*N}
K_{*,N}(z) = \left( \begin{array}{cc} \frac{c_* {\bf t}_*(z)}{1 - (c_* {\bf t}_*(z))^{2}} \Delta_*^{2} 
X_{p,N}^{*} {\bf T}_{\beta_*}(z) X_{p,N} & \frac{X_{p,N}^{-1} \Gamma_*^{*} X_{f,N}^{-*}}{(1 - (c_* {\bf t}_*(z))^{2}}  \\
\frac{X_{f,N}^{*}  {\bf T}_{\beta_*}(z) \, \Gamma_* {\bf T}_{\beta_*}(z) X_{p,N}}{1 - (c_* {\bf t}_*(z))^{2}} & \frac{c_* {\bf t}_*(z)}{1 - (c_* {\bf t}_*(z))^{2}} 
X_{f,N}^{*} {\bf T}_{\beta_*}(z) X_{f,N} \, \Delta_*^{2} \end{array} \right)
\end{equation}
Using (\ref{eq:convergence-XiN*XiN}) and (\ref{eq:XtimesDelta*}), we obtain that 
\begin{equation}
\label{eq:convergence-KN*-K*}
\left( \begin{array}{cc} X_{p,N} & 0 \\ 0 & X_{f,N} \end{array} \right) \, \left( I - K_{*,N}(z) \right) \left( \begin{array}{cc} X_{p,N}^{*} & 0 \\ 0 & X_{f,N}^{*} \end{array} \right) \rightarrow I - K_*(z)
\end{equation}
where $K_*(z)$ is defined by (\ref{eq:def-G*}). The solutions $y > \sqrt{x_{+,*}}$ of the equation 
$\mathrm{det}(I - K_{*}(y))=0$ are the $(y_{k,*})_{k=1, \ldots, s}$ introduced in Section \ref{sec:asym_beh_eigenval}. 
Using the arguments in \cite{benaych-nadakuditi-ann-math}
and \cite{chapon-couillet-hachem-mestre}, we obtain from (\ref{eq:convergence-KN*-K*}) that for $\delta > 0$ small enough, the equation
$\mathrm{det}(I - K_{*,N}(y))=0$, or equivalently the equation $\mathrm{det}(I + F_{*,N}(y))=0$ has $s$ solutions $(y_{k,N,*})_{k=1, \ldots, s}$ larger than $\sqrt{x_{+,*} + \delta}$ and verifying $y_{k,N,*} \rightarrow y_{k,*}$ for each $k=1, \ldots, s$. \cite{benaych-nadakuditi-ann-math} and \cite{chapon-couillet-hachem-mestre} imply that the equations 
$\mathrm{det}(I + F_{d,N}(y))=0$ and $\mathrm{det}(I + F_{N}(y))=0$ have also $s$ solutions larger than $\sqrt{x_{+,*} + \delta}$
and converging almost surely towards the $(y_{k,*})_{k=1, \ldots, s}$. This, in turn, shows that Theorem 
\ref{theo:convergence-spiked-eigenvalues} remains still valid when condition (\ref{eq:entries-Delta*-different}) does not hold.

\subsection{Particular cases and examples}\label{subsec:autocov_part_cases}
In order to get some insights on the number of eigenvalues $s$ that escape from $\mathcal{S}_N=[0, x_{+,N}]$ for each $N$ large enough, we first study informally the behaviour of $s$ when $c_N \rightarrow 0$. Intuitively, we should recover the results corresponding to the traditional regime, i.e. that $s = P$. For this, we use Corollary \ref{coro:evaluation-s-simple} and remark that $w_{+,N}$, which depends on $c_N$, satisfies $\phi_N^{'}(w_{+,N}) = 0$. Using $\phi_N^{'}(w_{+,N}) = 0$ and following the proof of Proposition 7.7 in \cite{loubaton-tieplova-2020} for $w_0 = w_{+,N}$ until Eq. (7.59), we obtain that $\frac{1}{M} \mathrm{Tr}(R_N(w_{+,N} I - R_N)^{-1}) < 1$. As $R_N > a I$ (see Assumption (\ref{eq:hypothesis-R-bis})), 
we obtain that 
$$
\frac{1}{M} \sum_{k=1}^{M} \frac{1}{w_{+,N} - \lambda_{k,N}} < \frac{1}{a}
$$
holds for each $c_N$. This implies that $\liminf_{c_N \rightarrow 0} w_{+,N} - \lambda_{1,N} > 0$, and that matrix 
$\left( \Theta_N^{*} (I_L \otimes (w_{+,N} I - R_N)^{-1}) \Theta_N \right)^{-1}$ remains bounded when $c_N \rightarrow 0$. 
As $x_{+,N}= \phi_N(w_{+,N})$, it is easy to check that $x_{+,N} = \mathcal{O}(c_N)$. Therefore, 
$\frac{c_N w_{+,N}}{\sqrt{x_{+,N}}} = \mathcal{O}(\sqrt{c_N})$, and $G_N \rightarrow 0$ 
when $c_N \rightarrow 0$. Therefore, when $c_N \rightarrow 0$, 
$$
H_N(\sqrt{x_{+,N}}) \rightarrow \left( \begin{array}{cc} 0 &  \Gamma_N^{*} \\ 
\Gamma_N & 0 \end{array} \right)
$$
As mentioned previously, matrix $\Gamma_N$ has rank $P \leq r$. Therefore, the eigenvalues of
matrix $\left( \begin{array}{cc} 0 &  \Gamma_N^{*} \\ 
\Gamma_N & 0 \end{array} \right)$ are $0$ with mutiplicity $2(r-P)$, $(\chi_k)_{k=1, \ldots, P}$
and $-(\chi_k)_{k=1, \ldots, P}$ where we recall that $(\chi_k)_{k=1, \ldots, P}$ represent the $P$ non zero
singular values of matrix $\Gamma_N$. Therefore, when $c_N \rightarrow 0$, $s$ converges towards $P$. 
This is in accordance with the traditional asymptotic regime where $N \rightarrow +\infty$ and $M$ is fixed. 
Indeed, in this context, matrix $\Sigma_{f,N} \Sigma_{p,N}^{*} \Sigma_{p,N} \Sigma_{f,N}^{*}$
converges towards the rank $P$ matrix $R^{L}_{f|p} \left(R^{L}_{f|p}\right)^{*}$, i.e. for $N$ large enough, 
matrix $\Sigma_{f,N} \Sigma_{p,N}^{*} \Sigma_{p,N} \Sigma_{f,N}^{*}$ has $P$ eigenvalues that are 
significantly larger the $M-P$ smallest ones. \\

When $c_N$ does not converge towards $0$, the presence of matrix $G_N$
in the expression (\ref{eq:expre-H(sqrt(x))}) in general deeply modifies the value of $s$.
In particular, the value of $s$ depends on the singular values $(\chi_{k,N})_{k=1, \ldots, P}$ of matrix
$\Gamma_N$, but also on the diagonal entries $(\delta_{k,N}^{2})_{k=1, \ldots, r}$ of matrix
$\Delta_N^{2}$, or equivalently, on the non zero eigenvalues of $R_{u,N}^{L} = \mathbb{E}(u_n^{L} u_n^{*L})$.
In contrast with the context of the usual spiked empirical covariance matrix 
models, $s$ may be larger than the number $P$ of non zero eigenvalues
of the true matrix $R_{f|p} R_{f|p}^{*}$. This implies that if $c_N$ is not small 
enough, then estimating the rank $P$ of matrix $R_{f|p} R_{f|p}^{*}$ by the number $s$ 
of eigenvalues of $\Sigma_{f,N} \Sigma_{p,N}^{*} \Sigma_{p,N} \Sigma_{f,N}^{*}$ that escape 
from $[0,x_{+,N}]$ does not lead to a consistent estimation scheme. \\

We now construct explicit examples where $R_N = \sigma^{2} I_M$ for some $\sigma^{2} > 0$,  $P=L=1$ and for which $s$ does not coincide with $P=1$. In particular, we now establish that for each $r \geq 2$, there exists useful signals $u$ for which $\mathrm{Rank}(R_{u,N}^{1}) = \mathrm{Rank}(\mathbb{E}(u_n u_n^{*})) = r$ and $s=2r-1$. Moreover, 
we show that the non zero eigenvalues of $R_{u,N}^{1}$ as well as the non zero singular 
value of $R_{f|p,N}^{1}$ can be arbitrarily large. In the following, matrices $R_{u,N}^{1}$ and 
$R_{f|p,N}^{1}$ will be denoted $R_{u,N}$ and $R_{f|p,N}$. We define $K=r-1$ and we consider a $M \times r$ matrix $\Theta_N = (C_N,D_{1,N}, \ldots, D_{K,N})$ verifying $\Theta_N^* \Theta_N = I_r$ and a $K$--dimensional white noise sequence $(i_n)_{n \in \mathbb{Z}}$ verifying $\mathbb{E}(i_n i_n^{*}) = I_K$. If $0 \leq a < 1$ and $b_1, \ldots, b_K$ are real numbers, we define the signal $(u_n)_{n \in \mathbb{Z}}$ by
\begin{align}
\nonumber
x_{n+1} & = a \, x_n + \sum_{k=1}^{K} b_k \, i_{k,n} \\
\label{eq:model-s-equal-2r-1}
u_n & = C_N \, x_n + \sum_{k=1}^{K} \delta_{k+1} \, D_{k,N} \, i_{k,n}
\end{align}
where $\delta_2, \ldots, \delta_r$ are strictly positive real numbers. As the state-space sequence is $1$--dimensional, $P$ coincides with $1$. We denote by $\delta_1$ the positive real number such that 
$$
\mathbb{E}(|x_n|^{2}) = \frac{\sum_{k=1}^{K} b_k^{2}}{1 - a^{2}} = \delta_1^{2}
$$
Then, if we denote by $\Delta^{2}$ the $r \times r$ diagonal matrix $\mathrm{Diag}(\delta_1^{2}, \ldots, \delta_r^{2})$, we obtain immediately that 
$$
R_{u,N}= \Theta_N \Delta^{2} \Theta_N^{*} = \delta_1^{2} \, C_N C_N^{*} + 
\sum_{k=1}^{K} \delta_{k+1}^{2} \, D_{k,N} D_{k,N}^{*}
$$
coincides with the eigenvalue/ eigenvector decomposition of $R_{u,N}$. We observe that the eigenvalues 
$(\delta_k^{2})_{k=1, \ldots, r}$ do not depend on $N$. Therefore, matrix $\Delta_N$ coincides for each $N$ with $\Delta=\Delta_*$. The matrix 
$R_{f|p,N} = \mathbb{E}(u_{n+1}u_n^{*})$ is given by 
$$
R_{f|p,N} = C_N \, \left( a \delta_1^{2} \, C_N^{*} + \sum_{k=1}^{K}b_k \delta_{k+1} \, D_{k,N}^{*} \right)
$$
Therefore, $R_{f|p,N}$ can we written as 
$$
R_{f|p,N} = \Theta_N \Gamma_* \Theta_N^{*}
$$
where $\Gamma_* = \Theta_N^{*} R_{f|p,N} \Theta_N$ is equal to 
$$
\Gamma_* = e_1 \, (a \delta_1^{2}, b_1 \delta_2, \ldots, b_K \delta_{K+1}) = \chi \, \Upsilon \, \tilde{\Upsilon}^*
$$
where $e_1$ is the first vector of the canonical basis of $\mathbb{C}^{r}$, $\Upsilon = e_1$, $\chi = \left(a \delta_1^{2})^{2} + \sum_{k=1}^{K}(b_k \delta_{k+1})^{2} \right)^{1/2}$ and $\tilde{\Upsilon}$ is the unit norm vector $\tilde{\Upsilon} = \frac{1}{\chi}  (a \delta_1^{2}, b_1 \delta_2, \ldots, b_K \delta_{K+1})^{T}$. $\chi$ thus represents the non zero singular value of rank 1 matrix $R_{f|p,N}$. 
Using (\ref{eq:expre-Rfp-Gamma}), we obtain immediately that  
matrix $\Gamma_*$ coincides with $\lim_{N \rightarrow +\infty} \Gamma_N$ where $\Gamma_N$ is defined 
by (\ref{eq:def-Gamma}). As $R_N = \sigma^{2}I_M$, and that $H_*(\sqrt{x_{+,*}}) = \lim_{N \rightarrow +\infty} H_N(\sqrt{x_{+,N}})$, it is easy to check using (\ref{eq:expre-H(sqrt(x))}) that $s$ coincides with the number of strictly negative eigenvalues of matrix $\left( \begin{array}{cc} G_* &  \Gamma_*^{*} \\ 
\Gamma_* & G_* \end{array} \right)$ where $G_*$ is defined by 
$$
G_* = \left( \frac{\sigma^{2} c_*}{w_{+,*} - \sigma^{2}(1-c_*)}\right)^{1/2} \left( (w_{+,*} - \sigma^{2}) \, I_r - \Delta^{2} \right)^{-1}
$$
Here, $w_{+,*} = \lim_{N \rightarrow +\infty} w_{+,N}$ is equal to 
$$
w_{+,*} =  \sigma^{2} \left(1 + \frac{1+\sqrt{1+8c_*}}{2} \right)
$$
(see the expression of $w_{+,N}$, Eq. (7-54) in \cite{loubaton-tieplova-2020}). It is easily checked that all the previous required Assumptions are verified by the present model. We now indicate how it is possible to choose the various parameters in order 
that $s$ coincides with $2r-1$. We first fix parameters $(\delta_k)_{k=1, \ldots, r}$ in such a way that $\delta_1^{2} \geq \delta_2^{2} \geq  \ldots \geq \delta_{r}^{2}$ and $\delta^{2}_k > (w_{+,*} - \sigma^{2})$ for each $k=1, \ldots, r$ 
while we consider in the following $(b_k)_{k=1, \ldots, K}$ verifying 
$$
1 - \frac{\sum_{k=1}^{K} b_k^{2}}{\delta_1^{2}} > 0
$$
and choose 
\begin{equation}
\label{eq:choice-a}
a = \left( 1 - \frac{\sum_{k=1}^{K} b_k^{2}}{\delta_1^{2}} \right)^{1/2}   
\end{equation}
Therefore, we of course have $\delta_1^{2} = \mathbb{E}(|x_n|^{2})$. We claim that with these set of 
parameters, $s = 2r -1$. For this, we first remark that matrix $G_* < 0$. 
As  $\mathrm{Rank}(\Gamma_*)=1$, the number $s$ of stricty negative eigenvalues of 
 $\left( \begin{array}{cc} G_* &  \Gamma_*^{*} \\ 
\Gamma_* & G_* \end{array} \right)$ is equal to $2r-1$ or to $2r$. Using the Schur complement 
trick twice, we obtain after some algebra that $s=2r - 1$ if and only 
$$
- \chi^{2} \tilde{\Upsilon}^{*} G_*^{-1} \tilde{\Upsilon} > - \left(\Upsilon^{*} G_*^{-1} \Upsilon\right)^{-1}
$$
a condition equivalent to
\begin{align}\label{ineq:s=2r-1}
    a^{2} > \frac{\sigma^{2} c_*}{\sigma^{2} c_* + w_{+,*} - \sigma^{2}} \, \left( 1 - \frac{w_{+,*} - \sigma^{2}}{\delta_1^{2}}\right)^{2} - \sum_{k=1}^{K} \frac{b_k^{2}}{\delta_1^{2}} \; \left( \frac{1 - \frac{w_{+,*} - \sigma^{2}}{\delta_1^{2}}}{1 -  \frac{w_{+,*} - \sigma^{2}}{\delta_{k+1}^{2}}} \right)
\end{align}

Using that $a^{2} =  1 - \frac{\sum_{k=1}^{K} b_k^{2}}{\delta_1^{2}}$ as well as $\delta^{2}_1 \geq \delta^{2}_{k+1}$ 
for each $k=1, \ldots, K$, we obtain that the above condition holds, and, 
therefore, that $s=2r-1$. We also mention that the condition $\delta_k^{2} > w_{+,*} - \sigma^{2}$ for each 
$k=1, \ldots, r$ does not induce any power limitation on the useful signal $u$. Moreover, using  $a^{2} = 1 - \frac{\sum_{k=1}^{K} b_k^{2}}{\delta_1^{2}}$, we obtain that the non zero singular value $\chi$ of $R_{f|p,N}$ 
is given by
$$
\chi^{2}= \delta_1^{4} \left( 1 - \frac{\sum_{k=1} b_k^{2}}{\delta_1^{2}}\right) + \sum_{k=1}^{K} b_k^{2} \delta_{k+1}^{2}
$$
As the $(\delta_k^{2})_{k=1, \ldots, r}$ can take any large values, the same property holds for $\chi$. In sum, 
even for powerful enough signals $u$ for which the largest singular value of $R_{f|p,N}$ is large, $s$ may be strictly larger than $P$. \\

\begin{figure}[ht!]
	\centering
	\par
	\includegraphics[scale=0.5]{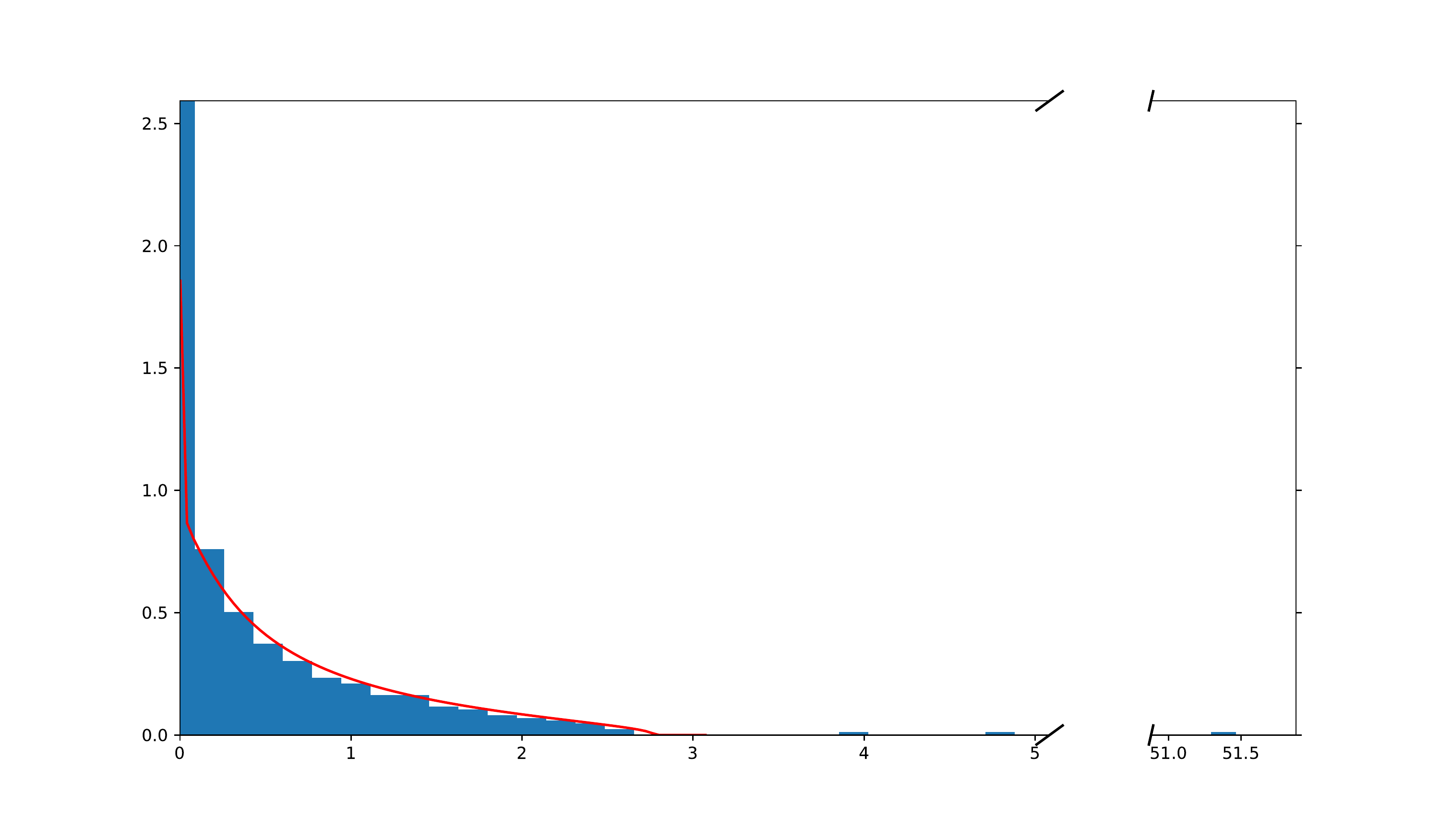}
	\caption{Histogram of the eigenvalues and graph of $g_N$, $r=2$, $s=3$}
	\label{fig:s=3}
\end{figure}

\begin{figure}[ht!]
	\centering
	\par
	\includegraphics[scale=0.5]{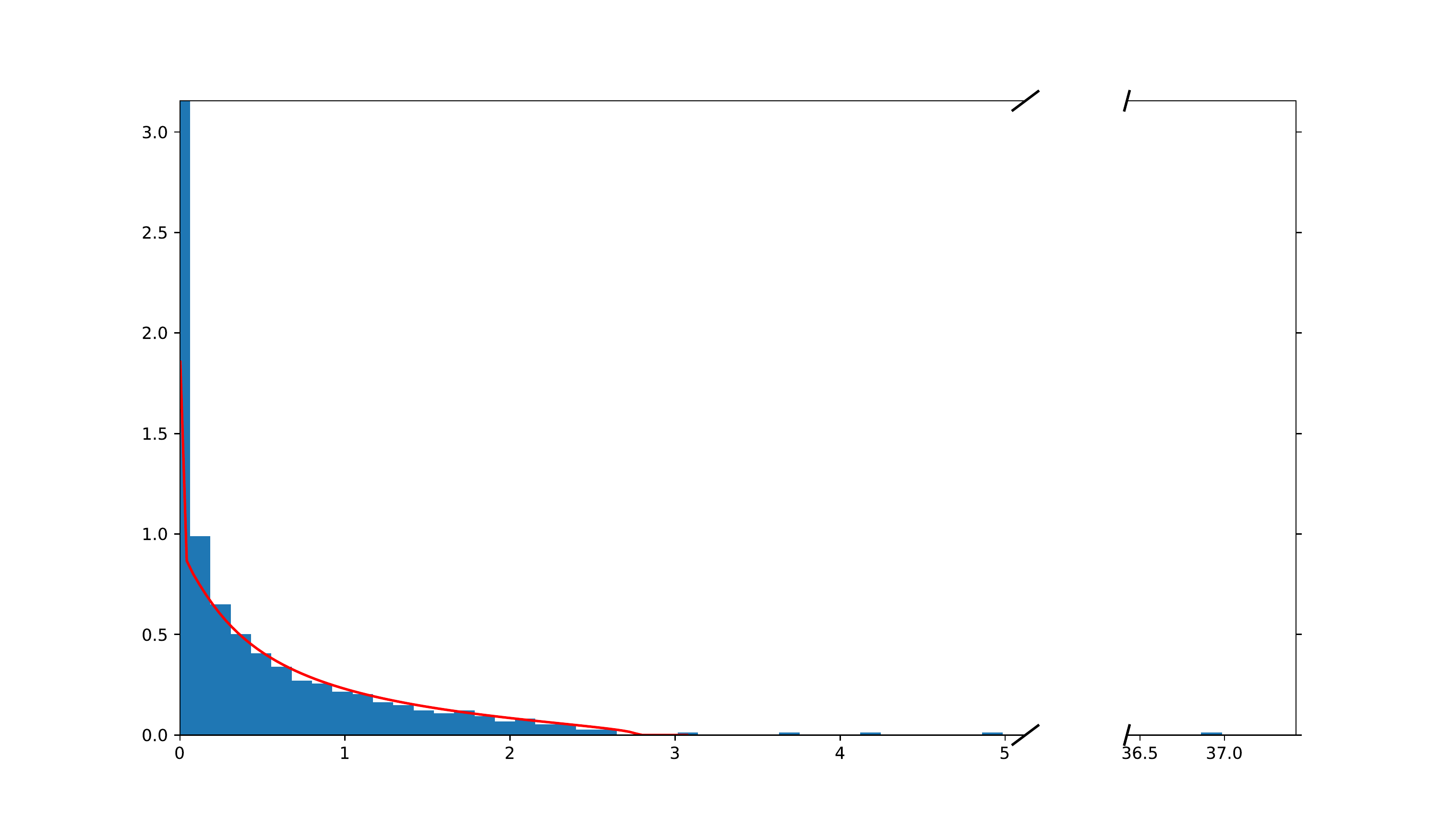}
	\caption{Histogram of the eigenvalues and graph of $g_N$, $r=3$, $s=5$}
	\label{fig:s=5}
\end{figure}

We illustrate the above analysis by numerical simulations in which $N=1200$ and $c_N = \frac{1}{2}$. The parameters of model  (\ref{eq:model-s-equal-2r-1}) are chosen as above by replacing $c_*$ by $c_N$. Figures \ref{fig:s=3} and \ref{fig:s=5} plot an histogram of the eigenvalues of a realization 
of matrix $\Sigma_{f,N} \Sigma_{p,N}^{*}  \Sigma_{p,N} \Sigma_{f,N}^{*}$, as well as the graph of 
the density $g_N$ of the deterministic equivalent measure $\nu_N$ of the empirical eigenvalue 
$\hat{\nu}_N$ of $W_{f,N} W_{p,N}^{*}  W_{p,N} W_{f,N}^{*}$. 
In the context of Fig. \ref{fig:s=3}, $r=2$ and it is seen that $s=3$ eigenvalues 
of  $\Sigma_{f,N} \Sigma_{p,N}^{*}  \Sigma_{p,N} \Sigma_{f,N}^{*}$ escape from the support of 
$\nu_N$. In the context of Fig. \ref{fig:s=5}, $r=3$ and $s=5$ as expected. We mention 
that in both figures, the largest eigenvalue, which, in some sense, is due to the useful signal, appears much larger than the other spurious ones. It can be checked that, as expected, for smaller values of $c_N$, the spurious eigenvalues that escape from the support of $\nu_N$ tend to become closer from $x_{+,N}$. This will be confirmed 
in Section \ref{sec:simulations} where more exhaustive Monte Carlo simulation results
evaluate the behaviour of two estimates of $s$
when $s=5$ and $c_N = \frac{1}{4}$. It will be seen that the estimates of $s$ belong to $\{2,3,4,5,6,7,8\}$, fail to detect $s=5$ very often, but never take the value $1$. \\

The above examples show that $s$ can take any odd value larger than $3$. We finally show that $s$ can also 
be equal to 2, and consider the following simple case. We assume that $P=K=1$, and that 
the scalar state-space sequence $(x_n)_{n \in \mathbb{Z}}$ is given by $x_{n+1} = a x_n + b i_n$
where $a \in ]0,1[$, $b > 0$, and $(i_n)_{n \in \mathbb{Z}}$ is scalar unit variance i.i.d. sequence. Moreover, $u_n$ is given by
\begin{equation}
\label{eq:case-jiang-feng}
u_n = \theta_N x_{n+1} = a \theta_N x_n + b \theta_N i_n
\end{equation}
where $\theta_N$ is a unit norm $M$--dimensional vector. Therefore, matrices $C_N$ and $D_N$ coincide with
vectors $a \theta_N$ and $b \theta_N$ respectively. We also consider the case where $L=1$. 
The covariance matrix $R_{u,N}= E(u_n u_n^{*})$ is of course equal to $\delta^{2} \, \theta_N \theta_N^{*}$ where $\delta^{2} = \mathbb{E}(|x_n|^{2}) = 
\frac{|b|^{2}}{1-a^{2}}$, so that $r=P=1$. We also mention that in the present case, 
$\delta^{2}$ does not depend on $N$. Moreover, $R_{f|p,N}= E(x_{n+1}x_{n}^{*}) = a \delta^{2} \theta_N \theta_N^{*} $. Therefore, matrix $\Gamma_*$ is reduced to the scalar $a \delta^{2}$, which also coincides with the non zero singular value $\chi$ of $R_{f|p,N}$. As $r=P=1$, $s$ may take the values $0,1,2$. In the following, we justify 
that it is possible to find $a$ and $b$ for which $s=2$. \\

It is easily seen that $s=2$ if $\delta^{2} > w_{+,*} - \sigma^{2}$ and
\begin{equation}
\label{eq:choice-a-s=2}
a^{2} < \frac{\sigma^{2} c_*}{\sigma^{2} c_* + (w_{+,*} - \sigma^{2})} \, \left(1 - \frac{(w_{+,*} - \sigma^{2}}{\delta^{2}}  \right)^{2}
\end{equation}
In order to find $a \in ]0,1[$ and $b$ for which these conditions hold, we fix $\delta^{2} > w_{+,*} - \sigma^{2}$, then choose
$a \in ]0,1[$ such that (\ref{eq:choice-a-s=2}) holds, and finally select $b$ in such a way that $|b|^{2} = \delta^{2}(1 - a^{2})$.  We again mention that $\delta^{2}$ and 
$\chi = a \delta^{2}$ can take arbitrarily large values, $\chi$ being however less than 
$\delta^{2}  \left( \frac{\sigma^{2} c_*}{\sigma^{2} c_* + (w_{+,*} - \sigma^{2})}\right)^{1/2} \, \left(1 - \frac{(w_{+,*} - \sigma^{2}}{\delta^{2}}  \right)$. \\

\begin{figure}[ht!]
	\centering
	\par
	\includegraphics[scale=0.8]{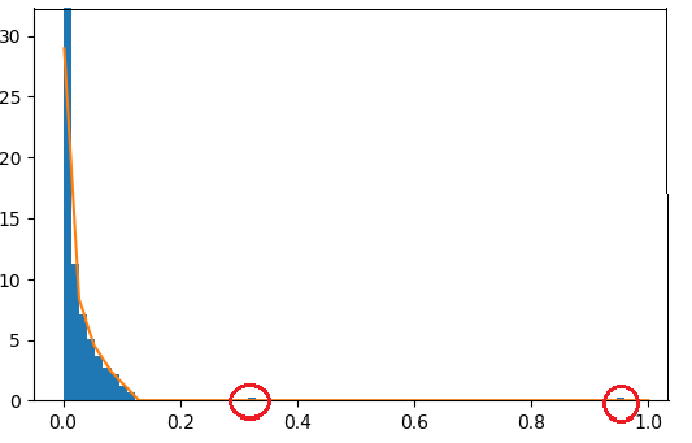}
	\caption{Histogram of the eigenvalues and graph of $g_N$, $r=1$, $s=2$}
	\label{fig:s=2}
\end{figure}

We again illustrate the above example by Fig. \ref{fig:s=2} obtained when $N = 1200$, $c_N = \frac{1}{2}$, and where it is seen that $s=2$.

\section{The canonical correlation coefficients between the past and the future}
\label{sec:cor_coeficients}
We showed in Section \ref{sec:autocov_signal} that the 
number of eigenvalues of $\hat{R}_{f|p}\hat{R}_{f|p}^*$ that escape from the interval $[0, x_{+,N}]$ is in general not a consistent estimator of the dimension $P$ of the minimal state space representation (\ref{eq:state-space}). In this section, we thus study the largest singular values of matrix  $(\frac{Y_f Y_f^{*}}{N})^{-1/2}\frac{Y_f Y_p^{*}}{N} (\frac{Y_p Y_p^{*}}{N})^{-1/2}$, or equivalently the largest eigenvalues of matrix 
$(\frac{Y_f Y_f^{*}}{N})^{-1/2}\frac{Y_f Y_p^{*}}{N} (\frac{Y_p Y_p^{*}}{N})^{-1} \frac{Y_p Y_f^{*}}{N} (\frac{Y_f Y_f^{*}}{N})^{-1/2}$. It is clear that apart $0$, the eigenvalues of the above matrix coincide with the eigenvalues of matrix $\Pi_{p,y} \Pi_{f,y}$ where for each $i=p,f$, $\Pi_{i,y}$ represents the orthogonal projection 
matrix on the row space of matrix $Y_i$, i.e. 
\begin{equation}
\label{eq:def-Pii}
\Pi_{i,y} = \frac{Y_i^{*}}{\sqrt{N}} \left( \frac{Y_i Y_i^{*}}{N}\right)^{-1} \frac{Y_i}{\sqrt{N}}
\end{equation}
We remark that the eigenvalues of $\Pi_{p,y} \Pi_{f,y}$ of course belong to $[0,1]$. 
We follow the same approach than in Section \ref{sec:autocov_signal}. We first study the eigenvalues 
of $\Pi_{p,v} \Pi_{f,v}$, where $\Pi_{i,v}$ is obtained from $\Pi_{i,y}$ by replacing $y$ by the noise $v$. 
Under certain assumptions on the useful signal $u$ (that appear simpler than in Section \ref{sec:autocov_signal}), 
we study the largest eigenvalues of $\Pi_{p,y} \Pi_{f,y}$ by remarking that  $\Pi_{p,y} \Pi_{f,y}$ is a low rank 
perturbation of $\Pi_{p,v} \Pi_{f,v}$, and use the approach developed in  \cite{benaych-nadakuditi-ann-math}, \cite{benaych-rao-2}, \cite{paul-2007}. We again mention that, while this general approach appears classical, as in Section \ref{sec:autocov_signal}, the complexity of the random matrix models that come into play makes the following 
results not obvious at all. \\

In the following, for the sake of simplicity, we will often use the same notations as in Section \ref{sec:autocov_signal} to represent different objects. This will not introduce any confusion because 
Section \ref{sec:cor_coeficients} and Section \ref{sec:autocov_signal} are independent. 
In particular, if $(\alpha_N)_{N \geq 1}$ is a sequence of positive numbers, we will say in this section that function $f_N(z)=\mathcal{O}_z(\alpha_N)$ on a domain  $\Omega \subset \mathbb{C} \setminus \mathbb{R}^{+}$ if there exists two nice polynomials $P_1$ and $P_2$ such that $|f_N(z)| \le \alpha_N P_1(|z|)P_2(\frac{1}{\rho(z)})$ for each $z \in \Omega$, where $\rho(z)=\dist(z,\mathbb{R^+})$. If $\Omega = \mathbb{C}\setminus\mathbb{R}^{+}$, we will just write
$f_N(z)=\mathcal{O}_z(\alpha_N)$ without mentioning the domain. For any diagonal $K\times K$ matrix $A(z)$, by $A(z)=\Ocal_z^K(\alpha_N)$, we mean that each diagonal element of $A(z)$ is $\Ocal_z(\alpha_N)$. Finally, we will use a lot the notation $f_N(z)=\mathcal{O}_{z^2}(\alpha_N)$ without mentioning the domain, which will mean that
$|f_N(z)| \leq \alpha_N  P_1(|z^{2}|)P_2(\frac{1}{\rho(z^{2})})$ for some nice polynomials $P_1,P_2$ when $z^{2} \in \mathbb{C} \setminus \mathbb{R}^{+}$, or equivalently, when $z\in\mathbb{C}\setminus\mathbb{R}$. 
We notice that 
if $P_1$, $P_2$ and $Q_1$, $Q_2$ are nice polynomials, then $P_1(|z|)P_2(\frac{1}{\rho(z)})+Q_1(|z|)Q_2(\frac{1}{\rho(z)})\le (P_1+Q_1)(|z|)(P_2+Q_2)(\frac{1}{\rho(z)})$, from which we conclude that if functions $f_1$ and $f_2$ are $\mathcal{O}_z(\alpha_N)$ then also $f_1(z)+f_2(z)=\mathcal{O}_z(\alpha_N)$. \\

\subsection{In the absence of signal}
\label{sec:zero_signal}
In this paragraph, we study the behaviour of the eigenvalues of $\Pi_{p,v} \Pi_{f,v}$. 
Due to the Gaussianity of the i.i.d. vectors $(v_n)_{n \geq 1}$, it exists 
i.i.d.  $\mathcal{N}_c(0,I_M)$ distributed vectors $(v_{iid,n})_{n \geq 1}$ such that $\mathbb{E}(v_{iid,n} v_{iid,n}^{*}) = I_M$ verifying $v_n = R_N^{1/2} v_{iid,n}$. It is clear that the row spaces of $V_p$ and $V_f$ coincide
with the row spaces of the block Hankel matrices $V_{p,iid}$
and $V_{f,iid}$ defined from vectors $(v_{n,iid})_{n=1,\ldots,N+2L-1}$. Therefore, the projection 
matrices $\Pi_{i,v}$ and $\Pi_{i,v_{iid}}$ coincide for $i=p,f$ and there is thus no restriction 
to assume in Section \ref{sec:zero_signal} that $R_N = I_M$. \\

As before, we denote by $ W_{p}, W_f$ the matrices defined by 
$W_p =  \frac{1}{\sqrt{N}} V_{p}$ and $W_f = \frac{1}{\sqrt{N}} V_{f}$. In order to simplify the 
notations of this Section, matrices $\Pi_{i,v}$, $i=p,f$ are denoted $\Pi_i$, $i=p,f$. Therefore, we have $\Pi_{p}=W_p^*(W_pW_p^*)^{-1}W_p$ and $\Pi_{f}=W_f^*(W_fW_f^*)^{-1}W_f$. We recall that $W_N$ is the $2ML \times N$ matrix
\begin{align}
\label{eq:def-WN-bis}
W_N=	\begin{pmatrix}
W_{p,N}\\
W_{f,N}
\end{pmatrix},
\end{align} 
As $R_N$ is supposed to be equal to $I_M$, the elements $(W_{i,j}^{m})_{i \leq 2L, j \leq N, m \leq M}$ of $W_N$ satisfy 
\begin{align}
  \label{eq:correlation-structure-W}
\mathbb{E}\{W_{i,j}^{m}W_{i^{\prime},j^{\prime}}^{m^{\prime}}\}= \frac{1}{N} \, \delta_{m-m^{\prime}}\delta_{i+j - (i^{\prime}+j^{\prime})}.
\end{align} 
where $W_{i,j}^m$ represents the element which  lies on the $(m+M(i-1))$-th line and $j$-th column for $1\le m\le M$, $1\le i\le 2L$ and $1\le j\le N$.   For each $j=1, \ldots, N$,$\{w_{j}\}_{j=1}^N, \{w_{p,j}\}_{j=1}^N$ and $\{w_{f,j}\}_{j=1}^N$ are the column of matrices $W, W_p$ and $W_f$ respectively. \\

We first verify that, almost surely, for $N$ large enough, matrices $W_{i,N} W_{i,N}^{*}$ are invertible, so that the orthogonal projection matrices  $\Pi_i$, $i=p,f$ are
well defined. For this, we mention that \cite{L:15} (see Theorem 1.1) established that 
the empirical eigenvalue distribution of $W_{i,N}W_{i,N}^{*}$ for $i=\{p,f\}$ converges towards the Marcenko-Pastur distribution with parameter $c_*$, and that almost surely, for $N$ greater 
than a random integer, its eigenvalues are located in a neighbourhood of $[(1 - \sqrt{c_*})^{2},(1 + \sqrt{c_*})^{2}]$. Therefore, almost surely, for $N$ large enough, matrices 
$W_{f,N} W_{f,N}^{*}$ and  $W_{p,N} W_{p,N}^{*}$ are invertible. Matrices $\Pi_{i,N}$ are thus well defined for $N$ large enough. \\

We next use again the results of \cite{L:15} to show the following Lemma which will be useful to 
establish Theorem \ref{th:no_eigen-2} below.  
\begin{lemma}
  \label{le:1eigenvaluePipPif}
  If $c_* > \frac{1}{2}$, then, almost surely, for $N$ large enough, $1$ is eigenvalue of $\Pi_{p,N} \Pi_{f,N}$ with multiplicity $2ML -N$
\end{lemma}
{\bf Proof.} It is clear that the eigenspace of $\Pi_{p,N} \Pi_{f,N}$ associated to the eigenvalue $1$ coincides with
$\mathrm{sp}_r(W_{p,N}) \cap \mathrm{sp}_r(W_{f,N})$, where for a matrix $A$, $\mathrm{sp}_r(A)$ represents the space generated by the rows of $A$. We have thus to verify
that if $c_* > 1/2$, then almost surely, for $N$ large enough,  $\mathrm{dim}\left(\mathrm{sp}_r(W_{p,N}) \cap \mathrm{sp}_r(W_{f,N}) \right) = 2ML -N$.
For this, we use again  \cite{L:15}. The eigenvalue distribution of $W_N W_N^{*}$ converges towards the Marcenko-Pastur distribution with
parameter $2 c_*$, and if $c_* > \frac{1}{2}$, i.e. if $2 c_* > 1$, then, for each $\epsilon > 0$,  $0$ is eigenvalue of $W_N W_N^{*}$ with multiplicity $2ML -N$ and the remaining $N$ eigenvalues
are located almost surely for each  $N$ large enough in  $[(1 - \sqrt{2c_*})^{2} - \epsilon,(1 + \sqrt{2c_*})^{2} + \epsilon]$. Therefore, we obtain that $\mathrm{dim}(\mathrm{sp}_r(W_N)) = N$
while we already know that $\mathrm{dim}(\mathrm{sp}_r(W_{p,N})) + \mathrm{dim}(\mathrm{sp}_r(W_{f,N})) = 2ML$. As $\mathrm{sp}_r(W_N) = \mathrm{sp}_r(W_{p,N}) + \mathrm{sp}_r(W_{f,N})$,
we obtain as expected that $\mathrm{dim}\left(\mathrm{sp}_r(W_{p,N}) \cap \mathrm{sp}_r(W_{f,N}) \right) = 2ML -N$. $\blacksquare$. \\

\subsubsection{Preliminary results}
In order to be able to use the perturbation approach developed in  \cite{benaych-nadakuditi-ann-math}, \cite{benaych-rao-2}, \cite{paul-2007}, it  appears necessary to evaluate the asymptotic behaviour of the 
resolvent of matrix $\Pi_p \Pi_f$. The corresponding results will also provide a characterization of the eigenvalues 
of $\Pi_p \Pi_f$. For this, we use in the following the integration by parts formula and the Poincar\'e-Nash inequality
(see Propositions \ref{prop:integration-by-parts}, \ref{prop:poincare}). The resolvent of $\Pi_p \Pi_f$ will be
interpreted as a function of the entries of matrix $W_N$. However, this approach needs some care because, 
considered as a function of the entries of $W_N$, matrices $\Pi_p$ and $\Pi_f$ are not differentiable
everywhere. In particular, for $i=p,f$, $\Pi_i$ is not differentiable when the rank of $W_{i,N}$ is less than $ML$. 
But, we have seen that almost surely, for $N$ large enough, matrices 
$W_f W_f^{*}$ and  $W_p W_p^{*}$ are invertible. In order to take benefit of this property, we use in the following a regularization term $\eta_N$ already introduced in \cite{hachem-loubaton-et-al-2012} in a different context. Another problem posed by the evaluation of the resolvent of $\Pi_p \Pi_f$ is due to the observation that, 
while matrix $\Pi_p \Pi_f$ has real eigenvalues that belong to $[0,1]$, it is not Hermitian. Some basic properties 
of the resolvent of $\Pi_p \Pi_f$ thus do not hold, in particular the upper bound (\ref{eq:upper-bound-norm-QA-Apositive}). 
In this paragraph, we first present the regularization term $\eta_N$ as well as some extra useful properties. 
\paragraph{Regularization term}
We define $\eta_N$ by
\begin{align}\label{eq:def_eta}
\eta_N=\det[\phi(W_{f,N}W_{f,N}^*)]\det[\phi(W_{p,N}W_{p,N}^*)],
\end{align}
where $\phi$ is a smooth function such that
\begin{eqnarray}
\label{eq:def-phi-regularization}
\phi(\lambda) & = & 1 \, \text{for } \lambda\in[(1-\sqrt{c_*})^2-\epsilon],\,[ (1+\sqrt{c_*})^2+\epsilon],\\
\nonumber
\phi(\lambda) & = & 0 \, \text{for } \lambda\in[-\infty,\,(1-\sqrt{c_*})^2-2\epsilon]\cup[(1+\sqrt{c_*})^2+2\epsilon,\,+\infty]
\end{eqnarray}
and $\phi(\lambda)\in(0,\,1)$ elsewhere. Here, $\epsilon$ verifies $(1 - \sqrt{c_*})^{2} - 2 \epsilon > 0$. Taking into account the almost sure behaviour of the eigenvalues of 
matrices $W_p W_p^{*}$ and $W_f W_f^{*}$, $\eta_N=1$  and 
\begin{align}\label{eq:bound_inver_WpWp*}
(W_{i,N}W_{i,N}^{*})^{-1}\eta_N\le\dfrac{I_{ML}}{(1-\sqrt{c_*})^2-2\epsilon}.
\end{align} 
almost surely for each $N$ larger than a random integer. We first mention the following useful property. 

\begin{lemma}
	For each $l,k\in\mathbb{N}$ it holds that
	\begin{align}\label{eq:moments_eta}
	\ex\{\eta_N^l\}=1+\Ocal\left(\dfrac{1}{N^k}\right)
	\end{align}
Moreover, if $X$ is a bounded random variable, we have for each integer $l \geq 1$
\begin{equation}
\label{eq:EetaX}
\mathbb{E}(\eta^{l} X) = \mathbb{E}(X) + \Ocal\left(\dfrac{1}{N^k}\right)
\end{equation}
for each integer $k$. 
\end{lemma}
{\bf Proof.} Denote
\begin{align}
\label{eq:def-EcalN}
\Ecal_N=\{\text{one of the eigenvalues of } W_pW_p^* \text{ or } W_fW_f^* \text{ escapes from the } [(1-\sqrt{c_*})^2-\epsilon,\, (1+\sqrt{c_*})^2+\epsilon]\}
\end{align} 
and define
another smooth function $\phi_0$ as
\begin{align*}
\phi_0(\lambda)=\begin{cases}
0 \, \text{for } \lambda\in[(1-\sqrt{c_*})^2,\, (1+\sqrt{c_*})^2],\\
1 \, \text{for } \lambda\in[-\infty,\,(1-\sqrt{c_*})^2-\epsilon] \cup[(1+\sqrt{c_*})^2+\epsilon,\,+\infty]
\end{cases}
\end{align*}
and $\phi_0(\lambda)\in(0,\,1)$ elsewhere. Then we have
\begin{align*}
P(\Ecal_N)\le P\left(\tr\phi_0(W_pW_p^*)\ge 1\right)\le\ex\left\{\left(\tr\phi_0(W_pW_p^*)\right)^{2k}\right\}
\end{align*}
for all $k\in\mathbb{N}$.
In order to evaluate $\ex\left\{\left(\tr\phi_0(W_pW_p^*)\right)^{2k}\right\}$, one can use the same steps as in the proof of Lemma~3.2 \cite{loubaton-tieplova-2020} and get immediately that 
$\ex\left\{\left(\tr\phi_0(W_pW_p^*)\right)^{2k}\right\}=\Ocal\left(\dfrac{1}{N^{2k}}\right)$
and therefore that $P(\Ecal_N)=\Ocal\left(\frac{1}{N^{2k}}\right)$ for each $k$. To show (\ref{eq:moments_eta}) we write 
\begin{multline*}
|\ex\{\eta_N^l-1\}|^2=|\ex\{(\eta_N-1)(1+\ldots+\eta_N^{l-1})\}|^2\le\ex\{(\eta_N-1)^2\}\ex\{(1+\ldots+\eta_N^{l-1})^2\}\\
\le\kappa\ex\{(\eta_N-1)^2{\bf 1}_{\Ecal_N}\}
\end{multline*}
because $\eta_N-1=0$ on $\Ecal_N^c$. Since by definition $\phi(\lambda)\in[0,1]$, we conclude that $0\le\eta_N\le1$ and $0\le(\eta_N-1)^2\le1$. This allows us to write that $\kappa\ex\{(\eta_N-1)^2{\bf 1}_{\Ecal_N}\}\le\kappa\ex\{{\bf 1}_{\Ecal_N}\}=\kappa P(\Ecal_N)=\Ocal\left(\dfrac{1}{N^{2k}}\right)$, which completes the proof. 
To verify (\ref{eq:EetaX}), we remark that 
\begin{align*}
|\ex\{(\eta_N^l-1)X\}|^2\le\ex\{(1-\eta_N^l)^2\}\ex\left\{|X|^2\right\}=\kappa \left(1-2 \left(1+\Ocal(N^{-k}) \right))+1+\Ocal(N^{-k})\right)=\Ocal\left(\dfrac{1}{N^k}\right).
\end{align*}
$\blacksquare$\\


\paragraph{Linearisation}
It is clear that almost surely,  
$\eta_N \Pi_{i,N} = \Pi_{i,N}$  for each $N$ large enough. Therefore, in order to evaluate the almost sure
behaviour of the resolvent of $\Pi_{p,N} \Pi_{f,N}$, it is sufficient to 
study the behaviour of the resolvent $Q_N(z)$ of $\eta_N \Pi_{p,N} \eta_N \Pi_{f,N}$  defined by 
\begin{equation*}
\label{eq:def-QN-section3}
Q_N(z) = \left(\eta_N \Pi_{p,N}\eta_N\Pi_{f,N} - z I \right)^{-1}
\end{equation*}
As the direct study of $Q_N(z)$ is not obvious, we rather use, as in Section \ref{sec:autocov_signal}, the linearisation trick and introduce the resolvent $\mathbf{Q}_N(z)$ of the $2N\times2N$ block matrix 
\begin{align*}
\begin{pmatrix}
0 & \eta_N\Pi_{p,N}\\
\eta_N\Pi_{f,N}&0
\end{pmatrix}.
\end{align*} 
which can be written as
\begin{equation}
\label{eq:expre-BQ-2}
\mathbf{Q}_N(z) =  \left( \begin{array}{cc} ({\bf \Q_{pp}})_{N}(z) &  ({\bf \Q_{pf}})_{N}(z) \\
 ({\bf \Q_{fp}})_{N}(z) &   ({\bf \Q_{ff}})_{N}(z) \end{array} \right) = \left( \begin{array}{cc} z Q_N(z^{2}) & Q_N(z^{2})\eta_N\Pi_{p,N} \\
\eta_N\Pi_{f,N} Q_N(z^{2}) &  z \hat{Q}_N(z^{2}) \end{array} \right)
\end{equation}
where $\hat{Q}_N(z)$ is the resolvent of matrix $\eta_N\Pi_{f,N}\eta_N\Pi_{p,N}$. Since $Q_N(z)$ and $\Q_N(z)$ are resolvents of non Hermitian matrices, the usual bound (\ref{eq:upper-bound-norm-QA-Apositive}) is not necessarily 
verified. A more specific control is thus needed.
\begin{lemma}\label{le:bound_Q}
	If $\im z\ne0$ (i.e. $z^2\in \mathbb{C}\setminus\mathbb{R}^+$), then 	$\|\Q(z)\|=\Ocal_{z^2}(1)$.
\end{lemma}
{\bf Proof.} It is sufficient to bound each of the four blocks of $\Q$. We start with ${\bf Q_{pf}}$. For this we use expression (\ref{eq:expre-BQ-2}) for ${\bf Q_{pf}}$, the fact that $\Pi_p=\Pi_p^2$ and that $(AB-x)^{-1}A=A(BA-x)^{-1}$ in the case $A = \eta \Pi_p, B = \eta \Pi_p \Pi_f$. This leads to 
\begin{equation}
\label{eq:expre-alter-Qpf}
{\bf Q_{pf}}
=(\eta_N^2\Pi_p\Pi_f-z^2)^{-1}\eta_N\Pi_p \Pi_p
=\eta_N\Pi_p(\eta_N^2\Pi_p\Pi_f\Pi_p-z^2)^{-1}\Pi_p.
\end{equation}
$(\eta_N^2\Pi_p\Pi_f\Pi_p-z^2)^{-1}$ is the resolvent of a positive Hermitian matrix evaluated at $z^{2} \in \mathbb{C} \setminus \mathbb{R}^{+}$, so that its norm can be bounded by $(\rho (z^2))^{-1}$ (see (\ref{eq:upper-bound-norm-QA-Apositive})). Since $\|\Pi_p\|\le 1$ and $\eta_N\le1$, we have 
\begin{align}
\|{\bf Q_{pf}}\|\le \dfrac{1}{\rho(z^2)}
\end{align}
It is easily seen that $\|{\bf Q_{fp}}\|$ can be evaluated similarly. In order to address ${\bf Q_{pp}}$, we use again (\ref{eq:expre-BQ-2}) and the resolvent identity (\ref{eq:resolvent-identity-QA}), and observe that:
\begin{align*}
{\bf Q_{pp}}=z(\eta_N^2\Pi_p\Pi_f-z^2)^{-1}=\dfrac{1}{z}(-I_N+\eta_N^2\Pi_p\Pi_f(\eta_N^2\Pi_p\Pi_f-z^2)^{-1})=\dfrac{1}{z}(-I_N+\eta_N\Pi_p{\bf Q_{fp}})
\end{align*}
It obviously holds that $\|-I_N+\eta_N\Pi_p{\bf Q_{fp}}\|\le 1+\frac{1}{\rho(z^2)}$. To show that $|z^{-1}|\le P(\rho(z^2)^{-1})$ for some nice polynomial $P$, we write
\begin{align}
\dfrac{1}{|z|^{2}}\le\dfrac{1}{\rho(z^2)}\le 1+ \dfrac{1}{ \rho(z^2)}\le\left(1+\dfrac{1}{\rho(z^2)}\right)^2
\end{align}
This brings us to the conclusion that $\|{\bf Q_{pp}}\|=\Ocal_{z^2}(1)$ and so for ${\bf Q_{ff}}$. This completes the proof of the Lemma. $\blacksquare$

\begin{remark}
	It is worth to remark that in the course of the proof, we obtained  that $\frac{1}{|z|}\Ocal_{z^2}(1)$ is still $\Ocal_{z^2}(1)$. Since $|z|\le\frac{1}{2}(1+|z|^2)$ holds, we also have $|z|\Ocal_{z^2}(1)=\Ocal_{z^2}(1)$.
\end{remark}

\begin{remark}
\label{re:TrQ-stieltjes-transform}
While $Q_N(z)$ is not the resolvent of an Hermitian matrix, 
$\frac{1}{N} \mathrm{Tr} Q_N(z)$ coincides with the Stieltjes transform of the empirical eigenvalue distribution $\hat{\nu}_N$ of matrix 
$\eta^{2} \Pi_p \Pi_f$, which, of course, is a probability measure carried by $[0,1]$, and thus by $\mathbb{R}^{+}$. Therefore, (\ref{eq:expre-BQ-2}) and property (\ref{eq:property-zs(z2)}) 
imply $\frac{1}{N} \Tr {\bf Q_{pp}}(z) = \frac{1}{N} \Tr {\bf Q_{ff}}(z)$ coincide with the Stieltjes transform of a probability measure carried by $[-1,1]$ which appears to be the eigenvalue distribution of  matrix 
$\begin{pmatrix}
0 & \eta_N\Pi_{p,N}\\
\eta_N\Pi_{f,N}&0
\end{pmatrix}$
\end{remark}

The proof of Lemma \ref{le:bound_Q} also leads to the following useful Corollary. 
\begin{corollary}\label{cor:tr_Qpf-Stj_transf}
	$N^{-1}\tr{\bf Q_{pf}}(z)$ and $N^{-1}\tr{\bf Q_{fp}}(z)$ coincide with the value taken  at $z^2$ by the Stieltjes transforms of some positive measures carried by $\mathbb{R}^+$. The same property holds for 
$\ex\{N^{-1}\tr{\bf Q_{pf}}(z)\}$ and $\ex\{N^{-1}\tr{\bf Q_{fp}}(z)\}$, and the mass of the corresponding measures can be written as $c_N+\Ocal(N^{-k})$ for each $k\in\mathbb{N}$.
\end{corollary}
{\bf Proof.}  We just establish the properties of $N^{-1}\tr{\bf Q_{pf}}(z)$. $(\eta_N^2\Pi_p\Pi_f\Pi_p-z^{2})^{-1}$ is the resolvent of a positive Hermitian matrix evaluated at point $z^{2}$. Therefore,  $N^{-1}\tr \eta_N\Pi_p(\eta_N^2\Pi_p\Pi_f\Pi_p-z^{2})^{-1}\Pi_p =  N^{-1} \tr {\bf Q_{pf}}(z)$ (see Eq. \ref{eq:expre-alter-Qpf})) coincides with the Stieltjes transform of a positive measure carried by $\mathbb{R}^+$ of total mass $N^{-1}\tr\eta_N\Pi_p^2=N^{-1}\tr\eta_N\Pi_p$ evaluated at $z^{2}$. This implies that $N^{-1}\ex\{\tr{\bf Q_{pf}}\}$ has the same property, 
and that the mass of the corresponding measure is equal to $N^{-1}\ex\{\tr\eta_N\Pi_p\}$. We claim that 
\begin{equation}
\label{eq:expre-E-trace-eta-Pip}
N^{-1}\ex\{\tr\eta_N\Pi_p\} = c_N+\Ocal(N^{-k})
\end{equation}
for each integer $k$. To justify (\ref{eq:expre-E-trace-eta-Pip}), we first use (\ref{eq:EetaX})
and obtain that $N^{-1}\ex\{\tr\eta_N\Pi_p\} = N^{-1}\ex\{\tr \Pi_p\} +\Ocal(N^{-k})$. 
It is clear that $N^{-1} \Tr \Pi_p = c_N$ on the event $\Ecal_N^{c}$, where we recall that 
$\Ecal_N$ is the set defined by (\ref{eq:def-EcalN}). Writing 
$$
N^{-1}\ex\{\tr \Pi_p\}  = N^{-1}\ex\{\tr \Pi_p \mathbf{1}_{\Ecal_N^{c}} \} +  N^{-1}\ex\{\tr\Pi_p  \mathbf{1}_{\Ecal_N}\} = c_N P(\Ecal_N^{c}) + N^{-1}\ex\{\tr\Pi_p  \mathbf{1}_{\Ecal_N}\}
$$
and using that $P(\Ecal_N) = \mathcal{O}(N^{-k})$ for each $k$, we obtain that $N^{-1}\ex\{\tr\Pi_p  \mathbf{1}_{\Ecal_N}\} = \mathcal{O}(N^{-k})$ for each $k$ and that $N^{-1}\ex\{\tr \Pi_p\} = c_N + \Ocal(N^{-k})$. 
This completes the proof of (\ref{eq:expre-E-trace-eta-Pip}). $\blacksquare$\\


\paragraph{Properties based on the invariance of the complex Gaussian distribution }
\label{subsec:invar}

\begin{lemma}
	\label{le:symetries-E(Q)-2}
	The matrix $\ex\{\eta_N(W_iW_i^*)^{-1}\}$ is block diagonal and matrices $\ex\{\eta_N\Pi_i\}$, $\ex\{\mathbf{Q_{ij}}\}$, $\ex\{\eta_N \mathbf{Q_{ij}}\}$, $\ex\{\eta_N\Pi_h\mathbf{Q_{ij}}\}$ and $\ex\{\eta_N\mathbf{Q_{ij}}W_h^*(W_hW_h^*)^{-2}W_h\}$ are diagonal, for $i,j,h=\{p,f\}$. Moreover, $\mathbb{E}(\eta_N \,  W_h^{*} (W_h W_h^{*})^{-1}) =
        \mathbb{E}(\eta_N \,  \mathbf{Q_{ij}}  \, W_h^{*} (W_h W_h^{*})^{-1}) =  \mathbb{E}(\eta_N \,  \Pi_k \, \mathbf{Q_{ij}}  \, W_h^{*} (W_h W_h^{*})^{-1}) = 0$
        for $i,j,h,k=\{p,f\}$. Finally, 
        if $i,j,h=\{p,f\}$,for each $n=1, \ldots, N$, we have
	\begin{align}
	\label{eq:symetry-entries-wrt-Nover2}
	&\ex\{(\mathbf{Q_{ij}})^{n,n}\}=\ex\{(\mathbf{Q_{\tilde{i}\tilde{j}}})^{N+1-n,N+1-n}\} \\
	\label{eq:symetry-entries-Pi-Q}
& \ex\{\eta_N(\Pi_h\mathbf{Q_{ij}})^{n,n}\}=\ex\{\eta_N(\Pi_{\tilde{h}}\mathbf{Q_{\tilde{i}\tilde{j}}})^{N+1-n,N+1-n}\} \\		
	\label{eq:tr_q-pp=tr_q-ff}
	& \tr\ex\{\mathbf{Q_{ij}}\}=\tr\ex\{\mathbf{Q_{\tilde{i}\tilde{j}}}\},\\
	  &\tr\ex\{\eta_N\Pi_h\mathbf{Q_{ij}}\}=\tr\ex\{\eta_N\Pi_{\tilde{h}}\mathbf{Q_{\tilde{i}\tilde{j}}}\},
          \label{eq:symetry-Pi-Q}
	\end{align}
	where ``$\;\tilde{\;}\;$'' changes index to opposite: $p\rightarrow f, f\rightarrow p$.
\end{lemma}
The proof is postponed to the Appendix. To establish the first statements of the Lemma, we remark 
that for each $\theta$, the probability distribution of $(v_n)_{n \in \mathbb{Z}}$ coincides 
with the probability distribution of $(z_n)_{n \in \mathbb{Z}}$ where $z$ is chosen as 
$z_n = v_n e^{-in\theta}$ for each $n$. We use the same trick when $z_n = v_{-n+N+2L}$ for each $n$ to prove (\ref{eq:symetry-entries-wrt-Nover2})~--(\ref{eq:symetry-Pi-Q}). \\

We now establish that the diagonal matrices $\ex\{\eta_N(W_iW_i^*)^{-1}\}$ and $\ex\{\eta_N\Pi_{i}\}$ are multiples of the identity matrix up to error terms.  
\begin{lemma}
\label{le:expectation-WW-inverse}
	For $i=\{p,f\}$, we have:
	\begin{align}
	&\ex\{\eta_N(W_iW_i^*)^{-1}\}=\dfrac{1}{1-c_N}I_{ML}+\Ocal^{ML}\left(\dfrac{1}{N^{3/2}}\right)\label{le:estimation_inver_WW*}\\
	&\ex\{\eta_N\Pi_{i}\}=c_NI_N+\Ocal^{N}\left(\dfrac{1}{N^{3/2}}\right)\label{le:expectation_Pi}.
	\end{align} 
	Moreover, $(ML)^{-1}\tr\ex\{\eta_N(W_iW_i^*)^{-1}\}=(1-c_N)^{-1}+\Ocal(\frac{1}{N^2})$.
\end{lemma}
The proof of Lemma \ref{le:expectation-WW-inverse} uses the integration by parts formula and the 
Poincar\'e-Nash inequality, and is provided in the Appendix. \\


\subsubsection{Expression of matrix $\ex\{\mathbf{Q}\}$ obtained using the integration by parts formula}
\label{subsec:ipp-2}
We now establish that matrices ${\bf Q_{ij}}$ are, up to error terms, multiples of $I_N$, and characterize 
the asymptotic behaviour of their common diagonal terms. For this, we state the following Proposition that is proved
by using the integration by parts formula and the Poincar\'e-Nash inequality.
\begin{proposition}
\label{prop:use-ipp-np-Q}
The following equalities hold for each $z \in \mathbb{C}^{+}$. 
\begin{align}
\label{eq:integ_by_part_QppPip}
\ex\Big\{{\bf Q_{pp}}\eta\Pi_p\Big\}  = & c_N\ex\Big\{{\bf Q}_{{\bf pp}}\Big\}
-(1-c_N)\ex\Big\{\eta{\bf Q_{pp}}W_p^*(W_pW_p^*)^{-2}W_p\Big\}\dfrac{1}{N}\ex\Big\{\tr\left(\eta\Pi^\perp_p{\bf Q_{fp}}\right)\Big\}+{\bf \Delta_{pp}} \\
\label{eq:Q_pfP_p}
\ex\{{\bf Q_{pf}}\eta\Pi_p\} = & c_N\ex\Big\{{\bf Q}_{{\bf pf}}\Big\}
-(1-c_N)\ex\Big\{\eta{\bf Q_{pp}}W_p^*(W_pW_p^*)^{-2}W_p\Big\}\dfrac{1}{N}\ex\Big\{\tr\left(\eta\Pi^\perp_p{\bf Q_{ff}}\right)\Big\}
+{\bf \Delta_{pf}}
\end{align}
\begin{align}
\label{eq:integ_by_parts_QppPi_f}
\ex\Big\{{\bf Q_{pp}}\eta\Pi_f\Big\}
= & c_N\ex\Big\{{\bf Q}_{{\bf pp}}\Big\}
-(1-c_N)\ex\Big\{\eta{\bf Q_{pf}}W_f^*(W_fW_f^*)^{-2}W_f\Big\}\dfrac{1}{N}\ex\Big\{\tr\left(\eta\Pi^\perp_f{\bf Q_{pp}}\right)\Big\}
+{\bf \Delta^1_{pp}} \\
\label{eq:integ_by_parts_QpfPi_f}
\ex\Big\{{\bf Q_{pf}}\eta\Pi_f\Big\}= & c_N\ex\Big\{{\bf Q}_{{\bf pf}}\Big\}
-(1-c_N)\ex\Big\{\eta{\bf Q_{pf}}W_f^*(W_fW_f^*)^{-2}W_f\Big\}\dfrac{1}{N}\ex\Big\{\tr\left(\eta\Pi^\perp_f{\bf Q_{pf}}\right)\Big\}
+{\bf \Delta^1_{pf}}
\end{align}
where matrices ${\bf \Delta_{pp}}, {\bf \Delta_{pf}}, {\bf \Delta^1_{pp}}, {\bf \Delta^1_{pf}}$ are diagonal matrices 
whose entries are $\mathcal{O}_{z^2}\left(N^{-3/2}\right)$ terms, and whose normalized traces
are $\Ocal_{z^2}\left(N^{-2}\right)$ terms. 
\end{proposition}
(\ref{eq:integ_by_part_QppPip}) is proved in the Appendix. (\ref{eq:Q_pfP_p}, \ref{eq:integ_by_parts_QppPi_f}, 
\ref{eq:integ_by_parts_QpfPi_f}) are established similarly. \\

In order to introduce the next result, we denote by $w_N(z)$ the function defined by
\begin{equation}
  \label{eq:def-wN-section3}
  w_N(z) = 1 + \frac{1}{N} \mathbb{E}\{ \Tr(\eta_N \Pi_p^{\perp} {\bf Q_{fp}}) \} =  1 + \frac{1}{N} \mathbb{E}\{ \Tr(\eta_N \Pi_f^{\perp} {\bf Q_{pf}}) \}
\end{equation}
where the equality between the second term and the third term in (\ref{eq:def-wN-section3}) comes from
(\ref{eq:symetry-Pi-Q}). We claim that
\begin{equation}
  \label{eq:inv-c2-zw2-O}
\frac{1}{c_N^{2} - z^{2} w_N^{2}} = \mathcal{O}_{z^{2}}(1)
\end{equation}
To verify (\ref{eq:inv-c2-zw2-O}), we first notice that (\ref{eq:expre-alter-Qpf}) implies that $\frac{1}{N} \mathbb{E}\{ \Tr(\eta_N \Pi_f^{\perp} {\bf Q_{pf}}\Pi_f^{\perp}) \} = \frac{1}{N} \mathbb{E}\{ \Tr(\eta_N \Pi_f^{\perp} {\bf Q_{pf}}) \}$ (the equality follows from $\Pi_f^{\perp} = (\Pi_f^{\perp})^{2}$) coincides with the value taken at point $z^{2}$ by the Stieltjes transform
of a positive measure carried by $\mathbb{R}^{+}$. Proposition 5.1, item 4 in \cite{hachem-loubaton-najim-2007}
thus implies that function $-\left(z^{2} w_N(z)\right)^{-1}$ has the same property. Moreover, the converse of (\ref{eq:properties-calS(R)}, \ref{eq:ImzS}) in Proposition 4.1 in \cite{loubaton-tieplova-2020} leads to the conclusion that $-\left(z^{2}(w_N -\frac{c_N^{2}}{z^{2} w_N})\right)^{-1}$ also coincides with the value taken at point $z^{2}$ by the Stieltjes transform of a positive measure carried by $\mathbb{R}^{+}$. Writing $|c_N^{2} - z^{2} w_N^{2}|^{-1}$ as
\begin{align*}
\left|\dfrac{1}{c_N^2-z^2w_N^2}\right|=\left|\dfrac{1}{z^2w_N\left(-\frac{c_N^2}{z^2w_N}+w_N\right)}\right|=\left|\dfrac{1}{z^2w_N}\right||z|^2\left|\dfrac{1}{z^2(-\frac{c_N^2}{z^2w_N}+w_N)}\right|
\end{align*}
leads to (\ref{eq:inv-c2-zw2-O}). We are in position to precise the behaviour of the diagonal
matrices $\mathbb{E}({\bf Q_{ij}})$.
\begin{proposition}
  \label{prop:multiple_I}
  For $i=p,f$, and for $i \neq j$, we have
  \begin{eqnarray}
    \label{eq:expre-E(Qii)-section-3}
   \mathbb{E}(\mathbf{Q_{ii}}(z)) & = & \frac{z w^{2}_N(z)}{c_N^{2} - (z w_N(z))^{2}} \, I_N + \, \mathcal{O}^{N}_{z^{2}}\left(\frac{1}{N^{3/2}}\right) \\
   \label{eq:expre-E(Qij)-section-3}
   \mathbb{E}(\mathbf{Q_{ij}}(z)) & = & \frac{c_N w_N(z)}{c_N^{2} - (z w_N(z))^{2}} \, I_N  + \, \mathcal{O}^{N}_{z^{2}}\left(\frac{1}{N^{3/2}}\right)
  \end{eqnarray}
  where the normalized traces of the $\mathcal{O}^{N}_{z^{2}}\left(\frac{1}{N^{3/2}}\right) $ error terms are $\mathcal{O}_{z^{2}}\left(\frac{1}{N^{2}}\right)$ terms.  
\end{proposition}
{\bf Proof.} We just establish (\ref{eq:expre-E(Qii)-section-3}) for $i=p$ and
(\ref{eq:expre-E(Qij)-section-3}) for $i=p, j=f$ because, due to (\ref{eq:symetry-entries-wrt-Nover2}),  
(\ref{eq:expre-E(Qii)-section-3}) and (\ref{eq:expre-E(Qij)-section-3}) for $i=f$ and
for $i=f, j=p$, respectively, are consequences of  (\ref{eq:expre-E(Qii)-section-3}) for $i=p$ and
(\ref{eq:expre-E(Qij)-section-3}) for $i=p, j=f$. We consider Proposition \ref{prop:use-ipp-np-Q}, and begin by showing 
that the use of (\ref{eq:integ_by_part_QppPip}) and (\ref{eq:Q_pfP_p}) allows to obtain the following relationship 
between $\mathbb{E}({\bf Q_{pp}})$ and  $\mathbb{E}({\bf Q_{pf}})$
\begin{equation}
\label{eq:first-relation-EQpp-EQfp}
z \mathbb{E}({\bf Q_{pf}}(z)) \, w_N(z) = c \mathbb{E}({\bf Q_{pp}}(z)) + \Ocal_{z^2}^N\left(\dfrac{1}{N^{3/2}}\right) 
\end{equation}
where the normalized trace of the $\mathcal{O}^{N}_{z^{2}}\left(\frac{1}{N^{3/2}}\right) $ error term is a $\mathcal{O}_{z^{2}}\left(\frac{1}{N^{2}}\right)$ term. To check (\ref{eq:first-relation-EQpp-EQfp}), we first notice that (\ref{eq:expre-BQ-2}) and (\ref{eq:resolvent-identity-QA}) lead to the equality
\begin{align}
\label{eq:expr_QffPpperp}
{\bf Q_{ff}}\eta\Pi^\perp_p=z(\eta^2\Pi_f\Pi_p-z^2)^{-1}\eta\Pi^\perp_p
=z\left(-\dfrac{1}{z^2}\eta\Pi^\perp_p+\dfrac{1}{z^2}(\eta^2\Pi_f\Pi_p-z^2)^{-1}\eta^3\Pi_f\Pi_p\Pi^\perp_p\right)
=-\dfrac{1}{z}\eta\Pi^\perp_p
\end{align}
(\ref{eq:expre-E-trace-eta-Pip}) implies that $N^{-1}\mathbb{E}\left( \tr \eta \Pi_p^{\perp} \right) =(1-c_N) + \Ocal(\frac{1}{N^k})$ for each $k$. Therefore, (\ref{eq:expr_QffPpperp}) leads to $\ex\{N^{-1}\tr{\bf Q_{ff}}\eta\Pi^\perp_p\}=-\frac{(1-c_N)}{z}+\Ocal(\frac{1}{N^k})$ for each $k$. 
Moreover,  (\ref{eq:expre-BQ-2}) and $\Pi_p^{2} = \Pi_p$ lead to $E({\bf Q_{pf}} \eta \Pi_p) = 
E(\eta {\bf Q_{pf}})$, which, using again (\ref{eq:EetaX}), can also be written as 
$E( {\bf Q_{pf}}) + \mathcal{O}^{N}_{z^{2}}(\frac{1}{N^{k}})$ for each $k$. Therefore, 
(\ref{eq:Q_pfP_p}) implies that 
\begin{align*}
\ex\Big\{\eta{\bf Q_{pp}}W_p^*(W_pW_p^*)^{-2}W_p\Big\}= & \dfrac{z}{1-c_N}\ex\{{\bf Q_{pf}}\}-\dfrac{z}{(1-c_N)^2}{\bf \Delta_{pf}} \, + \\
    & \Ocal^{N}_{z^2}\left(\dfrac{1}{N^k}\right) + \ex\Big\{\eta{\bf Q_{pp}}W_p^*(W_pW_p^*)^{-2}W_p\Big\}\Ocal_{z^2}\left(\dfrac{1}{N^k}\right).
\end{align*}
It is easily seen that $\ex\{\eta{\bf Q_{pp}}W_p^*(W_pW_p^*)^{-2}W_p\}\Ocal_{z^2}(\frac{1}{N^k}) = \Ocal_{z^2}^{N}(\frac{1}{N^k})$. We next notice that (\ref{eq:expre-BQ-2}) implies that ${\bf Q_{pp}} \eta \Pi_p = z {\bf Q_{pf}}$. Plugging 
this and the above expression of $\ex\Big\{\eta{\bf Q_{pp}}W_p^*(W_pW_p^*)^{-2}W_p\Big\}$ into 
(\ref{eq:integ_by_part_QppPip}), we obtain easily (\ref{eq:first-relation-EQpp-EQfp}). Moreover, the property of the normalized trace of the error term in (\ref{eq:first-relation-EQpp-EQfp}) follows immediately from
$\frac{1}{N} \Tr {\bf \Delta_{pp}} = \mathcal{O}_{z^{2}}\left(\frac{1}{N^{2}}\right)$ and $\frac{1}{N} \Tr {\bf \Delta_{pf}} = \mathcal{O}_{z^{2}}\left(\frac{1}{N^{2}}\right)$. \\

We now use in a similar way (\ref{eq:integ_by_parts_QppPi_f}) and (\ref{eq:integ_by_parts_QpfPi_f}) to obtain another relationship between $\mathbb{E}({\bf Q_{pp}})$ and  $\mathbb{E}({\bf Q_{pf}})$. We first notice 
that by (\ref{eq:symetry-Pi-Q}) and (\ref{eq:expr_QffPpperp}),
\begin{equation}
  \label{eq:expre-trace-E-Qpp-Pif-perp}
  \ex\{N^{-1}\tr{\bf Q_{pp}}\eta\Pi_f^\perp\}=\ex\{N^{-1}\tr{\bf Q_{ff}}\eta\Pi_p^\perp\}= - \frac{(1-c_N)}{z}+\Ocal(\frac{1}{N^k})
  \end{equation}
for each $k$. We then remark that (\ref{le:expectation_Pi}) implies that  
\begin{multline*}
\ex\{{\bf Q_{pp}}\eta\Pi_f\}=z\ex\{(-\frac{1}{z^2}+\frac{1}{z^2}(\eta_N^2\Pi_p\Pi_f-z^2)^{-1}\eta_N^2\Pi_p\Pi_f)\eta_N\Pi_f\}
=-\frac{1}{z}\ex\{\eta_N\Pi_f\}\\
+\frac{1}{z}\ex\{{\bf Q_{pf}}\eta\Pi_f\}+\Ocal^{N}_{z^2}(N^{-k})=-\frac{c_N}{z}I_N+\frac{1}{z}\ex\{{\bf Q_{pf}}\eta\Pi_f\}+\Ocal^{N}_{z^2}(N^{-3/2})
\end{multline*} 
for each $k$. Moreover, it holds that $\ex\{{\bf Q_{pf}}\eta\Pi_f\}=\ex\{(\eta^2\Pi_p\Pi_f-z^2)^{-1}\eta^2\Pi_p\Pi_f\}=I_N+z\ex\{{\bf Q_{pp}}\}$. (\ref{eq:integ_by_parts_QppPi_f}) thus allows to obtain that
$$
(1-c_N)\ex\{\eta{\bf Q_{pf}}W_f^*(W_fW_f^*)^{-2}W_f\}=I_N + z \mathbb{E}({\bf Q_{pp}})+\Ocal^N_{z^2}(N^{-3/2})
$$
(\ref{eq:integ_by_parts_QpfPi_f}) eventually leads to
\begin{equation}
\label{eq:second-relation-EQpp-EQfp}
\left(I_N + z \mathbb{E}({\bf Q_{pp}})\right) \, w_N(z) = c_N \mathbb{E}({\bf Q_{pp}}) + \Ocal^N_{z^2}(N^{-3/2})
\end{equation}
where the normalized trace of the $\mathcal{O}^{N}_{z^{2}}\left(\frac{1}{N^{3/2}}\right) $ error term is a $\mathcal{O}_{z^{2}}\left(\frac{1}{N^{2}}\right)$ term. 
(\ref{eq:expre-E(Qii)-section-3}), (\ref{eq:expre-E(Qij)-section-3}), and the property of the normalized traces of the error terms  then follow from (\ref{eq:first-relation-EQpp-EQfp}) and (\ref{eq:second-relation-EQpp-EQfp}).
$\blacksquare$ \\

Finally, to complete this paragraph, we denote
\begin{align}
\tilde{{\bs\alpha}}_N=\dfrac{1}{N}\ex\{\tr{\bf Q_{pp}}\}=\dfrac{1}{N}\ex\{\tr{\bf Q_{ff}}\}\label{def_bold_tild_alpha}\\
{\bs\alpha}_N=\dfrac{1}{N}\ex\{\tr{\bf Q_{pf}}\}=\dfrac{1}{N}\ex\{\tr{\bf Q_{fp}}\}\label{def:bold_alpha}
\end{align}
and remark that taking the normalized traces of (\ref{eq:expre-E(Qii)-section-3}) and (\ref{eq:expre-E(Qij)-section-3}) implies that $\tilde{{\bs\alpha}}_N(z) =  \frac{z w^{2}_N(z)}{c_N^{2} - (z w_N(z))^{2}} + \mathcal{O}_{z^{2}}\left(\frac{1}{N^{2}}\right)$ and ${\bs\alpha}_N(z) = \frac{c_N w_N(z)}{c_N^{2} - (z w_N(z))^{2}} + \, \mathcal{O}^{N}_{z^{2}}\left(\frac{1}{N^{2}}\right)$. We have thus shown the following Corollary. 
\begin{corollary}
\label{coro:EboldQ}
  For $i=p,f$, and for $i \neq j$, we have
  \begin{eqnarray}
    \label{eq:expre-E(Qii)-section-3-bis}
   \mathbb{E}(\mathbf{Q_{ii}}(z)) & = & \tilde{{\bs\alpha}}_N(z) \, I_N + \, \mathcal{O}^{N}_{z^{2}}\left(\frac{1}{N^{3/2}}\right) \\
   \label{eq:expre-E(Qij)-section-3-bis}
   \mathbb{E}(\mathbf{Q_{ij}}(z)) & = & {\bs\alpha}_N(z)\, I_N  + \, \mathcal{O}^{N}_{z^{2}}\left(\frac{1}{N^{3/2}}\right)
  \end{eqnarray}
  where the normalized traces of the $\mathcal{O}^{N}_{z^{2}}\left(\frac{1}{N^{3/2}}\right) $ error terms are $\mathcal{O}_{z^{2}}\left(\frac{1}{N^{2}}\right)$. 
\end{corollary}

  We now establish a relationship between ${\bs\alpha}_N$ and $\tilde{{\bs\alpha}}_N$ and take benefit of this to show that  $\tilde{{\bs\alpha}}_N$ is a solution of a
  perturbed degree 2 polynomial equation. We will deduce from this that $\mathbb{E}\left( \frac{1}{N} \mathrm{Tr} Q_N(z) \right)$ verifies a similar equation. This property will be useful to evaluate the limit eigenvalue distribution of $\Pi_p \Pi_f$. We notice that
  \begin{align*}
\dfrac{1}{N}\tr\eta\Pi_f^\perp{\bf Q_{pp}}=\dfrac{1}{N}\tr\left(\eta{\bf Q_{pp}}-\eta_N\Pi_fz(\eta^2\Pi_{p}\Pi_f-z^2)^{-1}\right)=\dfrac{1}{N}\tr\left(\eta{\bf Q_{pp}}-z{\bf Q_{fp}}\right) 
\end{align*}
Taking the expectation from the both sides, using (\ref{eq:tr_q-pp=tr_q-ff}) and replacing $\eta$ by $1$ in $\dfrac{1}{N}\tr\left(\eta{\bf Q_{pp}}\right)$, we get that 
\begin{align}\label{eq:tr_PifQpp}
\dfrac{1}{N}\ex\{\eta\tr\Pi_f^\perp{\bf Q_{pp}}\}=\tilde{{\bs\alpha}}-z{\bs\alpha}+\Ocal_{z^2}\left(\dfrac{1}{N^k}\right)
\end{align}
for each $k$. (\ref{eq:expre-trace-E-Qpp-Pif-perp}) thus implies that
\begin{align}
\label{eq:connection_bold_alph_tilde-alph}
{\bs\alpha}_N(z)=\dfrac{\tilde{{\bs\alpha}}_N(z)}{z}+\dfrac{1-c_N}{z^2}+\Ocal_{z^2}\left(\dfrac{1}{N^k}\right)
\end{align}
Taking the normalized trace of
(\ref{eq:first-relation-EQpp-EQfp}) leads to
\begin{equation}
  \label{eq:trace-first-relation-EQpp-EQfp}
  c_N \tilde{{\bs\alpha}}_N(z) = z {\bs \alpha}_N(z) w_N(z) + \Ocal_{z^2}\left(\dfrac{1}{N^2}\right)
  \end{equation}
We now express $w_N$ in terms of  ${\bs\alpha}_N$ and $\tilde{{\bs\alpha}}_N$. 
For this we use ${\bf Q_{fp}}=\eta\Pi_f(\eta^2\Pi_p\Pi_f-z^2)^{-1}$ and write
\begin{multline}\label{eq:Pi_pperpQfp}
N^{-1}\ex\{\tr(\eta\Pi_p^\perp{\bf Q_{fp}})\}=N^{-1}\ex\{\tr({\eta\bf Q_{fp}})\}-N^{-1}\ex\{\tr(\eta^2\Pi_p\Pi_f(\eta^2\Pi_p\Pi_f-z^2)^{-1})\}\\
={\bs\alpha}-1-zN^{-1}\ex\{\tr({\bf Q_{pp}})\}+\Ocal_{z^2}\left(\dfrac{1}{N^k}\right)={\bs\alpha}-1-z\tilde{{\bs\alpha}}+\Ocal_{z^2}\left(\dfrac{1}{N^k}\right)
\end{multline}
Therefore, we obtain that $w(z) = {\bs\alpha}(z)-z\tilde{{\bs\alpha}}(z)+\Ocal_{z^2}\left(\dfrac{1}{N^k}\right)$ for each $k$. Plugging this into (\ref{eq:trace-first-relation-EQpp-EQfp})
and using (\ref{eq:connection_bold_alph_tilde-alph}), we obtain after some algebra that
\begin{align}
  \label{eq:final_tilde-bold-alpha}
(1-z^2)\tilde{{\bs \alpha}}_N^2+\left(\dfrac{2(1-c_N)}{z}-z\right)\tilde{{\bs \alpha}}_N+\dfrac{(1-c_N)^2}{z^2}=\Ocal_{z^2}\left(\dfrac{1}{N^2}\right).
\end{align}
We define $\tilde{\alpha}_N(z)$ and $\alpha_N(z)$ by 
\begin{align}
\label{eq:def-nonboldalphatilde}
\tilde{\alpha}_N(z) = \frac{1}{N}\ex\{\tr Q_N(z)\} \\
\alpha_N(z) = \frac{1}{N} \ex\{\tr \eta\Pi_pQ_N(z)\}
\end{align}
for each $z \in \mathbb{C}\setminus \mathbb{R}^{+}$. (\ref{eq:expre-BQ-2}) implies that 
$\tilde{{\bs \alpha}}_N(z)=z\tilde{\alpha}_N(z^2)$ and ${\bs \alpha}_N(z)=\alpha_N(z^2)$ if 
$\Im z \neq 0$ or equivalently if $z^{2} \in \mathbb{C} \setminus \mathbb{R}^{+}$. Therefore, 
we deduce from (\ref{eq:final_tilde-bold-alpha}) that $\tilde{\alpha}_N(z)$ is a solution of 
the perturbed equation 
\begin{align*}
(1-z^2)z^2\tilde{ \alpha}_N^2(z^2)+\left(2(1-c_N)-z^2\right)\tilde{ \alpha}_N(z^2)+\dfrac{(1-c_N)^2}{z^2}=\Ocal_{z^2}\left(\dfrac{1}{N^2}\right).
\end{align*}
The l.h.s of this equation is a function of $z^2 \in  \mathbb{C} \setminus \mathbb{R}^{+}$, thus the error term at the r.h.s is also a function of $z^2$. By exchanging $z^2$ with $z$ we have
\begin{align}\label{eq:tilde_alpha}
(1-z)z\tilde{ \alpha}_N^2(z)+\left(2(1-c_N)-z\right)\tilde{ \alpha}_N(z)+\dfrac{(1-c_N)^2}{z}=\Ocal_{z}\left(\dfrac{1}{N^2}\right)
\end{align}
on $ \mathbb{C} \setminus \mathbb{R}^{+}$. Moreover, from (\ref{eq:connection_bold_alph_tilde-alph}), we obtain that
\begin{align}
  \label{eq:connection_alph_tilde-alph}
\alpha_N(z)=\tilde{\alpha}_N(z)+\dfrac{1-c_N}{z}+\Ocal_{z}\left(\dfrac{1}{N^k}\right).
\end{align}
on $\mathbb{C} \setminus \mathbb{R}^{+}$ for each integer $k \geq 1$. 
\begin{remark}
\label{eq:alpha-stieltjes}
	Corollary~\ref{cor:tr_Qpf-Stj_transf} implies that $\alpha_N$ is the Stieltjes transform of a positive measure carried by $\mathbb{R}^+$ with mass $c_N+\Ocal_{z}(N^{-k})$. This is in accordance with (\ref{eq:connection_alph_tilde-alph}) because $\tilde{\alpha}_N$ is the Stieltjes transform of a probability measure carried by $\mathbb{R}^{+}$ (i.e. the expectation of the empirical eigenvalue distribution of $\eta^{2} \Pi_p \Pi_f$) and $-\frac{1-c_N}{z}$ is the Stieltjes transform of measure $(1-c_N)\delta_0$.
\end{remark}

\subsubsection{Limiting distribution and almost sure localisation of the eigenvalues of $\Pi_p \Pi_f$}
\label{sec:stieltjes}
In this paragraph, we evaluate the almost sure asymptotic behaviour of the empirical eigenvalue distribution $\hat{\nu}_N$ of matrix $\Pi_p \Pi_f$. As $\eta_N = 1$ almost surely for $N$ large
enough, this can be done by evaluating the almost sure behaviour of $\frac{1}{N} \mathrm{Tr}(Q_N(z))$ where we recall that $Q_N(z)$ is the resolvent of the regularized
matrix $\eta_N^{2} \Pi_p \Pi_f$. We first notice that,  in conjunction with the Borel-Cantelli Lemma,  Lemma \ref{le:estim_var}, Eq. (\ref{eq:var_trQ}), applied for $i=j=p$ and $F=I$, implies immediately
that
\begin{equation}
  \label{eq:np-normalized-trace-resolvent-PipPif}
  \frac{1}{N} \mathrm{Tr}(Q_N(z)) - \mathbb{E}\left(\frac{1}{N} \mathrm{Tr}(Q_N(z))\right) \rightarrow 0 \; a.s.
\end{equation}
for each $z \in \mathbb{C} \setminus \mathbb{R}^{+}$. We are thus back to the evaluation of the asymptotic behaviour of $\tilde{\alpha}_N(z)= \mathbb{E}\left(\frac{1}{N} \mathrm{Tr}(Q_N(z))\right)$.
For this, we introduce the probability measure $\tilde{\nu}_N$ defined by
\begin{equation}
  \label{eq:def-convolutive-free-product}
\tilde{\nu}_N=(c_N\delta_1+(1-c_N)\delta_0)\boxtimes(c_N\delta_1+(1-c_N)\delta_0)  
\end{equation}
where  $\boxtimes$ represents the free multiplicative convolution product operator (see e.g. \cite{voiculescu-al-1992} Section 3.6). We recall that if $\Pi_1$ and $\Pi_2$ are orthogonal projection matrices onto the rows of
two mutually independent random Gaussian $ML \times N$ matrices with i.i.d. standard Gaussian entries, then the results of \cite{voiculescu-al-1992}
imply that the empirical eigenvalue distribution of $\Pi_1 \Pi_2$ has the same asymptotic behaviour than $\tilde{\nu}_N$. In the following,
we establish that, while $\Pi_p$ and $\Pi_f$ are not generated as $\Pi_1$ and $\Pi_2$, $\hat{\nu}_N$ behaves as $\tilde{\nu}_N$. \\

For this, we denote by $\tilde{t}_N$ the Stieltjes transform of $\tilde{\nu}_N$. The expression and the properties of $\tilde{t}_N$ and of
$\tilde{\nu}_N$ are well-known, see for example Example~3.6.7. \cite{voiculescu-al-1992}. If $z\in\mathbb{C^+}$, $\tilde{t}_N$ is given by
\begin{align}
\label{eq:expre-tN-section3}   
\tilde{t}_N(z)=\dfrac{z-2(1-c_N)+\sqrt{z(z-4c_N(1-c_N))}}{2(1-z)z},
\end{align}
where we define function $z\mapsto\sqrt{z}$ for $z=|z|e^{i\theta}$, $\theta\in[0,2\pi)$ as $\sqrt{z}=\sqrt{|z|}e^{i\theta/2}$. In particular, if $x\in\mathbb{R}^{+}$ and $z=xe^{i\theta}$ then $\sqrt{z}\xrightarrow[\theta\searrow0]{}\sqrt{x}$ and $\sqrt{z}\xrightarrow[\theta\nearrow2\pi]{}-\sqrt{x}$. Then one can easily obtain that  $\lim_{z\rightarrow x,z\in\mathbb{C^+}}\tilde{t}_N(z)$ exists for $x\in(-\infty,0)\cap(4c_N(1-c_N),+\infty)$ and $x\ne1$. This limit is still denoted $\tilde{t}_N(x)$, and 
\begin{align}\label{eq:formula_tilde_t}
\tilde{t}_N(x)=\begin{cases}
\dfrac{x-2(1-c_N)-\sqrt{x(x-4c_N(1-c_N))}}{2(1-x)x},\quad x<0\\
\dfrac{x-2(1-c_N)+\sqrt{x(x-4c_N(1-c_N))}}{2(1-x)x},\quad x>4c_N(1-c_N), x\ne1
\end{cases}
\end{align}
Moreover,  $\tilde{\nu}_N=(c_N\delta_1+(1-c_N)\delta_0)\boxtimes(c_N\delta_1+(1-c_N)\delta_0)$ is given by
\begin{align}
d\tilde{\nu}_N(\lambda)=\dfrac{\sqrt{\lambda(4c_N(1-c_N)-\lambda)}}{2\pi\lambda(1-\lambda)}{\bf 1}_{[0,4c_N(1-c_N)]}d\lambda+(1-c_N)\delta_{0}+\max(2c_N-1,0)\delta_{1}.
\end{align}
The support of  $\tilde{\nu}_N$, denoted by $\Scal_N$, is thus given by 
\begin{equation}
  \label{eq:def-Scal-section3}
        \Scal_N =   [0, 4c_N(1-c_N)]\cup \{ 1 \} \, {\bf 1}_{c_N>1/2}.
\end{equation}
Finally, $\tilde{t}_N$  satisfies the equation (\ref{eq:tilde_alpha}), but in which the term $\Ocal_z(N^{-2})$ is replaced by 0, i.e.
\begin{align}\label{eq:eq_tilde_t_N}
z(1-z)\tilde{t}_N^2(z)+(2(1-c_N)-z)\tilde{t}_N(z)+\dfrac{(1-c_N)^2}{z}=0
\end{align}
a property which suggests that $\tilde{\alpha}_N(z) - \tilde{t}_N(z) \rightarrow 0$. In order to establish this formally, we establish the
following Proposition.
\begin{proposition}
  \label{prop:behaviour-hatnuN-section3}
  $\tilde{\alpha}_N(z)$ can be written as
  \begin{equation}
    \label{eq:decomposition-haagerup-tildealpha}
    \tilde{\alpha}_N(z) = \tilde{t}_N(z) + \tilde{r}_N(z),
  \end{equation}
  where $\tilde{r}_N$ is holomorphic in $\mathbb{C} \setminus \mathbb{R}^{+}$, and verifies
  \begin{equation}
    \label{eq:evaluation-reminder-r}
  |r_N(z)| \leq \frac{1}{N^{2}} P_1(|z|) P_2\left(\frac{1}{\Im z}\right)
  \end{equation}
  for each $z \in \mathbb{C}^{+}$, where $P_1$ and $P_2$ are two nice polynomials.
  \end{proposition}
 The proof is given in the Appendix. \\

As $c_N \rightarrow c_*$, $\tilde{t}_N(z) \rightarrow \tilde{t}_*(z)$ where $\tilde{t}_*(z)$ is obtained 
from $\tilde{t}_N(z)$ by replacing $c_N$ by $c_*$ in Eq. (\ref{eq:expre-tN-section3}). $\tilde{t}_*$ is 
of course the Stieljes transform of the measure $\tilde{\nu}_*$ given by 
\begin{align}
d\tilde{\nu}_*(\lambda)=\dfrac{\sqrt{\lambda(4c_*(1-c_*)-\lambda)}}{2\pi\lambda(1-\lambda)}{\bf 1}_{[0,4c_*(1-c_*)]}d\lambda+(1-c_*)\delta_{0}+\max(2c_*-1,0)\delta_{1}
\end{align}
and the support $\Scal_*$ of $\tilde{\nu}_*$ is obtained by replacing $c_N$ by $c_*$ in 
(\ref{eq:def-Scal-section3}). 
Sequence $(\tilde{\nu}_N)_{N \geq 1}$ of course converges weakly towards the probability measure
$\tilde{\nu}_*$. We deduce from this and from Proposition \ref{prop:behaviour-hatnuN-section3} the following Theorem
which states that $(\hat{\nu}_N)_{N \geq 1}$ converges weakly almost surely towards $\tilde{\nu}_*$. 
Moreover, all the eigenvalues of $\Pi_p \Pi_f$ are almost surely localised in a neighbourhood of 
 $\Scal_*$.

\begin{theorem}
  \label{th:no_eigen-2}
	The empirical eigenvalue distribution $\hat{\nu}_N$ of $\Pi_{p,N}\Pi_{f,N}$ verifies 
	\begin{equation}
	\label{eq:weak-convergence-hatnu_2}
	\hat{\nu}_N \rightarrow \tilde{\nu}_*
	\end{equation}
	weakly almost surely. If $c_* < \frac{1}{2}$, for each $\epsilon > 0$, almost surely, for each $N$ large 
enough, all the eigenvalues of $\Pi_p \Pi_f$ belong to $[0, 4c_*(1-c_*) + \epsilon]$. If 
$c_* > \frac{1}{2}$, $1$ is eigenvalue of $\Pi_p \Pi_f$ with multiplicity $2ML -  N$, and for each $\epsilon > 0$, the $2(N-ML)$
remaining eigenvalues are almost surely located in $[0, 4c_*(1-c_*) + \epsilon]$ for $N$ large enough.
\end{theorem}
{\bf Proof.} (\ref{eq:np-normalized-trace-resolvent-PipPif}) and Proposition  \ref{prop:behaviour-hatnuN-section3} imply that
\begin{equation}
\label{eq:convergence-empirical-distribution_2}
\frac{1}{N} \mathrm{Tr}(Q_N(z)) -  \tilde{t}_N(z) \rightarrow 0 \; a.s.
\end{equation}
for each $z \in \mathbb{C}^{+}$. As $\tilde{t}_N(z) \rightarrow \tilde{t}_*(z)$ on $\mathbb{C}^{+}$, 
we obtain that $\frac{1}{N} \mathrm{Tr}(Q_N(z)) \rightarrow  \tilde{t}_*(z) \rightarrow 0$ almost surely 
on $\mathbb{C}^{+}$, and that (\ref{eq:weak-convergence-hatnu_2}) holds. \\

We remark that if $c_* = \frac{1}{2}$, the support  $\Scal_*$ of $\tilde{\nu}_*$ is equal to the whole interval 
$[0,1]$. As we know that the eigenvalues of $\Pi_p \Pi_f$ belong to $[0,1]$, the knowledge of $\Scal_*$ does not provide any valuable information of the almost sure location of these eigenvalues if $c_* = \frac{1}{2}$. If 
$c_* \neq \frac{1}{2}$, the almost sure localisation of the eigenvalues of $\Pi_p \Pi_f$ can be established using the Haagerup-Thornbjornsen approach (\cite{HT:05}) using decomposition
(\ref{eq:decomposition-haagerup-tildealpha}) of $\tilde{\alpha}_N(z)$. As the corresponding proof is rather standard, we just provide a sketch of proof. We first mention that (\ref{eq:decomposition-haagerup-tildealpha}) 
implies that if $\psi$ is a $\mathcal{C}_{\infty}$ function constant on the complementary of a compact subset, 
then, we have 
\begin{equation}
\label{eq:EtracepsiPipPif}
\mathbb{E}\left( \Tr(\psi(\Pi_p \Pi_f)) \right) = N \; \int_{\mathcal{S}_N} \psi(\lambda) \, d\tilde{\nu}_N(\lambda) + \mathcal{O}(\frac{1}{N})
\end{equation}
(see Proposition 6.2 in \cite{HT:05} or Proposition 4.6 in \cite{capitaine-donati-martin-feral-2009}). 
If moreover $\psi$ vanishes on $\mathcal{S}_N$, we obtain that 
\begin{equation}
\label{eq:EtracepsiPipPif-2}
\mathbb{E}\left( \Tr(\psi(\Pi_p \Pi_f)) \right) = \mathcal{O}(\frac{1}{N})
\end{equation}
while if $\psi'$ vanishes on  $\mathcal{S}_N$, the 
Poincar\'e-Nash inequality allows to establish that 
$ \Tr(\psi(\Pi_p \Pi_f) - \mathbb{E}\left( \Tr(\psi(\Pi_p \Pi_f)) \right)   \rightarrow 0$ almost surely. 
Therefore, (\ref{eq:EtracepsiPipPif}) implies that $\Tr(\psi(\Pi_p \Pi_f) - N \, \int_{\mathcal{S}_N} \psi(\lambda) \, d\tilde{\nu}_N(\lambda)\rightarrow 0$ almost surely if  $\psi'$ vanishes on  $\mathcal{S}_N$. 
We consider $\epsilon > 0$ small enough, and a function $\psi_1 \in \mathcal{C}_{\infty}$ that verifies:
\begin{align*}
\psi_1(\lambda) = & 1 \; \mbox{if $\lambda \in \left( [0, 4c_*(1-c_*) + \epsilon] \cup [1-\epsilon,1+\epsilon] \mathbf{1}_{c_*>1/2} \right)^{c}$} \\
\psi_1(\lambda) = & 0 \; \mbox{if $\lambda \in [0, 4c_*(1-c_*) + \epsilon/2] \cup [1-\epsilon/2,1+\epsilon/2] \mathbf{1}_{c_*>1/2}$} \\
\psi_1(\lambda) \in & [0,1] \; \mbox{elsewhere}
\end{align*}
As $c_N \rightarrow c_*$, $\psi_1$ (and therefore $\psi_1'$) vanishes on $\mathcal{S}_N$ for $N$ large enough, so that
$\Tr(\psi_1(\Pi_p \Pi_f)) \rightarrow 0$. 
The number of eigenvalues of $\Pi_p \Pi_f$ located into $ \in \left( [0, 4c_*(1-c_*) + \epsilon] \cup [1-\epsilon,1+\epsilon] \mathbf{1}_{c_*>1/2} \right)^{c}$ is clearly less than $\Tr(\psi_1(\Pi_p \Pi_f))$ which converges towards
$0$. Therefore, almost surely, for each $N$ large enough, all the eigenvalues of $\Pi_p \Pi_f$ 
belong to  $[0, 4c_*(1-c_*) + \epsilon] \cup [1-\epsilon,1+\epsilon] \mathbf{1}_{c_*>1/2}$. This completes 
the proof of Theorem \ref{eq:convergence-empirical-distribution_2} when $c_* < 1/2$. In order to 
address the case $c_* > 1/2$, we consider a function $\psi_2 \in \mathcal{C}_{\infty}$ satisfying
\begin{align*}
\psi_2(\lambda) = & 1 \; \mbox{if $\lambda \in [1-\epsilon, 1+\epsilon]$} \\
\psi_2(\lambda) = & 0 \; \mbox{if $\lambda \in [1-2 \epsilon, 1+ 2 \epsilon]^{c}$} \\
\psi_2(\lambda) \in & [0,1] \; \mbox{elsewhere}
\end{align*}
$\psi_2'$ vanishes of $\mathcal{S}_N$, and 
$\int_{\mathcal{S}_N} \psi_2(\lambda) \, d\tilde{\nu}_N(\lambda) = 2c_N - 1$. Therefore, we obtain that 
$ \Tr(\psi_2(\Pi_p \Pi_f)) - (2ML-N) \rightarrow 0$ almost surely. 
As there is no eigenvalue of $\Pi_p \Pi_f$ in $[1-2\epsilon,1-\epsilon)$, 
$ \Tr(\psi_2(\Pi_p \Pi_f))$ coincides with the number of eigenvalues of $\Pi_p \Pi_f$ 
located into $[1-\epsilon,1]$. As $ \Tr(\psi_2(\Pi_p \Pi_f)) -  (2ML-N) \rightarrow 0$ almost surely, 
we obtain that for $N$ large enough, $\Pi_p \Pi_f$ has $2ML - N$ eigenvalues located in 
$[1-\epsilon,1]$. Lemma \ref{le:1eigenvaluePipPif} implies that $1$ is eigenvalue of  $\Pi_p \Pi_f$ with multiplicity $2ML - N$, from which we
get that if $c_* > 1/2$, the eigenvalues of $\Pi_p \Pi_f$ belong to $[0, 4 c_*(1-c_*)+\epsilon] \cup \{ 1 \}$.  
$\blacksquare$ \\

In the following, it will be useful to introduce the measure $\nu_N$ defined by
\begin{equation}
  \label{eq:definition-tN}
  \nu_N=\frac{1}{c_N}\tilde{\nu}_N-\frac{1-c_N}{c_N}\delta_0 
=\dfrac{\sqrt{\lambda(4c_N(1-c_N)-\lambda)}}{2\pi c_N \lambda(1-\lambda)}{\bf 1}_{[0,4c_N(1-c_N)]}d\lambda+\max(2c_N-1,0)\delta_{1}
\end{equation}
It is easily seen that $\nu_N$ is the  probability measure carried by $\Scal_N$ with Stieltjes transform $t_N(z)$ 
defined on $\mathbb{C} \setminus \Scal_N$ by 
\begin{equation}
\label{eq:def-tN-section3}
t_N(z) = \frac{\tilde{t}_N(z)}{c_N}+\frac{1-c_N}{c_Nz}
\end{equation}
After some algebra, we obtain that
\begin{align}\label{eq:formula_t}
&t_N(z)=\dfrac{z(2c_N-1)+\sqrt{z(z-4c_N(1-c_N))}}{2c_N(1-z)z},\quad z\in\mathbb{C^+}\notag\\
&t_N(x)=\begin{cases}
\dfrac{x(2c_N-1)-\sqrt{x(x-4c_N(1-c_N))}}{2c_N(1-x)x},\quad x<0\\
\dfrac{x(2c_N-1)+\sqrt{x(x-4c_N(1-c_N))}}{2c_N(1-x)x},\quad x>4c_N(1-c_N), x\ne1
\end{cases}
\end{align}
We also define ${\bf\tilde{t}}_N(z)=z\tilde{t}_N(z^2)$ and $\t_N(z)=t_N(z^2)$ which are related by
\begin{align}
  \label{eq:connect_bolt_t_tilde-t}
  \t_N(z)=\dfrac{\tilde{ \t}_N(z)}{c_Nz}+\dfrac{1-c_N}{c_Nz^2}.
\end{align}
(\ref{eq:property-zs(z2)}) 
implies that ${\bf\tilde{t}}_N$ is the Stieljes transform of a probability measure whose support is clearly the set ${\bs\Scal}_N$
defined by
\begin{equation}
  \label{eq:def-boldS}
        {\bs\Scal}_N =   [-\sqrt{4c_N(1-c_N)},\sqrt{4c_N(1-c_N)}]\cup\{\pm1\}{\bf 1}_{c_N>1/2}
\end{equation}
While $\t_N$ is not a Stieltjes transform, we however mention that $\t_N$ is also holomorphic outside ${\bs\Scal}_N$.
Then, we deduce from (\ref{eq:connection_alph_tilde-alph}) and (\ref{eq:decomposition-haagerup-tildealpha}) the following obvious,
but useful properties.
\begin{corollary}
  \label{re:conver_bold}
The sequence $(\alpha_N(z))_{N \geq 1}$ verifies
	$$\alpha_N(z)-c_Nt_N(z)\rightarrow0$$
  for $z\in\mathbb{C} \setminus \mathbb{R}^{+}$. Moreover, we also have 
	\begin{align}
	&{\bs \alpha}_N(z)-c_N\t_N(z) \rightarrow 0 \\
	&\tilde{{\bs \alpha}}_N(z)-{\bf\tilde{t}}_N(z)\rightarrow 0
	\end{align}
        on $\mathbb{C}^{+}$.
\end{corollary}
We also denote by $\nu_*$ and $t_*(z)$ the limits of $\nu_N$ and $t_N(z)$ when $N \rightarrow +\infty$, i.e. 
their expressions are obtained by replacing $c_N$ by $c_*$ in (\ref{eq:definition-tN}) and 
(\ref{eq:formula_t}). We have of course $\tilde{\nu}_*  = c_N \nu_* + (1-c_*) \delta_0$. 
We also remark that if $\hat{\nu}_N^{\prime}$ represents the eigenvalue distribution of 
matrix $(W_p W_p^{*})^{-1/2} W_p W_f^{*} (W_f W_f^{*})^{-1} W_f W_p^{*} (W_p W_p^{*})^{-1/2}$, then $\hat{\nu}_N = c_N \hat{\nu}_N^{\prime} + (1-c_N) \delta_0$. Therefore, 
the relation $\tilde{\nu}_*  = c_N \nu_* + (1-c_*) \delta_0$ and the convergence result (\ref{eq:weak-convergence-hatnu_2}) imply that 
\begin{equation}
\label{eq:convergence-hatnuprime}
\hat{\nu}_N^{\prime} \rightarrow \nu_*
\end{equation}
almost surely. \\

In the following, we also denote by $\t_*(z)$ and ${\bf\tilde{t}}_*(z)$ the functions $t_*(z^{2})$ and $z\tilde{t}_*(z^2)$ respectively, that can also be
seen as the limits of $\t_N(z)$ and ${\bf\tilde{t}}_N(z)$ when $N \rightarrow +\infty$. ${\bf\tilde{t}}_*(z)$ is of course the
Stieltjes transform of a probability measure carried by the set $ {\bs\Scal}_*$ obtained by replacing $c_N$ by
$c_*$ in (\ref{eq:def-boldS}). \\

We finally conclude this section by a result which can be seen as the counterpart of Lemma \ref{le:extra-properties} derived in Section 
\ref{sec:autocov_signal}.
\begin{lemma}
  \label{le:limit_Qpp_Qpf}
	For each $z \in \mathbb{C}\setminus{\bs\Scal}_*$, $i\ne j\in\{p,f\}$ and for each 
	bounded sequences $(a_N, b_N)_{N \geq 1}$ of $N$--dimensional deterministic vectors, it holds that 
	\begin{align}
          \label{eq:convergence-sesquilinear-boldQii}
	  &	a_N^{*} \, ({\bf Q_{ii}})_N(z) \, b_N- {\bf\tilde{t}}_N(z)a_N^{*}b_N \rightarrow 0 \;  \mbox{almost surely} \\
           \label{eq:convergence-sesquilinear-boldQij}
	&	a_N^{*} \, ({\bf Q_{ij}})_N(z)  \, b_N - c_N{\bf t}_N(z)a_N^{*}b_N\rightarrow 0 \;  \mbox{almost surely}
	\end{align}
	Moreover, these convergences hold uniformly on each compact subset of $\mathbb{C}\setminus{\bs\Scal}_*$. The
	properties are still valid if $a_N,b_N$ are random bounded vectors that are 
	independent from the noise sequence $(v_n)_{n \geq 1}$. 
\end{lemma}
The proof is given in the Appendix.

\subsection{ In the presence of signal}\label{sec:cor_signal}

In this section  we assume that signal $(u_n)_{n \in \mathbb{Z}}$ is present, and evaluate its influence 
on the eigenvalues of matrix $\Pi_{p,y}\Pi_{f,y}$. For this,  we notice that  matrices
$\Pi_{p,y}$ and $\Pi_{f,y}$ are finite rank perturbation of matrices $\Pi_{p,v}$ and $\Pi_{f,v}$ due to the noise $(v_n)_{n \in \mathbb{Z}}$,. Therefore, 
$\Pi_{p,y} \Pi_{f,y}$ is itself a finite rank perturbation of $\Pi_{p,v} \Pi_{f,v}$ 
We can thus use the same approach as in the previous chapter. Since the  useful signal 
$(u_n)_{n \in \mathbb{Z}}$ is generated by the same minimal state-space representation (\ref{eq:state-space}), we  keep the notations from the Section~\ref{sec:signal_model}.  As before, we denote $\Sigma_{i,N}=\frac{Y_{i,N}}{\sqrt{N}}=W_{i,N}+ \frac{U_{i,N}}{\sqrt{N}}$. $\Pi_{i,y}$ and $\Pi_{i,v}$ are denoted respectively 
$\Pi_{i}$ and $\Pi_{i}^W$ for $i=p,f$ from now on. We remind that in the presence of signal, we cannot assume that $R_N=I_M$, thus $W_{i}=(I_L\otimes R_N)^{1/2}W_{i,iid}$
where matrix $W_{i,iid}$ is built from i.i.d. $\mathcal{N}_c(0,I_M)$ distributed random vector 
$(v_{n,iid})_{n=1, \ldots, n+2L-1}$. However, we recall that $\Pi_i^{W} = \Pi_i^{W_{iid}}$ for 
$i=p,f$. In the following, we will denote by $\eta_N$ (rather than $\eta_{N,iid}$) the regularization term defined by (\ref{eq:def_eta}) by replacing $W$ by $W_{iid}$ in order to simplify the notations. 

We also keep Assumptions~\ref{as:state-space-equation} and \ref{as:rank-O-H}, as well as Assumption~\ref{as:signal} on the limits of $\Delta_N$ and $\Theta_N^{*} \R_{f|p,N}^{L} \Theta_N$ related to the signal model. As in Section~\ref{sec:asym_beh_eigenval}, we derive the following results under condition (\ref{eq:entries-Delta*-different}), and briefly justify that Theorem \ref{th:fundamental-theorem-section3} remains valid if some of the entries of matrix 
$\Delta_*$ coincide. Finally, it appears that the more involved  Assumptions~\ref{as:convergence-alpha},  \ref{as:support-mu}, and ~\ref{as:signal-bruit} are not needed here and can be replaced with the following  milder one. \\

\begin{assumption}\label{as:theta_R_theta}
	$r\times r$ matrix $G_N=\Theta_N^*(I_L\otimes R_N^{-1})\Theta_N$ converge towards some matrix $G_*$.
\end{assumption} 

We now take benefit of Proposition \ref{prop:properties-RUhat-Rfphat} to evaluate the behaviour of the 
canonical correlation coefficients between the row spaces of matrices $U_{p,N}$ and $U_{f,N}$ when $N \rightarrow +\infty$. For this, we recall that $\Gamma_*$ represents the limit of $\Theta_N^{*} R_{f|p,N}^{L} \Theta_N$, as well as, under condition (\ref{eq:entries-Delta*-different}), the limit of
of $\Gamma_N = \Delta_N \tilde{\Theta}^{*}_{f,N} \Theta_{p,N} \Delta_N$ (see Eq. (\ref{eq:def-Gamma}) 
for the definition of $\Gamma_N$). As $\Delta_N \rightarrow \Delta_* > 0$, $\tilde{\Theta}^{*}_{f,N} \tilde{\Theta}_{p,N}$ 
converges towards the matrix $\Omega_*$ given by 
\begin{equation}
\label{eq:def-Omega*}
\Omega_* = \Delta_*^{-1} \Gamma_* \Delta_*^{-1}
\end{equation}
$\Omega_*$ of course verifies $\| \Omega_* \| \leq 1$ and $\mathrm{Rank}(\Omega_*) = P$. \\

We are now in position to formulate the main result of this Section. For this, we denote by $F_*$ the 
rank $P$ $r \times r$ matrix defined by 
\begin{equation}
\label{eq:def-F*}
F_*=\Omega^*_*(I_r+\Delta_*^{-1}G_*^{-1}\Delta_*^{-1})^{-1}\Omega_*(I_r+\Delta_*^{-1}G_*^{-1}\Delta_*^{-1})^{-1}
\end{equation}
As matrix $\Omega_*$ verifies $\| \Omega_* \| \leq 1$, matrix $F_*$ satisfies $\| F_* \| < 1$. 
Moreover, the eigenvalues of $F_*$ are real and belong to $[0,1)$. 
\begin{theorem}
\label{th:fundamental-theorem-section3}
\begin{itemize} 
\item The function $f_*(x)$ defined by 
\begin{equation}
\label{eq:def-f*}
f_*(x) =  x \, \left( \frac{\tilde{t}_*(x)}{(1-c_*)t_*(x)}\right)^{2}
\end{equation}
is strictly increasing on $[4c_*(1-c_*), 1]$, verifies $f_*(4c_*(1-c_*)) = \frac{c_*}{1-c_*}$, $f(1) = 1$
if $c_* < \frac{1}{2}$ and $f(1) = \left( \frac{c_*}{1-c_*} \right)^{2}$ if $c_* > \frac{1}{2}$. 
\item If $c_* \geq \frac{1}{2}$, the equation
\begin{equation}
\label{eq:equation-spiked-eigenvalues}
\mathrm{det} \left( f_*(x) \, I_r - F_* \right) = 0
\end{equation}
has no solution in $(4c_*(1-c_*), 1)$, and for each $\delta > 0$, almost surely, for $N$ large enough, all the eigenvalues of $\Pi_p \Pi_f$ belong to $[0, 4c_*(1-c_*)+\delta] \cup [1-\delta,1]$. Among the eigenvalues contained in $[1 -\delta, 1]$, $2ML -N + \mathcal{O}(1)$ are equal to 1, and, possibly,  $o(N)$ other eigenvalues 
converge towards $1$. 
\item If $c_* < \frac{1}{2}$, 
the equation (\ref{eq:equation-spiked-eigenvalues}) has $0 \leq s \leq P$ solutions that belong to $(4c_*(1-c_*),1)$ where $s$ is  the number of eigenvalues (taking into account the multiplicities) of $F_*$ that are striclty larger than $\frac{c_*}{1-c_*} < 1$. If $\rho_{1,*}, \ldots, \rho_{s,*}$ are the 
corresponding solutions, then the $s$ largest eigenvalues of $\Pi_p \Pi_f$ converge almost surely towards $\rho_{1,*}, \ldots, \rho_{s,*}$, and, for each $\delta > 0$, almost surely, for $N$ large enough, the remaining $N - s$ ones belong to $[0, 4c_*(1-c_*)+\delta]$.  
\end{itemize}
\end{theorem}
{\bf Proof.} The properties of function $f_*$ are proved in the Appendix. $x \in (4c_*(1-c_*),1)$ 
is solution of equation (\ref{eq:def-f*}) if and only $f_*(x)$ coincides with one of the eigenvalues of 
$F_*$. If $c_* \geq  \frac{1}{2}$, 
$f_*(x) \in [\frac{c_*}{1-c_*}, \left( \frac{c_*}{1-c_*} \right)^{2}]$ if $x \in [4c_*(1-c_*), 1]$. As $\frac{c_*}{1-c_*} \geq 1$ and the eigenvalues of $F_*$ belong to $[0,1)$, equation (\ref{eq:def-f*}) has no solution in $(4c_*(1-c_*), 1)$. If $c_* < \frac{1}{2}$, $f_*\left((4c_*(1-c_*), 1\right)$ coincides with the interval $(\frac{c_*}{1-c_*}, 1)$. Therefore, equation (\ref{eq:def-f*}) has $s$ solutions, where $s$ represents the number of 
eigenvalues of $F_*$ strictly larger than $\frac{c_*}{1-c_*}$. \\

We now establish the last statements of Theorem \ref{th:fundamental-theorem-section3} related to 
the possible eigenvalues of $\Pi_{p}\Pi_{f}$ that escape from  $\Scal_*=[0,4c_*(1-c_*)]\cup\{1\}{\bf 1}_{c_*>1/2}$. We first present the general approach 
of the proof. As before, we study the squares of the positive eigenvalues of the linearised version $\begin{pmatrix}
0&\Pi_p\\\Pi_f&0
\end{pmatrix}$ that escape form $[0,2\sqrt{c_*(1-c_*)}]\cup\{1\}{\bf 1}_{c_*>1/2}$. For this, for each $\delta >0$ small enough, we
consider  $y \in (\sqrt{4c_*(1-c_*)+\delta},1-\delta)$ if $c_* > \frac{1}{2}$ and 
$y \in (\sqrt{4c_*(1-c_*)+\delta},1]$ if $c_* < \frac{1}{2}$, which by Theorem \ref{th:no_eigen-2}, cannot be, almost surely, for $N$ large enough, an eigenvalue of matrix  $\begin{pmatrix}
0&\Pi_p^{W}\\\Pi_f^{W}&0
\end{pmatrix}$.  We take benefit of this property to express  $\mathrm{det}  \begin{pmatrix} -yI_{N}&\Pi_p\\ \Pi_f&-yI_{N} \end{pmatrix}$ in terms of $\mathrm{det}  \begin{pmatrix} -yI_{N}&\Pi_p^{W}\\ \Pi_f^{W} &-yI_{N} \end{pmatrix}$ and of the resolvent of matrix $\begin{pmatrix} 0 &\Pi_p^{W}\\ \Pi_f^{W} & 0 \end{pmatrix}$ evaluated at $y$, which is well defined. As we establish almost sure convergence results in the following, we notice that the regularisation term $\eta_N$ defined by (\ref{eq:def_eta}) by exchanging $W_i$ by $W_{i,iid}$, $i=p,f$, can be considered to be equal to 1. Therefore, the later resolvent coincides with ${\bf Q^{W}}(y) = {\bf Q^{W_{iid}}}(y)$ defined by (\ref{eq:expre-BQ-2}) for $z=y$. We then evaluate the asymptotic behaviour of  $\mathrm{det}  \begin{pmatrix} -yI_{N}&\Pi_p\\ \Pi_f&-yI_{N} \end{pmatrix}$, and deduce from this the last statements of Theorem \ref{th:fundamental-theorem-section3}. \\

The key point is to use
that $\begin{pmatrix} -yI_{N}&\Pi_p\\ \Pi_f&-yI_{N} \end{pmatrix}$ is a low rank perturbation of 
$\begin{pmatrix} -yI_{N}&\Pi_p^{W}\\ \Pi_f^{W} &-yI_{N} \end{pmatrix}$. In order to evaluate the 
corresponding low-rank matrix, we have first to evaluate $\Pi_i - \Pi_i^{W}$ for $i=p,f$. It is easy to see that $\Sigma_i\Sigma_i^*$ can be expressed as
\begin{align*}
\Sigma_i \Sigma_i^*=W_iW_i^*+(W_i\tilde{\Theta}_i\Delta_i,\Theta_i )\begin{pmatrix}
0&I_r\\
I_r&\Delta_p^2
\end{pmatrix}\begin{pmatrix}
\Delta_i\tilde{\Theta}_i^*W_i^*\\
\Theta^*
\end{pmatrix}
\end{align*}
where we recall that $\frac{U_i}{\sqrt{N}} = \Theta_i \Delta_i \tilde{\Theta}_i^{*}$ 
is the singular value decomposition of $\frac{U_i}{\sqrt{N}}$ (see Eq. (\ref{eq:svd-Up-Uf})). 

We first establish that, almost surely, for $N$ large enough, matrix $\Sigma_i \Sigma_i^*$ is invertible. 
For this, we need the following Lemma proved in the Appendix.
\begin{lemma}
\label{le:existence-Di}
We define $E_i$ as the $2r \times 2r$ matrix given by 
\begin{align}\label{eq:def_D}
E_i= \begin{pmatrix}
0&I_r\\
I_r&\Delta_i^2
\end{pmatrix}^{-1} \left(I_{2r}+\begin{pmatrix}
0&I_r\\
I_r&\Delta_i^2
\end{pmatrix} \begin{pmatrix}
\Delta_i\tilde{\Theta}_i^*\Pi_i^W\tilde{\Theta}_i\Delta_i&\Delta_i\tilde{\Theta}_i^*W_i^*(W_iW_i^*)^{-1}\Theta_i\\
\Theta_i^*(W_iW_i^*)^{-1}W_i\tilde{\Theta}_i\Delta_i&\Theta_i^*(W_iW_i^*)^{-1}\Theta_i
\end{pmatrix}\right) 
\end{align}
Then, we have 
\begin{equation}
\label{eq:convergence-Ei}
E_i-\begin{pmatrix}
		-(1-c_N)\Delta_N^2&I_r\\
		I_r&\frac{1}{1-c_N}\Theta_N^*(I_L\otimes R_N^{-1})\Theta_N
		\end{pmatrix}\rightarrow 0 \; \mbox{almost surely}
\end{equation}
\end{lemma}
The determinant of the second term of the left hand side of (\ref{eq:convergence-Ei}) is equal to 
$$
\mathrm{det}\left( -(1-c_N) \Delta_N^2 \right) \, \mathrm{det} \left( \frac{1}{1-c_N} ( \Theta_N^*(I_L\otimes R_N^{-1})\Theta_N + \Delta_N^{-2}) \right)
$$
and thus converges towards a non zero term. Therefore, almost surely, for $N$ large enough, 
matrix $E_i$ is invertible. In the following, we denote by $D_i = E_i^{-1}$ the inverse of $E_i$. 
The  Woodbury's identity implies that $\Sigma_i\Sigma_i^*$ is almost surely invertible for each 
$N$ large enough, and that 
\begin{align*}
(\Sigma_i\Sigma_i^*)^{-1}=(W_iW_i^*)^{-1}-((W_iW_i^*)^{-1}W_i\tilde{\Theta}_i\Delta_i,(W_iW_i^*)^{-1}\Theta_i)D_i\begin{pmatrix}
\Delta_i\tilde{\Theta}_i^*W_i^*(W_iW_i^*)^{-1}\\
\Theta_i^*(W_iW_i^*)^{-1}
\end{pmatrix},
\end{align*}
After some algebra, we obtain that
\begin{align*}
\Pi_i-\Pi_i^W=-\Acal_iD_i\Acal_i^*,
\end{align*}
where
\begin{align}\label{eq:def_Acal}
\Acal_i=(-\Pi_i^{W,\perp}\tilde{\Theta}_i\Delta_i,W_i^*(W_iW_i^*)^{-1}\Theta_i)
\end{align}
From this, we immediately get that
\begin{align}\label{eq:expression_matrix_PipPif}
\begin{pmatrix}
-yI_{N}&\Pi_p\\
\Pi_f&-yI_{N}
\end{pmatrix}
=\begin{pmatrix}
-yI_{N}&\Pi^W_p\\
\Pi^W_f&-yI_{N}
\end{pmatrix}-\begin{pmatrix}
\Acal_p&0\\
0&\Acal_f
\end{pmatrix}\begin{pmatrix}
D_p&0\\
0&D_f
\end{pmatrix}\begin{pmatrix}
0&\Acal_p^*\\
\Acal_f^*&0
\end{pmatrix}
\end{align}
or equivalently
\begin{align}\label{eq:expr_yPipPif}
\begin{pmatrix}
-yI_{N}&\Pi_p\\
\Pi_f&-yI_{N}
\end{pmatrix}
=\begin{pmatrix}
-yI_{N}&\Pi^W_p\\
\Pi^W_f&-yI_{N}
\end{pmatrix}\left(I_{2N}-{\bf Q^W}(y)\begin{pmatrix}
\Acal_p&0\\
0&\Acal_f
\end{pmatrix}\begin{pmatrix}
D_p&0\\
0&D_f
\end{pmatrix}\begin{pmatrix}
0&\Acal_p^*\\
\Acal_f^*&0
\end{pmatrix}\right)
\end{align}
Therefore, $y$ is an eigenvalue of $\begin{pmatrix}
0&\Pi_p\\
\Pi_f&0
\end{pmatrix}$ if and only if the determinant of the second term at the r.h.s. of  (\ref{eq:expr_yPipPif}) vanishes, or equivalently if
\begin{align}
\det\left(I_{2r}-\begin{pmatrix}
\Acal_p^*{\bf Q^W_{fp}}(y)\Acal_p&\Acal_p^*{\bf Q^W_{ff}}(y)\Acal_f\\
\Acal_f^*{\bf Q^W_{pp}}(y)\Acal_p&\Acal_f^*{\bf Q^W_{pf}}(y)\Acal_f
\end{pmatrix}\begin{pmatrix}
D_p&0\\
0&D_f
\end{pmatrix}\right)=0
\end{align}
or 
\begin{align}\label{eq:det_D-AQ}
\det\left(\begin{pmatrix}
E_p&0\\
0&E_f
\end{pmatrix}-\begin{pmatrix}
\Acal_p^*{\bf Q^W_{fp}}(y)\Acal_p&\Acal_p^*{\bf Q^W_{ff}}(y)\Acal_f\\
\Acal_f^*{\bf Q^W_{pp}}(y)\Acal_p&\Acal_f^*{\bf Q^W_{pf}}(y)\Acal_f
\end{pmatrix}\right)=0
\end{align}
We now establish that for each $y \in (\sqrt{4c_*(1-c_*)},1)$, the left hand side 
of (\ref{eq:det_D-AQ}) converges towards a deterministic term. In particular, we have the following result. 
\begin{lemma}\label{le:asympt}
	For each $z\in\mathbb{C}\setminus{\bs \Scal}_*$, where ${\bs \Scal}_*=(-2\sqrt{c_*(1-c_*)},\,2\sqrt{c_*(1-c_*)} )\cup\{\pm1\}  {\bf 1}_{c_*>1/2}$ and $i\ne j\in\{p,f\}$ we have:
	\begin{itemize}
		\item $\Acal_i^*{\bf Q^W_{ji}}\Acal_i
		-\begin{pmatrix}
		-\dfrac{(1-c_N)(1+z{\bf\tilde{t}}_N(z))}{z{\bf\tilde{t}}_N(z)+1-c_N}\Delta_N^2&0\\
		0&\dfrac{1+{\bf \tilde{t}}_N(z)z}{c_N(1-c_N)}\Theta_N^*(I_L\otimes R_N^{-1})\Theta_N
		\end{pmatrix}\rightarrow 0$ almost surely
		\item	$\Acal_f^*{\bf Q^W_{pp}}\Acal_p-\begin{pmatrix}
		-\dfrac{(1-c_N)^2}{z^2{\bf \tilde{t}}_N(z)}\Gamma_N&0\\0&0
		\end{pmatrix}\rightarrow 0$ almost surely
		\item $\Acal_p^*{\bf Q^W_{ff}}\Acal_f-\begin{pmatrix}
		-\dfrac{(1-c_N)^2}{z^2{\bf \tilde{t}}_N(z)}\Gamma^*_N&0\\0&0
		\end{pmatrix}\rightarrow 0$ almost surely
	\end{itemize}
	Moreover, almost surely, the three convergence items hold uniformly on each compact subset of $\mathbb{C}\setminus{\bs \Scal}_*$. 	
\end{lemma}
{\bf Proof.} The proof of this Lemma is postponed to the Section~\ref{sec:lemma_asympt}. \\

We remind that $\Theta_N^*(I_L\otimes R_N^{-1})\Theta_N$ is denoted by $G_N$. After trivial algebra,  Lemma~(\ref{le:asympt}) implies that asymptotically, for $N\rightarrow\infty$, the "limiting form" of  Eq. (\ref{eq:det_D-AQ}) is
\begin{align}
\det\begin{pmatrix}
\dfrac{(1-c_N)c_N}{z{\bf\tilde{t}}_N(z)+1-c_N}\Delta_N^{2}&I_r&\dfrac{(1-c_N)^2}{z^2{\bf \tilde{t}}_N(z)}\Gamma^*_N&0\\
I_r& -\dfrac{1-c_N+z{\bf\tilde{t}}_N(z)}{c_N(1-c_N)}G_N&0&0\\
\dfrac{(1-c_N)^2}{z^2{\bf \tilde{t}}_N(z)}\Gamma_N&0&\dfrac{(1-c_N)c_N}{z{\bf\tilde{t}}_N(z)+1-c_N}\Delta_N^{2}&I_r\\
0&0&I_r& -\dfrac{1-c_N+z{\bf\tilde{t}}_N(z)}{c_N(1-c_N)}G_N
\end{pmatrix}=0
\end{align}

Replacing  $z{\bf\tilde{t}}_N(z)+1-c_N$ by $z^2c_N\t_N(z)$ (see (\ref{eq:connect_bolt_t_tilde-t})) and taking the limits of the various terms when $N\rightarrow+\infty$ (due to Assumptions~\ref{as:rank-O-H}, \ref{as:signal}, \ref{as:theta_R_theta}), we can expect that the solutions of equation (\ref{eq:det_D-AQ}) tend to the solutions of the limiting equation, i.e. 
\begin{align}\label{eq:det_limit}
\det\begin{pmatrix}
\dfrac{1-c_*}{y^2\t_*(y)}\Delta_*^{2}&I_r&\dfrac{(1-c_*)^2}{y^2{\bf \tilde{t}}_*(y)}\Gamma^*_*&0\\
I_r& -\dfrac{y^2\t_*(y)}{1-c_*}G_*&0&0\\
\dfrac{(1-c_*)^2}{y^2{\bf \tilde{t}}_*(y)}\Gamma_*&0&\dfrac{1-c_*}{y^2\t_*(y)}\Delta_*^{2}&I_r\\
0&0&I_r& -\dfrac{y^2\t_*(y)}{1-c_*}G_*
\end{pmatrix}=0.
\end{align} 

We now study the solutions of  (\ref{eq:det_limit}). If we interchange the second and third row blocks and second  and third column blocks, the determinant will not change and using the Schur complement formula, the l.h.s. of (\ref{eq:det_limit}) becomes
\begin{multline*}
\det\begin{pmatrix}
-\dfrac{y^2\t_*(y)}{1-c_*}G_*&0\\
0& -\dfrac{y^2\t_*(y)}{1-c_*}G_*
\end{pmatrix} \times \\
\det\Bigg[\begin{pmatrix}
\dfrac{1-c_*}{y^2\t_*(y)}\Delta_*^{2}&\dfrac{(1-c_*)^2}{y^2{\bf \tilde{t}}_*(y)}\Gamma^*_*\\
\dfrac{(1-c_*)^2}{y^2{\bf \tilde{t}}_*(y)}\Gamma_*&\dfrac{1-c_*}{y^2\t_*(y)}\Delta_*^{2}
\end{pmatrix}
-\begin{pmatrix}
-\dfrac{y^2\t_*(y)}{1-c_*}G_*&0\\
0& -\dfrac{y^2\t_*(y)}{1-c_*}G_*
\end{pmatrix}^{-1}\Bigg]    
\end{multline*}

Since $\det\begin{pmatrix}
-\frac{y^2\t_*(y)}{1-c_*}G_*&0\\
0& -\frac{y^2\t_*(y)}{1-c_*}G_*
\end{pmatrix}\ne0$,  Eq. (\ref{eq:det_limit}) is equivalent to
\begin{align*}
\det\begin{pmatrix}
\dfrac{1-c_*}{y^2\t_*(y)}(\Delta_*^{2}+G_*^{-1})&\dfrac{(1-c_*)^2}{y^2{\bf \tilde{t}}_*(y)}\Gamma^*_*\\
\dfrac{(1-c_*)^2}{y^2{\bf \tilde{t}}_*(y)}\Gamma_*&\dfrac{1-c_*}{y^2\t_*(y)}(\Delta_*^{2}+G_*^{-1})\end{pmatrix}=0
\end{align*}
Using again the Schur complement formula, we obtain that the limiting form of Eq. (\ref{eq:det_D-AQ}) is 
\begin{align*}
\det\left(\dfrac{(1-c_*)^2}{y^4\t_*^2(y)}(\Delta_*^{2}+G_*^{-1})-\dfrac{(1-c_*)^4}{y^4{\bf \tilde{t}}^2_*(y)}\Gamma^*_*(\Delta_*^{2}+G_*^{-1})^{-1}\Gamma_*\right)=0,
\end{align*}
or equivalently,
\begin{align}\label{eq:det_bold}
\det\left(\dfrac{1}{(1-c_*)^2}\dfrac{{\bf \tilde{t}}^2_*(y)}{\t_*^2(y)}-\Gamma^*_*(\Delta_*^{2}+G_*^{-1})^{-1}\Gamma_*(\Delta_*^{2}+G_*^{-1})^{-1}\right)=0.
\end{align}
We write that ${\bf \tilde{t}}_*(y)=y\tilde{t}_*(y^2)$ and $\t_*(y)=t_*(y^2)$, and put $x = y^{2} \in (4c_*(1-c_*), 1)$. Then, using (\ref{eq:def-Omega*}), Eq. (\ref{eq:det_bold}) leads to equation (\ref{eq:equation-spiked-eigenvalues}). \\

In order to complete the proof of Theorem \ref{th:fundamental-theorem-section3}, it remains to resort to the stability arguments in \cite{benaych-nadakuditi-ann-math} and \cite{chapon-couillet-hachem-mestre}. For this, it is sufficient to use exactly the same arguments as in the proof of Corollary \ref{coro:final-result}. We thus omit 
the details. We just justify the statements related to the number of eigenvalues located 
into $[1-\delta, 1]$ when $c_* > \frac{1}{2}$. Lemma \ref{le:1eigenvaluePipPif} implies that 
$1$ is eigenvalue of $\Pi_p^{W} \Pi_f^{W}$  with multiplicity 
$2ML -N$. As $\Pi_p \Pi_f$ is a finite rank perturbation of $\Pi_p^{W} \Pi_f^{W}$, $1$ is eigenvalue of $\Pi_p \Pi_f$ with a multiplicity equal to  $2ML - N + \mathcal{O}(1)$. 
The stability arguments in \cite{benaych-nadakuditi-ann-math} and \cite{chapon-couillet-hachem-mestre} 
do not preclude the existence of other eigenvalues of $\Pi_p \Pi_f$ that converge towards $1$. As the eigenvalue 
distribution of $\Pi_p \Pi_f$ has the same limit as the eigenvalue distribution of  $\Pi^{W}_p \Pi^{W}_f$, i.e. measure $\tilde{\nu}_*$, for each $\delta > 0$ small enough, 
$\frac{1}{N} \# \{ \lambda_i(\Pi_p \Pi_f) \in [\delta,1] \} \rightarrow \tilde{\nu}_*([\delta,1]) = 2 c_* - 1$. 
Therefore, the number of remaining eigenvalues converging towards $1$ is a $o(N)$ term, as expected.  
$\blacksquare$ \\

Theorem \ref{th:fundamental-theorem-section3} allows to derive immediately the conditions under which it is possible to estimate consistently $P$ by the number of eigenvalues of $\Pi_p \Pi_f$ that escape from $\Scal_*$. 
\begin{corollary}
  \label{coro:consistent-estimation-P}
  $P$ coincides with the number of eigenvalues that escape from $\Scal_*$ if and only if $c_* < \frac{1}{2}$ and if the $P$ non zero eigenvalues of $F_*$  are strictly larger than $\frac{c_*}{1-c_*}$
\end{corollary}
The condition that the non zero eigenvalues of $F_*$ are bigger than $\frac{c_*}{1-c_*}$ implies that the singular values of $\Omega_*$ and the eigenvalues of $\Delta_*$ are large enough. In practice, 
this means that the canonical correlation coefficients between the past and the future of $u$ are large enough (thus making the singular values of $\Omega_*$ large) and the $r$ eigenvalues of $R_{u,N}^{L}$ are also large 
enough (thus making matrix $\Delta_*^{-1}$ small). It is interesting to notice that if $c_* > \frac{1}{2}$, the largest eigenvalues of $\Pi_p \Pi_f$ cannot be used to estimate $P$. \\

We finally mention that, as in the context of Corollary \ref{coro:final-result-finite-N}, Theorem \ref{th:fundamental-theorem-section3} can be formulated in terms of the finite $N$ equivalents of 
matrix $F_*$ and function $f_*(z)$ defined by 
\begin{equation}
    \label{eq:def-finite-N-version-F*}
    F_N = \Delta_N^{-1} \Gamma_N \Delta_N^{-1} (I_r + \Delta_N^{-1} G_N^{-1} \Delta_N^{-1})^{-1} \Delta_N^{-1} \Gamma_N \Delta_N^{-1} (I_r + \Delta_N^{-1} G_N^{-1} \Delta_N^{-1})^{-1} 
\end{equation}
and 
\begin{equation}
\label{eq:def-finite-N-version-f*}
f_N(x) = x \, \left( \frac{\tilde{t}_N(x)}{(1-c_N)t_N(x)}\right)^{2}
\end{equation}
It is easily seen that the properties of function $f_N$ are similar to the properties of $f_*$ stated in item (i) of Theorem \ref{th:fundamental-theorem-section3}, parameter $c_N$ replacing $c_*$. We thus have the following 
result.  
\begin{corollary}
\label{coro:final-result-finite-N-canonical-correlation}
If $c_* < \frac{1}{2}$, and if $\delta > 0$ is small enough, for $N$ large enough, $s$ coincides with the number of solutions of the equation $\mathrm{det}(f_N(x) - F_N) = 0$ that belong to 
$(4c_N(1-c_N)+\delta,1)$, as well as with the number of eigenvalues of $F_N$ that are strictly 
larger than $\frac{c_N}{1-c_N}+\kappa < 1$ for some $\kappa > 0$ small enough. If $\rho_{1,N}, \ldots, \rho_{s,N}$ are the corresponding solutions, then $\rho_{1,N}, \ldots, \rho_{s,N}$ converge almost surely towards $\rho_{1,*}, \ldots, \rho_{s,*}$. The 
$s$ largest eigenvalues of $\Pi_p \Pi_f$ have the same asymptotic behaviour than $\rho_{1,N}, \ldots, \rho_{s,N}$, 
and for each $\delta > 0$, almost surely, for $N$ large enough, the remaining $N-s$ ones belong to $[0,4c_N(1-c_N)+\delta]$.
\end{corollary}

\begin{figure}[ht!]
	\centering
	\par
	\includegraphics[scale=0.4]{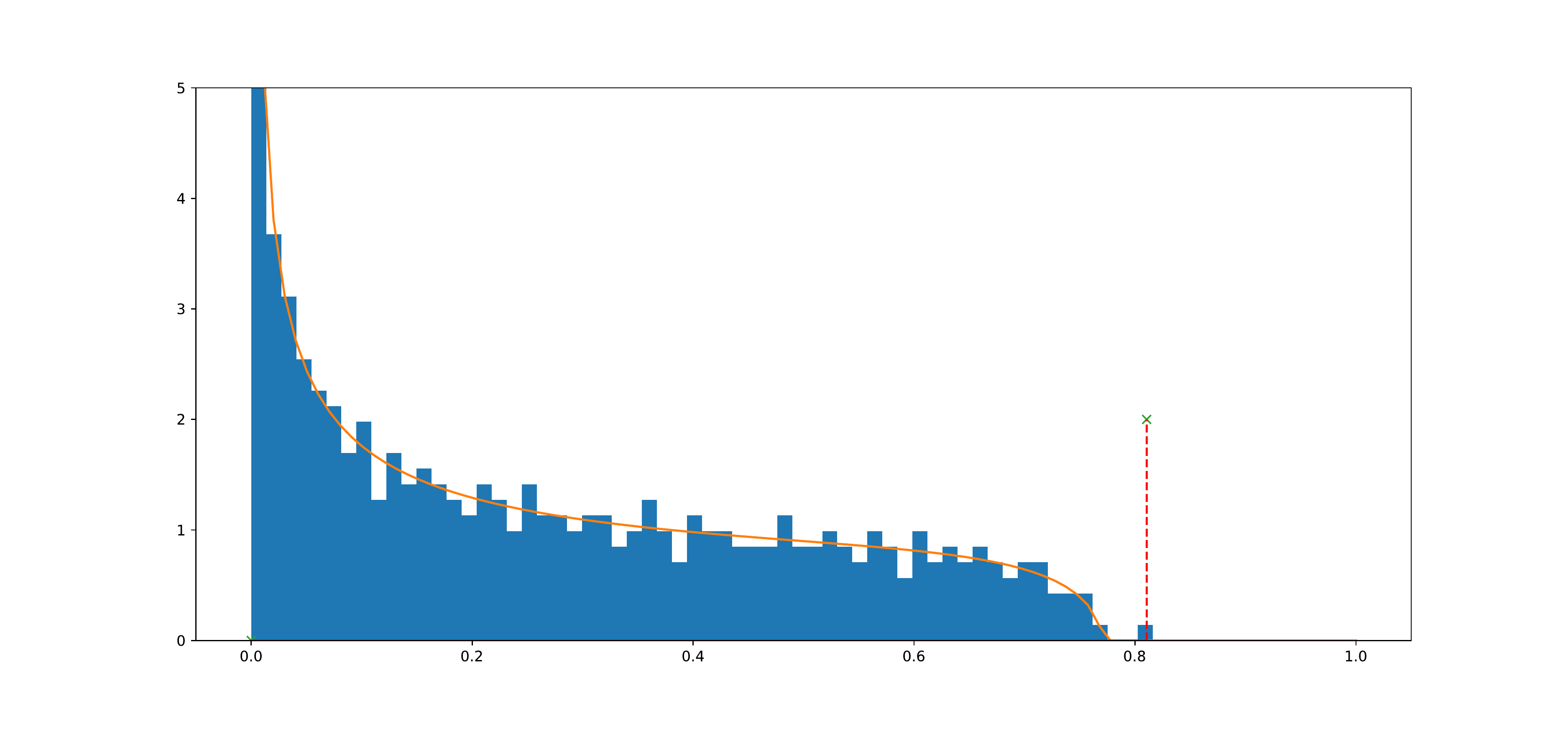}
	\caption{Histogram of the eigenvalues and graph of the density of $\nu_N$ with 1 outlier }
	\label{fig:out_1}
\end{figure}

\begin{figure}[ht!]
	\centering
	\par
	\includegraphics[scale=0.4]{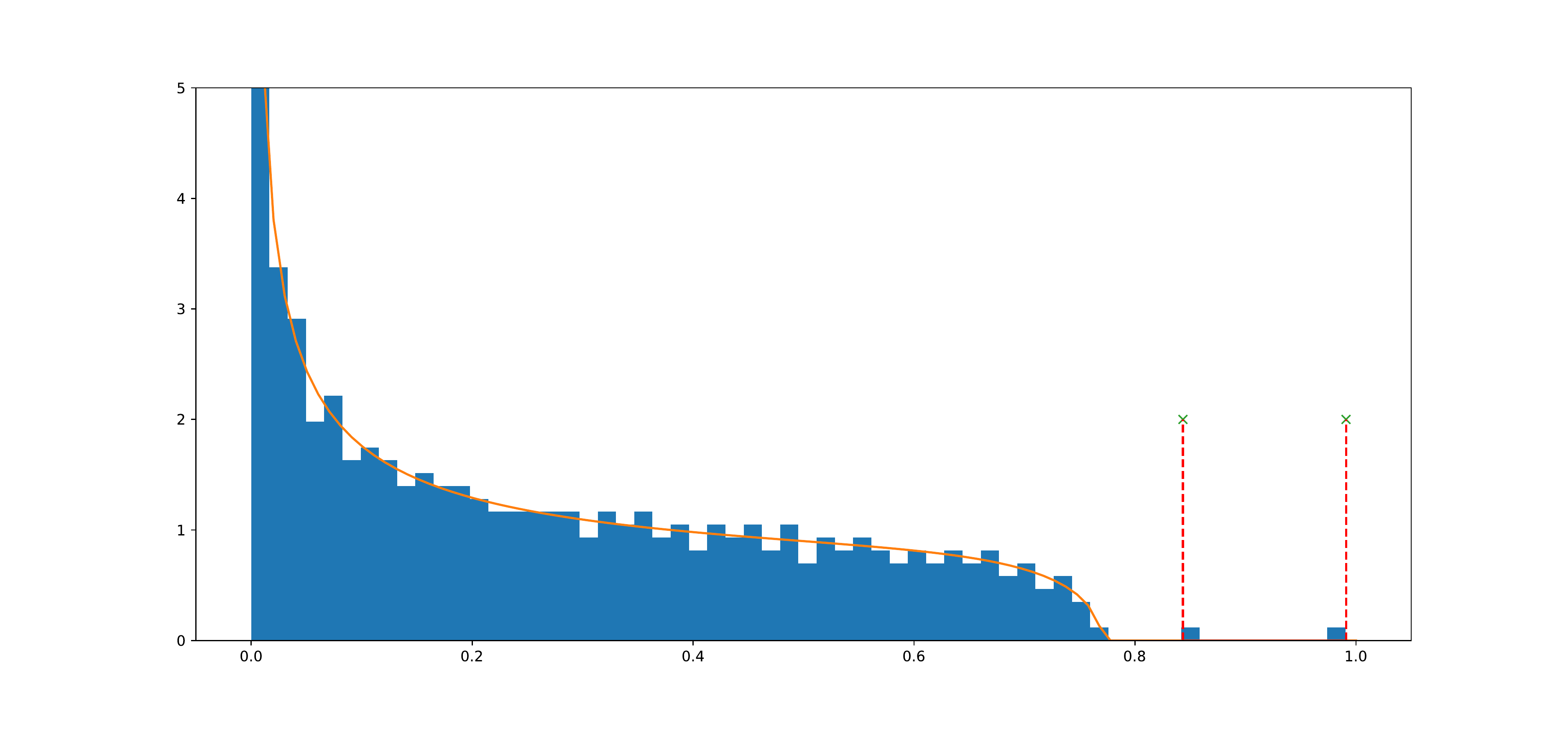}
	\caption{Histogram of the eigenvalues and graph of the density of $\nu_N$ with 2 outliers}
	\label{fig:out_2}
\end{figure}

We illustrate the above discussion by numerical experiments showing that eigenvalues outside the bulk  indeed tend to thesolutions of equation (\ref{eq:equation-spiked-eigenvalues}). We consider a simple case, when $P=2$, $K=1$ and $A$ is diagonal with eigenvalues $a_1$ and $a_2$. Figures \ref{fig:out_1}, \ref{fig:out_2} represent histograms 
of the eigenvalues of realizations of the matrix $(\hat{R}^{L}_{f,y})^{-1/2} \hat{R}^{L}_{f|p,y} (\hat{R}^{L}_{p,y})^{-1} \hat{R}_{f|p,y}^{L*} (\hat{R}^{L}_{f,y})^{-1/2}$, as well as the graph of the density of measure $\nu_N=\frac{1}{c_N}\tilde{\nu}_N-\frac{1-c_N}{c_N}\delta_0$ and the solutions of equation (\ref{eq:equation-spiked-eigenvalues}).

We take $N=2000$, $M=130$ and $L=4$, so $c_N=0.26$. The eigenvalues of matrix $R_N$ are defined by $\lambda_{k,N}  = 1/2 + \frac{\pi}{4} \cos\left(\frac{\pi (k-1)}{2M}\right)$ for $k=1, \ldots, M$, so that matrix $R_N$ verifies $\frac{1}{M} \mathrm{Tr}(R_N) \simeq 1$.
Figure  \ref{fig:out_1} corresponds to a choice of $(a_1,a_2)$ for which $s=1$, while  $s=2$ in the context of Figure   \ref{fig:out_2}. 

\subsection{When condition (\ref{eq:entries-Delta*-different}) does not hold.}
\label{subsec:without-condition-section3}
We briefly justify that Theorem \ref{th:fundamental-theorem-section3} remains valid when some of the entries of 
$\Delta_*$ coincide. For this, we use the same notations as in Section \ref{subsec:entries-Delta*}. The reader may check 
that when condition (\ref{eq:entries-Delta*-different}) does not hold, the limiting equation (\ref{eq:det_limit}) is replaced by
\begin{align}\label{eq:det_limit_nocondition}
\det\begin{pmatrix}
\dfrac{1-c_*}{y^2\t_*(y)}\Delta_*^{2}&I_r&\dfrac{(1-c_*)^2}{y^2{\bf \tilde{t}}_*(y)}X_{p,N}^{-1}\Gamma^*_*X_{f,N}^{-*}&0\\
I_r& -\dfrac{y^2\t_*(y)}{1-c_*}X_{p,N}^{*}G_*X_{p,N}&0&0\\
\dfrac{(1-c_*)^2}{y^2{\bf \tilde{t}}_*(y)}X_{f,N}^{-1}\Gamma_*X_{p,N}^{-*}&0&\dfrac{1-c_*}{y^2\t_*(y)}\Delta_*^{2}&I_r\\
0&0&I_r& -\dfrac{y^2\t_*(y)}{1-c_*}X_{f,N}^{*}G_*X_{f,N}
\end{pmatrix}=0.
\end{align} 
Following the same steps as in the proof of Theorem \ref{th:fundamental-theorem-section3}, we obtain that (\ref{eq:det_limit_nocondition}) 
is equivalent to 
\begin{align}\label{eq:det_bold_nocondition}
\det\left(\dfrac{1}{(1-c_*)^2}\dfrac{{\bf \tilde{t}}^2_*(y)}{\t_*^2(y)}-X_{p}^{-1}\Gamma^*_*X_{f}^{-*}(\Delta_*^{2}+(X_f^{*}G_*X_{f})^{-1})^{-1}X_f^{-1}\Gamma_*X_p^{-*}(\Delta_*^{2}+(X_p^{*}G_*X_p)^{-1})^{-1}\right)=0
\end{align}
or to 
\begin{align}\label{eq:det_bold_nocondition_bis}
\det\left(\dfrac{1}{(1-c_*)^2}\dfrac{{\bf \tilde{t}}^2_*(y)}{\t_*^2(y)}-\Gamma^*_*X_{f}^{-*}(\Delta_*^{2}+(X_f^{*}G_*X_{f})^{-1})^{-1}X_f^{-1}\Gamma_*X_p^{-*}(\Delta_*^{2}+(X_p^{*}G_*X_p)^{-1})^{-1}X_{p}^{-1}\right)=0
\end{align}
We remark that for $i=p,f$
$$
X_i^{-*}(\Delta_*^{2}+(X_i^{*}G_*X_{i})^{-1})^{-1}X_i^{-1} = \left(X_i \Delta^{2}_* X_{i}^* + G_*^{-1} \right)^{-1} = \left(\Delta^{2}_* X_i X_{i}^{*} + G_*^{-1} \right)^{-1}
$$
because we recall that $X_i \Delta_* = \Delta_* X_i$. As $X_{i,N} X_{i,N}^* \rightarrow I_r$ (see (\ref{eq:convergence-XiN*XiN})), it appears 
the limiting form of (\ref{eq:det_bold_nocondition}) is (\ref{eq:det_bold}), i.e. the final equation derived in the proof of 
Theorem \ref{th:fundamental-theorem-section3}. Using again the stability arguments in  \cite{benaych-nadakuditi-ann-math} and \cite{chapon-couillet-hachem-mestre}, we deduce that Theorem \ref{th:fundamental-theorem-section3} remains valid.

\subsection{Example.}
We now consider the particular models defined by (\ref{eq:model-s-equal-2r-1}) and assume 
that $R_N = \sigma^{2} I_M$. 
We use again the notations introduced to derive the properties of (\ref{eq:model-s-equal-2r-1}), and
evaluate the conditions under which $s=P=1$. For this, we have first 
to compute matrix $\Omega_*$. We notice that 
$$
(R_{u,N})^{\# 1/2} R_{f|p} (R_{u,N})^{\# 1/2} = \Theta_N \Delta_*^{-1} \Gamma_* \Delta_*^{-1} \Theta_N^{*} = \Theta_N  \Delta_*^{-1} \kappa \Upsilon \tilde{\Upsilon}^{*} \Delta_*^{-1} \Theta_N
$$
where we recall that $\Upsilon$ coincides with the first vector $e_1$ of the canonical basis of $\mathbb{C}^{r}$. Therefore, a simple calculation leads to the conclusion that matrix $\tilde{\Theta}_{f,N}^{*} \tilde{\Theta}_{p,N}$ converges towards $\Omega_*$ given by
$$
\Omega_* = \frac{1}{\delta_1} \,  e_1 \,  (a \delta_1^{2}, b_1 \delta_2, \ldots, b_K \delta_{K+1})\, \Delta_*^{-1} = e_1 \, \left(a, \frac{b_1}{\delta_1}, \ldots, \frac{b_K}{\delta_1} \right)
$$
The non zero singular value of $\Omega_*$ is thus equal $a^{2} + \frac{\|b\|^{2}}{\delta_1^{2}}$, 
which, by (\ref{eq:choice-a}), coincides with $1$. We notice that this is not surprising 
because it is easily seen that the intersection of the row spaces of matrices 
$U_p$ and $U_f$ is not reduced to $0$, and coincides with the one dimensional space 
generated by $(x_2, \ldots, x_{N+1})$. As $G_* = \frac{I}{\sigma^{2}}$, matrix $F_*$ is thus given by 
\begin{equation}
    \label{eq:expre-F*-example}
    F_* = \frac{1}{1+\frac{\sigma^{2}}{\delta_1^{2}}} \, \left( \begin{array}{c}
    a \\ \frac{b_1}{\delta_1} \\ \vdots \\ \frac{b_K}{\delta_{1}} \end{array} \right) \; 
    \left( a, \frac{b_1}{\delta_1}, \ldots, \frac{b_K}{\delta_{1}} \right) \, \left( I + \sigma^{2} \Delta_*^{-2} \right)^{-1}
\end{equation}
and the non zero eigenvalue $\lambda_1(F_*)$ of $F_*$ is given by
$$
\lambda_1(F_*) = \left( \frac{1}{1+\frac{\sigma^{2}}{\delta_1^{2}}} \right) \, 
\left[ \left( \frac{a^{2}}{1+\frac{\sigma^{2}}{\delta_1^{2}}} \right) + 
\sum_{k=1}^{K} \frac{b_k^{2}}{\delta_1^{2}} \;  \frac{1}{1+\frac{\sigma^{2}}{\delta_{k+1}^{2}}} \right]
$$
We conclude that $s=P=1$ if and only $c_* < \frac{1}{2}$ and $\lambda_1(F_*) > \frac{c_*}{1-c_*}$. In order to get more insights on this condition, we assume that 
the $(\delta_k)_{k=1, r}$ all coincide with $\delta$. In this context, the ratio 
$\frac{\delta^{2}}{\delta^{2}}$ can be interpreted as the signal to noise ratio. Then,  $\lambda_1(F_*) > \frac{c_*}{1-c_*}$ is equivalent to 
\begin{equation}
    \label{eq:condition-snr-s=1}
    \left( \frac{1}{1+\frac{\sigma^{2}}{\delta^{2}}} \right)^{2} > \frac{c_*}{1-c_*}
\end{equation}
or to 
\begin{equation}
    \label{eq:condition-snr-s=1-bis}
    \frac{\delta^{2}}{\sigma^{2}} > \frac{1}{\left( \frac{1-c_*}{c_*} \right)^{1/2} - 1} = \frac{\sqrt{c_*}}{\sqrt{1-c_*} - \sqrt{c_*}} =\frac{c_* + \sqrt{c_*(1-c_*)}}{1 -2 c_*}
\end{equation}
It is interesting to notice that for $i=p,f$, $Y_{i,N} = U_{i,N} + V_{i,N}$, where $\frac{U_{i,N}U_{i,N}^{*}}{N}$ is a rank $r$ matrix whose $r$ non zero eigenvalues converge towards $\delta^{2}$. Therefore, usual results related to spiked models imply that the $r$ largest eigenvalues of $\frac{Y_{i,N} Y_{i,N}^{*}}{N}$ escape from the support of the Marcenko-Pastur distribution $[\sigma^{2}(1-\sqrt{c_*})^{2}, \sigma^{2}(1+\sqrt{c_*})^{2}]$ if and only if the signal to noise ratio $\frac{\delta^{2}}{\sigma^{2}}$ is larger than the threshold $\sqrt{c_*}$. Not surprisingly,
condition (\ref{eq:condition-snr-s=1-bis}) appears stronger than $\frac{\delta^{2}}{\sigma^{2}} > \sqrt{c_*}$. However, if $c_*$ is small enough, $\frac{\sqrt{c_*}}{\sqrt{1-c_*} - \sqrt{c_*}} \simeq \sqrt{c_*}$ and the 2 conditions are nearly equivalent. 

\section{Monte Carlo Simulations}
\label{sec:simulations}
Our theoretical results allow to evaluate the number $s$ of eigenvalues of $\Sigma_f \Sigma_p^{*} \Sigma_p \Sigma_f^{*}$ and of $\Pi_{p}\Pi_{f}$ that escape from the support of the limit eigenvalue distribution  of $W_f W_p^{*} W_p W_f^{*}$ and $\Pi_p^{W} \Pi_f^{W}$ respectively. In this section, using Monte Carlo simulation results, we evaluate the behaviour of two estimates of $s$, and check whether the true value 
of $s$ is in practice well estimated. 
We still consider the simple model defined by (\ref{eq:model-s-equal-2r-1}), and choose the various parameters in such a way that $s=2r-1$ and $s=P=1$ in the context of matrices $\Sigma_f \Sigma_p^{*} \Sigma_p \Sigma_f^{*}$  and $\Pi_{p}\Pi_{f}$ respectively.  More precisely, we take $c_N=0.25$, $R_N=I_M$ (that is $\sigma=1$), $K=2$ and therefore $r=K+1=3$ and $s=5$. $a$ is chosen equal to $0.2$, and we choose 
 $\delta_1=\delta_2=\delta_3=\delta$ and $b_1=b_2=b=\frac{1}{\sqrt{2}}\delta(1-a^2)^{1/2}$.
 $\delta$ is chosen equal to $\delta=(w_{+,N}-\sigma^2)^{1/2}+0.3$ where  $w_{+,N}=\sigma^2\left(1+\dfrac{1+\sqrt{1+8c_N}}{2}\right)$, so that the signal to noise ratio 
 $\frac{\delta^{2}}{\sigma^{2}}$ is equal to $3.3 dB$. Our goal is twofold. While we know that  $s= 2r-1=5$, we first check that in the context of $\Sigma_f \Sigma_p^{*} \Sigma_p \Sigma_f^{*}$, the probability of estimating $s$ by $P=1$ is very low, thus confirming that estimating $P$ from the largest eigenvalues of $\Sigma_f \Sigma_p^{*} \Sigma_p \Sigma_f^{*}$ is irrelevant both theoretically and practically. Second, in the context of matrix $\Pi_p \Pi_f$, we evaluate the empirical probability that 
 the estimates of $s$ take the value $s=P=1$. \\
 
\begin{figure}[ht!]
	\centering
	\par
	\includegraphics[scale=0.45]{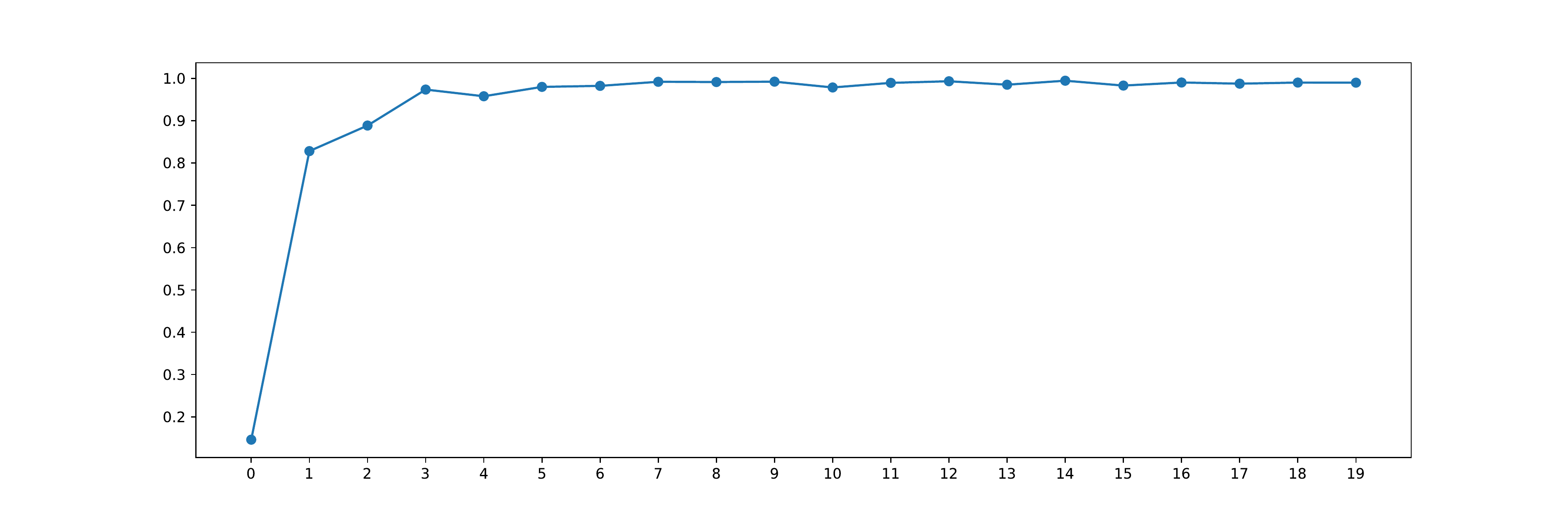}
	\caption{Ratios of eigenvalues $\lambda_{i+1}/\lambda_i$ of $\Sigma_f \Sigma_p^{*} \Sigma_p \Sigma_f^{*}$ w.r.t. $i-1$}
	\label{fig:ratios-autocorr}
\end{figure}

 1000 realisations of matrices  $\Sigma_f \Sigma_p^{*} \Sigma_p \Sigma_f^{*}$  and $\Pi_{p}\Pi_{f}$ were generated. Table ~\ref{table:proba_autocov} reports the results corresponding to the estimation of $s$ in the context of matrix $\Sigma_f \Sigma_p^{*} \Sigma_p \Sigma_f^{*}$. The first estimate $\tilde{s}$ of $s$ is the number of eigenvalues of $\Sigma_f \Sigma_p^{*} \Sigma_p \Sigma_f^{*}$ that are larger than $x_{+,N}(1+\epsilon_1)$ for $\epsilon_1 = 0.01$. The second estimate, $\hat{s}$, already used in \cite{hachem-2} and \cite{li-wang-yao-ann-stat-2016}, is defined by 
\begin{equation}
\label{eq:def-hats-autocov}
\hat{s}  = \underset{k}{\text{argmin}}\Big\{ \dfrac{\lambda_{k+1}}{\lambda_k} > 1 - \epsilon_2 \Big\} - 1
\end{equation}
for $\epsilon_2=0.05$. $\hat{s}$ appears to be more realistic than $\tilde{s}$ because, in practice, the noise variance $\sigma^{2}$, and thus $x_{+,N}$, are not necessarily known. 
Table~\ref{table:proba_autocov} provides the empirical probabilities that $\tilde{s}$ and $\hat{s}$ equal to $0,1,2,3,4,5,6,7,8$ for various values of $M$ and $N$ and Figure \ref{fig:ratios-autocorr} represents the ratios of eigenvalues $\lambda_{i+1}/\lambda_i$
of a realisation of $\Sigma_f \Sigma_p^{*} \Sigma_p \Sigma_f^{*}$ in terms of $i-1$ when $(M,N) = (600,2400)$. 
Figure \ref{fig:ratios-autocorr} indicates that the largest eigenvalue $\lambda_1$ is much larger than the next four ones because $\frac{\lambda_2}{\lambda_1} < 0.1$. Moreover, $\lambda_2$ is nearly equal to $1.3 \, \lambda_3$, and the next eigenvalues appear to be much closer one from each others. This confirms what we already noticed in the context of the numerical experiment of Section \ref{subsec:autocov_part_cases}: as $c_N = \frac{1}{4}$ is rather small, the largest eigenvalue corresponds to the useful signal, and appears much larger than the other 4 spurious 
outliers. Table~\ref{table:proba_autocov} tends to confirm that $(\lambda_i)_{i=3,4,5}$ are likely to be close from $x_{+,N}$ while $\lambda_2$ is very often significantly larger than 
$x_{+,N}$ thus explaining that $\tilde{s}$ and $\hat{s}$ do not take the value 1, and that 
$\tilde{s}$ and $\hat{s}$ take the values $2,3,4,5$ (and $\hat{s}$ sometimes $6,7,8$) . These experiments tend to indicate 
that the true value of $s$ is difficult to estimate, and more importantly, that the estimates are never equal to $P=1$. This confirms that $P$ cannot be estimated reliably from the largest eigenvalues of $\Sigma_f \Sigma_p^{*} \Sigma_p \Sigma_f^{*}$.

\begin{figure}[ht!]
	\centering
	\par
	\includegraphics[scale=0.45]{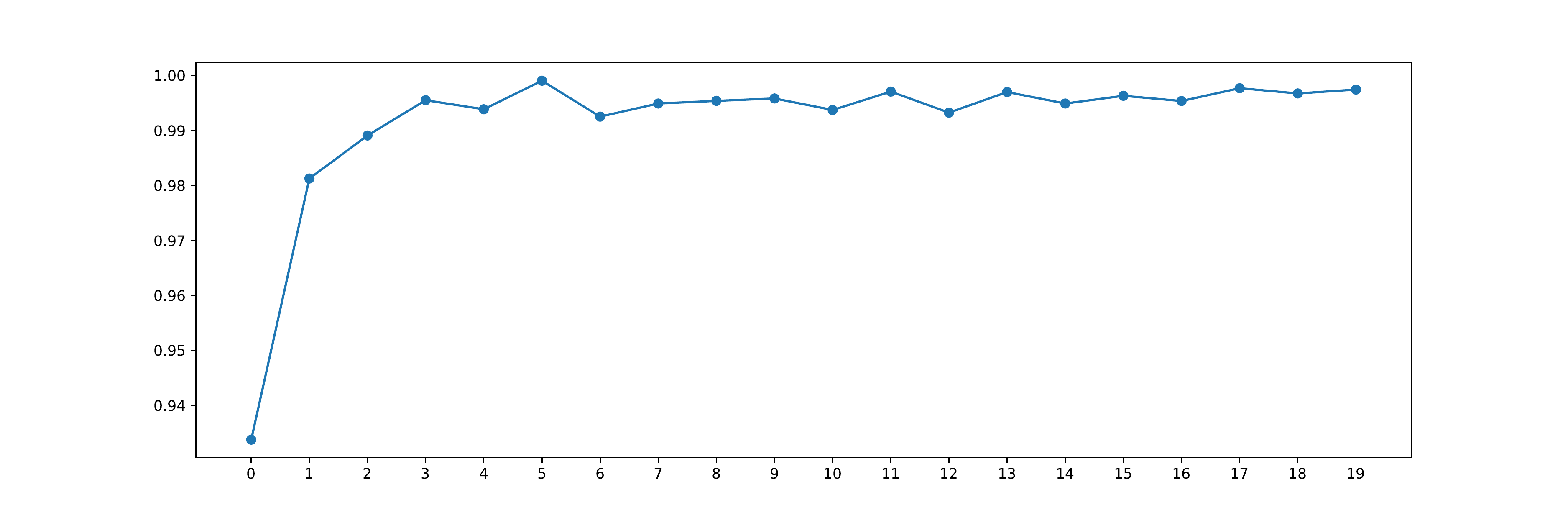}
	\caption{Ratios of eigenvalues $\lambda_{i+1}/\lambda_i$ of $\Pi_{p,y}\Pi_{f,y}$ w.r.t. $i-1$}
	\label{fig:ratios}
\end{figure}

\begin{table}[h]
\caption{Behaviour of $\tilde{s}$ and $\hat{s}$ for matrix $\Sigma_f \Sigma_p^{*} \Sigma_p \Sigma_f^{*}$}
\begin{center}
\begin{tabular}{l||c|c|c|c||ll||c|c|c|c}
    & \multicolumn{1}{l|}{\begin{tabular}[c]{@{}l@{}}M=100\\ N=400\end{tabular}} & \multicolumn{1}{l|}{\begin{tabular}[c]{@{}l@{}}M=200\\ N=800\end{tabular}} & \multicolumn{1}{l|}{\begin{tabular}[c]{@{}l@{}}M=400\\ N=1600\end{tabular}} & \multicolumn{1}{l||}{\begin{tabular}[c]{@{}l@{}}M=600\\ N=2400\end{tabular}} &                      &                      & \multicolumn{1}{l|}{\begin{tabular}[c]{@{}l@{}}M=100\\ N=400\end{tabular}} & \multicolumn{1}{l|}{\begin{tabular}[c]{@{}l@{}}M=200\\ N=800\end{tabular}} & \multicolumn{1}{l|}{\begin{tabular}[c]{@{}l@{}}M=400\\ N=1600\end{tabular}} & \multicolumn{1}{l}{\begin{tabular}[c]{@{}l@{}}M=600\\ N=2400\end{tabular}} \\ \hline\hline
$\tilde{s}=8$ & 0                                                                      & 0                                                                       & 0                                                                        & 0                                                                       & & $\hat{s}=8$ & 0.061                                                                       & 0                                                                       & 0                                                                        & 0                                                                       \\ \hline
$\tilde{s}=7$ & 0                                                                      & 0                                                                       & 0                                                                        & 0                                                                       & & $\hat{s}=7$ & 0.128                                                                       & 0                                                                       & 0                                                                        & 0                                                                       \\ \hline
$\tilde{s}=6$ & 0                                                                      & 0                                                                       & 0                                                                        & 0                                                                       & & $\hat{s}=6$ & 0.179                                                                       & 0.01                                                                       & 0                                                                        & 0                                                                       \\ \hline
$\tilde{s}=5$ & 0                                                                      & 0.005                                                                       & 0.09                                                                        & 0.27                                                                       & & $\hat{s}=5$ & 0.25                                                                       & 0.335                                                                       & 0.097                                                                        & 0                                                                       \\ \hline
$\tilde{s}=4$ & 0.235                                                                       & 0.56                                                                       & 0.86                                                                        & 0.72                                                                       & \multicolumn{1}{c}{} & $\hat{s}=4$ & 0.247                                                                       & 0.298                                                                       & 0.357                                                                        & 0.033                                                                       \\ \hline
$\tilde{s}=3$ & 0.745                                                                       & 0.425                                                                      & 0.05                                                                       & 0.01                                                                       & \multicolumn{1}{c}{} & $\hat{s}=3$ & 0.12                                                                       & 0.21                                                                       & 0.32                                                                        & 0.287                                                                       \\ \hline
$\tilde{s}=2$ & 0.02                                                                      & 0.01                                                                         & 0                                                                           & 0                                                                          &                      &       $\hat{s}=2$               & 0.005                                                                       & 0.147                                                                       & 0.226                                                                        & 0.68                                                                        \\ \hline
$\tilde{s}=1$ & 0                                                                          & 0                                                                          & 0                                                                           & 0                                                                          &                      &          $\hat{s}=1$            & 0.01                                                                          & 0                                                                          & 0                                                                           & 0                                                                          \\ \hline
$\tilde{s}$=0 & 0                                                                          & 0                                                                          & 0                                                                           & 0                                                                          &                      &          $\hat{s}=0$            & 0                                                                          & 0                                                                          & 0                                                                           & 0                                                                         
\end{tabular}
\end{center}
\label{table:proba_autocov}
\end{table}
Table~\ref{table:proba_cor_coef} is related to the estimation of $s$ in the context of matrix
$\Pi_p \Pi_f$. $\epsilon_1$ and $\epsilon_2$ being still equal to $0.01$ and $0.05$, $\tilde{s}$ represents this time the number of eigenvalues of  $\Pi_p \Pi_f$ that are larger than $4 c_N(1-c_N)(1 + \epsilon_1)$, while $\hat{s}$ is defined by
$$
\hat{s}  = \underset{k}{\text{argmin}}\Big\{ \dfrac{\lambda_{k+1}}{\lambda_k} > 1 - \epsilon_2 \Big\} - 1
$$
We also represent Fig. \ref{fig:ratios} the ratios of eigenvalues 
$\frac{\lambda_{i+1}}{\lambda_i}$ of a realisation of $\Pi_p \Pi_f$ in terms of $i-1$ when 
$(M,N) = (600, 2400)$. The largest eigenvalue appears significantly larger than $\lambda_2$, and the other eigenvalues are quite close one from each others. This behaviour is confirmed 
by the behaviour of $\tilde{s}$ and $\hat{s}$ which take the value 1 with high probability, 
thus confirming the relevance of the estimate of $P$ based on the largest eigenvalues of 
$\Pi_p \Pi_f$. We notice that in the context of matrix  $\Pi_p \Pi_f$, the estimate 
$\tilde{s}$ is in practice relevant because $4c_N(1-c_N)$ is of course known. 
\begin{table}[h]
\caption{Behaviour of $\tilde{s}$ and $\hat{s}$ for matrix $\Pi_{p}\Pi_{f}$}
\begin{center}
\begin{tabular}{l||c|c|c|c||ll||c|c|c|c}
    & \multicolumn{1}{l|}{\begin{tabular}[c]{@{}l@{}}M=100\\ N=400\end{tabular}} & \multicolumn{1}{l|}{\begin{tabular}[c]{@{}l@{}}M=200\\ N=800\end{tabular}} & \multicolumn{1}{l|}{\begin{tabular}[c]{@{}l@{}}M=400\\ N=1600\end{tabular}} & \multicolumn{1}{l||}{\begin{tabular}[c]{@{}l@{}}M=600\\ N=2400\end{tabular}} &  &  & \multicolumn{1}{l|}{\begin{tabular}[c]{@{}l@{}}M=100\\ N=400\end{tabular}} & \multicolumn{1}{l|}{\begin{tabular}[c]{@{}l@{}}M=200\\ N=800\end{tabular}} & \multicolumn{1}{l|}{\begin{tabular}[c]{@{}l@{}}M=400\\ N=1600\end{tabular}} & \multicolumn{1}{l}{\begin{tabular}[c]{@{}l@{}}M=600\\ N=2400\end{tabular}} \\ \hline\hline
$\tilde{s}=4$ & 0                                                                       & 0                                                                      & 0                                                                       & 0                                                                      &  & $\hat{s}=4$  & 0                                                                      & 0.007                                                                      & 0.007                                                                           & 0.008                                                                          \\ \hline
$\tilde{s}=3$ & 0                                                                       & 0                                                                      & 0                                                                       & 0                                                                      &  & $\hat{s}=2$  & 0                                                                      & 0.013                                                                      & 0.008                                                                           & 0.017                                                                          \\ \hline
$\tilde{s}=2$ & 0.001                                                                       & 0                                                                      & 0                                                                       & 0                                                                      &  & $\hat{s}=2$  & 0.07                                                                      & 0.012                                                                      & 0                                                                           & 0.01                                                                          \\ \hline
$\tilde{s}=1$ & 0.999                                                                       & 1                                                                      & 1                                                                       & 1                                                                      &  & $\hat{s}=1$  & 0.91                                                                      & 0.966                                                                      & 0.974                                                                           & 0.965                                                                          \\ \hline
$\tilde{s}=0$ & 0                                                                          & 0                                                                          & 0                                                                           & 0                                                                          &  &$\hat{s}=0$   & 0.02                                                                          & 0.002                                                                          & 0.011                                                                           & 0                                                                         
\end{tabular}
\end{center}
\label{table:proba_cor_coef}
\end{table}


\section{Conclusion}
\label{sec:concluding-remarks} 
In this paper, motivated by the problem of estimating consistently the minimal dimension $P$ of the state space realizations of the high-dimensional time series $y$, we have studied the behaviour of the largest singular values of
the empirical autocovariance matrix $\hat{R}_{f|p,y}^{L}$ as well as of its normalized version $(\hat{R}_{f,y}^{L}) ^{-1/2} \hat{R}_{f|p,y}^{L} (\hat{R}_{p,y}^{L})^{-1/2}$. In the high-dimensional asymptotic regime defined in Section \ref{ch:notations}, and under certain technical assumptions, we have shown that all the singular values  of $\hat{R}_{f|p,y}^{L}$ are less than a certain threshold, except a finite number $s$ of outliers. Unfortunately, $s$ is not related to $P$, and, when $P=1$, we have built simple examples for which 
$s$ can take any odd value. We also showed that the singular values of the normalized matrix  $(\hat{R}_{f,y}^{L})^{-1/2} \hat{R}_{f|p,y}^{L} (\hat{R}_{p,y}^{L})^{-1/2}$ lie almost surely in 
a neighbourhood of the interval $[0, 2 \sqrt{c_*(1-c_*)}]$, but this time, we proved that the number $s$ of outliers belong to $\{0,1, \ldots, P\}$, and that $s=P$ if $c_* < \frac{1}{2}$ and if the $P$ non zero eigenvalues of the rank $P$ matrix $F_*$ defined by (\ref{eq:def-F*}) are larger than 
$\frac{c_*}{1-c_*}$. Under this condition, which, in practice, means that the useful signal $u$
is powerful enough and its non zero canonical correlation coefficients between the past and the future 
are large enough, $P$ can be estimated consistently by the number of singular values of 
$(\hat{R}_{f,y}^{L})^{-1/2} \hat{R}_{f|p,y}^{L} (\hat{R}_{p,y}^{L})^{-1/2}$ that are larger than 
$2 \sqrt{c_*(1-c_*)}(1+\epsilon)$ for a certain parameter $\epsilon$ small enough. These results are established using a general approach already proposed in the literature in the context of
simple large random matrix models. However, the random matrix models considered in the present paper are quite complicated, and we needed to solve a number of non obvious new technical issues. 
We have also provided numerical simulation results that confirm the practical relevance of
our theoretical results. We finally mention that the existence of a consistent estimate of $P$
allows to consider the problem of estimating other parameters of the state space realizations 
of the useful signal $u$. This is a topic for future research.


\appendix

\section{Appendix}

\subsection{Proof of Lemma \ref{le:symetries-E(Q)-2}}
To prove that matrices $\ex\{\mathbf{Q_{ij}}\}$, $i,j=p,f$ are diagonal, we consider the new set of vectors $z_k=e^{-ik\theta}v_k$ and construct the matrices $Z_p$, $Z_f$ in the same way as $W_p$ and $W_f$. It is clear 
that sequence $(z_n)_{n \in \mathbb{Z}}$ has the same probability distribution 
that $(v_n)_{n \in \mathbb{Z}}$. $Z_p$ and $Z_f$ can be expressed as
\begin{align*}
&Z_p=\begin{pmatrix}
e^{-i\theta}I_M&\ldots&0\\
\vdots&\ddots&\vdots\\
0&\ldots&e^{-Li\theta}I_M
\end{pmatrix}W_p\begin{pmatrix}
1&\ldots&0\\
\vdots&\ddots&\vdots\\
0&\ldots&e^{-(N-1)i\theta}
\end{pmatrix},\\
&Z_f=e^{-Li\theta}\begin{pmatrix}
e^{-i\theta}I_M&\ldots&0\\
\vdots&\ddots&\vdots\\
0&\ldots&e^{-Li\theta}I_M
\end{pmatrix}W_f\begin{pmatrix}
1&\ldots&0\\
\vdots&\ddots&\vdots\\
0&\ldots&e^{-(N-1)i\theta}
\end{pmatrix}.
\end{align*}
Then \begin{align}
&Z_iZ_i^{*}=\begin{pmatrix}
e^{-i\theta}I_M&\ldots&0\\
\vdots&\ddots&\vdots\\
0&\ldots&e^{-Li\theta}I_M
\end{pmatrix}W_iW_i^*\begin{pmatrix}
e^{i\theta}I_M&\ldots&0\\
\vdots&\ddots&\vdots\\
0&\ldots&e^{Li\theta}I_M
\end{pmatrix},\\
&(Z_iZ_i^{*})^{-1}=\begin{pmatrix}
e^{-i\theta}I_M&\ldots&0\\
\vdots&\ddots&\vdots\\
0&\ldots&e^{-Li\theta}I_M
\end{pmatrix}(W_iW_i^*)^{-1}\begin{pmatrix}
e^{i\theta}I_M&\ldots&0\\
\vdots&\ddots&\vdots\\
0&\ldots&e^{Li\theta}I_M
\end{pmatrix}\label{eq:sym_inver_WW*}
\end{align}
as well as $\phi(Z_{f}Z_{f}^*)=\phi(W_fW_f^*)$ and $\phi(Z_{p}Z_{p}^*)=\phi(W_pW_p^*)$ where $\phi$ is defined
by (\ref{eq:def-phi-regularization}). Therefore, $\eta^z$ coincides with $\eta$. Next we define $\Pi^z_{i}=Z_i^*(Z_iZ_i^*)^{-1}Z_i$, $i=\{p,f\}$. The equality
\begin{align}\label{eq:sym_Pi}
\Pi^z_{i}=\begin{pmatrix}
1&\ldots&0\\
\vdots&\ddots&\vdots\\
0&\ldots&e^{(N-1)i\theta}
\end{pmatrix}\Pi_i\begin{pmatrix}
1&\ldots&0\\
\vdots&\ddots&\vdots\\
0&\ldots&e^{-(N-1)i\theta}
\end{pmatrix}
\end{align}
holds for $i=\{p,f\}$. 
We define matrix $
\mathbf{Q^Z}=\left(\begin{smallmatrix}
-zI_{ML}&\eta^z\Pi_p^z\\
\eta^z\Pi_f^z&-zI_{ML}
\end{smallmatrix}\right)^{-1}
$
and obtain immediately that 
\begin{align*}
\ex\{\mathbf{Q^Z}\}=\begin{pmatrix}
A&0\\
0&A
\end{pmatrix}\ex\{\mathbf{Q}\}\begin{pmatrix}
A^*&0\\
0&A^*
\end{pmatrix},
\ex\{ \eta_N \mathbf{Q^Z}\}=\begin{pmatrix}
A&0\\
0&A
\end{pmatrix}\ex\{\eta_N \mathbf{Q}\}\begin{pmatrix}
A^*&0\\
0&A^*
\end{pmatrix}
\end{align*}
where the $N\times N$ matrix $A$ is defined as
\begin{align*}
A=\begin{pmatrix}
1&\ldots&0\\
\vdots&\ddots&\vdots\\
0&\ldots&e^{(N-1)i\theta}
\end{pmatrix}
\end{align*}
Obviously for each $N\times N$ block $\ex\{\mathbf{Q_{ij}}^z\}$, $i,j=\{p,f\}$, we have
\begin{align*}
\ex\{\mathbf{Q_{ij}}^z\}= A \, \ex\{\mathbf{Q_{ij}}\} \, A^*, \ex\{\eta_N \mathbf{Q_{ij}}^z\}= A \, \ex\{\eta_N \mathbf{Q_{ij}}\} \, A^*
\end{align*}
and 
\begin{align*}
\ex\{\eta_N\Pi_h^z\mathbf{Q_{ij}}^z\}= A \, \ex\{\eta_N\Pi_h\mathbf{Q_{ij}}\} \, A^*
\end{align*}
for $h=\{p,f\}$.
Since $\ex\{\mathbf{Q^Z}\}=\ex\{\mathbf{Q}\}$,  $\ex\{\eta_N \mathbf{Q^Z}\}=\ex\{\eta_N \mathbf{Q}\}$, 
and $\ex\{\eta_N\Pi_h^z\mathbf{Q_{ij}}^z\} = \ex\{\eta_N\Pi_h\mathbf{Q_{ij}}\}$, for $1\le k,l\le N$ and $i,j,h=\{p,f\}$, we have
\begin{align*}
&\ex\{\mathbf{Q_{ij}}^{k,l}\}=e^{(k-1)i\theta}\ex\{\mathbf{Q_{ij}}^{k,l}\}e^{-(l-1)i\theta}=e^{(k-l)i\theta}\ex\{\mathbf{Q_{ij}}^{k,l}\}\\
& \ex\{\eta_N \mathbf{Q_{ij}}^{k,l}\}=e^{(k-1)i\theta}\ex\{\eta_N \mathbf{Q_{ij}}^{k,l}\}e^{-(l-1)i\theta}=e^{(k-l)i\theta}\ex\{\eta_N \mathbf{Q_{ij}}^{k,l}\} \\
&\ex\{\eta_N(\Pi_h\mathbf{Q_{ij}})^{k,l}\}=e^{(k-1)i\theta}\ex\{\eta_N(\Pi_h\mathbf{Q_{ij}})^{k,l}\}e^{-(l-1)i\theta}=e^{(k-l)i\theta}\ex\{\eta_N(\Pi_h\mathbf{Q_{ij}})^{k,l}\}
\end{align*}
This proves that $\ex\{\mathbf{Q_{ij}}^{k,l}\} = 0$,  $\ex\{\eta_N \mathbf{Q_{ij}}^{k,l}\} = 0$ and $\ex\{\eta_N(\Pi_h\mathbf{Q_{ij}})^{k,l}\}=0$ if $k \neq l$, as expected. We can prove similarly that matrices  $\ex\{\eta_N(W_iW_i^*)^{-1}\}$, $\ex\{\eta_N\Pi_i\}$ and $\ex\{\eta_N\mathbf{Q_{ij}}W_h^*(W_hW_h^*)^{-2}W_h\}$ are diagonal. We just verify that $\mathbb{E}(\eta_N W_{p}^{*} (W_p W_p)^{-1} ) = 0$, and omit the proof of $\mathbb{E}(\eta_N \,  \mathbf{Q_{ij}}  \, W_h^{*} (W_h W_h^{*})^{-1}) =  \mathbb{E}(\eta_N \,  \Pi_k \, \mathbf{Q_{ij}}  \, W_h^{*} (W_h W_h^{*})^{-1}) = 0$. It is clear that
$$
Z_p^{*} (Z_p Z_p^*)^{-1} = \begin{pmatrix}
1&\ldots&0\\
\vdots&\ddots&\vdots\\
0&\ldots&e^{i(N-1)\theta}
\end{pmatrix} W_p^{*} (W_p W_p^*)^{-1} \begin{pmatrix}
e^{i\theta}I_M&\ldots&0\\
\vdots&\ddots&\vdots\\
0&\ldots&e^{Li\theta}I_M
\end{pmatrix}
$$
The equalities $\eta^{z} = \eta$ and $\mathbb{E}(\eta^{z} Z_p^{*} (Z_p Z_p^*)^{-1}) = \mathbb{E}(\eta W_p^{*} (W_p W_p^*)^{-1})$
lead immediately to $\left(\mathbb{E}(\eta W_p^{*} (W_p W_p^*)^{-1})\right)_{n,l} = e^{i(n-1+l) \theta} \left(\mathbb{E}(\eta W_p^{*} (W_p W_p^*)^{-1})\right)_{n,l}$
for each $1 \leq n \leq N$ and $1 \leq l \leq L$. As $\theta$ can be chosen arbitrarily, we obtain that
$\left(\mathbb{E}(\eta W_p^{*} (W_p W_p^*)^{-1})\right)_{n,l} = 0$ as expected. \\

To prove (\ref{eq:symetry-entries-wrt-Nover2}) to (\ref{eq:symetry-Pi-Q}), we consider the sequence $(z_n)_{n \in \mathbb{Z}}$ defined by $z_n=v_{-n+N+2L}$ 
for each $n$. Again, the distributions of $z$ and $v$ coincide, and it is easy to see that for $i\in\{p,f\}$, $Z_i$ is given by 
\begin{align*}
Z_i=\begin{pmatrix}
0&\ldots&I_M\\
\vdots&&\vdots\\
I_M&\ldots&0
\end{pmatrix}W_{\tilde{i}}\begin{pmatrix}
0&\ldots&1\\
\vdots&&\vdots\\
1&\ldots&0
\end{pmatrix},
\end{align*}
and as consequence
\begin{align*}
Z_iZ_i^*=\begin{pmatrix}
0&\ldots&I_M\\
\vdots&&\vdots\\
I_M&\ldots&0
\end{pmatrix}W_{\tilde{i}}W_{\tilde{i}}^*\begin{pmatrix}
0&\ldots&I_M\\
\vdots&&\vdots\\
I_M&\ldots&0
\end{pmatrix}
\end{align*}
Therefore, $Z_iZ_i^*$ and $W_{\tilde{i}}W_{\tilde{i}}^*$ have the same eigenvalues, which implies that  $\phi(Z_iZ_i^*)=\phi(W_{\tilde{i}}W_{\tilde{i}}^*)$, and  that the new regularization term $\eta^z=\det\phi(Z_pZ_p^*)\det\phi(Z_fZ_f^*)$ will remain the same, i.e. $\eta^z=\eta$.
It is easy to see that $\Pi^z_p=A\Pi_fA$ and $\Pi^z_f=A\Pi_pA$, where this time $A=\begin{pmatrix}
0&\ldots&1\\
\vdots&&\vdots\\
1&\ldots&0
\end{pmatrix}$. From this, we obtain that  
\begin{align*}
\ex\{\mathbf{Q^Z}\}=\begin{pmatrix}
A&0\\
0&A
\end{pmatrix}\ex\Big\{\begin{pmatrix}
-zI_N&\eta\Pi_f\\
\eta\Pi_p&-zI_N
\end{pmatrix}^{-1}\Big\}\begin{pmatrix}
A&0\\
0&A
\end{pmatrix}.
\end{align*}
Using the inverse block matrix formula and the fact that  $\ex\{\mathbf{Q^Z}\} = \ex\{\mathbf{Q}\}$, we obtain that $\ex\{\mathbf{Q_{pp}}\} = A\ex\{\mathbf{Q_{ff}}\}A$ and $\ex\{\mathbf{Q_{pf}}\} = A\ex\{\mathbf{Q_{fp}}\}A$. This immediately implies that for every $1\le k\le N$ and $h,i,j=\{p,f\}$ we have $\ex\{(\mathbf{Q_{ij}})^{k,k}\}=\ex\{(\mathbf{Q_{\tilde{i}\tilde{j}}})^{N+1-k,N+1-k}\}$ and $\ex\{\eta_N(\Pi_h\mathbf{Q_{ij}})^{k,k}\}=\ex\{\eta_N(\Pi_{\tilde{h}}\mathbf{Q_{\tilde{i}\tilde{j}}})^{N+1-k,N+1-k}\}$. As a consequence, $\ex\{\tr\mathbf{Q_{ij}}\}=\ex\{\tr\mathbf{Q_{\tilde{i}\tilde{j}}}\}$ and $\ex\{\eta_N\Pi_h\mathbf{Q_{ij}}\}=\ex\{\eta_N\Pi_{\tilde{h}}\mathbf{Q_{\tilde{i}\tilde{j}}}\}$ as expected.

\subsection{Proof of Lemma \ref{le:expectation-WW-inverse}}

The lemma is established using the integration by parts formula and the Poincar\'e-Nash inequality. As the partial derivatives of $\eta$ with respect to elements of $W_p,W_f$ will appear, we first state the following useful lemma. We recall that 
$\phi$ and $\Ecal_N$ are defined respectively by (\ref{eq:def-phi-regularization}) and 
(\ref{eq:def-EcalN}). 
\begin{lemma}\label{le:estimation_eta}
	Let $\Omega$ be the event defined by:
	\begin{align}
          \label{eq:def-set-Omega}
	\Omega=\Ecal_N	\cap\{\text{all eigenvalues of } W_pW_p^* \text{ and } W_fW_f^* \in\supp(\phi)\}.
	\end{align}
	Then it holds that 
	\begin{align}
	\dfrac{\partial \eta_N}{\partial W_{i,j}^m}=0 \text{ on } \Omega^c
	\end{align}
	and
	\begin{align}
	\ex\left\{\left|\dfrac{\partial \eta_N}{\partial W_{i,j}^m}\right|^2\right\}=\Ocal\left(\dfrac{1}{N^k}\right)
	\end{align}
	for all $1\le m\le M$, $1\le i\le 2L$, $1\le j\le N$ and each $k$.
\end{lemma}
The proof of the 
lemma is an adaptation of Lemma~11 and calculations from Proposition 4 of \cite{hachem-loubaton-et-al-2012}. \\

We just prove Lemma \ref{le:expectation-WW-inverse} for $i=p$. In the following, we drop index $i$ and  denote $G=(WW^*)^{-1}$. To prove the lemma, we apply the integration by parts formula (\ref{integr}) to $\eta_NG_{i_1i_2}^{m_1m_2}W_{i_2,j_2}^{m_2}\bar{W}_{j_1,i_3}^{m_3}$ considered as a function of the entries of
the $2ML \times N$ matrix $\overline{W}_N$ whose elements are the complex conjugates of those of $W_N$.
We recall that the correlation structure of the elements of
$W_N$ is given by (\ref{eq:correlation-structure-W}).
\begin{multline}\label{eq:GWW*}
\ex\{\eta_NG_{i_1i_2}^{m_1m_2}W_{i_2,j_2}^{m_2}\bar{W}_{j_1,i_3}^{m_3}\}=
\sum_{m^\prime,i^\prime,j^\prime}
\ex\{\bar{W}_{j_1,i_3}^{m_3}W_{i^\prime,j^\prime}^{m^\prime}\}\\
\times\left(\ex\left\{\dfrac{\partial\eta_N}{\partial W_{i^\prime,j^\prime}^{m^\prime}}G_{i_1i_2}^{m_1m_2}W_{i_2,j_2}^{m_2}\right\}+
\ex\left\{\eta_N\dfrac{\partial G_{i_1i_2}^{m_1m_2}}{\partial W_{i^\prime,j^\prime}^{m^\prime}}W_{i_2,j_2}^{m_2}\right\}+
\ex\left\{\eta_NG_{i_1i_2}^{m_1m_2}\dfrac{\partial W_{i_2,j_2}^{m_2}}{\partial W_{i^\prime,j^\prime}^{m^\prime}}\right\}\right)
\end{multline}
Lemma~\ref{le:estimation_eta} implies that the first term of the r.h.s of (\ref{eq:GWW*}) is of order $\Ocal(N^{-k})$ for each $k$. Indeed,
\begin{align*}
\ex\left\{\dfrac{\partial\eta_N}{\partial W_{i^\prime,j^\prime}^{m^\prime}}G_{i_1i_2}^{m_1m_2}W_{i_2,j_2}^{m_2}\right\}=
\ex\left\{{\bf 1}_{\Omega}\dfrac{\partial\eta_N}{\partial W_{i^\prime,j^\prime}^{m^\prime}}G_{i_1i_2}^{m_1m_2}W_{i_2,j_2}^{m_2}\right\}
\end{align*}
where we recall that $\Omega$ is defined by (\ref{eq:def-set-Omega}). The Schwartz inequality leads to
\begin{align}
\left|\ex\left\{{\bf 1}_{\Omega}\dfrac{\partial\eta_N}{\partial W_{i^\prime,j^\prime}^{m^\prime}}G_{i_1i_2}^{m_1m_2}W_{i_2,j_2}^{m_2}\right\}\right|^2\le
\ex\left\{\left|\dfrac{\partial\eta_N}{\partial W_{i^\prime,j^\prime}^{m^\prime}}\right|^2\right\}\ex\left\{\left|{\bf 1}_{\Omega}G_{i_1i_2}^{m_1m_2}W_{i_2,j_2}^{m_2}\right|^2\right\}
\end{align}
On $\Omega$, the eigenvalues of $WW^*$ belong to $((1-\sqrt{c_*})^2-2\epsilon,(1+\sqrt{c_*})^2+2\epsilon)$, so that $\|G{\bf 1}_{\Omega}\|$ and $\|W{\bf 1}_{\Omega}\|$ are bounded by a nice constant. Therefore  $\left|{\bf 1}_{\Omega}G_{i_1i_2}^{m_1m_2}W_{i_2,j_2}^{m_2}\right|$ has the same property. After some calculations, (\ref{eq:GWW*}) becomes
\begin{multline}\label{eq:GWW*-2}
\ex\{\eta_NG_{i_1i_2}^{m_1m_2}W_{i_2,j_2}^{m_2}\bar{W}_{j_1,i_3}^{m_3}\}=\dfrac{1}{N}
\sum_{m^\prime,i^\prime,j^\prime}\delta_{m^\prime,m_3}\delta_{i_3+j_1,i^{\prime}+j^{\prime}}\\
\times
\left(-\ex\left\{\eta_NG_{i_1i^\prime}^{m_1m^\prime}(W^*G)_{j^\prime,i_2}^{m_2}W_{i_2,j_2}^{m_2}\right\}+
\ex\{\eta_NG_{i_1i_2}^{m_1m_2}\delta_{m^\prime,m_2}\delta_{i_2,i^\prime}\delta_{j_2,j^\prime}\}\right)+
\Ocal\left(\dfrac{1}{N^k}\right)
\end{multline}
Defining $l=i_3-i^\prime=j^\prime-j_1$ which runs from $-L+1$ to $L-1$ and taking into account (\ref{eq:def-J}), we get $\delta_{m^\prime,m_3}\delta_{i_3+j_1,i^{\prime}+j^{\prime}}=(J^{(l)}_L\otimes I_M)_{i^\prime i_3}^{m^\prime m_3}(J^{(l)}_N)_{j_1j^\prime}$. Then, after summing over $i^\prime, j^\prime$ and $m^\prime$, (\ref{eq:GWW*-2}) becomes
\begin{multline*}
\ex\{\eta_NG_{i_1i_2}^{m_1m_2}W_{i_2,j_2}^{m_2}\bar{W}_{j_1,i_3}^{m_3}\}=
-\dfrac{1}{N}\ex\left\{\eta_N\left(G(J^{(l)}_L\otimes I_M)\right)_{i_1i_3}^{m_1m_3}(J^{(l)}_NW^*G)_{j_1,i_2}^{m_2}W_{i_2,j_2}^{m_2}\right\}\\
+\dfrac{1}{N}\ex\{\eta_NG_{i_1i_2}^{m_1m_2}(J^{(l)}_L\otimes I_M)_{i_2 i_3}^{m_2 m_3}(J^{(l)}_N)_{j_1j_2}\}+
\Ocal\left(\dfrac{1}{N^k}\right)
\end{multline*}
Summing both sides over $i_2,m_2$, we obtain that
\begin{multline}\label{eq:integ_parts_etaGWW*}
\ex\{\eta_N(GW)_{i_1j_2}^{m_1}\bar{W}_{j_1,i_3}^{m_3}\}=
-\dfrac{1}{N}\sum_{l=-(L-1)}^{L-1}\ex\left\{\eta_N(G(J^{(l)}_L\otimes I_M))_{i_1 i_3}^{m_1 m_3}(J^{(l)}_N\Pi)_{j_1j_2}\right\}\\
+\dfrac{1}{N}\sum_{l=-(L-1)}^{L-1}\ex\left\{\eta_N(G(J^{(l)}_L\otimes I_M))_{i_1 i_3}^{m_1 m_3}(J^{(l)}_N)_{j_1j_2}\right\}
+\Ocal\left(\dfrac{1}{N^k}\right).
\end{multline}
At this point, in order to prove (\ref{le:estimation_inver_WW*}), we take $j_2=j_1$ and sum over this index. Since $GWW^*=I_{ML}$, we have
\begin{align*}
\ex\{\eta_N\}I_{ML}=-\sum_{l=-(L-1)}^{L-1}\ex\left\{\eta_NG(J^{(l)}_L\otimes I_M)\dfrac{1}{N}\tr(J^{(l)}_N\Pi)\right\}
+\sum_{l=-(L-1)}^{L-1}\ex\left\{\eta_NG(J^{(l)}_L\otimes I_M)\dfrac{1}{N}\tr J^{(l)}_N\right\}
+\Ocal\left(\dfrac{1}{N^k}\right)
\end{align*}
Obviously $\frac{1}{N}\tr J^{(l)}_N$ is equal to 0 for $l\ne0$ and to 1 if $l=0$, and, as was discussed above, we can replace $\ex\{\eta_N\}$ by 1 on the l.h.s. and $\eta_N$ by $\eta_N^2$ on the first term of the r.h.s. while adding a $\Ocal(N^{-k})$ term. Then
\begin{multline}
\label{eq:preliminary-etaG}
I_{ML}=-\sum_{l=-(L-1)}^{L-1}\ex\left\{\eta_NG(J^{(l)}_L\otimes I_M)\right\}\ex\left\{\dfrac{1}{N}\tr(\eta_NJ^{(l)}_N\Pi)\right\}-\sum_{l=-(L-1)}^{L-1}\ex\left\{\eta_NG(J^{(l)}_L\otimes I_M)\dfrac{1}{N}\tr(\eta_NJ^{(l)}_N\Pi)^\circ\right\}\\
+\ex\left\{\eta_NG\right\}
+\Ocal\left(\dfrac{1}{N^k}\right)
\end{multline}
Lemma~\ref{le:symetries-E(Q)-2} implies that $\ex\{\eta_N\Pi\}$ is diagonal, so $\ex\left\{\dfrac{1}{N}\tr(\eta_NJ^{(l)}_N\Pi)\right\}=0$ for all $l\ne0$ and moreover since $\frac{1}{N}\tr\Pi=c_N$ it is easy to see that $\ex\left\{\frac{1}{N}\tr(\eta_N\Pi)\right\}=c_N+\Ocal(N^{-k})$ for each $k$. Thus, 
(\ref{eq:preliminary-etaG}) leads to 
\begin{multline}\label{eq:eta-G}
\ex\left\{\eta_N(WW^*)^{-1}\right\}=\dfrac{1}{1-c}I_{ML}
+\dfrac{1}{1-c}\sum_{l=-(L-1)}^{L-1}\ex\left\{\eta_NG(J^{(l)}_L\otimes I_M)\dfrac{1}{N}\tr(\eta_NJ^{(l)}_N\Pi)^\circ\right\}
+\Ocal\left(\dfrac{1}{N^k}\right)
\end{multline}
Finally, we show that each element of matrix  $\sum\ex\left\{\eta_NG(J^{(l)}_L\otimes I_M)\frac{1}{N}\tr(J^{(l)}_N(\eta_N\Pi)^\circ)\right\}$ is of order $\Ocal(N^{-3/2})$. For this, we 
apply the Schwartz inequality: 
\begin{align*}
\left|\ex\left\{(\f_{i_1}^{m_1})^*\eta_NG(J^{(l)}_L\otimes I_M)\f_{i_2}^{m_2}\frac{1}{N}\tr(\eta_NJ^{(l)}_N\Pi^\circ)\right\}\right|
\le\left(\var\left((\f_{i_1}^{m_1})^*\eta_NG(J^{(l)}_L\otimes I_M)\f_{i_2}^{m_2}\right)\var\left(\frac{1}{N}\tr(\eta_NJ^{(l)}_N\Pi)\right)\right)^{1/2}
\end{align*} 
In  order to evaluate these variances, one should follow the steps of the proof of Proposition~3.1 in \cite{L:15}.  In \cite{L:15}, matrix $\eta G$ is replaced
by the resolvent of $W W^{*}$ evaluated at $z \in \mathbb{C}^{+}$. The proof of Proposition~3.1 in \cite{L:15} uses the fact that the norm of this resolvent is bounded by $\frac{1}{\im z}$,
a result that is of course not true in the present context. However, the above upper bound is replaced by  $\eta_NG\le \kappa I_N$  (see (\ref{eq:bound_inver_WpWp*})). This
allows to  obtain the same estimations as in Proposition~3.1 in \cite{L:15}:
\begin{align*}
&\var\left((\f_{i_1}^{m_1})^*\eta_NG(J^{(l)}_L\otimes I_M)\f_{i_2}^{m_2}\right)=\Ocal\left(\dfrac{1}{N}\right)\\
&\var\left(\frac{1}{N}\tr(\eta_NJ^{(l)}_N\Pi)\right)=\Ocal\left(\dfrac{1}{N^2}\right)\\
&\var\left(\dfrac{1}{ML}\tr\eta_NG(J^{(l)}_L\otimes I_M)\right)=\Ocal\left(\dfrac{1}{N^2}\right)
\end{align*}
and to conclude that (\ref{le:estimation_inver_WW*}) holds. 

To estimate the expectation of $(ML)^{-1}\tr\eta_N(WW^*)^{-1}$ we take the normalized trace from both sides of (\ref{eq:eta-G}) and use again the Schwartz inequality:
\begin{multline*}
\left|\ex\left\{\frac{1}{ML}\tr(\eta_NG(J^{(l)}_L\otimes I_M))\frac{1}{N}\tr(\eta_NJ^{(l)}_N\Pi^\circ)\right\}\right|
\le\left(\var\left(\frac{1}{ML}\tr\eta_NG(J^{(l)}_L\otimes I_M)\right)\var\left(\frac{1}{N}\tr(\eta_NJ^{(l)}_N\Pi)\right)\right)^{1/2}\\
=\Ocal\left(\dfrac{1}{N^2}\right)
\end{multline*}
Then we get immediately $(ML)^{-1}\tr\ex\{\eta_N(W_iW_i^*)^{-1}\}=(1-c_N)^{-1}+\Ocal(\frac{1}{N^2})$.

Finally, to prove (\ref{le:expectation_Pi}) we return to equation  (\ref{eq:integ_parts_etaGWW*}) but this time we take $m_1=m_3$, $i_1=i_3$ and sum both sides over these indexes:
\begin{multline*}
\ex\{\eta_N\Pi\}=
-c_N\sum_{l=-(L-1)}^{L-1}\ex\left\{\dfrac{1}{ML}\tr(\eta_NG(J^{(l)}_L\otimes I_M))(J^{(l)}_N\Pi)\right\}\\
+c_N\sum_{l=-(L-1)}^{L-1}\ex\left\{\dfrac{1}{ML}\tr(\eta_NG(J^{(l)}_L\otimes I_M))J^{(l)}_N\right\}
+\Ocal\left(\dfrac{1}{N^k}\right)
\end{multline*}
Analogous to what we have seen above, we replace $\eta_N$ by $\eta_N^2$ in the first term of the r.h.s. and  remark that $\ex\{\tr(\eta_NG(J^{(l)}_L\otimes I_M))\}=0$ for all $l\ne0$, since $\ex\{\eta_NG\}$ is block diagonal. Moreover $\ex\{(ML)^{-1}\tr(\eta_NG)\}=(1-c_N)^{-1}+\Ocal(\frac{1}{N^2})$, so that, after trivial algebra, we get
\begin{align*}
\ex\{\eta_N\Pi\}=
c_NI_N+\Ocal\left(\frac{1}{N^2}\right)+\sum_{l=-(L-1)}^{L-1}\ex\left\{\dfrac{1}{ML}\tr(\eta_NG(J^{(l)}_L\otimes I_M))^\circ\eta_NJ^{(l)}_N\Pi\right\}
\end{align*}
The Schwartz inequality allows to obtain  (\ref{le:expectation_Pi}).

\subsection{Proof of Proposition \ref{prop:use-ipp-np-Q}}

We just establish (\ref{eq:integ_by_part_QppPip}). For this, we evaluate each entry of
$\mathbb{E}\left( {\bf Q_{pp}}\eta\Pi_p \right)$ by using the integration by parts formula
(\ref{integr}). In this formula, each entry $\ex\{\left({\bf Q_{pp}}\eta\Pi_p\right)_{rs}\}$
of $\mathbb{E}\left( {\bf Q_{pp}}\eta\Pi_p \right)$ is considered as a function of the entries of
the $2ML \times N$ matrix $\overline{W}_N$ whose elements are the complex conjugate of those of
$W_N$. 


\begin{multline}
\ex\{\left({\bf Q_{pp}}\eta\Pi_p\right)_{rs}\}=\sum_{t=1}^{N}\sum_{i_1,i_2=1}^{L}\sum_{m_1,m_2=1}^{M}\ex\Big\{{\bf Q}_{{\bf pp}}^{rt}\eta \bar{W}_{p,i_1t}^{m_1}\left((W_pW_p^*)^{-1}\right)_{i_1i_2}^{m_1m_2}W_{p,i_2s}^{m_2}\Big\}=\sum\ex\{\bar{W}_{p,i_1t}^{m_1}W_{i_3u}^{m_3}\}\\
\times\ex\Bigg\{\dfrac{\partial\left({\bf Q}_{{\bf pp}}^{rt}\eta \left((W_pW_p^*)^{-1}\right)_{i_1i_2}^{m_1m_2}W_{p,i_2s}^{m_2}\right)}{\partial W_{i_3u}^{m_3}}\Bigg\}
=\dfrac{1}{N}\sum \ex\Bigg\{\delta_{m_1,m_3}\delta_{i_1+t,i_3+u}
{\bf Q}_{{\bf pp}}^{rt}\eta \left((W_pW_p^*)^{-1}\right)_{i_1i_2}^{m_1m_2}\dfrac{\partial W_{p,i_2s}^{m_2}}{\partial W_{i_3u}^{m_3}}\\
+{\bf Q}_{{\bf pp}}^{rt}\eta \dfrac{\partial\left((W_pW_p^*)^{-1}\right)_{i_1i_2}^{m_1m_2}}{\partial W_{i_3u}^{m_3}} W_{p,i_2s}^{m_2}
+\dfrac{\partial{\bf Q}_{{\bf pp}}^{rt}}{\partial W_{i_3u}^{m_3}}\eta \left((W_pW_p^*)^{-1}\right)_{i_1i_2}^{m_1m_2} W_{p,i_2s}^{m_2}
+{\bf Q}_{{\bf pp}}^{rt}\dfrac{\partial\eta}{\partial W_{i_3u}^{m_3}} \left((W_pW_p^*)^{-1}\right)_{i_1i_2}^{m_1m_2} W_{p,i_2s}^{m_2}\Bigg\}
\label{eq:exp-bigipp}
\end{multline}
Here we take the derivative with respect to each element of $W = (W_p^{T},W_f^{T})^{T}$, so index $i_3$ takes values from $1$ to $2L$.  We denote each term of the r.h.s. of (\ref{eq:exp-bigipp}) without expectation by $(T_1)_{rs}$, $(T_2)_{rs}$, $(T_3)_{rs}$, $(T_4)_{rs}$ respectively and treat them  separately. In order to simplify the notations, for $i=1,2,3,4$, we denote $(T_i)_{rs}$ by $T_i$ in the following calculations. 
\begin{align*}
T_1=\dfrac{1}{N}\sum \delta_{m_1,m_3}\delta_{i_1+t,i_3+u}{\bf Q}_{{\bf pp}}^{rt}\eta \left((W_pW_p^*)^{-1}\right)_{i_1i_2}^{m_1m_2}\dfrac{\partial W_{p,i_2s}^{m_2}}{\partial W_{i_3u}^{m_3}}
=\dfrac{1}{N}\sum \delta_{m_1,m_2}\delta_{i_1+t,i_2+s}{\bf Q}_{{\bf pp}}^{rt}\eta \left((W_pW_p^*)^{-1}\right)_{i_1i_2}^{m_1m_2}
\end{align*}
We define $l=i_1-i_2$ and rewrite $\delta_{i_1+t,i_2+s}=\delta_{i_1-i_2,l}\delta_{s-t,l}=(J^{(l)}_M)_{i_2i_1}(J^{(l)}_N)_{ts}$. Taking into account (\ref{eq:def-J}), we obtain
\begin{align}
T_1=\dfrac{1}{N}\sum (J^{(l)}_{N})_{ts}(J^{(l)}_L\otimes I_M)_{i_2i_1}^{m_2m_1}{\bf Q}_{{\bf pp}}^{rt}\eta \left((W_pW_p^*)^{-1}\right)_{i_1i_2}^{m_1m_2}
=\sum_{l=-(L-1)}^{L-1}\left({\bf Q}_{{\bf pp}}J_N^{(l)}\right)_{rs}
\dfrac{1}{N}\tr\left((J^{(l)}_L\otimes I_M)\eta(W_pW_p^*)^{-1}\right)
\end{align}
We take the expectation and obtain
\begin{align*}
\ex\{T_1\}
=\sum_{l=-(L-1)}^{L-1}\ex\Big\{\left({\bf Q}_{{\bf pp}}J_N^{(l)}\right)_{rs}
\Big\}\dfrac{1}{N}\ex\Big\{\tr\left((J^{(l)}_L\otimes I_M)\eta(W_pW_p^*)^{-1}\right)\Big\}\\
+\sum_{l=-(L-1)}^{L-1}\ex\Big\{\left({\bf Q}_{{\bf pp}}^\circ J_N^{(l)}\right)_{rs}
\dfrac{1}{N}\tr\left((J^{(l)}_L\otimes I_M)\eta(W_pW_p^*)^{-1}\right)\Big\}
\end{align*}
We denote the second term of the r.h.s by $T_1^{\mathcal{E}}$. According to (\ref{le:estimation_inver_WW*}), $\ex\{(ML)^{-1}\tr\eta(W_pW_p^*)^{-1}\}=\frac{1}{1-c_N}+\mathcal{O}(\frac{1}{N^2})$. Therefore, if $l=0$ we have 
$$
\dfrac{1}{N}\ex\{\tr(\eta(W_pW_p^*)^{-1})\}=\dfrac{c_N}{(1-c_N)}+\mathcal{O}\left(\dfrac{1}{N^2}\right)
$$
and if $l\ne0$, we have  $\frac{1}{N}\ex\Big\{\tr\left((J^{(l)}_L\otimes I_M)\eta(W_pW_p^*)^{-1}\right)\Big\}=0$ by  Lemma~\ref{le:symetries-E(Q)-2}. Lemma~\ref{le:bound_Q} thus leads to
\begin{align}\label{eq:exp_t_1}
\ex\{T_1\}=\dfrac{c_N}{1-c_N}\ex\Big\{\left({\bf Q}_{{\bf pp}}\right)_{rs}\Big\}+\mathcal{O}_{z^2}\left(\dfrac{1}{N^2}\right)+T_1^{\mathcal{E}}.
\end{align}

For second term, we have 
\begin{align*}
T_2=-\dfrac{1}{N}\sum\delta_{m_1,m_3}\delta_{i_1+t,i_3+u}{\bf Q}_{{\bf pp}}^{rt}\eta\left((W_pW_p^*)^{-1}\right)_{i_1i_3}^{m_1m_3}\left(W_p^*(W_pW_p^*)^{-1}\right)_{ui_2}^{m_2}W_{p,i_2s}^{m_2}
\end{align*}
We define $l=i_1-i_3$. Then,  $\delta_{i_1+t,i_3+u}=\delta_{i_1-i_3,l}\delta_{u-t,l}=(J^{(l)}_M)_{i_3i_1}(J^{(l)}_N)_{tu}$. This gives us
\begin{align}
T_2=-\sum_{l=-(L-1)}^{L-1}\left(\eta{\bf Q}_{{\bf pp}}J_N^{(l)}\Pi_p\right)_{rs}
\dfrac{1}{N}\tr\left((J^{(l)}_L\otimes I_M)(W_pW_p^*)^{-1}\right)
\end{align}
Taking the expectation and replacing $\eta$ by $\eta^2$, we have for each $k \geq 1$,
\begin{multline*}
\ex\{T_2\}=-\sum_{l=-(L-1)}^{L-1}\ex\Big\{\left(\eta{\bf Q}_{{\bf pp}}J_N^{(l)}\Pi_p\right)_{rs}\Big\}
\dfrac{1}{N}\ex\Big\{\tr\left(\eta(J^{(l)}_L\otimes I_M)(W_pW_p^*)^{-1}\right)\Big\}\\
-\sum_{l=-(L-1)}^{L-1}\ex\Big\{\left(\eta{\bf Q}_{{\bf pp}}J_N^{(l)}\Pi_p\right)_{rs}^\circ
\dfrac{1}{N}\tr\left(\eta(J^{(l)}_L\otimes I_M)(W_pW_p^*)^{-1}\right)\Big\}+\Ocal_{z^2}\left(\dfrac{1}{N^k}\right)
\end{multline*}
As previously, we denote the second term of the r.h.s. by $T_2^{\mathcal{E}}$ and notice that in the first term of the r.h.s.,  according to Lemma~\ref{le:symetries-E(Q)-2},  all the terms corresponding to $l \neq 0$ are zeros, and $\ex\{(ML)^{-1}\tr\eta(W_pW_p^*)^{-1}\}=\frac{1}{1-c_N}+\mathcal{O}(\frac{1}{N^2})$. Therefore, we obtain that 
\begin{align}\label{eq:exp_t_2}
\ex\{T_2\}=-\dfrac{c_N}{1-c_N}\ex\Big\{\left(\eta{\bf Q}_{{\bf pp}}\Pi_p\right)_{rs}\Big\}+T_2^{\mathcal{E}}+\Ocal_{z^2}\left(\dfrac{1}{N^2}\right)
\end{align} 
To deal with the third term, $T_3$, we first should find the derivatives of the resolvent w.r.t. the entries of $W$. For this, we write
\begin{align}\label{{eq:der_Q}}
\partial\Q=-\Q\partial\begin{pmatrix}
0&\eta\Pi_p\\
\eta\Pi_f&0
\end{pmatrix}\Q=-\begin{pmatrix}
{\bf Q_{pf}}\partial(\eta \Pi_f){\bf Q_{pp}}+{\bf Q_{pp}}\partial(\eta \Pi_p){\bf Q_{fp}}&{\bf Q_{pf}}\partial(\eta \Pi_f){\bf Q_{pf}}+{\bf Q_{pp}}\partial(\eta \Pi_p){\bf Q_{ff}}\\
{\bf Q_{ff}}\partial(\eta \Pi_f){\bf Q_{pp}}+{\bf Q_{fp}}\partial(\eta \Pi_p){\bf Q_{fp}}&{\bf Q_{ff}}\partial(\eta \Pi_f){\bf Q_{pf}}+{\bf Q_{fp}}\partial(\eta \Pi_p){\bf Q_{ff}}
\end{pmatrix}
\end{align}
and evaluate the derivative with respect to the element $ W^{m_3}_{i_3u}$. As was discussed before, since   $\|\Q\|$ and $\|\Pi_{i}\|$, $i=p,f$, are bounded (see Lemma~\ref{le:bound_Q}), the expectation of the entries of the terms containing $\frac{\partial \eta}{\partial W^{m_3}_{i_3u}}$ are $\mathcal{O}_{z^2}(N^{-k})$ terms for each $k \geq 1$. This justifies that we gather all this terms together in a matrix, denoted $\mathcal{E}$, whose entries are  $\mathcal{O}_{z^2}(N^{-k})$ for any $k$.
We also need to evaluate the derivative of projectors $\Pi_{p}$ and $\Pi_f$. For this, we use classical perturbation theory results (\cite{kato-book}, see also Theorem 6 in  \cite{abe-lou-mou-ieeeit-1997} for the statement of the result), and obtain
\begin{align}\label{eq:der_proj}
\delta\Pi_p=\Pi_p^\perp\delta(W_p^*W_p)(W_p^*W_p)^\#+(W_p^*W_p)^\#\delta(W_p^*W_p)\Pi_p^\perp\,
\end{align}
where $(W_p^*W_p)^\#$ is the pseudoinverse of $W_p^*W_p$, which, in this case, is equal to $W_p^*(W_pW_p^*)^{-2}W_p$. The expression of 
$\delta\Pi_f$ is similar. The derivative with respect to $ W^{m_3}_{i_3u}$ is thus given by
\begin{align*}
\dfrac{\partial\Pi_p}{\partial  W^{m_3}_{i_3u}}
=\left(\Pi_p^\perp W_p^*\f_{i_3}^{m_3}\e_uW_p^*(W_pW_p^*)^{-2}W_p+W_p^*(W_pW_p^*)^{-2}W_pW_p^*\f_{i_3}^{m_3}\e_u\Pi_p^\perp\right){\bf 1}_{i_3\le L}.
\end{align*}
In this context, $\f_{i_3}^{m_3}$ is a vector of the canonical basis of $\mathbb{C}^{ML}$ rather than of $\mathbb{C}^{2ML}$. Since $\Pi_p^\perp W_p^*=0$ the first term disappears and we obtain 
\begin{align*}
\dfrac{\partial\Pi_p}{\partial  W^{m_3}_{i_3u}}=W_p^*(W_pW_p^*)^{-1}\f_{i_3}^{m_3}\e^*_u\Pi_p^\perp{\bf 1}_{i_3\le L}
\end{align*}
For $\Pi_f$ the formula is analogous, but  $\f_{i_3}^{m_3}$ is replaced by $\f_{i_3-L}^{m_3}$:
\begin{align*}
\dfrac{\partial\Pi_f}{\partial  W^{m_3}_{i_3u}}=W_f^*(W_fW_f^*)^{-1}\f_{i_3-L}^{m_3}\e^*_u\Pi_f^\perp{\bf 1}_{i_3> L}
\end{align*}
Putting these expressions in (\ref{{eq:der_Q}}), we get that
\begin{multline}\label{eq:final_der_Q}
\dfrac{\partial\Q}{\partial  W^{m_3}_{i_3u}}=-\eta{\bf 1}_{i_3\le L}
\begin{pmatrix}
{\bf Q_{pp}}(W_p^*(W_pW_p^*)^{-1}\f_{i_3}^{m_3}\e^*_{u}\Pi^\perp_p{\bf Q_{fp}}&
{\bf Q_{pp}}(W_p^*(W_pW_p^*)^{-1}\f_{i_3}^{m_3}\e^*_{u}\Pi^\perp_p{\bf Q_{ff}}\\
{\bf Q_{fp}}(W_p^*(W_pW_p^*)^{-1}\f_{i_3}^{m_3}\e^*_{u}\Pi^\perp_p{\bf Q_{fp}}&{\bf Q_{fp}}(W_p^*(W_pW_p^*)^{-1}\f_{i_3}^{m_3}\e^*_{u}\Pi^\perp_p{\bf Q_{ff}}
\end{pmatrix}\\
-\eta{\bf 1}_{i_3> L}
\begin{pmatrix}
{\bf Q_{pf}}(W_f^*(W_fW_f^*)^{-1}\f_{i_3-L}^{m_3}\e^*_{u}\Pi^\perp_f{\bf Q_{pp}}&
{\bf Q_{pf}}(W_f^*(W_fW_f^*)^{-1}\f_{i_3-L}^{m_3}\e^*_{u}\Pi^\perp_f{\bf Q_{pf}}\\
{\bf Q_{ff}}(W_f^*(W_fW_f^*)^{-1}\f_{i_3-L}^{m_3}\e^*_{u}\Pi^\perp_f{\bf Q_{pp}}&{\bf Q_{ff}}(W_f^*(W_fW_f^*)^{-1}\f_{i_3-L}^{m_3}\e^*_{u}\Pi^\perp_f{\bf Q_{pf}}
\end{pmatrix}
+\mathcal{E}
\end{multline}
We are now ready to address the term $T_3$. We first we sum over $i_2,m_2$, and obtain that $T_3$ can be written as
\begin{multline}
T_3=-\dfrac{1}{N}\sum\delta_{m_1,m_3}\delta_{i_1+t,i_3+u}\eta^2
\left({\bf Q_{pp}}W_p^*(W_pW_p^*)^{-1}\right)_{ri_3}^{m_3}\left(\Pi^\perp_p{\bf Q_{fp}}\right)_{ut}\left((W_pW_p^*)^{-1}W_p\right)_{i_1s}^{m_1}{\bf 1}_{i_3\le L}\\
- \dfrac{1}{N}\sum\delta_{m_1,m_3}\delta_{i_1+t,i_3+u}\eta^2
\left({\bf Q_{pf}}W_f^*(W_fW_f^*)^{-1}\right)_{ri_3-L}^{m_3}\left(\Pi^\perp_f{\bf Q_{pp}}\right)_{ut}\left((W_pW_p^*)^{-1}W_p\right)_{i_1s}^{m_1}{\bf 1}_{i_3> L}+\mathcal{E}_3
\label{eq:expre-bis-T3}
\end{multline}
where $\mathcal{E}_3$ represents the contribution of matrix $\mathcal{E}$ to $T_3$. In order to express the first term of the r.h.s. of 
(\ref{eq:expre-bis-T3}),  we again define $l=i_1-i_3$ which belongs to $\{-(L-1),\ldots, L-1\}$ and notice that $\delta_{i_1+t,i_3+u}=\delta_{i_1-i_3,l}\delta_{u-t,l}=(J_L^{(l)})_{i_3i_1}(J_N^{(l)})_{tu}$. As for second term of the r.h.s. of (\ref{eq:expre-bis-T3}), we first exchange  $i_3 > L$ by $i_3 -L$ which runs from $1$ to $L$. The second term becomes 
$$\frac{1}{N}\sum\delta_{m_1,m_3}\delta_{i_1+t,i_3+L+u}\eta^2
({\bf Q_{pf}}W_f^*(W_fW_f^*)^{-1})_{ri_3}^{m_3}(\Pi^\perp_f{\bf Q_{pp}})_{ut}((W_pW_p^*)^{-1}W_p)_{i_1s}^{m_1}{\bf 1}_{i_3<L}$$
We again put $l=i_1-i_3$ and remark that $\delta_{i_1+t,i_3+L+u}=\delta_{i_1-i_3,l}\delta_{u-t,l-L}=(J_L^{(l)})_{i_3,i_1}(J_N^{(l-L)})_{tu}$. Therefore, we obtain that $T_3$ is equal to
\begin{multline*}
T_3=-\sum_{l=-(L-1)}^{L-1}\eta^2\left({\bf Q_{pp}}W_p^*(W_pW_p^*)^{-1}(J^{(l)}_L\otimes I_M)(W_pW_p^*)^{-1}W_p\right)_{rs}\dfrac{1}{N}\tr\left(\Pi^\perp_p{\bf Q_{fp}}J^{(l)}_N\right)\\
-\sum_{l=-(L-1)}^{L-1}\eta^2\left({\bf Q_{pf}}W_f^*(W_fW_f^*)^{-1}(J^{(l)}_L\otimes I_M)(W_pW_p^*)^{-1}W_p\right)_{rs}\dfrac{1}{N}\tr\left(\Pi^\perp_f{\bf Q_{pp}}J^{(l-L)}_N\right)+\mathcal{E}_3
\end{multline*} 
Taking the expectation, we obtain that 
\begin{multline*}
\ex\{T_3\}=-\sum_{l=-(L-1)}^{L-1}\ex\Big\{\eta\left({\bf Q_{pp}}W_p^*(W_pW_p^*)^{-1}(J^{(l)}_L\otimes I_M)(W_pW_p^*)^{-1}W_p\right)_{rs}\Big\}\dfrac{1}{N}\ex\Big\{\tr\left(\eta\Pi^\perp_p{\bf Q_{fp}}J^{(l)}_N\right)\Big\}\\
-\sum_{l=-(L-1)}^{L-1}\ex\Big\{\eta\left({\bf Q_{pf}}W_f^*(W_fW_f^*)^{-1}(J^{(l)}_L\otimes I_M)(W_pW_p^*)^{-1}W_p\right)_{rs}\Big\}\dfrac{1}{N}\ex\Big\{\tr\left(\eta\Pi^\perp_f{\bf Q_{pp}}J^{(l-L)}_N\right)\Big\}\\
+\ex\{\mathcal{E}_3\}+T_3^{\mathcal{E}},
\end{multline*}
where, as above, $T_3^{\mathcal{E}}$ is defined by 
\begin{multline*}
T_3^{\mathcal{E}} = \sum_{l} \mathbb{E} \left[ \left(\eta{\bf Q_{pp}}W_p^*(W_pW_p^*)^{-1}(J^{(l)}_L\otimes I_M)(W_pW_p^*)^{-1}W_p\right)_{rs}^\circ  \,   
\dfrac{1}{N}\tr\left(\Pi^\perp_p{\bf Q_{fp}}J^{(l)}_N\right)^{\circ} \right] + \\
\sum_{l} \mathbb{E} \left[ \left(\eta{\bf Q_{pf}}W_f^*(W_fW_f^*)^{-1}(J^{(l)}_L\otimes I_M)(W_pW_p^*)^{-1}W_p\right)_{rs}^\circ \, 
\dfrac{1}{N}\tr\left(\Pi^\perp_f{\bf Q_{pp}}J^{(l-L)}_N\right)^{\circ} \right]
\end{multline*}
According to Lemma~\ref{le:symetries-E(Q)-2}, $\ex\{\eta\Pi^\perp_p{\bf Q_{fp}}\}$ and $\ex\{\eta\Pi^\perp_f{\bf Q_{pp}}\}$ are diagonal. 
Therefore, the traces of these matrices multiplied by $J_N^{(k)}$ for $k\ne0$ are zeros. This leads to
\begin{align}\label{eq:exp_t_3}
\ex\{T_3\}=-\ex\Big\{\eta\left({\bf Q_{pp}}W_p^*(W_pW_p^*)^{-2}W_p\right)_{rs}\Big\}\dfrac{1}{N}\ex\Big\{\tr\left(\eta\Pi^\perp_p{\bf Q_{fp}}\right)\Big\}
+\ex\{\mathcal{E}_3\}+T_3^{\mathcal{E}}.
\end{align}
Finally, the various terms of $T_4$ contain the terms $(\frac{\partial \eta}{\partial W^{m_3}_{i_3u}})_{i_3=1, \ldots, 2L, m_3=1, \ldots, M} $. Therefore, $T_4$ is denoted $\mathcal{E}_4$, and $\mathbb{E}(T_4) = \mathbb{E}(\mathcal{E}_4) = \Ocal_{z^2}(N^{-k})$ for each $k$. \\

Combining (\ref{eq:exp_t_1}), (\ref{eq:exp_t_2}), (\ref{eq:exp_t_3}), we have thus obtained that 
\begin{multline*}
\ex\Big\{({\bf Q_{pp}}\eta\Pi_p)_{rs}\Big\}=\dfrac{c_N}{1-c_N}\ex\Big\{({\bf Q}_{{\bf pp}})_{rs}\Big\}
-\dfrac{c_N}{1-c_N}\ex\Big\{\eta({\bf Q}_{{\bf pp}}\Pi_p)_{rs}\Big\}\\
-\ex\Big\{\eta({\bf Q_{pp}}W_p^*(W_pW_p^*)^{-2}W_p)_{rs}\Big\}\dfrac{1}{N}\ex\Big\{\tr\left(\eta\Pi^\perp_p{\bf Q_{fp}}\right)\Big\}
+\frac{1}{1-c_N} ({\bf \Delta_{pp}})_{rs}
\end{multline*}
where $({\bf \Delta_{pp}})_{rs}$ represents the term 
$$
({\bf \Delta_{pp}})_{rs} = (1-c_N) \, \left[ \ex\{T_1^{\mathcal{E}}+T_2^{\mathcal{E}}+T_3^{\mathcal{E}} \}+\ex\{\mathcal{E}_3\} + \ex\{\mathcal{E}_4\} + \Ocal_{z^{2}}(\frac{1}{N^2}) \right]
$$
obtained by gathering the various error terms defined in the evaluation of $(T_i)_{i=1,2,3,4}$. Therefore, we 
eventually get the following expression of matrix $\ex\Big\{{\bf Q_{pp}}\eta\Pi_p\Big\}$:
\begin{multline*}
\ex\Big\{{\bf Q_{pp}}\eta\Pi_p\Big\}=\dfrac{c_N}{1-c_N}\ex\Big\{{\bf Q}_{{\bf pp}}\Big\}
-\dfrac{c_N}{1-c_N}\ex\Big\{\eta{\bf Q}_{{\bf pp}}\Pi_p\Big\}\\
-\ex\Big\{\eta{\bf Q_{pp}}W_p^*(W_pW_p^*)^{-2}W_p\Big\}\dfrac{1}{N}\ex\Big\{\tr\left(\eta\Pi^\perp_p{\bf Q_{fp}}\right)\Big\}
+\frac{1}{1-c_N} \, {\bf \Delta_{pp}}
\end{multline*}
which leads immediately to (\ref{eq:integ_by_part_QppPip}). \\


It remains to establish the properties of matrix ${\bf \Delta_{pp}}$.  According to Lemma~\ref{le:symetries-E(Q)-2}, $\ex\{{\bf Q_{pp}}\eta\Pi_p\}, \ex\{{\bf Q_{pp}}\}$ and $\ex\{\eta{\bf Q_{pp}}W^*_p(W_pW^*_p)^{-2}W_p\}$ are diagonal. Therefore, (\ref{eq:integ_by_part_QppPip}) implies that ${\bf \Delta_{pp}}$ is also diagonal. In order to evaluate the order of magnitude of the entries of ${\bf \Delta_{pp}}$
and of $\frac{1}{N} \mathrm{Tr}({\bf \Delta_{pp}})$, we 
 first prove the next lemma which is based on the Poincar\'e-Nash inequality.

\begin{lemma}\label{le:estim_var}
	Let $(F_N)_{N \geq 1}$ and $(G_N)_{N \geq 1}$ be sequences of deterministic $N \times N$ matrices such that $\sup_N \|F_N\|$, $\sup_N \|G_N\| \le \kappa$, and consider sequences of deterministic $N$--dimensional vectors $(a_{1,N})_{N \geq 1}$, $(a_{2,N})_{N \geq 1}$ such that $sup_N \|a_{i,N}\| \le\kappa$ for $i=1,2$. Then, for each $z \in \mathbb{C}^+$ and $i,j,h=\{p,f\}$, it holds that
	\begin{align}\label{eq:var_trQ}
	&\var\left\{\dfrac{1}{N}\tr F\mathbf{Q_{ij}}\right\}=\Ocal_{z^2}\left(\dfrac{1}{N^{2}}\right),\\
	&\var\left\{\dfrac{1}{N}\tr \mathbf{Q_{ij}}F\eta_N\Pi_hG\right\}=\Ocal_{z^2}\left(\dfrac{1}{N^{2}}\right),\label{eq:var_trQPi}\\
	&\var\left\{\dfrac{1}{N}\tr \mathbf{Q_{ij}}F\eta_N\Pi^\perp_hG\right\}=\Ocal_{z^2}\left(\dfrac{1}{N^{2}}\right),\label{eq:var_trQPi_perp}\\	
	&\var\left\{\dfrac{1}{N}\tr \eta {\bf Q_{ij}}W_h^* (W_hW_h^*)^{-1}F (W_kW_k^*)^{-1}W_k  \right\} = \Ocal_{z^2}\left(\dfrac{1}{N}\right), \label{eq:var_tr_QW*(WW*)F(WW*)W}\\
	&\var\left(a_1^*\eta{\bf Q_{ij}}W_h^* (W_hW_h^*)^{-1}F (W_kW_k^*)^{-1}W_ka_2\right)=\Ocal_{z^2}\left(\dfrac{1}{N}\right),\label{eq:var_QW*(WW*)F(WW*)W}\\
	  &\var\left\{a_1^*\mathbf{Q_{ij}}a_2\right\}=\Ocal_{z^2}\left(\dfrac{1}{N}\right),\label{eq:var_Q}\\
          & \var \left[\left( a_1^*\mathbf{Q_{ij}}a_2 - \mathbb{E}(a_1^*\mathbf{Q_{ij}}a_2)\right)^{2} \right] = \Ocal_{z^2}\left(\dfrac{1}{N^{2}}\right),\label{eq:var_Q_as}\\
	&\var\left\{a_1^*\mathbf{Q_{ij}}F\eta_N\Pi_ha_2\right\}=\Ocal_{z^2}\left(\dfrac{1}{N}\right).\label{eq:var_QPi}
	\end{align}
\end{lemma}
{\bf Proof.} 
We just prove (\ref{eq:var_trQ}) for ${\bf Q_{pp}}$. The proofs of the other items are omitted. We denote by $\xi$ the term $\xi=\frac{1}{N}\tr F\mathbf{Q_{pp}}$. The Poincar\'e-Nash inequality (\ref{prop:poincare}) leads to
\begin{align*}
\var\{\xi\}&\le \sum_{\substack{i_1,j_1,m_1\\i_2,j_2,m_2}}\ex\left\{\left(\dfrac{\partial\xi}{\partial \overline{W}_{i_1,j_1}^{m_1}}\right)^*\ex\{W_{i_1,j_1}^{m_1}\overline{W}_{i_2,j_2}^{m_2}\}\dfrac{\partial\xi}{\partial\overline{W}_{i_2,j_2}^{m_2}}\right\}\\
&+\sum_{\substack{i_1,j_1,m_1\\i_2,j_2,m_2}}\ex\left\{\dfrac{\partial\xi}{\partial W_{i_1,j_1}^{m_1}}\ex\{W_{i_1,j_1}^{m_1}\overline{W}_{i_2,j_2}^{m_2}\}\left(\dfrac{\partial\xi}{\partial W_{i_2,j_2}^{m_2}}\right)^*\right\}.
\end{align*}
We just evaluate the second term of the r.h.s., denoted by $\phi$.
The derivative of $\xi$ can be found with the help of (\ref{eq:final_der_Q}):
\begin{multline*}
\dfrac{\partial\xi}{\partial W_{i_1j_1}^{m_1}}
=-\dfrac{\eta}{N}\tr F
{\bf Q_{pp}}W_p^*(W_pW_p^*)^{-1}\f_{i_1}^{m_1}\e^*_{j_1}\Pi^\perp_p{\bf Q_{fp}}{\bf 1}_{i_1\le L}\\
-\dfrac{\eta}{N}\tr F
{\bf Q_{pf}}W_f^*(W_fW_f^*)^{-1}\f_{i_1-L}^{m_1}\e^*_{j_1}\Pi^\perp_f{\bf Q_{pp}}{\bf 1}_{i_1> L}
+\Ocal\left(\dfrac{1}{N^{k}}\right)
\end{multline*}
$\phi$ is clearly the sum of four similar terms. We  just evaluate
\begin{align}
\dfrac{1}{N^3}\sum_{\substack{i_1,j_1,m_1\\i_2,j_2,m_2}}\delta_{m_1,m_2}\delta_{i_1+j_1,i_2+j_2}\ex\left\{\eta^2\e_{j_1}^*\Pi^\perp_p{\bf Q_{fp}}F
{\bf Q_{pp}}W_p^*(W_pW_p^*)^{-1}\f_{i_1}^{m_1}\f_{i_2}^{m_2*}(W_pW_p^*)^{-1}W_p{\bf Q^*_{pp}}F{\bf Q^*_{fp}}\Pi^\perp_p\e_{j_2}\right\}
\label{eq:expre-NP-xi-appendix}
\end{align}
where $1\le i_1,i_2\le L$. Now we again denote $l=i_1-i_2=j_2-j_1$ which lies in $(-L+1,L-1)$ and remark that  $\sum_{m_1,m_2,i_1,i_2}\delta_{m_1,m_2}\delta_{i_1-i_2,l}\f_{i_1}^{m_1}\f_{i_2}^{m_2*}=(J^{(l)}_L\otimes I_M)$ as well as $\sum_{j_1,j_2}\delta_{j_2-j_1,l}\e_{j_2}\e_{j_1}^*=J^{(l)}_N$. This allows to rewrite (\ref{eq:expre-NP-xi-appendix}) as
\begin{align}
\dfrac{1}{N^3}\sum_{l=-(L-1)}^{L-1}\ex\left\{\eta^2\tr \Pi^\perp_p{\bf Q_{fp}}F
{\bf Q_{pp}}W_p^*(W_pW_p^*)^{-1}(J^{(l)}_L\otimes I_M)(W_pW_p^*)^{-1}W_p{\bf Q^*_{pp}}F^*{\bf Q^*_{fp}}\Pi^\perp_pJ^{(l)}_N\right\}
\end{align}
For each $N \times ML$ matrices $A$ and $B$, the Schwartz inequality and the inequality between arithmetic and geometric means lead to
\begin{align*}
\left| \dfrac{1}{N}\tr A(I_M\otimes J^{*(l)}_{L})B^{*}J^{*(l)}_N \right| \le \dfrac{1}{2N}\tr A(I_M\otimes J^{*(l)}_LJ^{(l)}_L)A^{*}+\dfrac{1}{2N}\tr BJ^{*(l)}_NJ^{(l)}_NB^{*}.
\end{align*}
Therefore, since $I_M\otimes J^{*(l)}_LJ^{(l)}_L \le I_{ML}$ and $J^{*(l)}_NJ^{(l)}_N\le I_N$, the inequality
\begin{align}\label{eq:ineq_AJBJ}
\left| \dfrac{1}{N} \tr A(I_M\otimes J^{*(l)}_{L})B^{*}J^{*(l)}_N \right| \le\dfrac{C}{2N}(\tr A^*A+\tr B^*B).
\end{align}
holds. We take $A=B= \Pi^\perp_p{\bf Q_{fp}}F
{\bf Q_{pp}}W_p^*\eta(W_pW_p^*)^{-1}$, and have to check that $N^{-1}\ex\{\tr AA^*\} = \Ocal_{z}^2(1)$. For this, 
we remark that $\eta W_p^{*} W_p \leq ((1+\sqrt{c_*})^{2} + 2 \epsilon) I_N$ and $\eta^2(W_pW_p^*)^{-2}\le ((1-\sqrt{c_*})^2-2\epsilon)^{-2}I_{ML}$ (see (\ref{eq:bound_inver_WpWp*})). Therefore, $W_p^*\eta^2(W_pW_p^*)^{-2}W_p \leq \kappa I_N$, and 
\begin{align}\label{eq:eval_var_exemp}
\dfrac{1}{N}\ex\left\{\tr \Pi^\perp_p{\bf Q_{fp}}F
{\bf Q_{pp}}W_p^*\eta^2(W_pW_p^*)^{-2}W_p{\bf Q^*_{pp}}F^*{\bf Q^*_{fp}}\Pi^\perp_p\right\} = \Ocal_{z}^2(1)
\end{align} 
as expected. This completes the proof of (\ref{eq:var_trQ}). $\blacksquare$ \\

We return to the evaluation of the (diagonal) entries of ${\bf \Delta_{pp}}$. As the terms $\mathbb{E}(\mathcal{E}_3)$ and $\mathbb{E}(\mathcal{E}_4)$ 
are $\Ocal_{z}^2(\frac{1}{N^{k}})$ for each $k$, it remains to 
evaluate the order of magnitude of the terms $(T_i^{\Ecal})_{rr}$ for $i=1,2,3$
defined by (\ref{eq:exp_t_1},\ref{eq:exp_t_2},\ref{eq:exp_t_3}) respectively when $r=s$.  
We start with $(T_1^{\Ecal})_{rr}$ and use Schwartz inequality:
\begin{multline*}
|(T_1^{\Ecal})_{rr}|=\Bigg|\sum_{l=-(L-1)}^{L-1}\ex\left\{\left({\bf Q^{\circ}_{pp}}J^{(l)}_N\right)_{rr}\dfrac{1}{N}\tr\left((I_M\otimes J^{(l)}_M)\eta (W_pW_p^*)^{-1}\right)\right\}\Bigg|\\
\le\sum_{l=-(L-1)}^{L-1}\left(\var\left(\left({\bf Q_{pp}}J^{(l)}_N\right)_{rr}\right)\var\left(\dfrac{1}{N}\tr\left((I_M\otimes J^{(l)}_M)\eta (W_pW_p^*)^{-1}\right)\right)\right)^{1/2}
\end{multline*} 
We apply (\ref{eq:var_Q}) for $a_1=\e_r$ and $a_2=J^{(l)}_N\e_r$  and take into account that $\var(\frac{1}{N}\tr((I_M\otimes J^{(l)}_M)\eta (W_pW_p^*)^{-1}))=\Ocal(N^{-2})$. Then
\begin{align}
|(T_1^{\Ecal})_{rr}|\le\Ocal_{z}^2\left(\dfrac{1}{N^{3/2}}\right)
\end{align}
As for $(T_2^{\Ecal})_{rr}$, we have
\begin{multline*}
|(T_2^{\Ecal})_{rr}|=\Bigg|\sum_{l=-(L-1)}^{L-1}\ex\left\{\left(\eta{\bf Q_{pp}}J^{(l)}_N\Pi_p\right)^{\circ}_{rr}\dfrac{1}{N}\tr\left((I_M\otimes J^{(l)}_M)\eta (W_pW_p^*)^{-1}\right)\right\}\Bigg|\\
\le\sum_{l=-(L-1)}^{L-1}\left(\var\left(\left(\eta{\bf Q_{pp}}J^{(l)}_N\Pi_p\right)_{rr}\right)\var\left(\dfrac{1}{N}\tr\left((I_M\otimes J^{(l)}_M)\eta (W_pW_p^*)^{-1}\right)\right)\right)^{1/2}
\end{multline*}
From (\ref{eq:var_QPi}) we get immediately
\begin{align}
|(T_2^{\Ecal})_{rr}|=\Ocal_{z}^2\left(\dfrac{1}{N^{3/2}}\right)
\end{align}
For $T_3^{\Ecal}$ we obtain
\begin{multline*}
|(T_3^{\Ecal})_{rr}|=\Bigg|\sum_{l=-(L-1)}^{L-1}\ex\left\{\left(\eta{\bf Q_{pp}}W_p^* (W_pW_p^*)^{-1}(I_M\otimes J^{(l)}_M) (W_pW_p^*)^{-1}W_p \right)^{\circ}_{rr}\dfrac{1}{N}\tr\left(\eta J^{(l)}_N\Pi^\perp_p{\bf Q_{fp}}\right)\right\}\\
+\sum_{l=-(L-1)}^{L-1}\ex\left\{\left(\eta{\bf Q_{pf}}W_f^* (W_fW_f^*)^{-1}(I_M\otimes J^{(l)}_M) (W_pW_p^*)^{-1}W_p \right)^{\circ}_{rr}\dfrac{1}{N}\tr\left(\eta J^{(l-L)}_N\Pi^\perp_f{\bf Q_{pp}}\right)\right\}\Bigg|\\
\le\sum_{l=-(L-1)}^{L-1}\left(\var\left(\left(\eta{\bf Q_{pp}}W_p^* (W_pW_p^*)^{-1}(I_M\otimes J^{(l)}_M) (W_pW_p^*)^{-1}W_p \right)_{rr}\right)\var\left(\dfrac{1}{N}\tr\left(\eta J^{(l)}_N\Pi^\perp_p{\bf Q_{fp}}\right)\right)\right)^{1/2}\\
+\sum_{l=-(L-1)}^{L-1}\left(\var\left(\left(\eta{\bf Q_{pf}}W_f^* (W_fW_f^*)^{-1}(I_M\otimes J^{(l)}_M) (W_pW_p^*)^{-1}W_p \right)_{rr}\right)\var\left(\dfrac{1}{N}\tr\left(\eta J^{(l)}_N\Pi^\perp_f{\bf Q_{pp}}\right)\right)\right)^{1/2}
\end{multline*}
from what, using again (\ref{eq:var_QW*(WW*)F(WW*)W}) and (\ref{eq:var_trQPi_perp}), we immediately get 
\begin{align}
|(T_3^{\Ecal})_{rr}|=\Ocal_{z}^2\left(\dfrac{1}{N^{3/2}}\right)
\end{align}
To evaluate the normalized trace of ${\bf\Delta_{pp}}$, we still use the Schwartz inequality, and take benefit of the 
estimates (\ref{eq:var_trQ})-(\ref{eq:var_tr_QW*(WW*)F(WW*)W}) to improve the rate of convergence of $\frac{1}{N} \mathrm{Tr} {\bf\Delta_{pp}}$.

\subsection{Proof of Proposition \ref{prop:behaviour-hatnuN-section3}}
In order to  evaluate  $\tilde{ \alpha}_N-\tilde{t}_N$, it is natural to take the difference between equations (\ref{eq:tilde_alpha}) and (\ref{eq:eq_tilde_t_N}):
\begin{align*}
(\tilde{ \alpha}_N-\tilde{t}_N)\left((1-z)z(\tilde{ \alpha}_N+\tilde{t}_N)+2(1-c_N)-z \right)=\Ocal_{z}\left(\dfrac{1}{N^2}\right)
\end{align*}
We remind that $\tilde{ \alpha}_N=\alpha_N-\frac{1-c_N}{z}+\Ocal_{z}(N^{-k})$ (see (\ref{eq:connection_alph_tilde-alph})) and rewrite  the last equation as
$$
(\tilde{ \alpha}_N-\tilde{t}_N)\left((1-z)z \alpha_N-(1-z)(1-c_N)+(1-z)z\tilde{t}_N+2(1-c_N)-z+\Ocal_{z}(N^{-k})\right)=\Ocal_{z}\left(\dfrac{1}{N^2}\right)
$$
or equivalently as
$$
(\tilde{ \alpha}_N-\tilde{t}_N)\left((1-z)z \alpha_N-(1-z)(1-c_N)+(1-z)z\tilde{t}_N+2(1-c_N)-z \right)+(\tilde{ \alpha}_N-\tilde{t}_N)\Ocal_{z}\left(\dfrac{1}{N^{k}}\right)=\Ocal_{z}\left(\dfrac{1}{N^2}\right)
$$
Since $\tilde{\alpha}_N$, $\tilde{t}_N$ and $\alpha_N$ (see Remark \ref{eq:alpha-stieltjes}) are the Stieltjes transforms of a positive measures carried by $\mathbb{R}^{+}$, we obtain that 
$\alpha_N=\Ocal_{z}(1)$, $\tilde{ \alpha}_N=\Ocal_z(1)$, $\tilde{t}_N=\Ocal_z(1)$, and that $(\tilde{ \alpha}_N-\tilde{t}_N)\Ocal_{z}(N^{-k})=\Ocal_{z}(N^{-k})$. $(\tilde{ \alpha}_N-\tilde{t}_N)$ can thus be written as 
\begin{align*}
\tilde{\alpha}_N-\tilde{t}_N
=\dfrac{\Ocal_{z}\left(N^{-2}\right)}{(1-z)z \alpha_N-(1-z)(1-c_N)+(1-z)z\tilde{t}_N+2(1-c_N)-z}
\end{align*}
We now evaluate the denominator for $z \in \mathbb{C}^{+}$. For this, we return to (\ref{eq:eq_tilde_t_N}) and write:
\begin{align*}
(1-z)z\tilde{t}_N+2(1-c_N)-z=-\dfrac{(1-c_N)^2}{z\tilde{t}_N}
\end{align*}
Moreover, since $\tilde{t}_N$ is the Stieltjes transform of a positive measure carried by $\mathbb{R^+}$, $\im z\tilde{t}_N>0$ for $z\in\mathbb{C^+}$ (see (\ref{eq:ImzS})) and  $\im((1-z)z\tilde{t}_N)=\im z-\im \dfrac{(1-c_N)^2}{z\tilde{t}_N}>\im z$. We rewrite the denominator as
\begin{align*}
(1-z)z \alpha_N-(1-z)(1-c_N)+(1-z)z\tilde{t}_N+2(1-c_N)-z=(1-z)\left(z \alpha_N-(1-c_N)-\dfrac{(1-c_N)^2}{(1-z)z\tilde{t}_N}\right)
\end{align*}
 $\im z\alpha_N>0$ for $z\in\mathbb{C^+}$ because $\alpha_N$ is the Stieltjes transform of a positive measure curried by $\mathbb{R^+}$. Thus 
\begin{align*}
|(1-z)z \alpha_N-(1-z)(1-c_N)+(1-z)z\tilde{t}_N+2(1-c_N)-z|\ge|1-z|\im\dfrac{-(1-c_N)^2}{(1-z)z\tilde{t}_N}=\dfrac{(1-c_N)^2\im((1-z)z\tilde{t}_N)}{|1-z||z|^2|\tilde{t}_N|^2}
\end{align*}
As $\im((1-z)z\tilde{t}_N)>\im z$ and $|\tilde{t}_N(z)|\le(\im z)^{-1}$ on $\mathbb{C}^{+}$, and that
$$
\left|\Ocal_{z}\left(\dfrac{1}{N^2}\right)\right| \leq \dfrac{1}{N^2} \, P_1(|z|) P_2\left(\frac{1}{\Im z}\right)
$$
on $\mathbb{C}^{+}$ (because $\frac{1}{\rho(z)} \leq \frac{1}{\Im z}$ on $\mathbb{C}^{+}$), we obtain that 
\begin{equation}
  \label{eq:estimation_tilde_alpha-tilde_t}
|\tilde{\alpha}_N(z)-\tilde{t}_N(z)| \leq  \dfrac{1}{N^2} \, P_1(|z|) P_2\left(\frac{1}{\Im z}\right)
\end{equation}
for each $z \in \mathbb{C}^{+}$. This completes the proof of Proposition \ref{prop:behaviour-hatnuN-section3}.

\subsection{Proof of Lemma \ref{le:limit_Qpp_Qpf}}
We first justify  (\ref{eq:convergence-sesquilinear-boldQii}) and  (\ref{eq:convergence-sesquilinear-boldQij}) when $z \in \mathbb{C}^{+}$. For this, we  mention that (\ref{eq:var_Q}) and (\ref{eq:var_Q_as}) imply that the fourth order moments of  $	a_N^{*} \, ({\bf Q_{ii}})_N(z) \, b_N - \mathbb{E}(a_N^{*} \,({\bf Q_{ii}})_N(z) \, b_N)$ and  $	a_N^{*} \, ({\bf Q_{ij}})_N(z) \, b_N - \mathbb{E}(a_N^{*} \,({\bf Q_{ij}})_N(z) \, b_N)$ are $\mathcal{O}_{z^{2}}\left( \frac{1}{N^{2}} \right)$ terms. Borel-Cantelli's Lemma thus implies that
    $	a_N^{*} \, ({\bf Q_{ii}})_N(z) \, b_N - \mathbb{E}(a_N^{*} \,({\bf Q_{ii}})_N(z) \, b_N)$ and  $	a_N^{*} \, ({\bf Q_{ij}})_N(z) \, b_N - \mathbb{E}(a_N^{*} \,({\bf Q_{ij}})_N(z) \, b_N)$ converge almost surely towards $0$ . (\ref{eq:expre-E(Qii)-section-3-bis}) and
    (\ref{eq:expre-E(Qij)-section-3-bis}) as well as Corollary \ref{re:conver_bold} complete the proof of  (\ref{eq:convergence-sesquilinear-boldQii}) and  (\ref{eq:convergence-sesquilinear-boldQij}) on $\mathbb{C}^{+}$. In order to extend the convergence to  $\mathbb{C}\setminus{\bs\Scal}_*$,
    we remark that Theorem \ref{th:no_eigen-2} implies that almost surely, for $N$ large enough, $a_N^{*} \, ({\bf Q_{ii}})_N(z) \, b_N- {\bf\tilde{t}}_N(z)a_N^{*}b_N$ and $a_N^{*} \, ({\bf Q_{ij}})_N(z)  \, b_N - c_N{\bf t}_N(z)a_N^{*}b_N$ are holomorphic on  $\mathbb{C}\setminus{\bs\Scal}_*$, and bounded on each compact subset of  $\mathbb{C}\setminus{\bs\Scal}_*$.
    Therefore, Montel's theorem implies that  (\ref{eq:expre-E(Qii)-section-3-bis}) and
    (\ref{eq:expre-E(Qij)-section-3-bis}) holds for each $z \in \mathbb{C}\setminus{\bs\Scal}_*$ and uniformly on the compact
    subsets of $\mathbb{C}\setminus{\bs\Scal}_*$. The extension to the context of random vectors $(a_N,b_N)$ is justified using 
    the arguments used in the course of the proof of Lemma \ref{le:extra-properties}. 

\subsection{Proof of the properties of function $f_*$ defined by (\ref{eq:def-f*})}

(\ref{eq:def-tN-section3}) implies that $c_* x t_*(x) = x \tilde{t}_*(x) + 1-c_*$ for each $x \in (4c_*(1-c_*),1)$. As $\tilde{t}_*$ is the Stieltjes transform of a positive measure supported by $\Scal_*$, 
function $x \rightarrow x \tilde{t}_*(x)$ is increasing on $(4c_*(1-c_*),1)$. As we also have
\begin{align}\label{eq:tilde_t_by_t}
\dfrac{\tilde{t}_*(x)}{t_*(x)}=c_*-\dfrac{1-c_*}{xt_*(x)}.
\end{align}
we obtain that $x \rightarrow \frac{\tilde{t}_*(x)}{t_*(x)}$ is increasing on $(4c_*(1-c_*),1)$.
Using (\ref{eq:formula_t}), we check that $x t_*(x)$ and $\dfrac{\tilde{t}_*(x)}{t_*(x)}$ are well defined at $4c_*(1-c_*)$ and that 
\begin{align*}
(xt_*(x))\Big|_{x=4c_*(1-c_*)}=\dfrac{4c_*(1-c_*)(2c_*-1)}{2c_*(2c_*-1)^2}=\dfrac{2(1-c_*)}{2c_*-1}\Rightarrow \dfrac{\tilde{t}_*(x)}{t_*(x)}\Big|_{x=4c_*(1-c_*)}=c_*-\dfrac{2c_*-1}{2}=\dfrac{1}{2}
\end{align*}
This shows that $\frac{\tilde{t}_*(x)}{t_*(x)}$ is positive  on $[4c_*(1-c_*),1)$ and that 
$x \rightarrow \left( \frac{\tilde{t}_*(x)}{t_*(x)} \right)^{2}$, and thus 
$x \rightarrow f_*(x)$, are increasing on $[4c_*(1-c_*),1)$. Moreover, it is easily checked that 
$f_*(4c_*(1-c_*)) = \frac{c_*}{1-c_*}$. It remains to show that $f_*$ is well defined at $1$, 
and to evaluate $f_*(1)$. For this, 
we remark that if $c_*<1/2$, then due to (\ref{eq:formula_tilde_t}) and (\ref{eq:formula_t}), we have
\begin{align*}
\dfrac{\tilde{t}_*(x)}{t_*(x)}\Big|_{x=1}=\lim_{y\rightarrow1}\dfrac{c_*4(1-c_*)^2(1-y)\left(x(2c_*-1)-\sqrt{x(x-4c_*(1-c_*))}\right)}{x(1-x)4c_*(1-c_*)\left(x-2(1-c_*)-\sqrt{x(x-4c_*(1-c_*))}\right)}=1-c_*
\end{align*}
and for $c_*\ge1/2$
\begin{align*}
\dfrac{\tilde{t}_*(x)}{t_*(x)}\Big|_{x=1}=\lim_{x\rightarrow1}\dfrac{c_*\left(x-2(1-c_*)+\sqrt{x(x-4c_*(1-c_*))}\right)}{x(2c_*-1)+\sqrt{x(x-4c_*(1-c_*))}}=c_*
\end{align*}
This leads to $f_*(1) = 1$ if $c_* < \frac{1}{2}$ and $f_*(1) = \left( \frac{c_*}{1-c_*} \right)^{2}$ 
if $c_* \geq 1/2$. 

\subsection{Proof of Lemma~\ref{le:existence-Di}}
\label{sec:lemma-existence-Di}
We express $E_i$ as 
\begin{align}
E_i=\begin{pmatrix}
-\Delta_i^{2}&I_r\\
I_r&0
\end{pmatrix}+\begin{pmatrix}
\Delta_i\tilde{\Theta}_i^*\Pi_i^W\tilde{\Theta}_i\Delta_i&\Delta_i\tilde{\Theta}_i^*W_i^*(W_iW^*_i)^{-1}\Theta_i\\
\Theta_i^*(W_iW_i^*)^{-1}W_i\tilde{\Theta}_i\Delta_i&\Theta_i^*(W_iW_i^*)^{-1}\Theta_i
\end{pmatrix}
\end{align}
We remind that, as $\eta_N = 1$ almost surely for $N$ large enough, it is possible to 
introduce $\eta_N$ whenever it is useful without modifying the almost sure behaviour of 
the various terms. Lemma~\ref{le:symetries-E(Q)-2} implies that$\ex\{ \eta W_i^*(W_iW^*_i)^{-1}\}=\ex\{ \eta (W_iW_i^*)^{-1}W_i\}=0$ for $i=p,f$ while  (\ref{le:estimation_inver_WW*})-(\ref{le:expectation_Pi}) lead to $\ex\{ \eta \Pi_i^W\}=c_NI_r+\Ocal(N^{-k})$ and $\ex\{\eta (W_iW_i^*)^{-1}\}=(1-c_N)^{-1}(I_L\otimes R_N^{-1})+\Ocal(N^{-k})$ for each $k\in\mathbb{N}$. Combining these evaluations with the Poincar\'e-Nash inequality, we obtain immediately
(\ref{eq:convergence-Ei}).

\subsection{Proof of Lemma~\ref{le:asympt}}\label{sec:lemma_asympt}
We just provide a brief justification of the first item of Lemma ~\ref{le:asympt}. For this, we notice that the $2r \times 2r$ matrix $\Acal_i^*{\bf Q^W_{ji}}\Acal_i$ is given by
\begin{equation}
    \label{eq:expre-item1-le-asympt}
\Acal_i^*{\bf Q^W_{ji}}\Acal_i = \left( \begin{array}{c} -\Delta_i \tilde{\Theta}_i^{*} \Pi_{i}^{W,\perp} \\
\Theta_i^{*} (W_i W_i^{*})^{-1} W_i^{*} \end{array} \right)  {\bf Q_{ji}^{W}} \left( - \Pi_{i}^{W,\perp}
\tilde{\Theta}_i \Delta_i, W_i^{*} (W_i W_i^{*})^{-1}  \Theta_i \right)
\end{equation}
We recall that, as $\eta_N = 1$ for each $N$ large enough, we can add $\eta_N$ everywhere in the 4 $r \times r$ blocks of $\Acal_i^*{\bf Q^W_{ji}}\Acal_i$ without modifying their almost sure behaviour. 
We first justify that the two $r \times r$ non diagonal blocks of  $\Acal_i^*{\bf Q^W_{ji}}\Acal_i$ converge almost surely towards $0$. For this, we first notice that, using the same arguments as in Lemma~\ref{le:symetries-E(Q)-2}, it can be easily shown that 
$\ex\{\Pi_i^{W}{\bf Q^W_{ji}} \eta W_i^*(W_iW_i^*)^{-1}\}=\ex\{{\bf Q^W_{ji}} \eta W_i^*(W_iW_i^*)^{-1}\}=0$. Using the Poincar\'e-Nash inequality, it is possible to prove that 
$\var \{ a_N^{*} \Pi_{i}^{W,\perp} {\bf Q_{ij}^{W}} \eta W_i^{*} (W_i W_i^{*})^{-1} b_N \} = \Ocal_{z^{2}}(\frac{1}{N})$ and 
$\var \{ \left(a_N^{*} \Pi_{i}^{W,\perp} {\bf Q_{ij}^{W}} \eta W_i^{*} (W_i W_i^{*})^{-1} b_N\right)^{2} \} = \Ocal_{z^{2}}(\frac{1}{N^{2}})$, where $a_N$ (resp.  $b_N$) is a $N$ dimensional (resp. $ML$-dimensional) deterministic vector such that $\sup_{N} \|a_N\| < +\infty$
(resp.  $\sup_{N} \|b_N\| < +\infty$). This immediately implies that 
$\mathbb{E} \left| a_N^{*} \Pi_{i}^{W,\perp} {\bf Q_{ij}^{W}} \eta W_i^{*} (W_i W_i^{*})^{-1} b_N \right|^{4} = \Ocal_{z^{2}}(\frac{1}{N^{2}})$ and that 
$a_N^{*} \Pi_{i}^{W,\perp} {\bf Q_{ij}^{W}} \eta W_i^{*} (W_i W_i^{*})^{-1} b_N$, and thus 
$a_N^{*} \Pi_{i}^{W,\perp} {\bf Q_{ij}^{W}} W_i^{*} (W_i W_i^{*})^{-1} b_N$ converge almost surely towards $0$. 
The extension of these properties to the context of bounded random vectors $(a_N,b_N)$ independent from 
the sequence $(v_n)_{n \in \mathbb{Z}}$ (see the proof of Lemma \ref{le:extra-properties}) leads to the conclusion that the two $r \times r$ non diagonal blocks of  $\Acal_i^*{\bf Q^W_{ji}}\Acal_i$ converge almost surely towards $0$ as expected. \\

We now evaluate the almost sure behaviour of the two $r \times r$ diagonal blocks of $\Acal_i^*{\bf Q^W_{ji}}\Acal_i$, and consider the case  $i=p$, $j=f$ without loss of generality. In the expression of   $(\Acal_p^*{\bf Q^W_{fp}}\Acal_p)_{11} = 
\Delta_p \tilde{\Theta}_p^{*}  \Pi^{W,\perp}_p{\bf Q^W_{fp}}\Pi^{W\perp}_p \tilde{\Theta}_p \Delta_p$, it is possible to replace $\Pi^{W,\perp}_p = I - \Pi_p^{W}$ by $I - \eta \Pi_p^{W}$. Using  (\ref{eq:expre-BQ-2}) and the resolvent identity
\begin{equation}
\label{eq:resolvent-identity-bfQ-appendix}  
I + z {\bf Q^{W}} = {\bf Q^{W}} \left( \begin{array}{cc} 0 & \eta \Pi_p^{W} \\ \eta \Pi_f^{W} & 0 
\end{array} \right) =  \left( \begin{array}{cc} 0 & \eta \Pi_p^{W} \\ \eta \Pi_f^{W} & 0 
\end{array} \right) {\bf Q^{W}}
\end{equation}
we obtain easily that 
\begin{align*}
(I-\eta \Pi^{W}_p){\bf Q^W_{fp}}(I-\eta \Pi^{W}_p) = {\bf Q^W_{fp}}-I_N-z{\bf Q^W_{pp}}-I_N-z{\bf Q^W_{ff}}+\eta \Pi^{W}_p+z^2{\bf Q^W_{pf}}
\end{align*}
Using the Poincar\'e-Nash inequality, it is easy to check that if $a_N$ and $b_N$ are two deterministic
$N$--dimensional vectors for which $\sup_{N} \|a_N\| < +\infty$ and $\sup_{N} \|b_N\| < +\infty$, then, it holds that $\Var\{a_N^{*} \eta \Pi^{W}_p b_N\} = \Ocal(\frac{1}{N})$ and 
$\Var\{\left(a_N^{*} \eta \Pi^{W}_p b_N - \mathbb{E}(a_N^{*} \eta \Pi^{W}_p b_N)\right)^{2}\} =  \Ocal(\frac{1}{N^{2}})$. Therefore, $\mathbb{E}\left(a_N^{*} \eta \Pi^{W}_p b_N - \mathbb{E}(a_N^{*} \eta \Pi^{W}_p b_N)\right)^{4} = \Ocal(\frac{1}{N^{2}})$, so that $a_N^{*} \eta \Pi^{W}_p b_N - \mathbb{E}(a_N^{*} \eta \Pi^{W}_p b_N) \rightarrow 0$ almost surely. (\ref{le:expectation_Pi}) thus leads to the conclusion that
$$
a_N^{*} \eta \Pi^{W}_p b_N -c_N \, a_N^{*} b_N \rightarrow 0 \; a.s.
$$
The use of Lemma~\ref{le:limit_Qpp_Qpf} implies that  $a_N^*\Pi^{W,\perp}_p{\bf Q_{fp}}\Pi^{W,\perp}_pb_N-((1+z^2)c_N\t_N(z)-1-2z{\bf\tilde{t}}_N(z)-(1-c_N))a_N^*b_N\rightarrow0$. Moreover, 
this property also holds when $(a_N,b_N)$ are bounded random vectors $(a_N,b_N)$ independent from 
the sequence $(v_n)_{n \in \mathbb{Z}}$. We deduce from this that 
$$
(\Acal_p^*{\bf Q^W_{fp}}\Acal_p)_{11} - \left( (1+z^2)c_N\t_N(z)-1-2z{\bf\tilde{t}}_N(z)-(1-c_N) \right) \, \Delta_N \rightarrow 0 \; a.s.
$$
holds. In order to obtain the expression stated in the Lemma, we refer to  (\ref{eq:connect_bolt_t_tilde-t}) and replace $c_N \t_N(z)$ by $c_N\t_N(z)=\frac{{\bf\tilde{t}}_N(z)}{z}+\frac{1-c_N}{z^2}$:  
\begin{align}\label{eq:for_last}
(1+z^2)c_N\t_N(z)-1-2z{\bf\tilde{t}}_N(z)-(1-c_N)={\bf\tilde{t}}_N(z)\left(\dfrac{1}{z}-z\right)+\dfrac{1-c_N}{z^2}-1
\end{align}
Let us remind that ${\bf\tilde{t}}_N$ satisfies Eq. (\ref{eq:final_tilde-bold-alpha}) but in which term $\Ocal_{z^2}(N^{-2})$ is replaced with 0, i.e.
\begin{align}\label{eq:bold_tilde_t}
(1-z^2){\bf\tilde{t}}_N^2(z)+\left(\dfrac{2(1-c_N)}{z}-z\right){\bf\tilde{t}}_N(z)+\dfrac{(1-c_N)^2}{z^2}=0
\end{align}
In order to simplify (\ref{eq:for_last}) we rewrite Eq. (\ref{eq:bold_tilde_t}) as
\begin{align*}
(z{\bf\tilde{t}}_N(z)+(1-c_N))\left({\bf\tilde{t}}_N(z)\left(\dfrac{1}{z}-z\right)+\dfrac{1-c_N}{z^2}-1\right) +(1-c_N)+z(1-c_N){\bf\tilde{t}}_N(z)=0
\end{align*}
and get immediately that the r.h.s. of (\ref{eq:for_last}) is equal to $-\frac{(1-c_N)(1+z{\bf\tilde{t}}_N(z))}{z{\bf\tilde{t}}_N(z)+(1-c_N)}$. This establishes the expression stated in the Lemma.\\

We finally evaluate the behaviour of $(\Acal_p^*{\bf Q^W_{fp}}\Acal_p)_{22} = \Theta_p^{*} (W_pW_p^*)^{-1}W_p{\bf Q^W_{fp}}W_p^*(W_pW_p^*)^{-1} \Theta_p$. We recall that $W_i = (I \otimes R^{1/2}) W_{i,iid}$ for $i=p,f$, so that
$$
\Theta_p^{*} (W_pW_p^*)^{-1}W_p{\bf Q^W_{fp}}W_p^*(W_pW_p^*)^{-1} \Theta_p = 
\Theta_p^{*} (I \otimes R^{-1/2})  (W_{p,iid}W_{p,iid} ^*)^{-1}W_{p,iid}{\bf Q^W_{iid,fp}}W_{p,iid}^*(W_{p,iid}W_{p,iid}^*)^{-1}  (I \otimes R^{-1/2})  \Theta_p
$$
because ${\bf Q^W_{iid,fp}} = {\bf Q^W_{fp}}$. It is thus sufficient to study 
the behaviour of 
$$
a_N^{*} \eta_N (W_{p,iid}W_{p,iid} ^*)^{-1}W_{p,iid}{\bf Q^W_{iid,fp}}W_{p,iid}^*(W_{p,iid}W_{p,iid}^*)^{-1} b_N
$$
where $a_N,b_N$ are deterministic $ML$--dimensional vectors such that  $\sup_{N} \|a_N\| < +\infty$ and $\sup_{N} \|b_N\| < +\infty$. We also recall that the regularization term $\eta_N$ is built from matrix $W_{iid}$. In order to simplify the notations, we prefer to denote $W_{i,iid}$ by $W_{i}$ in the following. After some calculations, the Poincar\'e-Nash inequality leads to 
$$
a_N^{*} \eta_N (W_pW_p^*)^{-1}W_p{\bf Q^W_{fp}} \eta W_p^*(W_pW_p^*)^{-1} b_N - \mathbb{E}\left( a_N^{*} \eta_N (W_pW_p^*)^{-1}W_p{\bf Q^W_{fp}} \eta W_p^*(W_pW_p^*)^{-1} b_N \right) \rightarrow 0 \; a.s.
$$
It is thus sufficient to evaluate  $\ex\{(\eta W_pW_p^*)^{-1}W_p{\bf Q^W_{fp}} \eta W_p^*(W_pW_p^*)^{-1}\}$ 
using the integration by parts formula. By Lemma~\ref{le:symetries-E(Q)-2}, this matrix is diagonal, and we therefore consider its diagonal elements. For this, we need to repeat the calculations of Section~\ref{subsec:ipp-2}. In order to avoid to reproduce another tedious and similar calculations, we provide only the ideas and main steps. It is first necessary to  apply the integration by parts formula for $\sum_{r,t,m_2,i_2}\ex\{{\bf \eta Q^W_{fp}}_{rt}\bar{W}^{m_1}_{p,i_1t}((W_pW_p^*)^{-1})^{m_1m_2}_{i_1i_2}W^{m_2}_{p,i_2r}\}$ and follow the calculations of Section~\ref{subsec:ipp-2} using similar arguments. We first obtain
\begin{multline*}
\ex\{\left(\eta(W_pW_p^*)^{-1}W_{p}{\bf Q^W_{fp}}W_p^*\right)^{m_1m_1}_{i_1i_1}\}
=\ex\{\eta ((W_pW_p^*)^{-1})^{m_1m_1}_{i_1i_1}\}\dfrac{1}{N}\left(\ex\{\tr{\bf Q^W_{fp}}\}-\ex\{\tr{\bf Q^W_{fp}}\eta \Pi_p^W\}\right)\\
-
\ex\left\{\eta \left((W_pW_p^*)^{-1}W_p{\bf Q^W_{fp}}\eta W_p^*(W_pW_p^*)^{-1}\right)^{m_1m_1}_{i_1i_1}\right\}\dfrac{1}{N}\ex\{\tr \eta \Pi^{W,\perp}_p{\bf Q^W_{fp}}\}+\Ocal_{z^2}^{N}(N^{-3/2})
\end{multline*}
Since $\ex\{\eta (W_pW_p^*)^{-1}\}=(1-c_N)^{-1}I_N+\Ocal^{N}(N^{-3/2})$ and that the equality ${\bf Q^W_{fp}} \eta \Pi_p^W=I_N+z{\bf Q^W_{ff}}$ holds (see (\ref{eq:resolvent-identity-bfQ-appendix})), we can simplify the r.h.s. of the last equation: 
\begin{multline}\label{eq:exp_last_lem-1}
\ex\{\eta \left((W_pW_p^*)^{-1}W_{p}{\bf Q^W_{fp}}W_p^*\right)^{m_1m_1}_{i_1i_1}\}=\dfrac{1}{1-c_N}({\bs\alpha}_N-1-z\tilde{{\bs \alpha}}_N)\\
-
\ex\left\{\eta \left((W_pW_p^*)^{-1}W_p{\bf Q^W_{fp}} \eta W_p^*(W_pW_p^*)^{-1}\right)^{m_1m_1}_{i_1i_1}\right\}({\bs\alpha}_N-1-z\tilde{{\bs \alpha}}_N)+\Ocal_{z^2}(N^{-3/2})
\end{multline}

 We express $\ex\{\eta \left((W_pW_p^*)^{-1}W_{p}{\bf Q^W_{ff}}W_p^*\right)^{m_1m_1}_{i_1i_1}\}$ similarly:
\begin{multline*}
\ex\{\eta \left((W_pW_p^*)^{-1}W_{p}{\bf Q^W_{ff}}W_p^*\right)^{m_1m_1}_{i_1i_1}\}
=\ex\{\eta ((W_pW_p^*)^{-1})^{m_1m_1}_{i_1i_1}\}\dfrac{1}{N}\left(\ex\{\tr{\bf Q^W_{ff}}\}-\ex\{\tr{\bf Q^W_{ff}} \eta \Pi_p^W\}\right)\\
-
\ex\left\{\eta \left((W_pW_p^*)^{-1}W_p{\bf Q^W_{ff}} \eta W_p^*(W_pW_p^*)^{-1}\right)^{m_1m_1}_{i_1i_1}\right\}\dfrac{1}{N}\ex\{\tr \eta \Pi^{W,\perp}_p{\bf Q^W_{ff}}\}+\Ocal_{z^2}(N^{-3/2})
\end{multline*}
We remark that $N^{-1}\ex\{\tr{\bf Q^W_{ff}}\eta \Pi^{W,\perp}_p\}=-z^{-1}\ex\{N^{-1} \eta \tr\Pi^{W,\perp}_p\}=-\frac{1-c_N}{z} + \Ocal_{z^2}(N^{k})$  for each integer $k$ (see (\ref{eq:expr_QffPpperp})). The last equation can thus be rewritten as
\begin{multline}\label{eq:exp_last_lem-2}
\ex\{\eta \left((W_pW_p^*)^{-1}W_{p}{\bf Q^W_{ff}}W_p^*\right)^{m_1m_1}_{i_1i_1}\}=-\dfrac{1}{z}\\
+\dfrac{1-c_N}{z}
\ex\left\{ \eta \left((W_pW_p^*)^{-1}W_p{\bf Q^W_{fp}} \eta W_p^*(W_pW_p^*)^{-1}\right)^{m_1m_1}_{i_1i_1}\right\}+\Ocal_{z^2}(N^{-3/2})
\end{multline}
Moreover,  using the resolvent identity, $(W_pW_p^*)^{-1}W_{p}{\bf Q^W_{ff}}W_p^*$ can be rewritten as $(W_pW_p^*)^{-1}W_{p}(-z^{-1}I_N+z^{-1}(\Pi^W_f\Pi^W_p-z^2)^{-1}\Pi^W_f\Pi^W_p)W_p^*=-z^{-1}I_N+z^{-1}(W_pW_p^*)^{-1}W_{p}{\bf Q^W_{fp}} \eta \Pi_p^{W} W_p^*$. Using the obvious identity 
$\Pi_p^{W} W_p^{*}= W_p^{*}$ and comparing (\ref{eq:exp_last_lem-1}) and (\ref{eq:exp_last_lem-2}) we that:
\begin{multline*}
\ex\{\left( \eta (W_pW_p^*)^{-1}W_{p}{\bf Q^W_{fp}} \eta W_p^*(W_pW_p^*)^{-1}\right)^{m_1m_1}_{i_1i_1}\}
=\dfrac{{\bs\alpha}_N-1-z\tilde{{\bs \alpha}}_N}{(1-c_N)((1-c_N)+{\bs\alpha}_N-1-z\tilde{{\bs \alpha}}_N)}+\Ocal_{z^2}(N^{-3/2})
\end{multline*}
As we can see, all diagonal elements of $\ex\{ \eta (W_pW_p^*)^{-1}W_{p}{\bf Q^W_{fp}}\eta W_p^*(W_pW_p^*)^{-1}\}$ are equal up to an error term. Therefore, the matrix $\ex\{\eta (W_pW_p^*)^{-1}W_{p}{\bf Q^W_{fp}} \eta W_p^*(W_pW_p^*)^{-1}\}$ is a multiple of $I_N$ up to an error term. We thus conclude that 
\begin{align}\label{eq:asympt_ff}
a_N^*(W_pW_p^*)^{-1}W_{p}{\bf Q^W_{fp}}W_p^*(W_pW_p^*)^{-1}b_N-\frac{c_N{\bf t}_N-1-z\tilde{{\bf t}}_N}{(1-c_N)((1-c_N)+c_N{\bf t}_N-1-z\tilde{{\bf t}}_N)}a_N^*b_N\rightarrow0.
\end{align}
After replacing $c_N\t_N$ with $\frac{{\bf\tilde{t}}_N(z)}{z}+\frac{1-c_N}{z^2}$ we find that $c_N{\bf t}_N-1-z\tilde{{\bf t}}_N={\bf\tilde{t}}_N(z)(\frac{1}{z}-z)+\frac{1-c_N}{z^2}-1$ which is also the expression obtained in (\ref{eq:for_last}). We remind  that
\begin{align*}
{\bf\tilde{t}}_N(z)\left(\dfrac{1}{z}-z\right)+\dfrac{1-c_N}{z^2}-1=-\frac{(1-c_N)(1+z{\bf\tilde{t}}_N(z))}{z{\bf\tilde{t}}_N(z)+(1-c_N)}
\end{align*}
Plugging this expression into (\ref{eq:asympt_ff}), and remarking that (\ref{eq:asympt_ff}) still holds when $(a_N,b_N)$ are random bounded vectors independent from $(v_n)_{n \geq 1}$, we  obtain the asymptotic behaviour $(\Acal_p^*{\bf Q^W_{fp}}\Acal_p)_{22}$ stated in the Lemma.\\

\end{document}